\documentclass[conference,compsoc,letterpaper]{IEEEtran}

\IEEEoverridecommandlockouts
\usepackage{tabularx}

\usepackage{lineno}

\ifCLASSOPTIONcompsoc
  \usepackage[nocompress]{cite}
\else
  \usepackage{cite}
\fi
\usepackage[normalem]{ulem}

\usepackage{hyperref}

\ifCLASSINFOpdf
  \usepackage[pdftex]{graphicx}
  \graphicspath{{../pdf/}{../jpeg/}}
  \DeclareGraphicsExtensions{.pdf,.jpeg,.png}
\else
\fi

\usepackage{amsmath}
\usepackage{xcolor}
\usepackage{algorithm}
\usepackage{algpseudocode} 

\usepackage{makecell}

 \usepackage[caption=false,font=footnotesize,labelfont=sf,textfont=sf]{subfig}

\usepackage{amssymb}
\usepackage{booktabs} 
\usepackage{multirow} 
\usepackage{threeparttable}

\usepackage{url}

\usepackage{listings}
\begin{document}
%
\title{AutoPYara: Next-Gen YARA Rule Generator for Malware Family Clustering%
\thanks{This manuscript is an extended version of the paper accepted at IEEE ACSAC 2026, featuring supplementary appendices.}}

\author{
\IEEEauthorblockN{
Mabon Ninan$^{*}$,
Nhat Nguyen$^{*}$,
Soumyajyoti Dutta,
Sidharth Anil,
and Marcus Botacin
}
\IEEEauthorblockA{
\textit{Texas A\&M University}\\
\{ninanmm, nmnguy29, soumyajyoti1998, sid.anil, botacin\}@tamu.edu
}

\IEEEauthorblockA{\footnotesize{
$^{*}$ Equally Contribution}
}
}


\maketitle

\begin{abstract}
Millions of malware variants threaten users daily, such that clustering these samples is key for fast incident response. AutoYara is the current state-of-the-art for automatic YARA rule generation, but the task remains challenging on realistic datasets, where intra-family diversity and noisy clustering can degrade rule quality. We present AutoPYara, a solution armored with: (1) precise cluster centroid selection based on similarity hashing, and (2) an improved heuristic for estimating subclusters within a malware family using historical VirusTotal data. We implement AutoPYara as an open-source Python package \texttt{pip install autopyara} to support reproducibility and practical use. We show that AutoPYara, (1) increases rule coverage by 14 percentage points; (2) reduces variance by 7.1 percentage points, comparing the average variance AutoPYara exhibits to that of AutoYara; and (3) improves threat-hunting accuracy by 10\%. AutoPYara produces robust and effective YARA rules than prior automatic approaches in practical settings.

\end{abstract}



%
\IEEEpeerreviewmaketitle

\section{Introduction}

\noindent Among millions of malware samples threatening users daily, most are variants of the same family. Handling these samples individually does not scale~\cite{miller2016_integration_performanc_measurement_malware_detection}; thus, clustering samples into families is key for fast incident response, as it allows the same remediation to be applied to all family samples~\cite{9223163,9141989}. Multiple approaches have been proposed to cluster unknown samples into families, from signatures~\cite{10.1007/978-3-540-70500-0_25} to Machine Learning (ML)~\cite{8376209}. Despite their individual success, a one-size-fits-all solution for malware family identification remains an open problem. The current State-of-the-Art (SOTA) for this task is AutoYara~\cite{raff2020_auto_yara}, a hybrid approach that uses ML for determining Indicator of Compromise (IoCs) that uniquely identify the malware, and YARA~\cite{yara} signatures for the actual matching, being it the industry-standard. It extracts unique N-grams of bytes from goodware and malware to create an extensive database that is further used with a biclustering~\cite{dhillon2001_spectralcoclustering} algorithm to identify unique byte sequences that are only seen in malware. This allows the automatic derivation of rules to match these unique patterns.
Although AutoYara was a step ahead in automating rule generation, we identified key blind spots: \textbf{(1)} \uline{AutoYara was evaluated only on cases with known ground truth}. In real-world cases, complete ground truth is not available (attackers do not disclose their templates), such that ground truth is just approximated by AntiVirus (AV) companies via labeling strategies. Different labeling strategies lead to different ``groundtruths'', hence it is essential to evaluate their impact to understand how sensitive rule generation frameworks (previous and ours) are to these classification uncertainties; \textbf{(2)} \uline{AutoYara's evaluations only considered samples with known patterns} aligned to the pre-trained Bloom filters, ignoring that most in-the-wild samples have unknown patterns caused by concept evolution and byte distribution shift, causing rule generation to fail; \textbf{(3)} \uline{AutoYara was only evaluated on large cluster sizes}, which tend to be more stable. In contrast, in-the-wild samples might be collected at different family sizes, and their impact on rule generation is unknown.

To characterize the impact of these factors on automatic rule generation, we conducted exhaustive experiments on a dataset of $\approx$~71,000 samples collected over 8 years by an AV company in 3 countries: the US, Brazil, and Japan. We compared the results under different strategies to approximate ground truth: the application of two similarity hashing functions (SSdeep~\cite{kornblum2006identifying} and J-SDhash~\cite{botacin2021_ssdeephashing}); and VirusTotal (VT) labels~\cite{DBLP:journals/corr/abs-2006-10615}. However, both strategies introduce distinct trade-offs regarding precision and data volume. Despite producing homogeneous clusters, similarity hashing limits sample coverage and introduces methodological variables like the similarity threshold. AV labels are comprehensive and cover most of the dataset, but result in heterogeneous clusters with increased False Positives (FPs). 

Our results highlight the need for: \textbf{(i)} methodical standards to evaluate rule generation with high variance and uncertain labels, which we address by introducing guidelines and best practices for experiments; and \textbf{(ii)} algorithmic solutions to address the inefficient handling of heterogeneous data with high variance. Our analysis of AutoYara revealed that: \textbf{(1)} most of the variability in the results arises from the randomness of the original algorithm, which initiates clustering from arbitrary points that may not align well with heterogeneous distributions observed in real scenarios; \textbf{(2)} the AutoYara algorithm underestimates the number of subclusters a malware family has due to the use of the Variational Bayesian Gaussian Mixture Model (VBGMM)~\cite{bishop2006pattern} algorithm as the guesser for the number of clusters.

We propose solving the problem \textbf{(1)} by adding a crafted centroid selection step to the rule generation process. This ensures that the generation always starts from the best possible centroid to cover a heterogeneous distribution of samples, and in \textbf{(2)} we develop a heuristic to replace VBGMM and increase the number of sub-clusters that the malware families are fitted into, allowing heterogeneous samples to be classified in the same family. The heuristic estimates the number of sub-clusters based on historical malware family data obtained from the VirusTotal oracle. To evaluate our contributions, we created AutoPYara, a solution that implements our proposed algorithm. AutoPYara is as efficient as AutoYara, as it shares the same Java backend. However, it is more user-friendly, as it is implemented as a Python front-end and distributed as a PyPI package. Applied to the same dataset, AutoPYara outperforms AutoYara by up to 14\%~\footnote{Presented results represent absolute gains. \% often means \% points.} in its best configuration on real-world samples. Our results demonstrate the impact of centroid selection and reopen the possibility of using VT labels for rule generation—not as a labeling strategy, which we show is problematic, but as a heuristic, which we show is powerful.

\vspace{-4.0mm}

\section{Background}
\label{sec:Background}
\subsection{An Overview of AutoYara Design}
\label{subsec:OverView}

\begin{figure}[!htpb]
    \centering
    \includegraphics[width=1\linewidth]{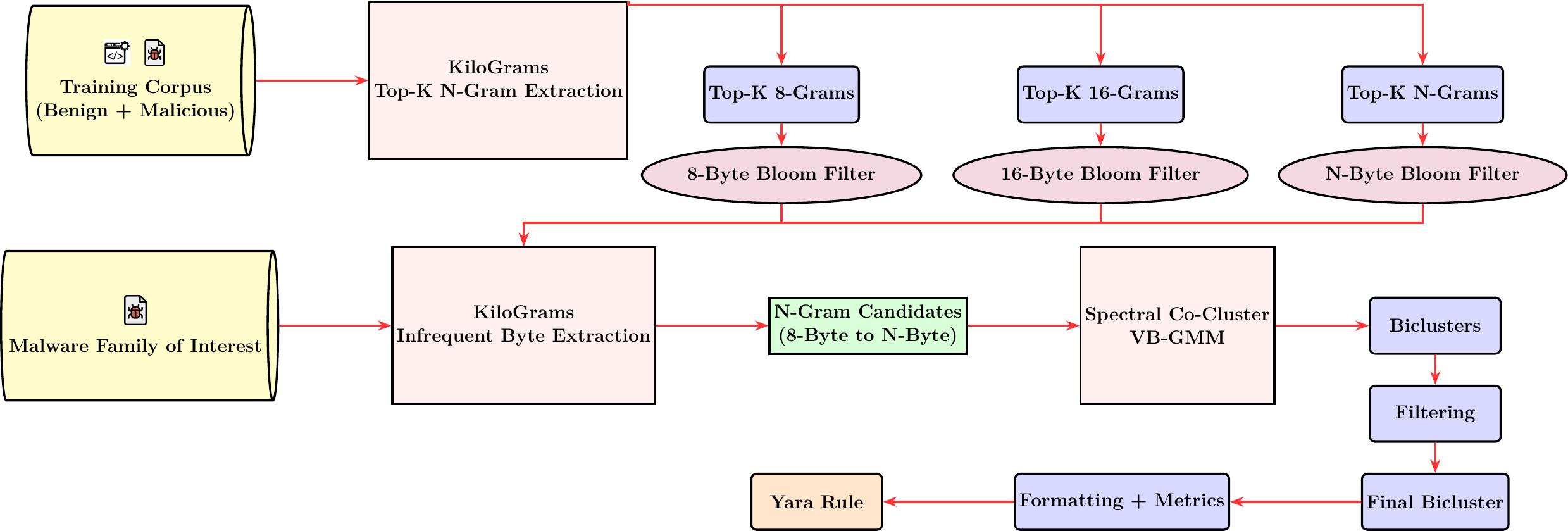}
    \vspace{-4.0mm}
    \caption{The AutoYara system pipeline.}
    \label{fig:yara pipeline}
\end{figure}

\noindent AutoYara~\cite{raff2020_auto_yara} generates rules based on byte patterns that are not frequent in goodware samples (to avoid FPs) and are frequent among the malware samples (to fingerprint them). AutoYara implements this logic via two steps: \textbf{(1)} creating a blacklist of goodware byte patterns during training; \textbf{(2)} identifying frequent malware-family patterns that are disjoint from previously seen goodware patterns during generation. Figure~\ref{fig:yara pipeline} provides a representation of the framework.
AutoYara's training step uses KiloGrams~\cite{raff2019_kilograms} to extract the top-K byte N-grams from the dataset, producing Bloom filters~\cite{bloom1970_bloomfilter} to store the byte blacklists. Next, the generation step extracts new N-grams from the input files and biclusters using Spectral Co-Clustering~\cite{dhillon2001_spectralcoclustering}
(details in Appendix~\ref{app:biclustering})
 with VBGMM~\cite{corduneanu2001_vbgmm}. This reveals distinct byte groups that are filtered using heuristics (e.g., coverage, cluster size, low FP, etc.) and merged into a disjunction of \textit{soft} conjunctions. The disjunctions have the form \textit{at least X conditions match}, where each string corresponds to an extracted N-gram and each condition to a bicluster. AutoYara ships pretrained BFs for the top 1M frequent bytes from EMBER~\cite{anderson2018emberopendatasettraining}.

\subsection{A critical review of AutoYara Clustering}

\noindent AutoYara employs unsupervised clustering~\cite{jain1999_clustering_review} to group byte patterns, reducing FPs by requiring simultaneous signature matches. It aims to organize N-gram candidates into the fewest clusters covering the most samples. While the literature offers various Hard~\cite{lloyd1982_kmeans, ergun2022_learningaugmentedkmeansclustering, ester1996_DBSCAN} and Soft~\cite{bezdek1984_fuzzycmeans, reynolds2009_gaussianmixture, corduneanu2001_vbgmm} clustering approaches, AutoYara utilizes Gaussian Mixture Models (GMMs) as its soft clustering solution. Table~\ref{tab:clustering_comparison} contrasts AutoYara with other literature methods and our proposed approach. While AutoYara addresses the manual $k$ selection of K-Means using GMMs, it remains sensitive to initialization and assumes Gaussian distributions (often underfitting malware biclusters). Other methods like Augmented K-Means~\cite{ergun2022_learningaugmentedkmeansclustering} offer noise resistance, and DBSCAN~\cite{ester1996_DBSCAN} handles non-spherical data without a preset $k$. 

\begin{table}[!htbp]
\caption{Comparison of Clustering Methods}
\vspace{-6mm}
\begin{center}
\resizebox{\columnwidth}{!}{
\begin{tabular}{|l|c|c|c|c|}
\hline
\textbf{Method} & \textbf{Type} & \textbf{Auto-$k$} & \textbf{Non-Gauss.$^\dagger$} & \textbf{Biclusters} \\ 
\hline
K-Means~\cite{lloyd1982_kmeans} & Hard & $\times$ & $\times$ & $\times$ \\
C-Means~\cite{bezdek1984_fuzzycmeans} & Soft & $\times$ & $\times$ & $\times$ \\
DBSCAN~\cite{ester1996_DBSCAN} & Density & \textcolor{green}{\checkmark} & \textcolor{green}{\checkmark} & $\times$ \\
\textbf{AutoYara} & \textbf{Soft} & \textcolor{green}{\checkmark} & $\times$ & \textcolor{green}{\checkmark} \\ 
\hline
\textbf{AutoPYara (base)} & \textbf{Soft} & $\times$ & \textcolor{green}{\checkmark} & \textcolor{green}{\checkmark} \\
\textbf{AutoPYara (heuristic)} & \textbf{Soft} & \textcolor{green}{\checkmark} & \textcolor{green}{\checkmark} & \textcolor{green}{\checkmark} \\
\hline

\multicolumn{5}{l}{$^\dagger$Handles non-spherical/non-Gaussian distributions.}
\end{tabular}
}
\label{tab:clustering_comparison}
\end{center}
\vspace{-4mm}
\end{table}

\noindent \textbf{Hypothesis:} We propose combining the benefits of Soft clustering (Fuzzy C-Means~\cite{bezdek1984_fuzzycmeans}) with the density-based automatic $k$ determination of DBSCAN to overcome the initialization and distribution limitations of AutoYara.
\section{AutoPYara}
\label{sec:Methodology}
\subsection{Design}

\noindent\textbf{Overview:} Based on the review of AutoYara in Section~\ref{sec:Background}, we identified two key aspects that could be improved: \textbf{(1)} the centroid selection process; \textbf{(2)} the $k$ selection process; and we propose refactoring AutoYara into AutoPYara to integrate these changes. Compared to AutoYara's architecture visualized in Figure~\ref{fig:yara pipeline}, the AutoPYara pipeline remains the same, but the internal biclustering algorithm is replaced by our algorithm.

\noindent \textbf{Key Insights} We recognize multiple positive aspects in AutoYara, but highlight that some project decisions must be rethought. Improving AutoYara's clustering is not as simple as extending it. Rather, it requires replacing algorithms that work in a paradigm with each other (Section~\ref{sec:Background}). We adopted a deconstruction-plus-reconstruction approach, where we first downgraded AutoYara's algorithms to base algorithms from another paradigm, and then extended this paradigm to surpass the original ones via our observations.

\noindent \textbf{To improve the $k$ selection:}

\noindent \textbf{1)} We believe that external threat intelligence (e.g., similarity hashing) can be used as additional context for clustering algorithms to improve biclustering performance. The intuition is that by implementing external threat intelligence and domain-specific context, we have a robust framework that is difficult to evade.\newline
\noindent \textbf{2)} To add external threat intelligence into the clustering process, we need the algorithm to consider global structures and properties of the handled data. \newline
\noindent \textbf{3)} To consider global structures, we must fully control the clustering process to input this data to the algorithms.\newline 
\noindent \textbf{4)} We cannot control the clustering structure of VBGMM, which auto-determines $k$. We should default back to K-Means~\cite{lloyd1982_kmeans} to control centroids and $k$.\newline
\noindent \textbf{5)} Since K-Means produces hard labels~\cite{bezdek1984_fuzzycmeans}, it is not comparable to VBGMM, which uses soft labels. We must replace the traditional K-Means with Fuzzy C-Means to use soft labels. In Fuzzy C-Means, a large $m$ will evenly spread out the probability distribution of each sample, leading to all samples having the same probability for all other clusters. A small $m$ will create a fast probability drop-off, which is similar to hard clustering. It provides a mechanism to create overlapping biclusters that are better at handling noise in real-world samples.\newline
\noindent \textbf{6)} The final challenge is that we don't have $k$ for clustering (it was previously auto-determined by VBGMM).To control $k$ and improve clustering quality, we devise heuristics grounded in family labels from AvClass~\cite{DBLP:journals/corr/abs-2006-10615} and the frameworks’ historical behavior on these families. Each sample is queried against VirusTotal using its unique file hash (SHA-256 by default). The resulting raw AV engine scan reports are passed directly to AVClass, which handles internal vendor naming conflicts, normalizes generic terms, and yields a single canonical family label per binary. Empirically, VBGMM underestimates $k$, so we adjust it upward to enhance clustering performance.

\noindent\textbf{To improve the initial centroid selection:}

\noindent \textbf{1)} Fuzzy C-Means is controllable and relies on soft labels; we extend it to consider external threat intelligence in the form of provided labels. It becomes AugmentedK-Means Soft (AKMS), our algorithmic contribution, extending Augmented K-Means~\cite{ergun2022_learningaugmentedkmeansclustering} to operate on soft labels.\newline
\noindent \textbf{2)} Given AKMS, we can also benefit from its heuristic predictor to select the best cluster centroids. The intuition is that for real-world applications, external information can be leveraged to produce better clusters. This external information could be provided from SSdeep~\cite{KORNBLUM200691_ssdeep}, J-SDhash~\cite{botacin2021_ssdeephashing} hashing, or VirusTotal~\cite{virustotal}. \newline
\noindent \textbf{3)}
AKMS might ingest external information to guide its heuristic by clustering the provided hashes or labels with DBSCAN~\cite{ester1996_DBSCAN}, yielding \textit{``hint''} clusters whose centroids can initialize the algorithm. As DBSCAN does not make spherical or Gaussian assumptions and automatically clusters points based on their density, this model's real-world samples are accurate due to its arbitrary cluster-shaping and consideration of noisy outliers. No user-specified $k$ is needed; Augmented K-means infers it from labels.

\vspace{-2.0mm}

\subsection{Algorithmic Challenges}
\label{subsec:algChal}

\noindent \textbf{Solving the Alignment Problem.}  In\-te\-grat\-ing Aug\-ment\-edK-Means with Spectral Co-Clustering presents a dimensional alignment challenge. The AKMS algorithm requires a robust initialization and an error-ceiling hyperparameter $\alpha$. The following algorithms detail the core logic and the coordinate-wise estimation subroutine used to achieve a $(1 + O(\alpha))$-approximate clustering. Algorithm~\ref{alg:learning-kmeans} outlines the AugmentedKMeans proposed by Ergun et al.~\cite{ergun2022_learningaugmentedkmeansclustering}.
(implementation: src/clustering/augmented\_kmeans.py)

In the original AugmentedK-Means, the predictor label vector $\Pi$ directly aligns with the samples in the matrix input as shown in Algorithm.~\ref{alg:learning-kmeans}.

\begin{algorithm}[H]
\caption{Learning-augmented k-means clustering}
\label{alg:learning-kmeans}
\begin{algorithmic}[1]
\Require A point set $X$ with labels given by a predictor $\Pi$ with label error rate $\lambda$
\Ensure $(1 + O(\alpha))$-approximate $k$-means clustering of $X$
\For{$i = 1$ to $k$}
    \State Let $Y_i$ be the set of points with label $i$
    \State Run \textsc{CRDEST} for each of the $d$ coordinates of $Y_i$
    \State Let $C_i'$ be the coordinate-wise outputs of \textsc{CRDEST}
\EndFor
\State \Return $C_1', \ldots, C_k'$
\end{algorithmic}
\end{algorithm}

If there were 8 samples, we should expect the size of $\Pi$ to be 8. However, Spectral Co-Clustering transforms our original input $S\times{}F$ matrix into a $(S + F)\times{}L$ Z matrix, where both samples and features are represented in the same latent space.  To maintain the theoretical guarantees of AugmentedK-Means when clustering in this transformed Z space, we must extend $\Pi$ to cover all rows of Z. Let $\Pi_{S} = [\pi_1, \pi_2, ..., \pi_S]$ be the initial predictor labels for S samples. For each feature $f_j$, we define its label as: $\pi_{s+j} = mode(\{\pi_i | sample\ i\ contains\ feature\ j\})$.

Each feature inherits the most frequent label among the samples that contain it. E.g., if feature $f_1$ are in samples $S_1, S_3, S_5, S_6$ and the samples are labeled $[0, 0, 1, 2]$ respectively, feature $f_1$ would be assigned label $0$ since it represents the most samples in this label. For size reference, an example of the full dimension transformation is extending $\Pi_{S} = [0, 3, 1, 3, 2]$ to $\Pi_{S + F} = [0, 3, 1, 3, 2, 0, 2, 2, 1]$ given the sample count of $S = 5$ and feature count of $F=4$. It preserves information from the AugmentedK-Means predictor while guiding the centroid placement in the transformed spectral Z space.

 This extension respects both the original predictor information and the structural relationship revealed by Spectral Co-Clustering. E.g., consider an extreme scenario in the Z latent space where sample points are close, feature points are close, and both the sample and feature \textit{``clouds''} are far. In this scenario, VBGMM might create two biclusters: one with all samples and no features, and one with no samples and all features. This would not be usable for rule generation. AugmentedK-Means recognizes that features are connected to samples and tries to place its centroids between the two \textit{``clouds''} to bridge the gap rather than place its centroids inside each sample/feature cloud. No prior work addresses this, which will be hereafter called \textbf{mode extension} since we are extending the size of $\Pi$ by filling in the feature labels with the label mode.

\noindent \textbf{Solving the Calibration Problem.} AugmentedK-Means requires an $\alpha$ hyperparameter as the error ceiling of the predictor. We followed the methodology of Ergun et al.~\cite{ergun2022_learningaugmentedkmeansclustering} to get $\alpha$. 
(implementation:  src/clustering/crdest.py)

To do so, we ``\textit{iterate over a small range of possible $\alpha$ from .01 to .15 (Algorithm~\ref{alg:crdest}) with a step size of 0.01 and select the clustering that results in the lowest objective cost}~\cite{ergun2022_learningaugmentedkmeansclustering}. We minimized the sum of squared distances as the cost function by using the Coordinate-wise Estimation (CRDEST) to perform this estimation efficiently; a breakdown of its key properties and algorithmic steps is provided in Appendix~\ref{app:aug}.

\begin{algorithm}[H]
\caption{Coordinate-wise estimation (CRDEST)}
\label{alg:crdest}
\begin{algorithmic}[1]
\Require Points $x_1, \ldots, x_{2m} \in \mathbb{R}$, corruption level $\lambda \leq \alpha$
\For{each trial}
    \State Randomly partition the points into two groups $X_1, X_2$ of size $m$
    \State Let $I = [a, b]$ be the shortest interval containing $m(1 - 5\alpha)$ points of $X_1$
    \State $Z \gets X_2 \cap I$
    \State $z \gets \tfrac{1}{|Z|}\sum_{x \in Z} x$
    \State \Return $z$
\EndFor
\end{algorithmic}
\end{algorithm}

\subsection{Heuristics To Estimate k}

\noindent Our working hypothesis is that VT reports provide an informative signal that positions each sample within our search space better than chance, because vendor family labels carry intrinsic information. We hypothesize that knowledge transfers directly from the AV label domain to the similarity hashing domain. Specifically, diversity knowledge should transfer: if an AV family is \textbf{(i)} more or less homogeneous, \textbf{(ii)} larger or more compact, or (\textbf{iii)} contains more or fewer sub-variants in the AV label domain, we force this same structural diversity into the hash domain to bias $k$ to match the real-world scenario. We can leverage VT to estimate the number of clusters $k$. However, a key challenge is that VT vendors often disagree and produce noisy labels; hence, VT is not a perfect oracle. A perfect oracle would exist only if all VT engines agreed on a family label for every sample, which we do not observe. Empirically, clusters rarely achieve unanimity. Within a cluster of $N$ samples, we typically observe multiple distinct candidate family labels across vendors rather than a single consensus (and the number of unique candidates generally increases with $N$). This observed label diversity both informs our estimation of $k$ and gives us a larger range of candidates to estimate an appropriate value. Accordingly, we normalize VT reports using AVClass2 \cite{DBLP:journals/corr/abs-2006-10615} and build a per-family history (an oracle) of $k$ values observed in prior runs. For each new batch, we extract VT-derived family labels and query this oracle to select $k$. We assume family-level information propagates consistently across samples, so that diversity, heterogeneity, and spread-not label identity-are what AV vendors implicitly preserve, even under vendor disagreement. Concretely, when we query the oracle for a sample, it returns the family's historical $k$ values, how samples from that family have clustered in the past. It is this structural signal, rather than label identity, that we leverage to estimate $k$. We tested three approaches:\newline
\noindent \textbf{1)} \textbf{Max $k$:} For a given family, choose the largest historically observed $k$. If that value exceeds the number of available samples, cap it at the largest feasible $k$. We hypothesize that the original algorithm systematically underestimates the optimal $k$ value, leading to over-generalized rules that fail to capture true diversity within malware families. By using the maximum historically observed $k$, we ensure that the clustering process accounts for the full complexity and variability that have been documented for each family, preventing the generation of overly broad rules that may produce FPs. \newline
\noindent \textbf{2)} \textbf{Statistical Mode $k$:} For a given family, we choose the most frequently observed $k$ across prior runs, assuming this reflects a stable choice. This approach is based on the hypothesis that the most frequently occurring $k$ value represents the consensus optimal clustering structure for a malware family, filtering out both underestimated and overestimated $k$ values that may result from data anomalies or algorithmic instability. The statistical mode captures the "typical" complexity level of the family by averaging out temporary fluctuations caused by unusual batches, incomplete data, or algorithmic edge cases. \newline
\noindent \textbf{3)} \textbf{Random $k$:} For a given family, sample $k$ uniformly at random from the historical values observed for that family, serving as a variance-injecting baseline.

The VT-based heuristics serve as a practical optimization, selecting $k$ using family histories during operation. In turn, our raw-$k$ baselines (Sec.~\ref{subsec:AutoPYaraKEval}) are used solely for evaluation and are not employed to drive $k$ selection.

\vspace{-2.0mm}
\subsection{Implementation}

\noindent Our implementation was designed to achieve \textbf{efficiency} and \textbf{usability}. To make AutoPYara efficient, we kept the original JAVA engine as the base, patching it with our proposed algorithm modifications. To increase usability, we provided a Python API for accessing the JAVA methods, allowing AutoPYara integration in multiple pipelines. This required breaking the original AutoYara architecture from a monolithic solution into a JAVA backend and Python frontend (details in Figure~\ref{fig:pyara interface} of Appendix~\ref{app:framework}). The Java-to-Python communication is performed via JPype \cite{jpype}, and the generated rules are checked for validity via their compilation with Yara-Python libraries~\cite{yaramod, yara_python}. AutoPYara (Sample Rule in Appendix~\ref{app:yararule}) is released as a library on PyPI.

\section{Experimental Setting} 
\label{sec:exp}
\subsection{Dataset}

\noindent We run our experiments on (malware and goodware) Windows (PE) files collected by an AV company from infected customers over 8 years, representing a realistic sample of in-the-wild malware. This dataset has been characterized in a previous study~\cite{botacin2021_bbsample} and includes samples collected by the same company and the same engine in 3 different countries: the USA, Brazil, and Japan. Using the same vendor in the same timeframe makes the samples comparable. Collection in different countries adds diversity and increases external generalization, as malware originating from different countries often displays distinct characteristics~\cite{10.1145/3678890.3678893}. We used the malware for rule generation and the goodware for Bloom Filter training, ensuring no FP arises from biases like data collected in distinct places and times. In total, we leverage 71k malware samples and 22k goodware samples for the experimental results. Unlike studies that focus on specific attack strategies~\cite{10.1145/3708821.3736206,10460021}, we \textbf{do not remove repeated samples} from our experiments, as they may appear across different malware campaigns and thus better reflect real-world conditions. We also do not perform any additional pre-processing on the samples to evaluate them as seen by the AV companies in the wild.

\subsection{Experimental Methodology}
\label{subsec:expmeth}

\noindent Our experiments follow a methodical strategy where malware samples are clustered into families, which serve as input for both AutoYara (SOTA) and AutoPYara (our solution) in the comparative analysis. The same set of goodware files is provided to eliminate any goodware bias. We then analyze the generated rules and, to minimize variability from inherent stochasticity, report each setting as the mean over multiple runs. We applied a convergence algorithm to determine the number of runs required to precisely represent each cluster. The framework is set to generate $N=10$ candidate rules per cluster, but adaptively determines early-stop conditions. After the generation of each rule, it evaluates its performance considering the Accuracy. If the improvement between consecutive evaluations falls below the threshold $\theta$, the algorithm halts rule generation early. The choice of a $5\%$ threshold reflects a balance between sensitivity and stability and is meant to prevent termination due to small fluctuations, ensuring that additional rules are generated if they provide meaningful improvement.

\noindent\textbf{Experimental Setup.} We ran our experiments on a high-performance server (Intel Xeon Gold 5318Y, 256GB RAM, running Ubuntu 22.04.3 LTS). We were also able to run AutoPYara on a consumer-grade laptop (Intel i7-1255U, 16GB RAM), showing that it is practical on end-user machines. AutoPYara is hardware-agnostic, requiring minimal resources (1 core, 1GB RAM) and supporting Python 3.9–3.12, Java 11+, and Linux/Windows.

\subsection{The Clustering Task.} 
\label{sec:Clustering Task}
\vspace{-4mm}

\begin{figure}[htbp]
\centerline{\includegraphics[
    width=\linewidth,
    trim=0cm 8cm 0cm 2.5cm,
    clip
]{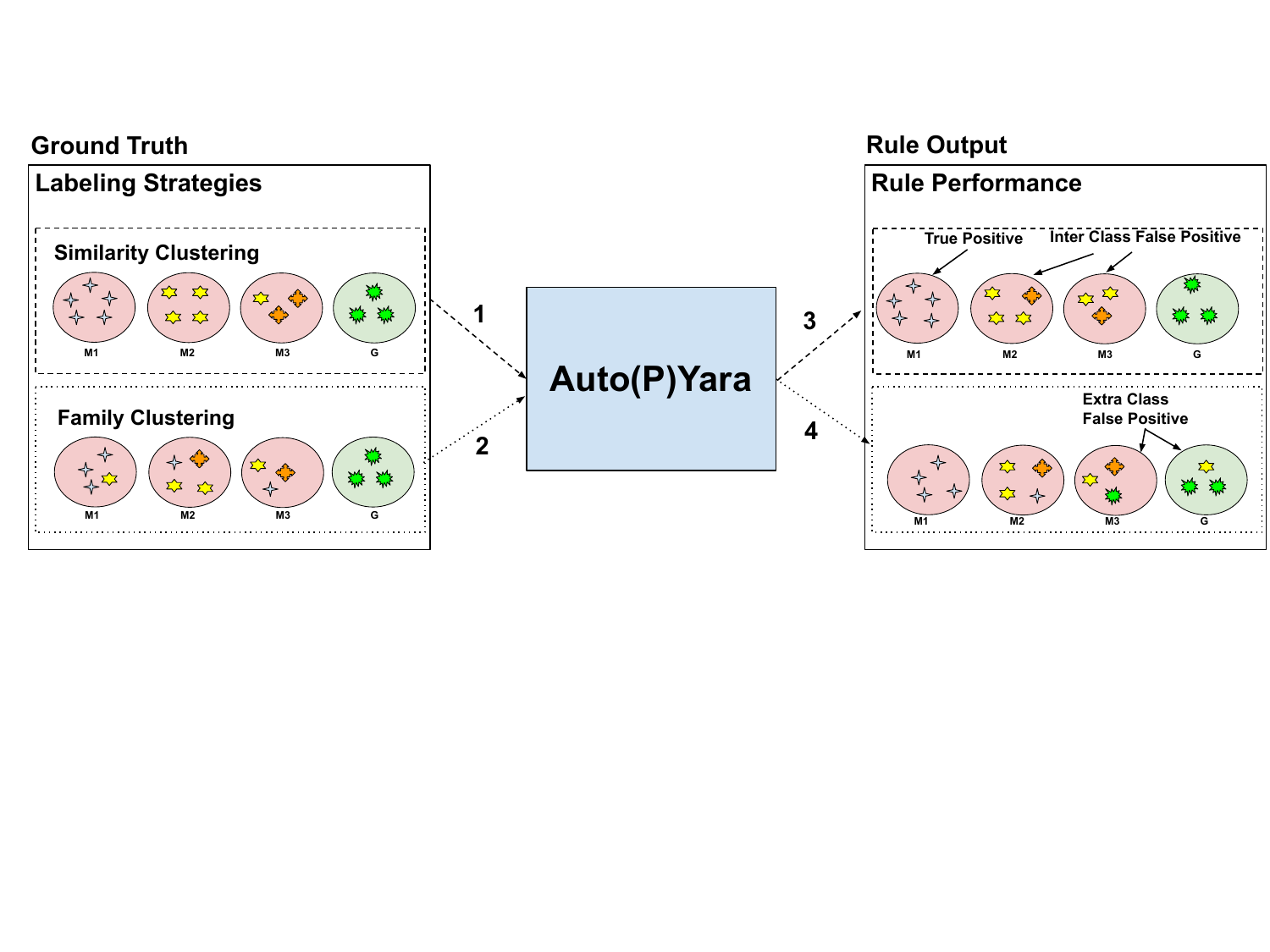}}
\caption{Overview of the rule generation task.}
\vspace{-4.0mm}

\label{fig:YaraFWCycle}
\end{figure}

 \noindent Unlike a typical ML clustering task, which assigns unlabeled samples to clusters, generating YARA rules for malware families is different. Its goal is to generate rules for \uline{already-labeled} samples. This happens when an AV company has already identified a malware family and wants to generate YARA rules for it. Figure~\ref{fig:YaraFWCycle} exemplifies this process, where a set of labeled malware and a reference set of goodware samples (Path 1) are input to AutoYara/AutoPYara, and a set of rules (Path 3) is generated as output. This process is naturally subject to variations in the input clusters. If a different cluster distribution is input (Path 2) to the rule generator, a different output (Path 4) is observed.

 While accuracy is naturally a relevant metric for this task, a full understanding of the tool's performance can only be achieved by breaking it down into its components, as they present very different characteristics. On the one hand, the tool aims to maximize the \underline{True Positive Rate (TPR)} for each cluster, which represents the rule's coverage, i.e., the portion of samples in a cluster that the rule can actually match. On the other hand, its goal is to minimize the False Positive Rate (FPR). It is key to note that not all mistakes (FPs) are the same. FPs can occur in two ways: First, as \underline{Extra-class False Positive Rate}, where a rule misclassifies goodware as malware; and also as \underline{Inter-class False Positive Rate}, where a rule incorrectly classifies a sample from a different malware family. The operational consequences of these two types are fundamentally different: intra-cluster FPs obscure malware boundaries, while extra-cluster FPs directly jeopardize goodware and normal operations. This work measures YARA rules using these three metrics. We consider it equally important to ensure that benign software continues to run without interruption and to achieve a high TP rate. An effective YARA rule must achieve a high TP rate while keeping both FP rate extremely low, ideally close to zero.

\vspace{-3mm}
\subsection{Threat Model}
\vspace{-2mm}

\noindent We evaluated AutoPYara across realistic environments with two distinct threat models based on operational priorities:

\noindent\textbf{1. Detection (Conservative):} We aim to prevent operational disruption and alert fatigue in production pipelines. FPs matter most here, so we require strict FP minimization even if it reduces coverage.

\noindent\textbf{2. Hunting (Aggressive):} We aim to discover unseen malware, prioritizing maximum TP rate while accepting higher FP tolerance for analyst triage.

These requirements are often conflicting: strict FP suppression and maximum TP coverage typically trade off against each other. We measure this conflict directly, then show how AutoPYara reconciles the two, addressing conservative detection and aggressive hunting together within a single framework.

\vspace{-3.0mm}

\section{Evaluation}
\label{sec:eval}
\vspace{-2.0mm}

\noindent We evaluated AutoYara and AutoPYara in two conditions. First (Subsec~\ref{subsec:AutoYaraEval} to ~\ref{subsec:AutoPYaraKEval}), we assessed their \uline{maximum performance potential} by covering the entire dataset. Second (Subsec~\ref{subsec:threathunting}), we evaluate their \uline{generalization} via train-test splits. The first condition is sub-divided into: \textbf{(i)} the effects that the original AutoYara solution was subject to (Subsec~\ref{subsec:AutoYaraEval}), acting as a guideline for evaluating automatic rule generation procedures, and \textbf{(ii)} the individual contribution of each AutoPYara component (Subsec~\ref{subsec:AutoPYaraEval}) to mitigate these effects, working as an ablation study for our tool.
\subsection{Characterizing AutoYara \& Its Effects}
\label{subsec:AutoYaraEval}

\noindent To evaluate whether AutoPYara is superior to the SOTA AutoYara, we have to establish clear Criteria (Cs) for comparison. Although the original paper describes the foundational approaches, it does not analyze how key parameters affect the tool’s performance. Identifying these factors allows us to establish fair benchmarks for comparison, ensuring we can quantify improvements and characterize their performance in the real world malware samples.

\noindent \textbf{\underline{C1}. The Impact of Labeling Strategy: }A natural first step to evaluate the tool's performance is its TP rate. Since our dataset consists of in-the-wild samples, it lacks family labels for comparison, requiring the derivation of our own labeling strategy for the experiments~\cite{JOYCE2021_AGTR}. But \textit{What is the best labeling strategy?} In the absence of standards, we adopt two different approaches to evaluate their impact on the framework: \textbf{(1)} \textbf{similarity-based} clustering via the hashing functions SSdeep and J-SDhash; and \textbf{(2)} \textbf{label-based} clustering, built on VirusTotal-reported family labels (normalized by AVClass2 \cite{DBLP:journals/corr/abs-2006-10615}). These clustering methods are used as they are the most popular and widely adopted strategies to group samples in the malware analysis literature~\cite{JOYCE2021_AGTR,lee2019dexofuzzy,10.1016/j.cose.2022.102921,kornblum2006identifying}. They are fundamentally different, with the first focusing on the binary characteristics of the samples and the second on their high-level behavior~\cite{vitale2025polymorphism}. It is plausible to hypothesize that generated labels might vary~\cite{botacin2021_ssdeephashing}, impacting the rule generation. We applied these labeling methods to all samples and generated the different clusters. We discarded singleton clusters (i.e., clusters containing only a single file) because AutoYara is designed to derive generalized rules from recurring patterns, not from singleton outliers.

\begin{table}[ht]
\centering
\scriptsize
\setlength{\tabcolsep}{1.5pt}
\vspace{-2mm}
\caption{Cluster Statistics Across Method}
\label{tab:ClusterStats}
\vspace{-3mm}
\begin{tabular}{lccccccccccc}
\toprule
\multirow{2}{*}{Metric}
& \multicolumn{5}{c}{SSdeep}
& \multicolumn{5}{c}{J-SDhash}
& VT \\
\cmidrule(lr){2-6}
\cmidrule(lr){7-11}
& 50 & 60 & 70 & 80 & 90
& 50 & 60 & 70 & 80 & 90
& \\
\midrule
\#Clusters & 8807 & 8874 & 8874 & 8761 & 8588 & 7753 & 7921 & 8169 & 8135 & 8158 & 1121 \\
Max Size & 1129 & 1126 & 708 & 708 & 466 & 10649 & 8673 & 2447 & 1391 & 1262 & 5558 \\
Avg Size & 4.87 & 4.61 & 4.36 & 4.18 & 3.89 & 6 & 5.59 & 5.12 & 4.89 & 4.45 & 38.89 \\
\bottomrule
\end{tabular}

\vspace{-2mm}

\end{table}

Table~\ref{tab:ClusterStats} shows that different labeling strategies yield datasets with varying sizes and internal structures. These differences alter underlying distribution shapes within the dataset, affecting the evaluation context and influencing the perceived performance. The variation across cluster counts, maximum sizes, and average distributions highlights our central hypothesis: Different labeling methods fundamentally alter how clusters are defined, affecting the task's difficulty and the conclusions one can draw from the results, leading to conflicting and unfair interpretations. 

Labeling Strategy variation has operational consequences: it can fragment large clusters into many smaller ones, inflating the apparent dataset size and compounding the computational burden of detection. Every additional cluster corresponds to a new set of rules or signatures that security systems must maintain and apply. Massive clusters may oversimplify detection rules, trading off specificity for scalability and potentially introducing both FPs and missed detections.  In practice, this means that the structure of the labels governs the balance between precision, scalability, and operational overhead. Evaluating tools or clustering methods on a single dataset without acknowledging these differences risks drawing conclusions that do not generalize to real-world deployment. We measure this impact via the TP rates across these variations and show how changes in dataset construction (e.g., size and structure) propagate into changes in rule quality and detection outcomes, skewing overall performance.

\begin{figure}[htbp]
    \centering
    \subfloat[Without Failed Clusters\label{fig:BaseLineAutoYaraBoxOGBF}]{%
        \includegraphics[width=0.48\linewidth]{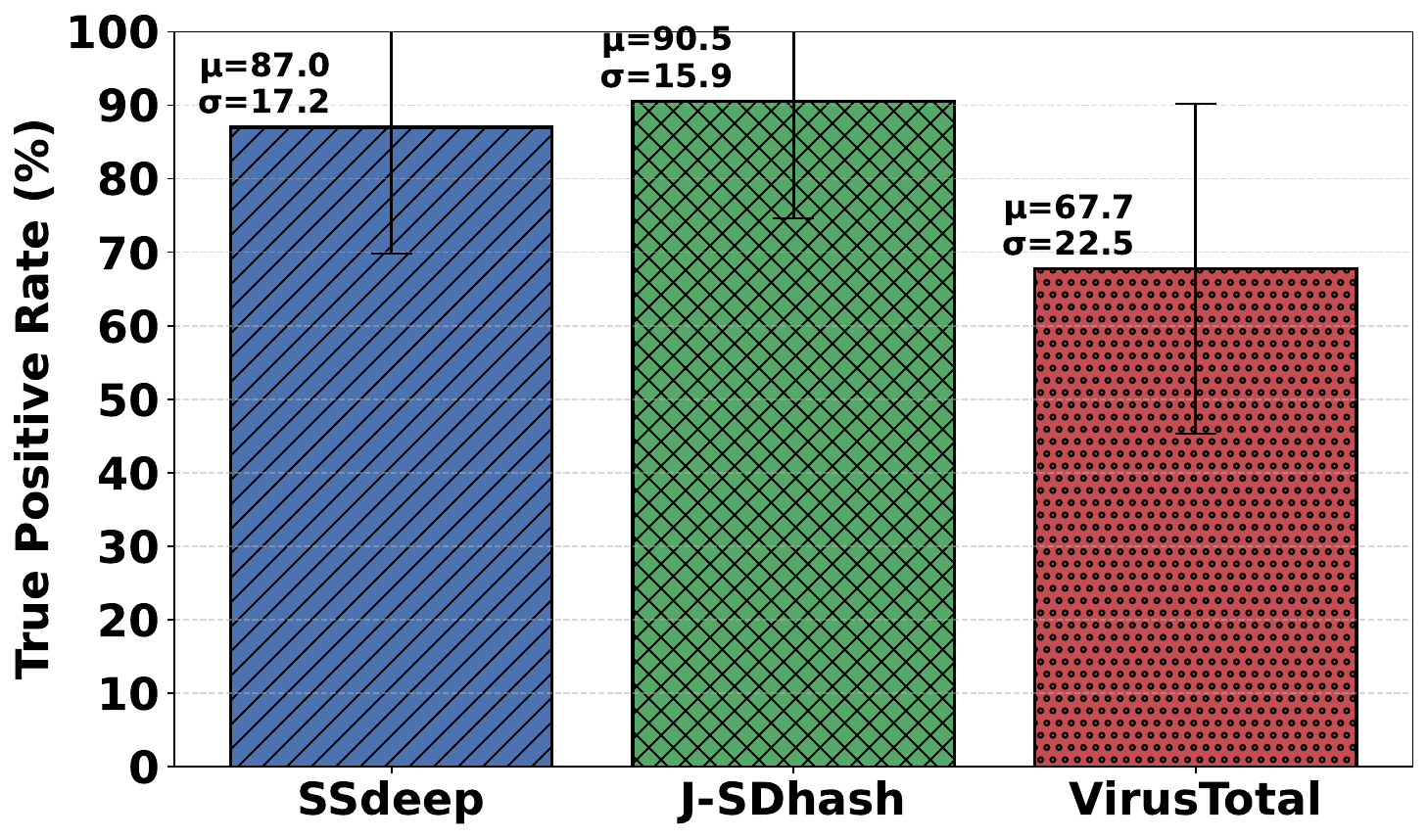}
    }\hfill
    \subfloat[With Failed Clusters\label{fig:BaseLineAutoYaraBoxOGBF_True}]{%
        \includegraphics[width=0.48\linewidth]{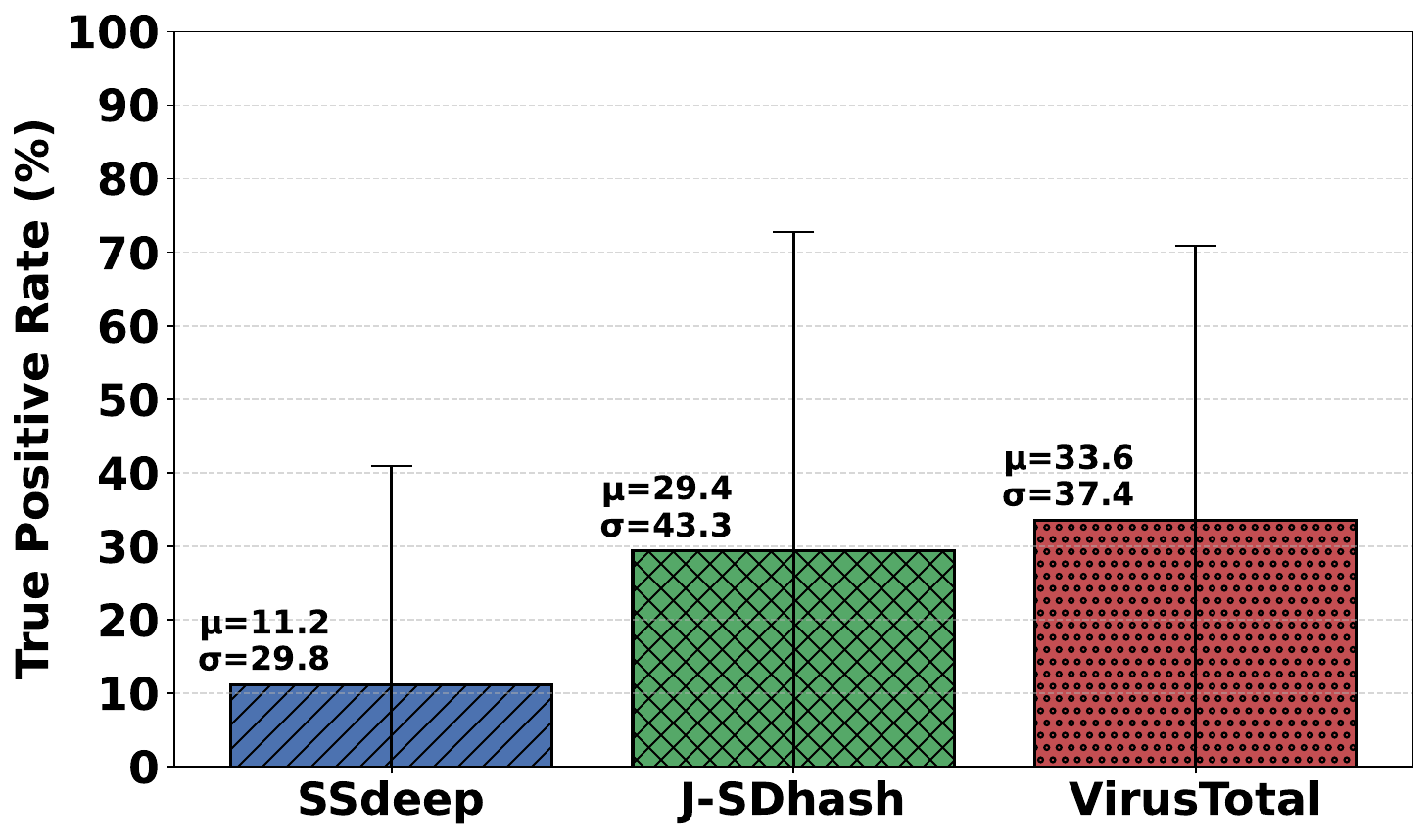}
    }
\caption{Comparison of AutoYara performance: (a) baseline excluding failed clusters; (b) observed performance including failed clusters.}    \label{fig:EmberBloomFilterYara}
\end{figure}

 \noindent Figure~\ref{fig:BaseLineAutoYaraBoxOGBF} shows the impact of different labeling approaches, which create varying clusters that yield markedly different approximations even when applied to the same data. This divergence underscores how the definition of a labeling strategy is not fixed but shaped by the algorithmic choices made during its construction. This variation arises because SSdeep and J-SDhash group samples by \textit{intrinsic code-level similarity}, capturing functional resemblance even across unrelated labels, whereas the label-based approach depends on external categorizations that overlook such relationships. The first method generally surpasses the label-based ones in \textbf{TPs}, as it relies on more coherent and internally consistent similarity criteria than on labels that group samples according to heterogeneous standards.

\noindent\textbf{\underline{C2}. The Impact of Bloom Filters:} 
We next examine the role of Bloom Filters within our framework and hypothesize that, like the labeling strategy, they may also influence its behavior. At first glance, the results shown in Figure~\ref{fig:BaseLineAutoYaraBoxOGBF} suggest that AutoYara has great performance with an average TP rate between 67\% and 90.5\%. This is comparable to the results presented by the original paper despite using a different dataset. However, there is a methodological oversight in this result, as it only accounts for the clusters that succeeded in producing rules. The true AutoYara performance, considering the failure rates, is shown in Figure~\ref{fig:BaseLineAutoYaraBoxOGBF_True}, and demonstrates a significant difference in the real-world performance of the tool. The variation arises because the provided Bloom Filters do not contain byte patterns present in malware found in the wild. The original paper only reported results for the successful clusters, while we considered both successful and unsuccessful cases, as we had more control over the dataset. Note that different strategies cause impacts at different levels, with the hashing-based labels experiencing a significant drop in performance compared to the family-based one. Fig.~\ref{fig:EmberBloomFilterYara} shows that the family clustering task is more challenging in the real world and that the shift in performance between the cases cannot be neglected.

\uline{Recovering BF performance.} Outdated Bloom Filters create an initial performance gap, but we can retrain them to improve coverage. We retrain the Bloom Filters and repeat the experiment to compare how coverage and performance change throughout our evaluation.

\begin{figure}[htbp]
    \centering
    \includegraphics[
        width=\columnwidth,
        height=0.22\textheight,
        keepaspectratio
    ]{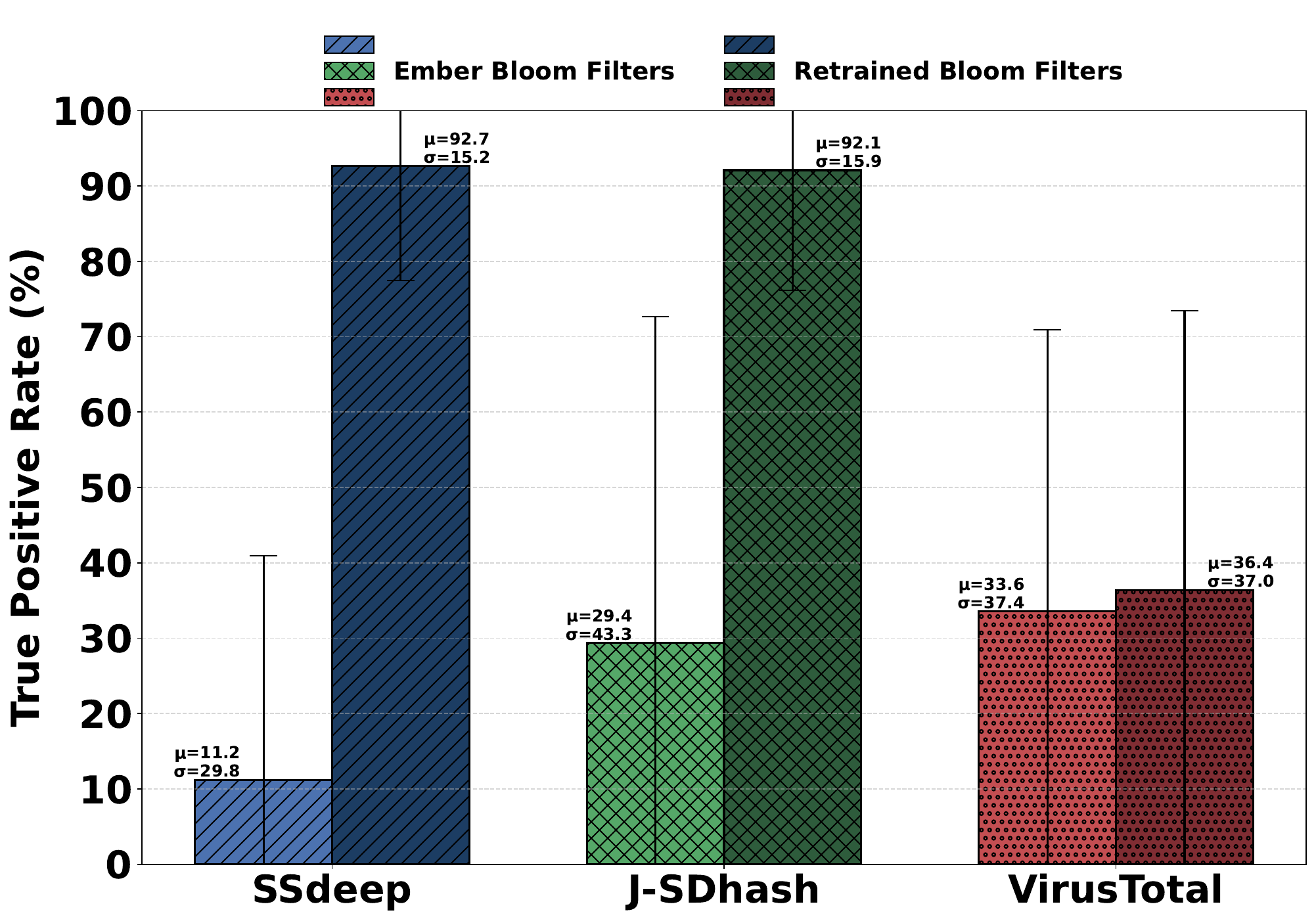}
\vspace{-2mm}

    \caption{Impact of Bloom Filter retraining on AutoYara, showing recovery of true positive rate compared to the baseline failed cluster configuration.}
    \label{fig:BaseLineAutoYaraBoxRTBFVSOGBF_True}
    \vspace{-2mm}

\end{figure}

Figure~\ref{fig:BaseLineAutoYaraBoxRTBFVSOGBF_True} quantifies the impact of this retraining step. We observe a clear improvement in TP rate performance. Retraining increases the number of generated rules across clusters by incorporating additional N-gram patterns, which reduces the number of failures. 
This improvement after retraining the Bloom Filters varies across the labeling strategies, given their fundamental differences. Byte-based clustering (SSdeep and J-SDhash) responds strongly to changes at the byte or function level, so retraining the Bloom Filters that use these hashes directly affects performance. The failures typically occur due to insufficient byte-level similarity information captured by the original Bloom Filter configuration. In contrast, VirusTotal labels aggregate samples using broader family concepts, which we find are less sensitive to fine-grained changes in the underlying code. The challenges here are fundamentally different, as VT label-based clustering struggles not from a lack of byte-level information, but from the inherent difficulty of reconciling heterogeneous label assignments. When samples carry conflicting or incompatible family labels, it becomes practically impossible to satisfy all constraints simultaneously, regardless of the underlying binary similarity. The extent of gains from retraining BFs depends on how closely the label criterion tracks true binary similarity rather than external taxonomic grouping.

\noindent\textbf{\underline{C3}. The Intrinsic High Variance:} In the same way that Bloom Filters can impact results, other factors may have a significant effect. The VBGMM clustering step is inherently non-deterministic because it does not use a fixed random seed. Random initialization and probabilistic convergence can yield different cluster assignments even when the input data is identical, sometimes converging to local rather than global optima. This stochastic behavior introduces variability into the rule generation process, leading to rules that differ in coverage, specificity, and quality across runs. To assess the magnitude of this effect, we measured how much variability appears in both the clustering output and the corresponding rule sets. We apply an early-stop algorithm to generate and evaluate rules across multiple runs, and report the cumulative metric. This provides a dynamic view of performance over time, highlighting long-term trends and changes, and reduces the influence of outliers. In Figure~\ref{fig:BaseLineAutoYaraBoxRTBFVSOGBF_True}, we highlight that the variance in the AutoYara framework is $\approx$30\% (very high).

\begin{table}[H]
\vspace{-3mm}

\centering
\scriptsize
\setlength{\tabcolsep}{3pt}

\caption{Extra Class False Positive Rate (\%)}
\label{tab:Merged-False-Positives}
\vspace{-2mm}

\begin{tabular}{lccc}
\toprule
\textbf{Method} & \textbf{SSdeep} & \textbf{J-SDhash} & \textbf{VT} \\
\midrule
AY (EBF)   & $0.99 \pm 0.89$ & $1.19 \pm 0.91$ & $48.66 \pm 33.34$ \\
AY (RBF)   & $1.23 \pm 0.83$ & $1.13 \pm 0.95$ & $20.36 \pm 12.79$ \\
AutoPYara  & $1.10 \pm 0.81$ & $1.01 \pm 0.91$ & $54.28 \pm 22.05$ \\
BestK      & $0.20 \pm 0.17$ & $0.20 \pm 0.17$ & $12.53 \pm 4.95$ \\
WorstK     & $0.22 \pm 0.16$ & $0.22 \pm 0.16$ & $3.42 \pm 2.49$ \\
\midrule
\multicolumn{4}{l}{\scriptsize AY: AutoYara, EBF/RBF: Ember/Retrained BF, VT: VirusTotal}
\end{tabular}
\vspace{-2mm}
\end{table}

\noindent\textbf{\underline{C4}. Not All Mistakes Are the Same:} As per Sec.~\ref{sec:Clustering Task}, TP gains should not come at the expense of FPs; thus, we also evaluated AutoYara via \textbf{Intra} and \textbf{Extra} class FPRs. Our experiments produced an Intra-Class FPR below 2\% for the original BFs (see Table~\ref{tab:FPR-Single-Col} of Appendix~\ref{app:fpa}). The rate is zeroed after retraining the BFs. This reinforces the importance of Bloom Filters and the careful selection of labeling strategies. Table~\ref{tab:Merged-False-Positives} presents the Extra-Class FPR across different labeling strategies.
SSdeep and J-SDhash exhibit lower Extra-Class FPR compared to the family label-based clustering. The inherent nature of similarity-based clustering is responsible for rules with an Extra-Class FPR. AV label clusters from VT are independent of internal patterns, which results in rules that are forced to be more generic, given files that have conflicting patterns. This causes rules to be more stringent with detection, resulting in high Extra-Class FPR. The VT approach is impractical where goodware is present, risking legitimate software operations. In malware‑only workflows, searching within a curated repository or triaging a lab quarantine feed rather than scanning a user’s hard drive, this limitation is less severe. Similarity hashing–based clustering provides robust performance, maintaining low Intra- and Extra-class FPR.

\noindent\textbf{\underline{C5}. The Impact of Threshold Variation:} Given similarity hashing as an effective labeling strategy, we now characterize its threshold variation. This clustering method allows us to adjust a similarity threshold, which dictates how selectively the algorithm groups the samples. In practice, it controls the balance between precision and inclusiveness, shaping the quality of the derived labels.

\begin{figure}[htbp]
    \centering
    {%
        \includegraphics[width=0.70\linewidth]{
            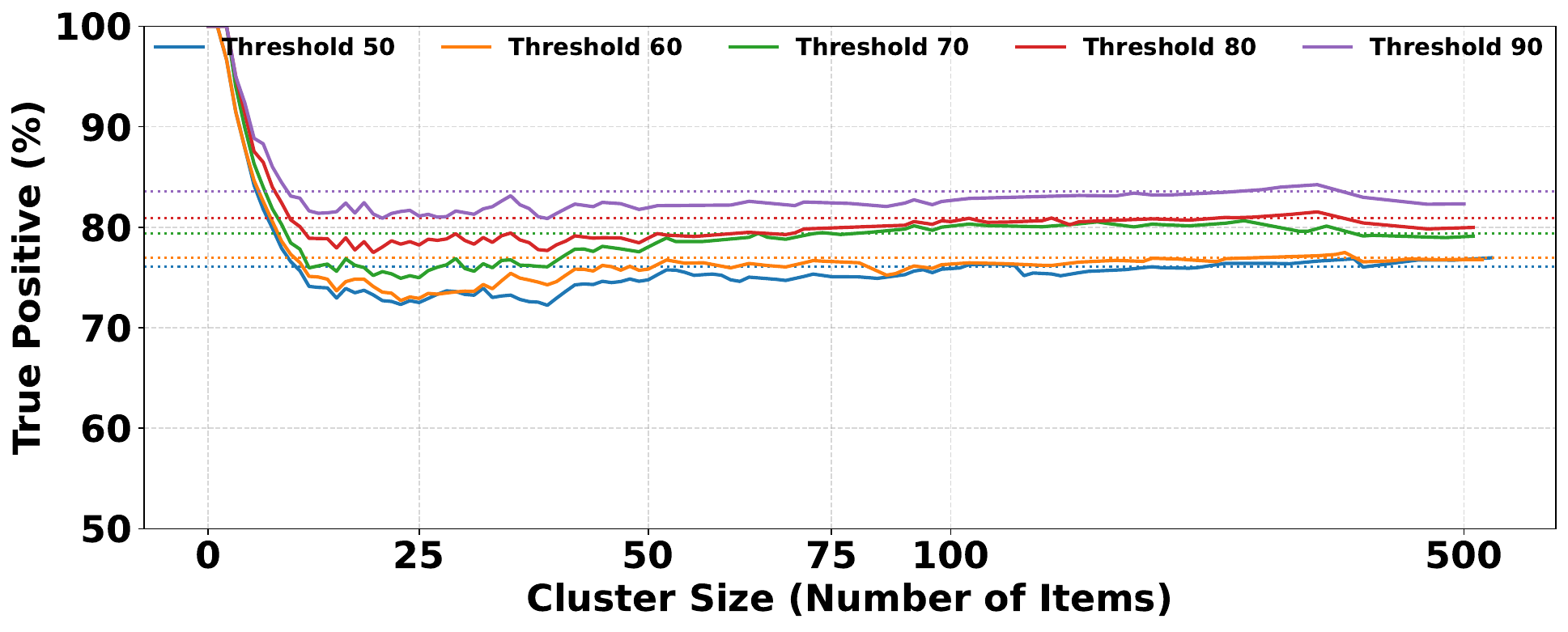
        }
    }

    \vspace{-2mm}

    {%
        \includegraphics[width=0.75\linewidth]{
            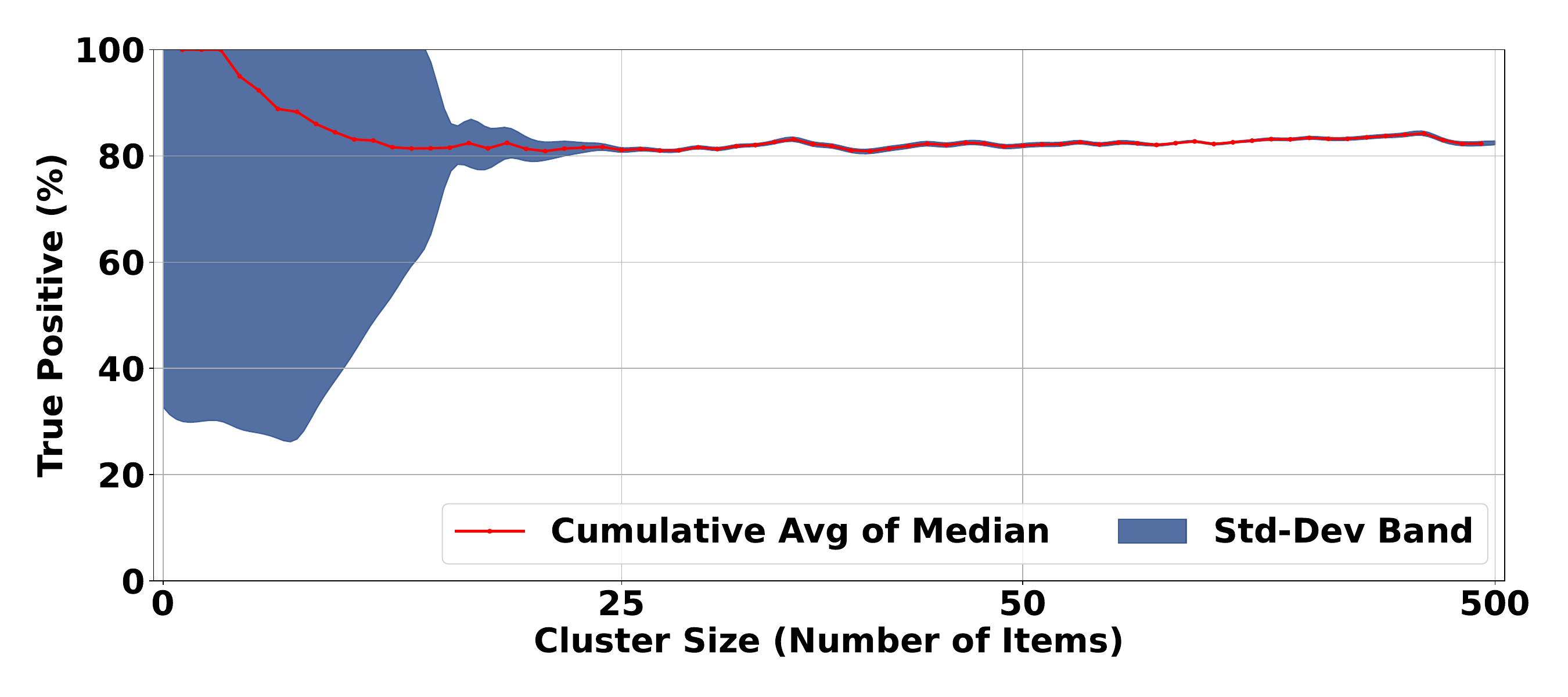
        }
    }

    \vspace{-3mm}
    \caption{AutoYara across clustering configurations. (a) Difference in performance between clusters derived using different SSDeep similarity thresholds. (b) Variation in performance between small and large clusters.}
    \label{fig:AutoYaraSSdeepComparison}
    \vspace{-4.0mm}

\end{figure}

Figure~\ref{fig:AutoYaraSSdeepComparison}.a highlights a 10\% difference in the threshold range from 50 to 90 in AutoYara and illustrates its impact on the results. This variation occurs because tighter thresholds produce purer clusters—in other words, the files are more closely matched at a byte level, resulting in better clusters. 
Figure~\ref{fig:AutoYaraSSdeepComparison}.a demonstrates that all thresholds exhibit similar variations, suggesting an inherent characteristic of malware clustering, rather than an artifact of our specific approach. This fundamental property indicates that the relationship between clustering parameters and detection performance follows predictable patterns inherent to the nature of similarity analysis. Stricter thresholds generally improve rule quality and effectiveness, albeit at the cost of reduced coverage. This reduces the impact of automatic rule generation, as each unclustered sample must be processed separately by a human analyst. The balance between cluster purity and diversity becomes critical when tuning thresholds for optimal performance and representativeness. When sample sets are well characterized, prioritizing rule purity yields highly precise, specific signatures; by contrast, hunting for unknown threats benefits from greater rule diversity to broaden detection coverage.

\noindent\textbf{\underline{C6}. The Impact of Cluster Size:} We previously demonstrated how data-aggregated analyses can obscure effects by combining results across clusters into a single overall score. Specifically, aggregation may mask variations in performance across different clusters and hide meaningful intra-cluster differences. We show that aggregation can lead to incomplete results, where some cluster configurations perform well in specific areas of the dataset, but underperform in others. Similar issues arise when aggregate performance is evaluated w.r.t. cluster size, potentially overlooking how size-specific configurations can influence results.

 The distribution of cluster sizes can impact the framework’s performance: a few large clusters may inflate scores, while more granular clusters can reveal weaknesses that would otherwise remain hidden \cite{10.1007/978-3-642-15512-3_13,CORDEIRODEAMORIM2021114971}. Larger clusters represent rules that must generalize across more samples. The hypothesis is that rules derived from small, homogeneous clusters should be highly effective within those narrow bounds but might lack generalization. Figure~\ref{fig:AutoYaraSSdeepComparison}.b shows the variation in the TP rate of the rules for a given threshold using AutoYara. Small clusters exhibit higher mean TP rate at the cost of larger variance. This reinforces that finding a few stable patterns among two similar samples is inherently easier than identifying patterns across $N$ more diverse samples. Cluster size directly influences rule effectiveness: performance often starts high for small clusters and declines as cluster size and diversity increase due to broader, less specific patterns. Our experiments show that variance is inherent to the family clustering problem, rather than to AutoYara alone, and can risk misinterpreting the results. Figure~\ref{fig:AutoYaraSSdeepComparison}.b highlights this variance, underscoring the need for more fine-grained analysis. We observed these patterns across all tested thresholds for both SSDeep and JC-SDhash ground truth. Relying on a single summary metric can be misleading; Adopting a fine-grained approach to rule evaluation, we can isolate peak performance and generalization across the dataset, which is key for practical deployment.

\subsection{Impact of Centroid Selection: AutoPYara [F1]}
\label{subsec:AutoPYaraEval}

 \noindent In the previous section, we identified key factors and metrics that influence rule generation, and now we will characterize and evaluate the AutoPYara framework. We compare its behavior directly with the original AutoYara framework under the same considerations. To assess AutoPYara's strengths, we follow the same strategy as in the earlier analysis and break down the results by component. We first evaluate centroid selection methods while keeping other parameters constant, then examine the impact of $k$ values, thereby isolating variables to understand how each component shapes the framework's performance. To conduct a fair comparison, we extract the $k$ value from the VBGMM step of the AutoYara framework and apply this same $k$, with only the centroid selection differing between them. We isolate variables between the two frameworks, avoiding comparisons that fail to control key parameters.

\noindent \textbf{\underline{C1}. The Impact of Labeling Strategies:} We recreate the previous experiment across different labeling strategies. We use AutoPYara with centroid selection, to verify the impact identified on the original AutoYara framework.

\begin{figure}[htbp]
    \centering
    \includegraphics[
        width=\columnwidth,
        height=0.22\textheight,
        keepaspectratio
    ]{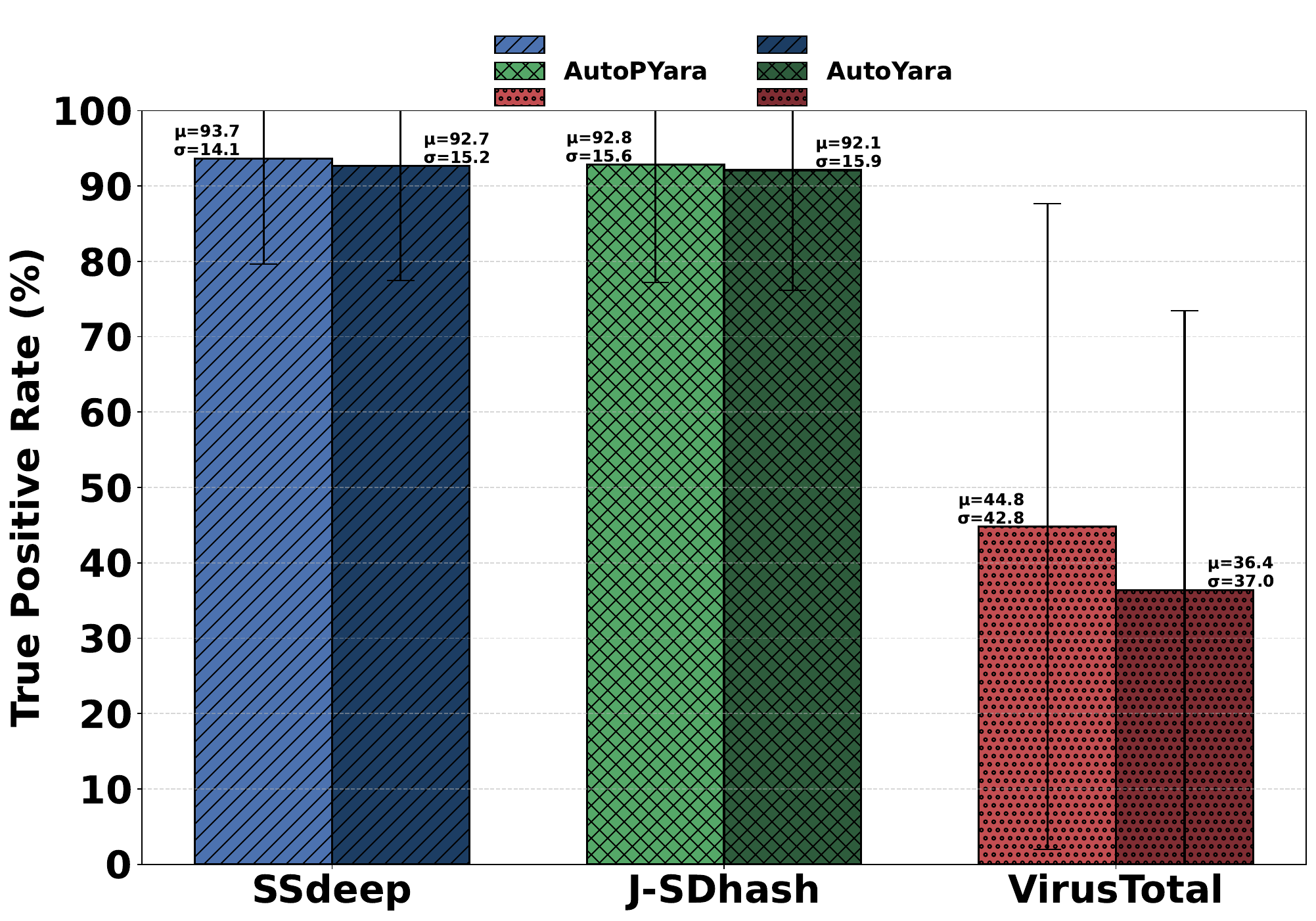}

    \vspace{-2mm}

    \caption{Comparison of labeling strategies between AutoYara and AutoPYara, highlighting performance differences under different configurations.}
    \label{fig:compare_autoyara_autopyara_rtbf}

    \vspace{-2mm}
\end{figure}
\begin{figure*}[t]
    \centering

    {%
        \includegraphics[width=0.32\textwidth]{Figures/AutoYara/RTBF/VairiationPlots/Merged/SSdeepAVG_Zoom.pdf}
    }
    \hfill
    {%
        \includegraphics[width=0.32\textwidth]{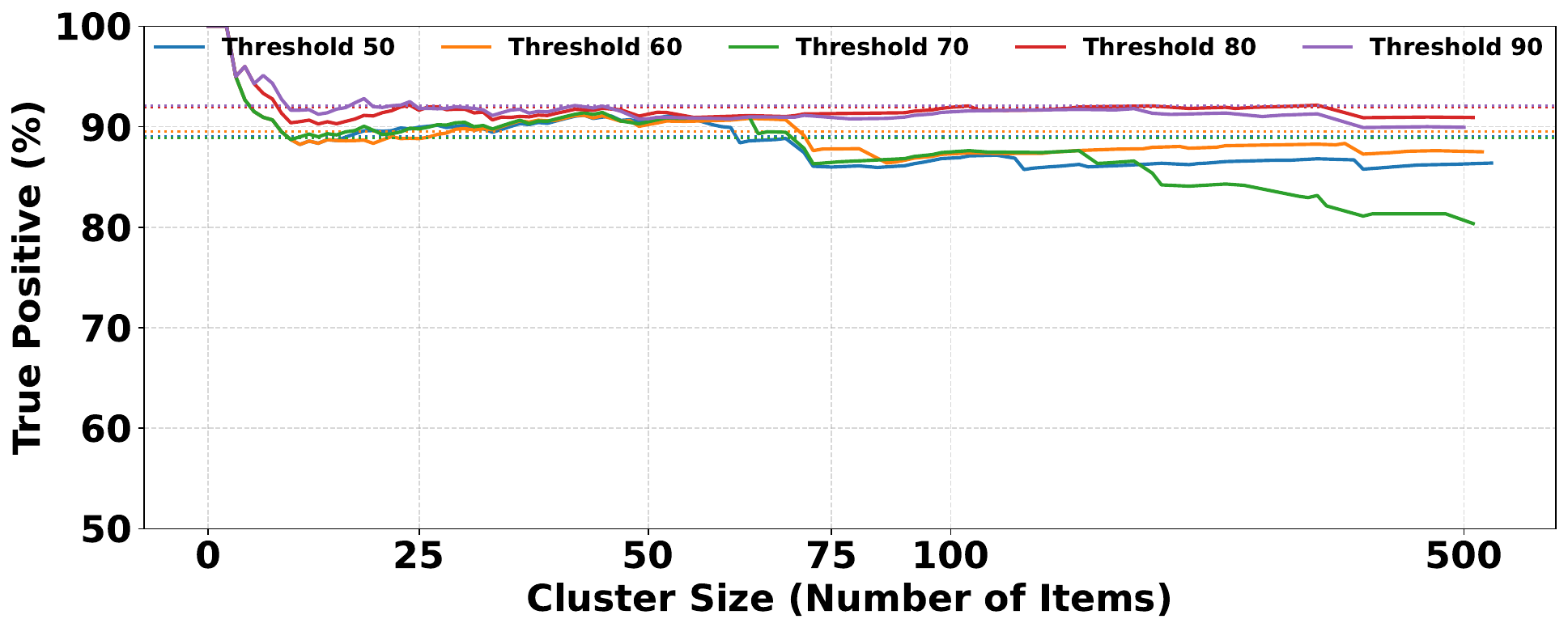}
    }
    \hfill
    {%
        \includegraphics[width=0.33\textwidth]{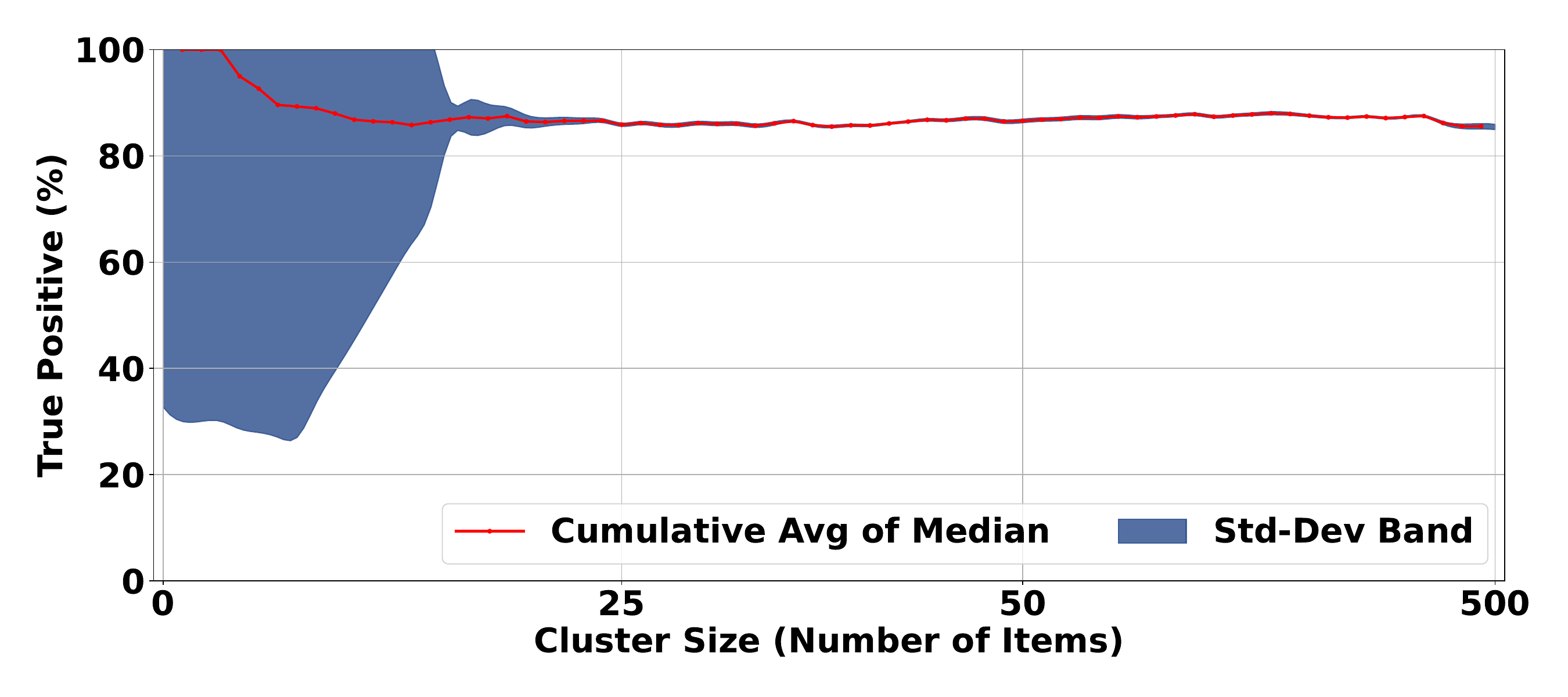}
    }

    \vspace{-3mm}
    \caption{Comparison of SSdeep-based labeling and cluster size variations: (a) threshold variation analysis of \textbf{AutoYara}, (b) threshold variation analysis of \textbf{AutoPYara}, and (c) cluster size impact on AutoPYara using SSdeep (TH90).}
    \label{fig:AutoYara_Thresholds_Comparison}
    \vspace{-5mm}
\end{figure*}

\noindent Figure~\ref{fig:compare_autoyara_autopyara_rtbf} shows the performance differences across the three labeling strategies. AutoPYara performs, on average, 3.4\% better than AutoYara on the same datasets and pre-clusters. Although 3.4\% may appear modest at first glance, this gain translates to an average of 2400 additional detections across malware samples in the dataset, which represents a significant absolute improvement that would be highly valuable in the real-world operations of an AV company. The improvement is most pronounced on the VT clusters, which exhibit far greater heterogeneity compared to baselines derived from similarity hashing–based clustering algorithms. In this setting, precise centroid selection in AutoPYara is critical; picking centroids that capture the dominant, stable subpatterns yields clusters with higher intra-cluster coherence and clearer inter-cluster separation. The rule generator can then assemble subpatterns correctly—promoting shared anchors within a cluster and separating variant features across clusters—so conditions are formed in a more disjoint manner rather than overlapping. This reduces rule interference and confusion, enabling rules that generalize to unseen samples from the same family while avoiding spurious matches. Note that, given its nature, the average size of the cluster generated using the AVClass labels is 8 times greater than that of the similarity-based clustering methods. AutoPYara’s centroid selection produces larger gains precisely where heterogeneity is highest.

\noindent \textbf{\underline{C2}. The Impact of Bloom Filters:}
We analyzed how retraining Bloom Filters affects AutoPYara when using centroid selection. Like AutoYara, we consider all clusters (including those that fail) when we evaluate the performance of centroid selection in AutoPYara. These failed clusters cause a noticeable drop in performance, highlighting the impact of Bloom Filters within the framework. 

\begin{figure}[htbp]
    \centering
    \includegraphics[
        width=\columnwidth,
        height=0.22\textheight,
        keepaspectratio
    ]{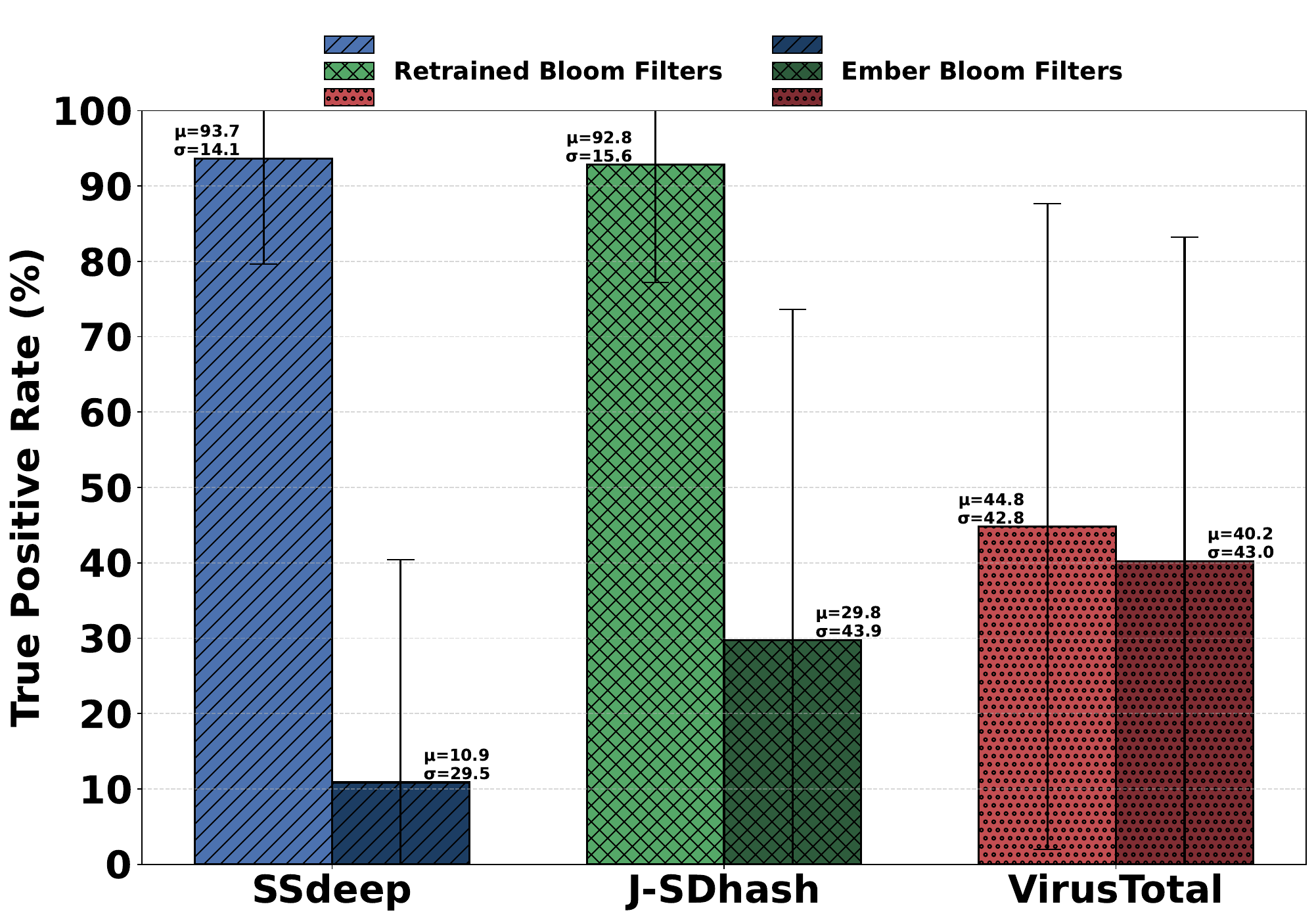}

    \vspace{-2mm}

    \caption{Impact of Bloom Filter retraining on AutoPYara, showing recovery of TPR compared to the baseline failed cluster configuration.}
    \label{fig:BaseLineAutoYaraBoxRTBFVSOGBF_TrueS}

    \vspace{-2mm}
\end{figure}

Figure~\ref{fig:BaseLineAutoYaraBoxRTBFVSOGBF_TrueS} highlights AutoPYara’s performance based on the Bloom Filters. Consistent with our previous findings, retraining Bloom Filters on new samples results in a clear performance improvement. Considering SSdeep clusters, AutoPYara with Retrained Bloom Filters achieves an absolute gain of 82.8\%, compared to 81.5\% for AutoYara. Using J-SDhash as the labeling strategy, the gain is 63.0\% compared to 62.7\% with AutoYara. For the VT clusters, AutoPYara yields a 4.6\% gain compared to 2.8\% with AutoYara. Centroid selection mitigates heterogeneity but lacks sufficient statistical power to address all clustering challenges, particularly when compensating for missing data (e.g., absent n-grams) that can significantly impact cluster representation. This limitation positions Bloom Filter retraining as an essential component of the pipeline. The increased sensitivity of retrained filters to centroid selection (AutoPYara) stems from the fact that retraining typically produces more clusters, introducing granularity which magnifies the influence of centroid selection on overall detection.

\noindent\textbf{\underline{C3}. The Intrinsic High Variance:}
AutoPYara inherits the lack of determinism of AutoYara. As a result, the framework exhibits performance variations across different runs, consistent with the behavior of the original framework. Figure~\ref{fig:compare_autoyara_autopyara_rtbf} illustrates this considerable variance through the error bars. If both frameworks display high non-determinism, it indicates that this behavior stems from the underlying problem of family clustering rather than the specific framework. Our results demonstrate notable consistency improvements for SSdeep clusters, where AutoPYara reduces standard deviation from 15.2\% to 14.1\% and 15.9\% to 15.6\% for J-SDhash clusters. Conversely, VT clusters exhibit increased variability, with standard deviation rising from 37\% to 42.8\%. This trend persists and reflects the heterogeneous nature of the VT clusters combined with AutoPYara’s centroid selection.

\noindent\textbf{\underline{C4}. Not All Mistakes Are the Same:} We evaluated AutoPYara's FPRs as before. For SSdeep and J-SDhash, AutoPYara achieves an intra-class FPR of 0\% (Table~\ref{tab:FPR-Single-Col} in Appendix~\ref{app:fpa}) and an Extra-class FPR below 2\%. These results highlight its potential for real-world deployment, since practical malware detectors must prioritize keeping FPR's low. For VT clusters, it has a large Extra-class FPR of 54.28\% compared to 20.36\% for AutoYara, as shown in Table~\ref{tab:Merged-False-Positives}. Note that this FPR increase does not occur for SSdeep and J-SDhash clusters because AutoPYara’s centroid-based approach makes selection more precise for homogeneous clusters, which is not the case for VT labels.

While centroids enable AutoPYara to generalize across heterogeneous VT clusters, the resulting rules also capture broader patterns that were previously unnoticed. These generalized patterns increase coverage but inadvertently also match goodware. Thus, addressing heterogeneity solely at the centroid level is insufficient. The extra detections are beneficial when only malware families are considered, but the rules are too broad for practical hunting. Given precise clusters, as in SSdeep and J-SDhash, AutoPYara outperforms AutoYara. When clusters are noisy and less precise, centroid selection amplifies that noise.

\noindent \textbf{\underline{C5}. The Impact of Threshold Variation:} Adjusting similarity thresholds directly shapes the diversity of samples in each cluster, which affects the quality of centroids in AutoPYara. A higher threshold creates smaller, more uniform clusters, while a lower threshold results in larger, more diverse groups. If centroids remain effective amid greater heterogeneity, decreasing the threshold should help AutoYara generate broader, more general rules. Using a higher threshold may improve precision by keeping clusters purer, and we test this by varying the threshold and measuring how changes impact rule performance. AutoYara shows a threshold variation of 10 pp, as shown in Figure~\ref{fig:AutoYara_Thresholds_Comparison}(a). AutoPYara significantly reduces this variation to 2.9 pp, as shown in Figure~\ref{fig:AutoYara_Thresholds_Comparison}(b), a 7.1 pp reduction in variation.
\vspace{-2mm}

\subsection{Impact of k on AutoPYara }
\label{subsec:AutoPYaraKEval}
\vspace{-2mm}

\begin{figure*}[htbp]

    \centering
    \subfloat[AutoPYara Centroids Only\label{fig:autopyara_ssdeep_centroids}]{%
        \includegraphics[width=0.32\linewidth]{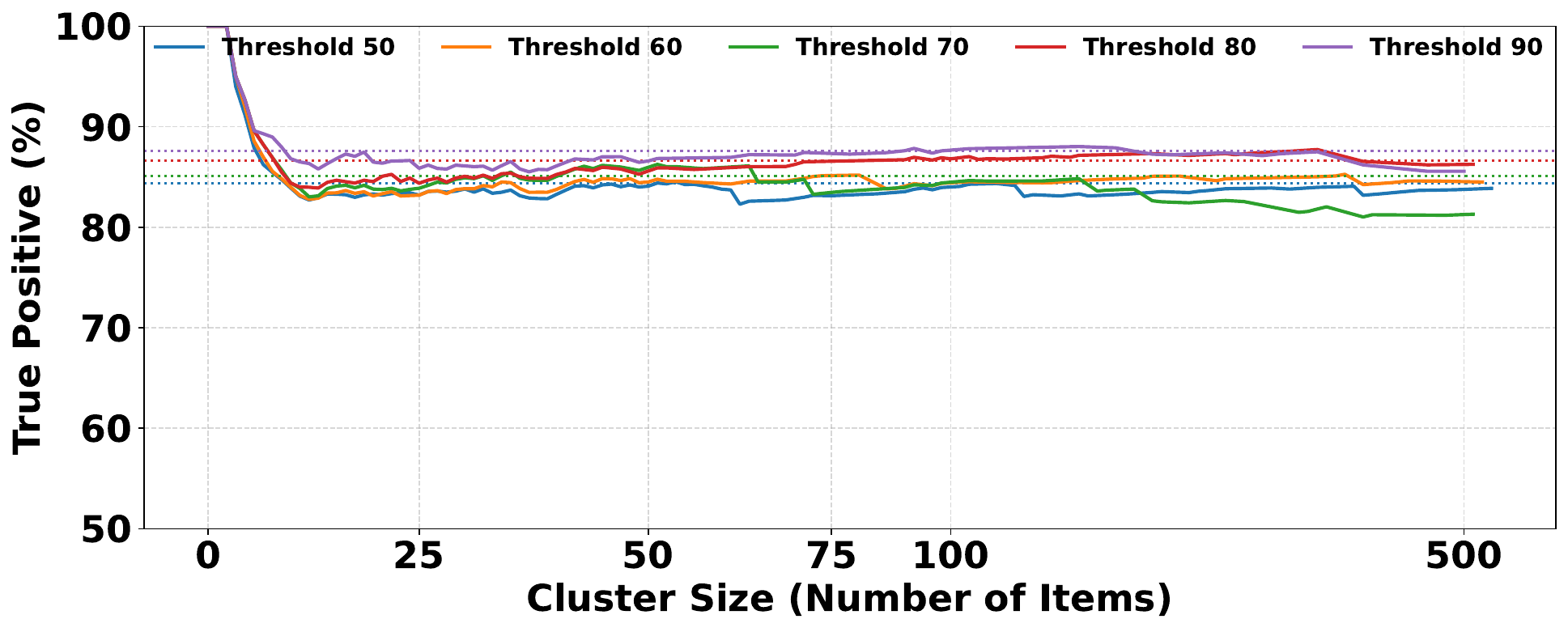}
    }\hfill
    \subfloat[AutoPYara Best K\label{fig:autopyara_ssdeep_best_k}]{%
        \includegraphics[width=0.32\linewidth]{Figures/AutoPYara/RTBF/VairiationPlots/Merged/BestK_SSdeepAVG_Zoom.pdf}
    }\hfill
    \subfloat[AutoPYara Worst K\label{fig:autopyara_ssdeep_worst_k}]{%
        \includegraphics[width=0.32\linewidth]{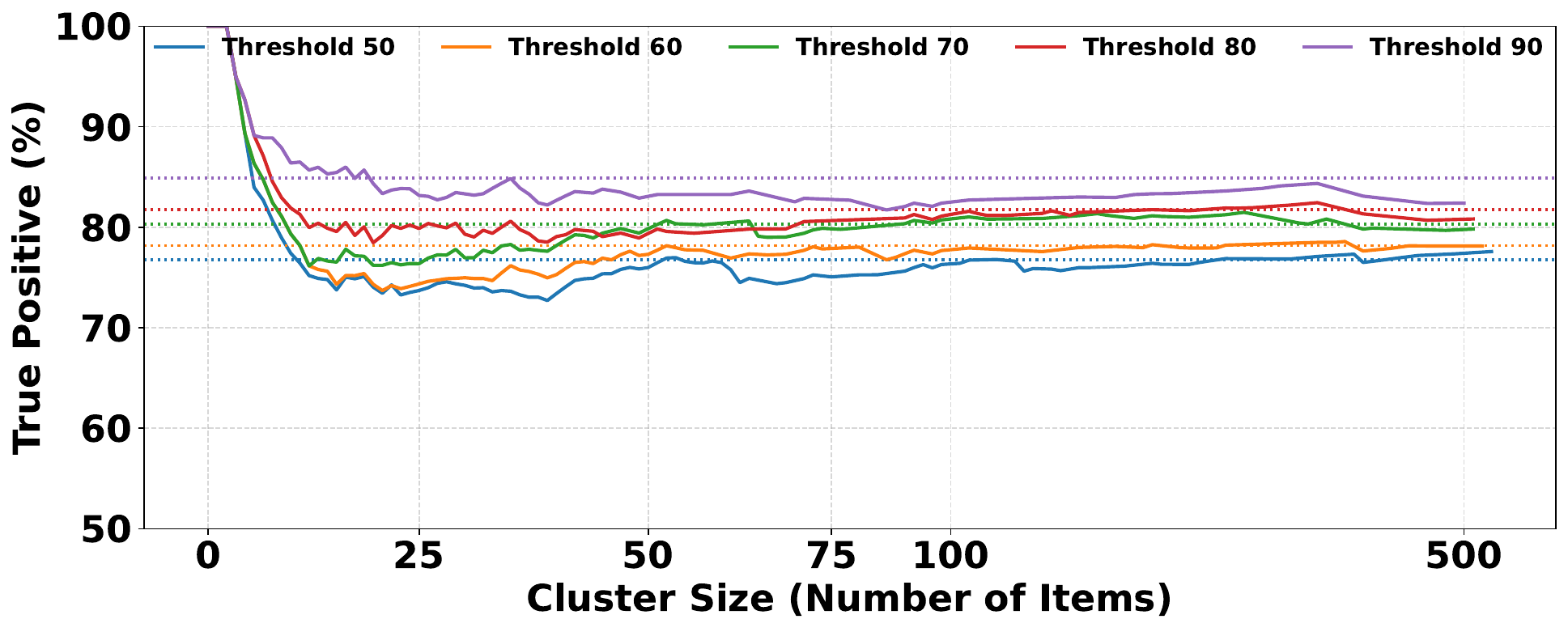}
    }
    
\vspace{-3mm}

    \subfloat[Oracle Based Heuristic: MODE\label{fig:autopyara_ssdeep_heuristic_mode}]{%
        \includegraphics[width=0.32\linewidth]{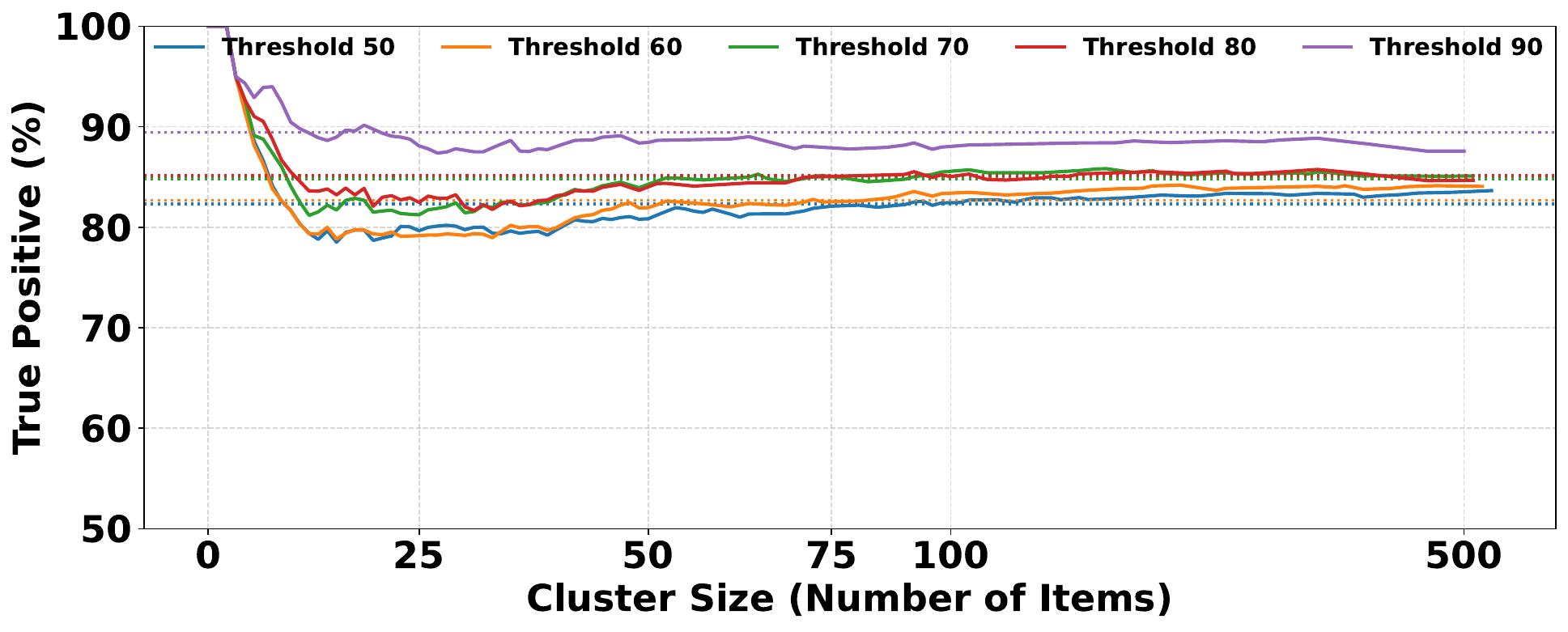}
    }\hfill
    \subfloat[Oracle Based Heuristic: MAX\label{fig:autopyara_ssdeep_heuristic_max}]{%
        \includegraphics[width=0.32\linewidth]{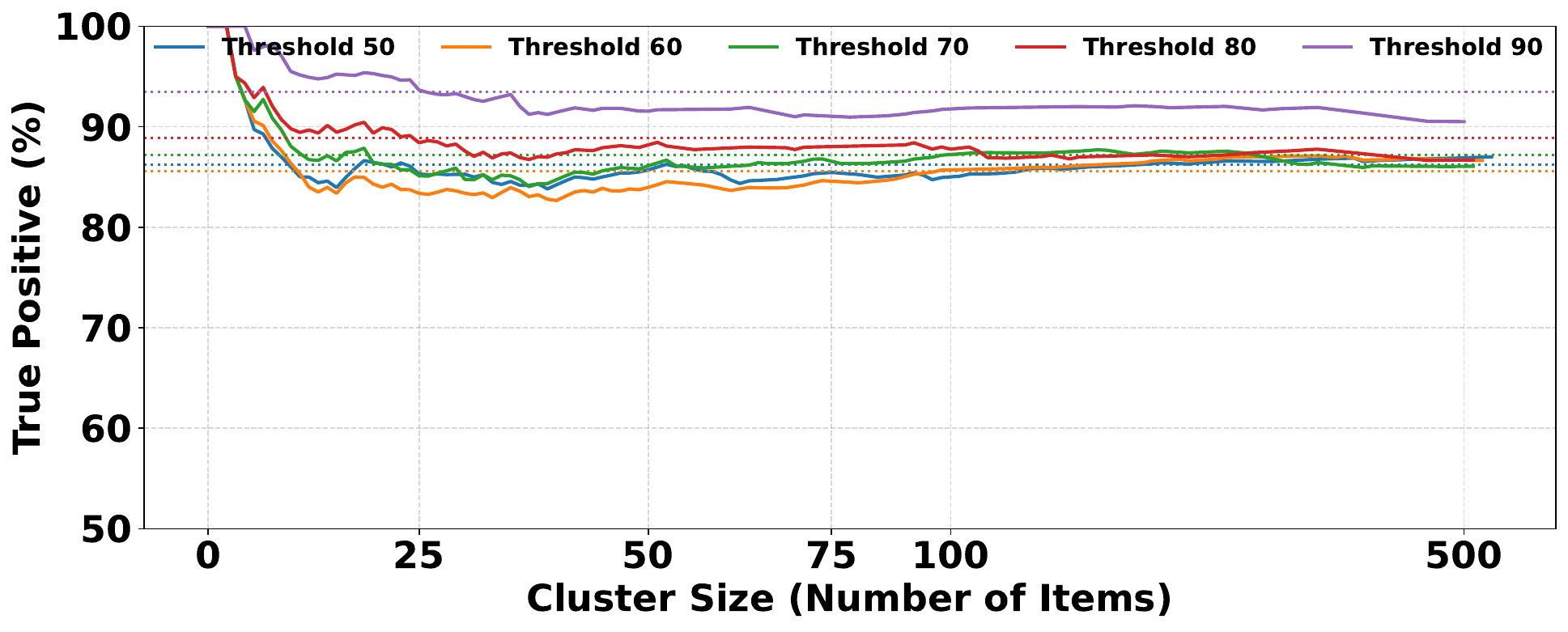}
    }\hfill
    \subfloat[Oracle Based Heuristic: Random\label{fig:autopyara_ssdeep_heuristic_random}]{%
        \includegraphics[width=0.32\linewidth]{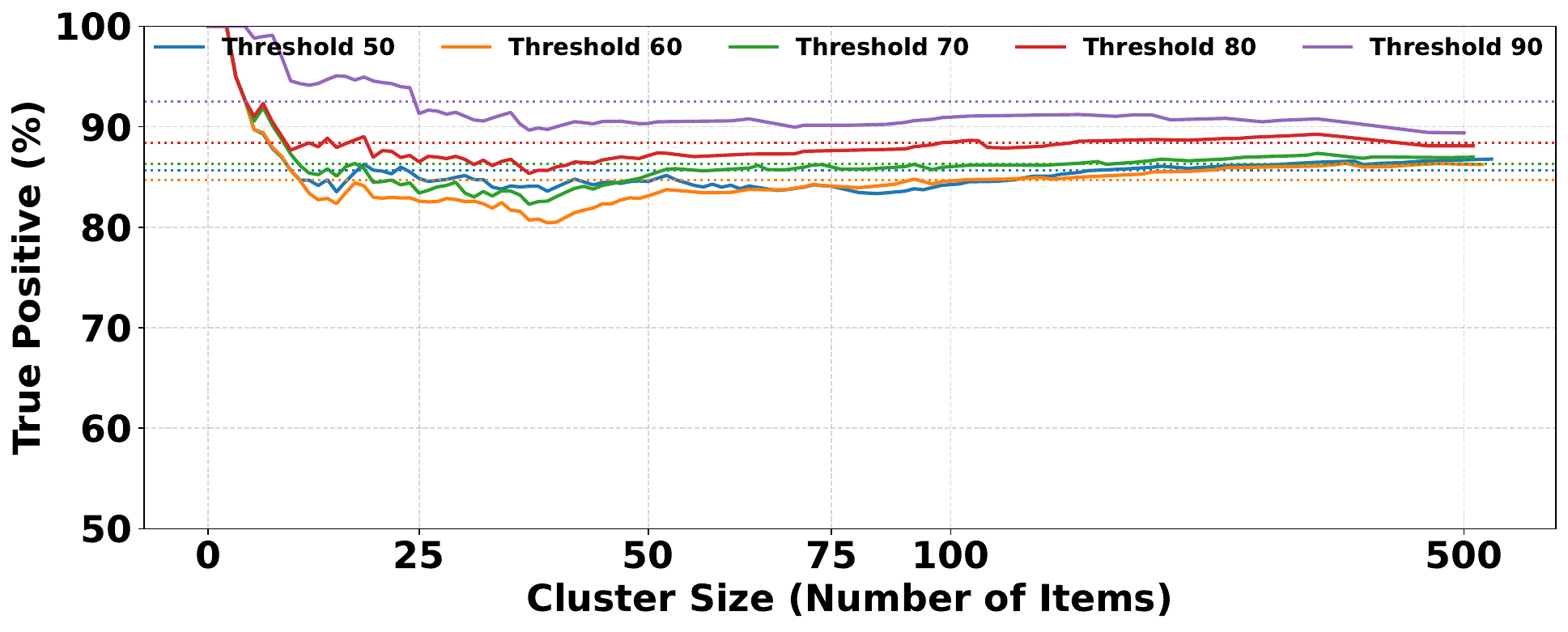}
    }

    \vspace{-3mm}

    \subfloat[Sub-Optimal Heuristic: Mean\label{fig:autopyara_ssdeep_nonheuristic_MEAN}]{%
        \includegraphics[width=0.32\linewidth]{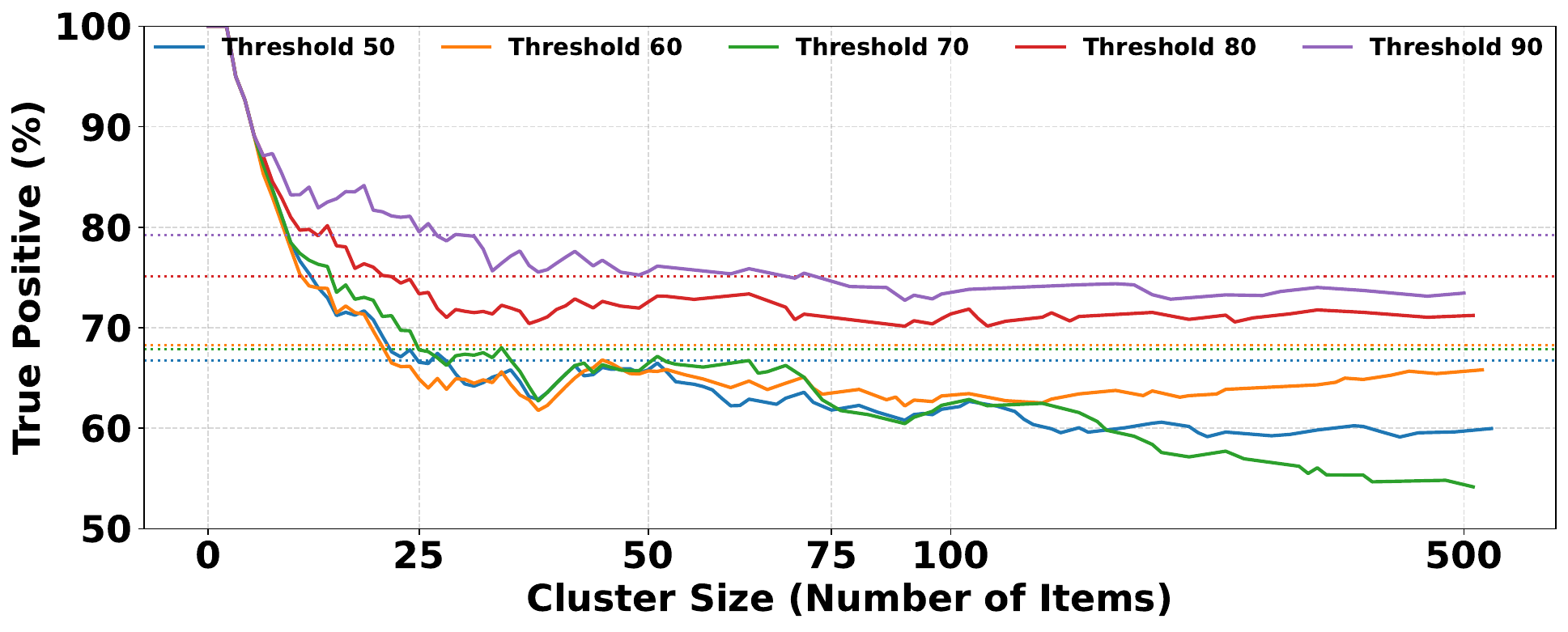}
    }\hfill
    \subfloat[Sub-Optimal Heuristic: MAX\label{fig:autopyara_ssdeep_nonheuristic_max}]{%
        \includegraphics[width=0.32\linewidth]{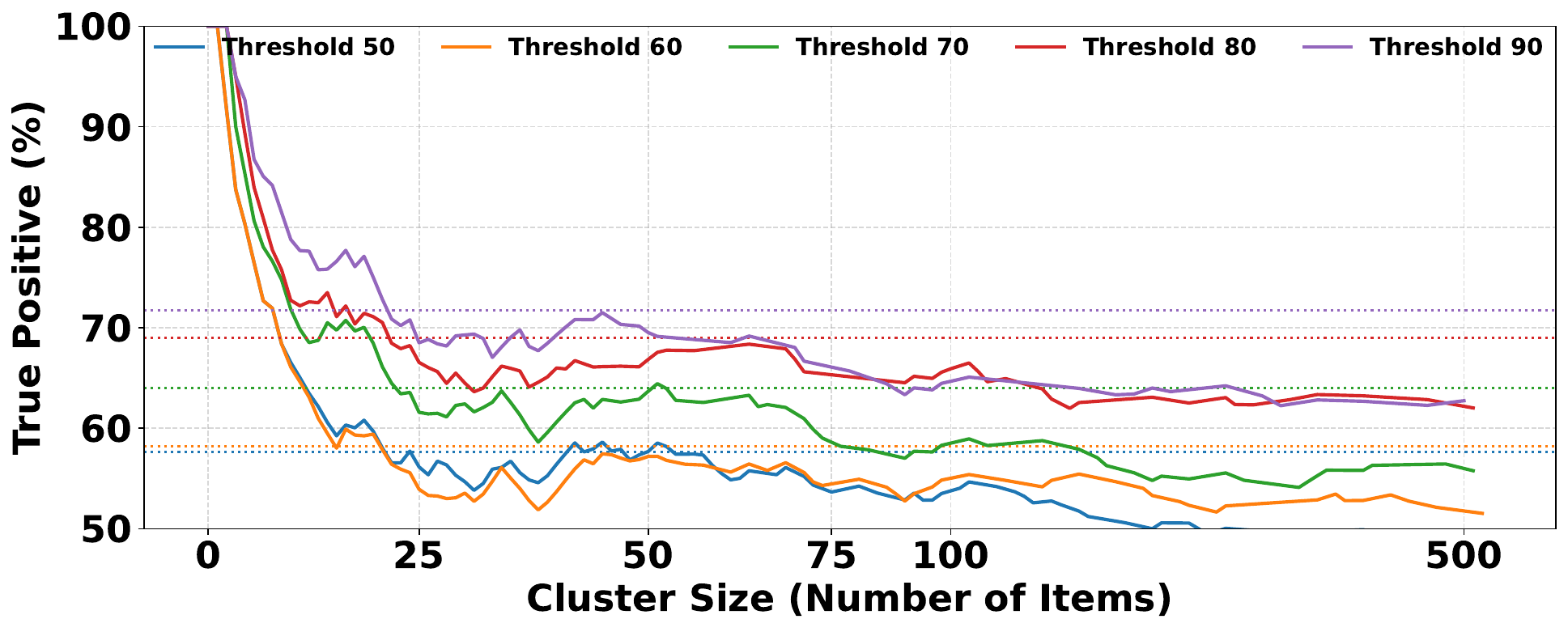}
    }\hfill
    \subfloat[Sub-Optimal Heuristic: Random\label{fig:autopyara_ssdeep_random}]{%
        \includegraphics[width=0.32\linewidth]{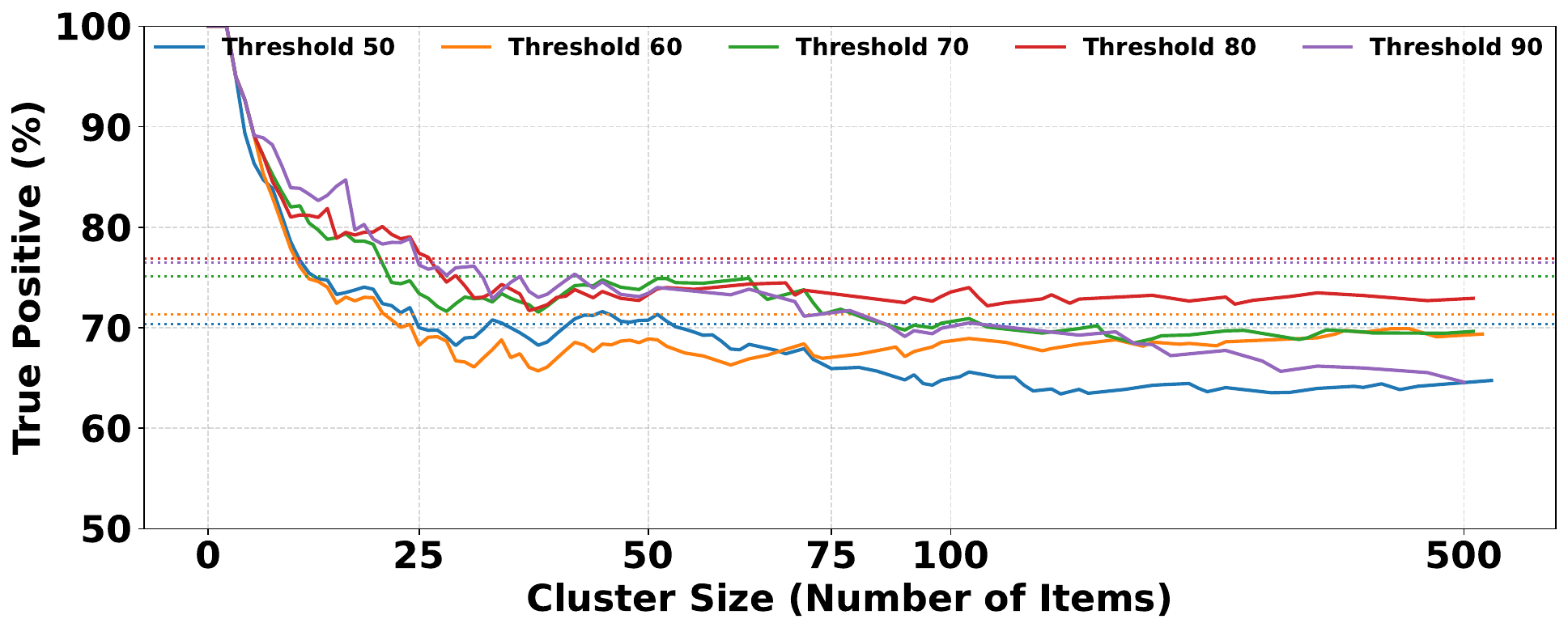}
    }
    
    \caption{Comparing AutoPYara in Different Configuration demonstrating the overall trend and gains using SSdeep  as labeling strategy}
    \label{fig:autopyara_ssdeep_Kimpact}
        \vspace{-5mm}

\end{figure*}

\noindent Comparing Figure~\ref{fig:AutoYaraSSdeepComparison}.b to Figure~\ref{fig:AutoYara_Thresholds_Comparison}.c, we show that AutoPYara consistently outperforms AutoYara by roughly 4\% across the samples. Smaller clusters tend to be more homogeneous, making it easier for the framework to derive rules that capture clear and representative patterns. As clusters increase in size and diversity, shared characteristics become diluted, reducing rule specificity and performance. AutoPYara’s centroid-based strategy offsets this by anchoring rules around representative points, enabling more robust generalization across samples.

 So far, we use $k$ derived from the VBGMM step of AutoYara as our reference point in previous experiments. This approach allowed us to validate our framework's improvements under controlled conditions, but it reflects a lab setting. We now transition to a more realistic setup where AutoPYara independently determines the $k$ value through its own clustering analysis. In this setup, we verify whether there is room to improve performance through $k$ selection and, if so, whether this process can be carried out automatically in a systematic approach. This methodological shift from externally-referenced to internally-determined $k$ values allows us to evaluate system performance under genuine operational conditions, optimize the K-selection algorithm, and demonstrate the framework's robustness without external clustering guidance. This transition is crucial toward developing an independent and optimized AutoPYara. All experiments in this phase isolate the variable of interest by fixing the centroid-based clustering as the baseline while systematically shifting only the $k$ parameter.

\noindent\textbf{[F2] Evaluation of Ideal $k$ Values:} Our first test investigates whether there is room for performance improvement through $k$-selection. To establish an upper bound, we assume access to an oracle, which we simulate via brute-force search. The oracle provides the ideal $k$ for each case, thereby revealing the maximum performance AutoPYara could achieve under perfect conditions.

\noindent\underline{\textit{(1) Exhaustive Search for Optimal $k$:}} We refine AutoYara’s initial estimate $k$ by exhaustively searching its neighborhood. For each candidate $k$, we evaluate performance and select the value that produces the best results. This brute-force search is not intended as a practical deployment strategy but rather as a methodological tool to approximate an oracle, allowing us to simulate access to the ideal $k$ values and establish the upper-bound performance AutoYara could achieve. We provide additional information regarding the search space in Appendix~\ref{app:note}.

\noindent\underline{\textit{(2) Worst-Case $k$:}} We also evaluate the worst-case $k$ as a lower-bound baseline to verify that the observed gains from AutoYara using VBGMM and AutoPYara reflect genuine improvements. To that end, we repeated the above exhaustive search to identify $k$ with the worst rule quality.

\begin{figure}[H]
    \centering
    \subfloat{%
        \includegraphics[width=0.95\linewidth]{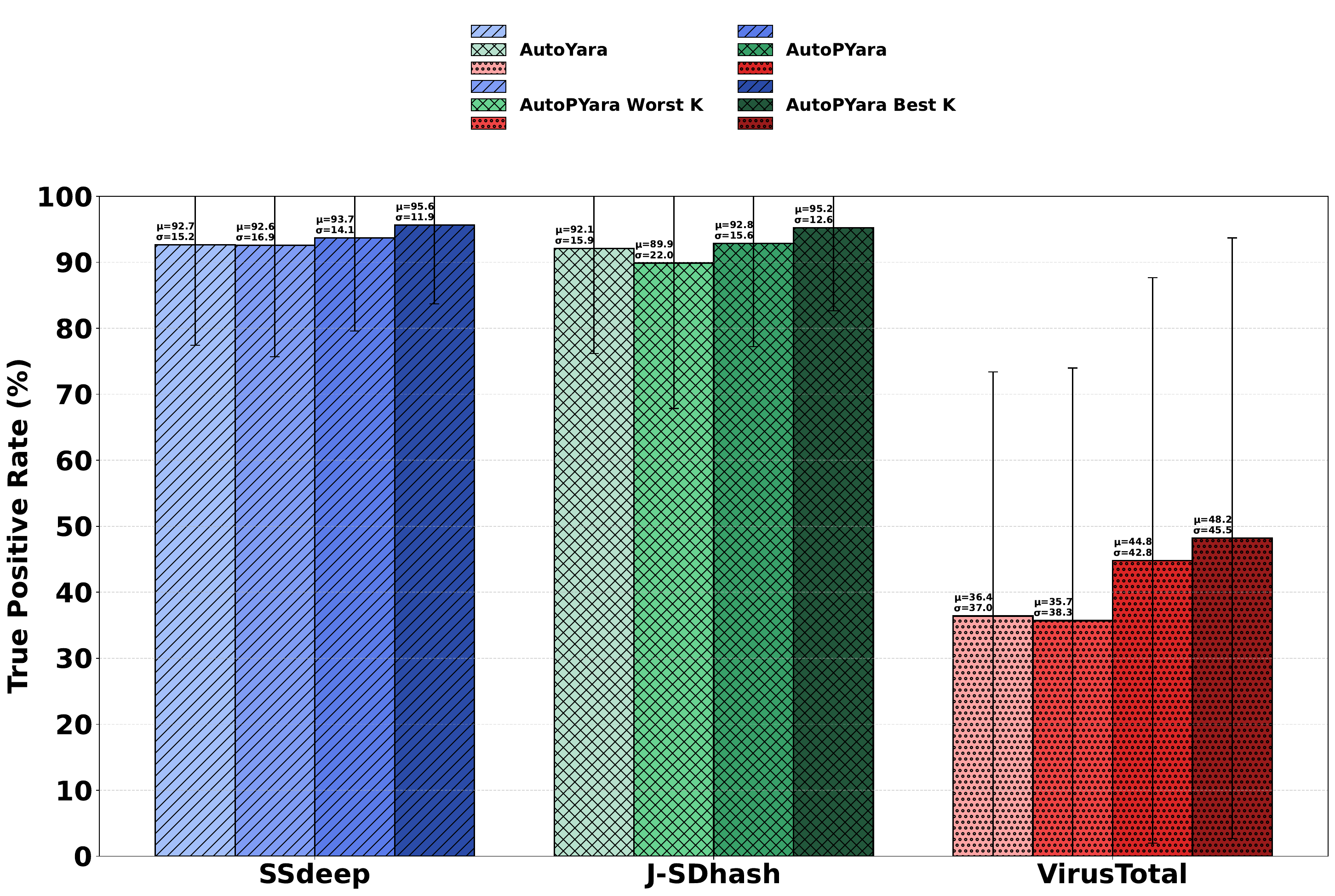}
    }

    \caption{Highlight the impact of $k$ on AutoPYara Using Extreme $k$ Values across different labeling strategies}

    \label{fig:AutoYaraVSAutoPYaraSpread}
    \vspace{-2mm}
\end{figure}
 \noindent The value of $k$ clearly affects AutoPYara’s performance. Fig.~\ref{fig:AutoYaraVSAutoPYaraSpread} shows that on average, choosing an appropriate $k$ increases the TPR by 3\% for SSdeep clusters, 5.3\% for J-SDhash clusters, and 12.5\% for VirusTotal-derived clusters when comparing the best configuration to its suboptimal values of $k$. Although these gains may look marginal at first glance, remember that smaller clusters skew the averages. Similarity-based clustering tends to contain a large frequency of clusters with just 2-3 samples. These limit the range of $k$ and thus the achievable gains. VT labeling, in turn, by not relying on intrinsic code-level similarity, produces clusters with significantly more files per family, resulting in larger cluster sizes, with bigger room for improvement. Fig.~\ref{fig:AutoYaraVSAutoPYaraSpread} shows the gap between the best and a suboptimal $k$ configuration is $\approx$~12.5\% in this setting. These results highlight the importance of selecting $k$ for the framework and the difference between the $k$ used in AutoYara and the optimal $k$ for AutoPYara. While VBGMM performs better than the worst-case baseline as intended, yielding real gains, it remains suboptimal (w.r.t. the best case), typically underestimating $k$, presenting room for improvement.

\noindent\textbf{[F3] Heuristic Approachs to Estimate $k$:}

\noindent\textbf{(1) Uninformed Heuristic to Estimate $k$:}
To establish a principled approach to estimate $k$ without relying on VBGMM, we first explored three naive strategies that do not incorporate any semantic understanding of sample types. These heuristics are intended as a substitute for exhaustive search. The three are not exhaustive, but they are representative of different approaches. The three methods are: \textbf{(i)} set $k$ to the total number of files in a cluster; \textbf{(ii)} set $k$ to the mean length to equally divide the data; and \textbf{(iii)} randomly guess an arbitrary value of $k$. These strategies are naive because they ignore the data’s structure and yield unstable, biased clusters.
Figures~\ref{fig:autopyara_ssdeep_nonheuristic_MEAN},~\ref{fig:autopyara_ssdeep_nonheuristic_max},~\ref{fig:autopyara_ssdeep_random} reinforce the impact of $k$ selection and highlight how the impact is more pronounced in larger clusters, where there are fewer chances for randomly sampling a proper $k$ value. This confirms that a stronger approach for $k$ definition is warranted; thus, our proposal is an informed heuristic.

\noindent\textbf{(2) Informed Heuristic to Estimate $k$:} Although we demonstrated the potential of a $k$ oracle to improve rule generation performance, true oracles for $k$ do not exist in reality. To approximate an oracle, we require reliable $k$ values—without exhaustive search, better than the original VBGMM estimation, and more effective than a random guess. As per Sec.~\ref{sec:Methodology}, we adopt a systematic heuristic that leverages historical AutoPYara runs indexed by VT family labels (via AVClass~\cite{DBLP:journals/corr/abs-2006-10615}) to estimate $k$ for new samples from the same family. We do not rely on VT for cluster labeling; instead, we repurpose them as a practical proxy to bypass expensive hash-based clustering and to anchor $k$ estimation in observed family trends. While VT’s family assignments can be inconsistent and not always semantically meaningful, this usage still improves practical reliability over VBGMM alone and avoids the cost of brute force. VT now serves a different role; it enables family-level grouping to drive our heuristic for estimating $k$, while hashes are still used for the centroids.  Our core hypothesis is that we can leverage historical family information tied to VT's labels to position our search range for optimal $k$ values in regions with a higher probability of success compared to random initialization, thus gaining efficiency and grounding it in practical malware evolution trends. Figures~\ref{fig:autopyara_ssdeep_heuristic_mode},~\ref{fig:autopyara_ssdeep_heuristic_max}, and~\ref{fig:autopyara_ssdeep_heuristic_random} show that our approach performs comparably to the Best configuration within a 2\% error margin. It also performs 10\% better than AutoYara on the same samples. For larger clusters, our informed choice of $k$ achieves superior performance relative to the heuristic and random settings. Together, these results \textbf{rescue} the application of VT to family clustering, which had previously been viewed as impractical.

\begin{figure*}[htbp]
    \centering

    {%
        \includegraphics[width=0.31\textwidth]{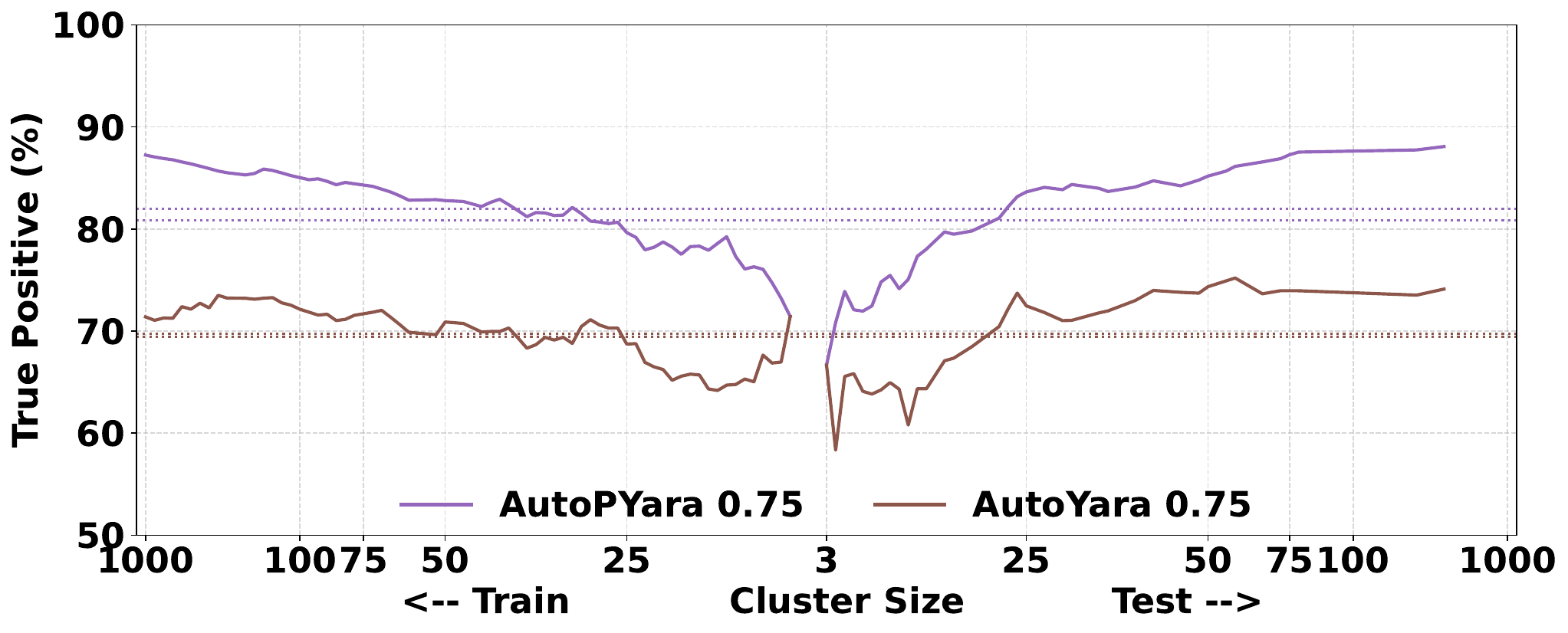}
    }
    \hfill
    {%
        \includegraphics[width=0.31\textwidth]{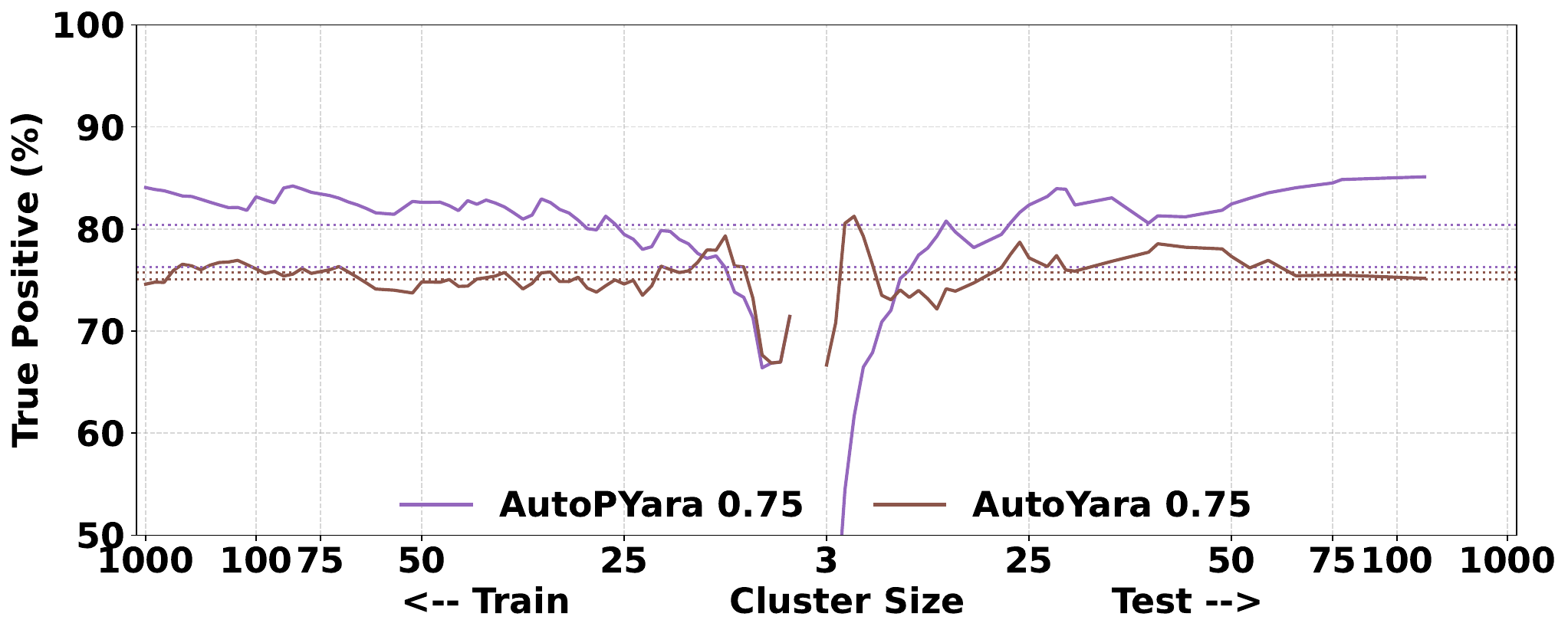}
    }
    \hfill
    {%
        \includegraphics[width=0.31\textwidth]{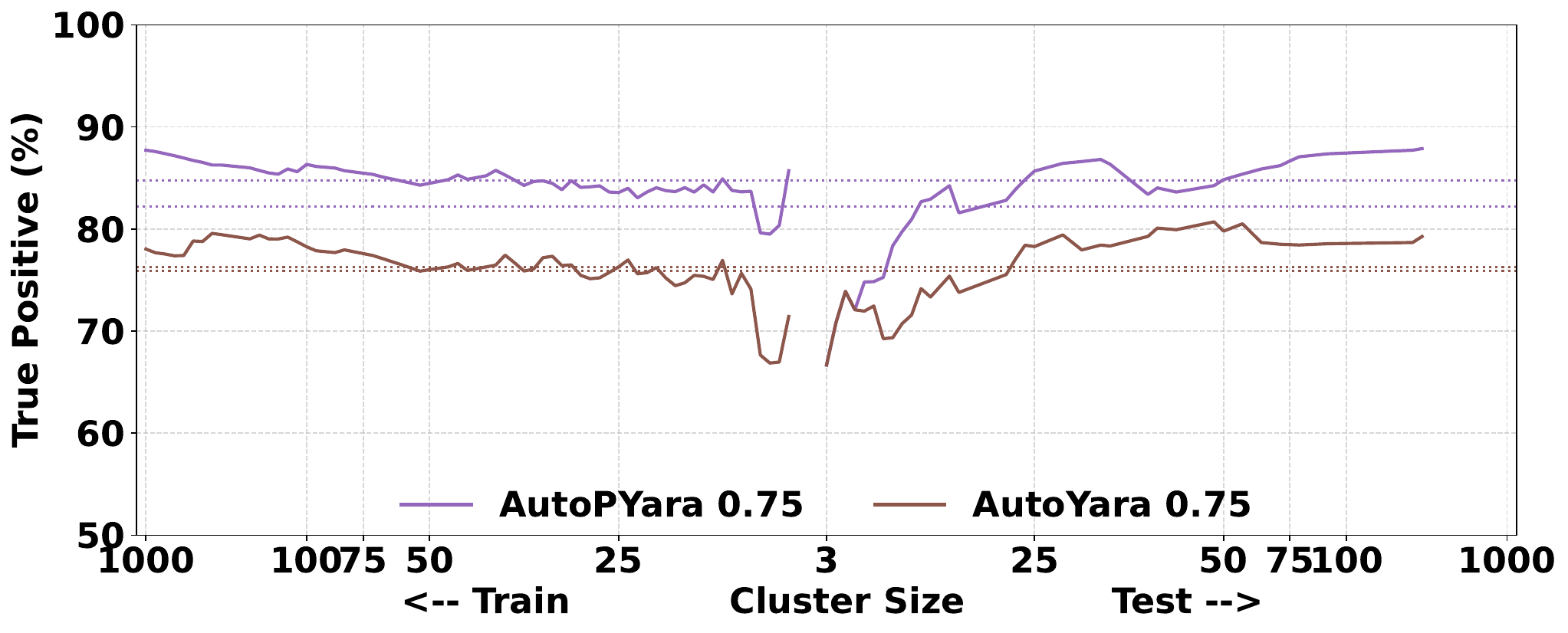}
    }

    \vspace{-3mm}
    \caption{Threat hunting comparison between AutoYara and AutoPYara under different availability assumptions: (a) oracle-based ideal configuration, (b) no-oracle setting, and (c) real-world scenario.}
    \label{fig:ThreatHuntingComparisonCombined}
        \vspace{-5mm}

\end{figure*}

\vspace{-2mm}

\subsection{The Qualitative AutoPYara Impact}
\label{subsec:QualitativeEval}
\vspace{-2mm}

\begin{figure*}[htbp]
    \centering
    \includegraphics[width=1\linewidth]{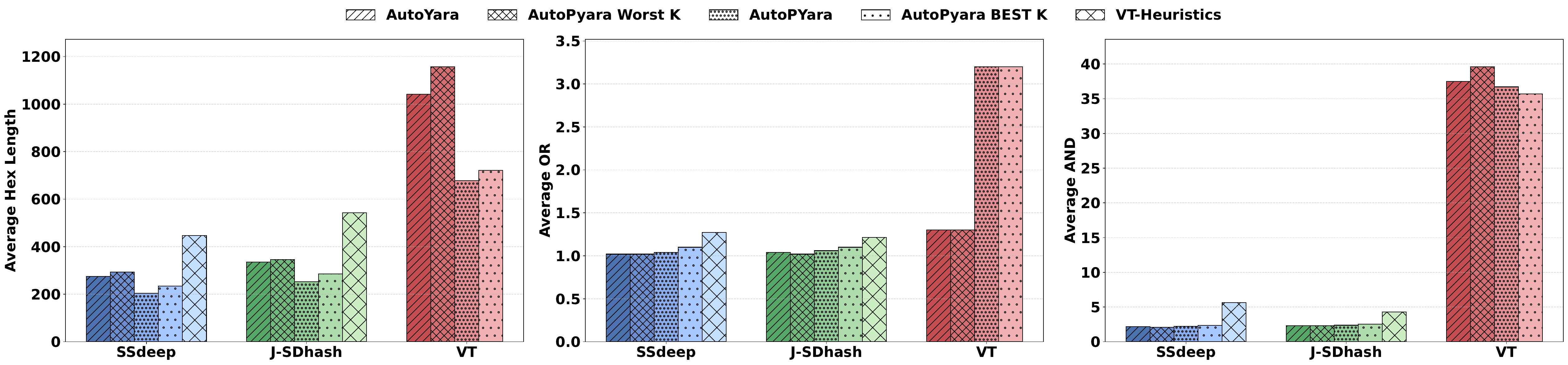}
        \vspace{-.5cm}
    \caption{Qualitative Analysis. Comparison of structural rule metrics, (a) hex string length and boolean logic complexity:(b) AND, (c) OR counts}
    \label{fig:QualitativeAnalysis}
\end{figure*}

\noindent AutoPYara's centroid selection and better $k$ values affect the generated rule's quality, which is a known~\cite{botacin2024_malwareanalysts} essential aspects for human analysts. We evaluated rule quality via three metrics: \textbf{1)} \textit{Number of Strings}; \textbf{2)} \textit{Hex Length}; and \textbf{(3)} \textit{Conditions Count}; These metrics compose the \textit{Compactness} metric. Given two rules with similar TP and FP rates, compactness is preferable.
Figure~\ref{fig:QualitativeAnalysis} provides a high-level overview of the overall distribution of metrics for the generated rules, comparing AutoYara and AutoPYara across four different configurations. AutoPYara creates smaller, more complex rules with greater coverage than the original framework, thus being superior to AutoYara both quantitatively and qualitatively. 
We provide more details in Appendix~\ref{app:rme}. Further, in Appendix~\ref{app:yararule}, we provide two rules to demonstrate how AutoPYara derives more compact signatures with superior coverage compared to AutoYara.

\subsection{AutoPYara as a predictor for Threat Hunting} 
\label{subsec:threathunting}

\noindent After characterizing the full potential of AutoYara and AutoPYara using the entire dataset, we split each cluster into training and test sets to evaluate their generalization on the threat-hunting task. This allows for an assessment by creating detection rules from limited samples and deploying them to identify unknown variants of the same malware families. The training portion consists of limited samples that analysts have access to when crafting rules, while the test portion consists of unseen samples that arrive in operational networks. The evaluation is performed at the cluster level using a split ratio of 75/25. The splits are non-overlapping to prevent data leakage, keeping train and test samples from each cluster strictly isolated. \\
We evaluated the task in 3 scenarios:

\noindent\textbf{1. When VT labels are available} 
The best scenario for AutoPYara operation is when VT labels are available in the oracle for \textbf{all} the targeted families, allowing better $k$ estimation via the \uline{informed heuristic}. Figure~\ref{fig:ThreatHuntingComparisonCombined}.a shows that AutoPYara achieves a 20\% gain when leveraging the informed heuristic compared to the baseline. By utilizing the gt labels from the VT oracle, the system more accurately estimates the optimal number of clusters ($k$), effectively minimizing both over-clustering and family fragmentation.

\noindent\textbf{2. When VT labels are not available}
The worst threat hunting scenario for AutoPYara operation is when VT labels are not available in the oracle for \textbf{any} of the targeted families, limiting it to operate only with the \uline{uninformed} $k$ estimation heuristic. Figure~\ref{fig:ThreatHuntingComparisonCombined}.b shows that AutoPYara's performance remains resilient, maintaining a 5\% lead over the baseline even when operating under the uninformed heuristic. While the lack of VT labels removes the ability to fine-tune $k$ based on known family distributions, the system manages to outperform generic clustering methods by relying on structural code similarities.

\noindent\textbf{3. When VT labels are partially available} In real hunting cases, VT oracle labels will be available to only part of the samples, limiting the benefit of the improved heuristic to them. We simulate this by using the real sample proportions in our dataset with VT labels, applying the informed heuristic to them, and falling back to the uninformed heuristics when needed. Figure~\ref{fig:ThreatHuntingComparisonCombined}.c, we highlight that AutoPYara outperforms AutoYara by an absolute gain of $\approx$~10\% in both the training and threat hunting scenarios.

 We also evaluated different splits, and AutoPYara consistently outperformed AutoYara. For each split ratio, samples from each cluster are randomly selected for rule generation, and the remaining are used to evaluate detection of the generated rules. This experiment measures how effectively rules derived from varying amounts of training data detect and classify previously unseen test samples.
\section{Benchmarking Issues \& Contributions}
\label{sec:benchmarkfinal}

\begin{figure}[htbp]
\vspace{-2mm}
    \centering
        {%
        \includegraphics[width=0.95\linewidth]{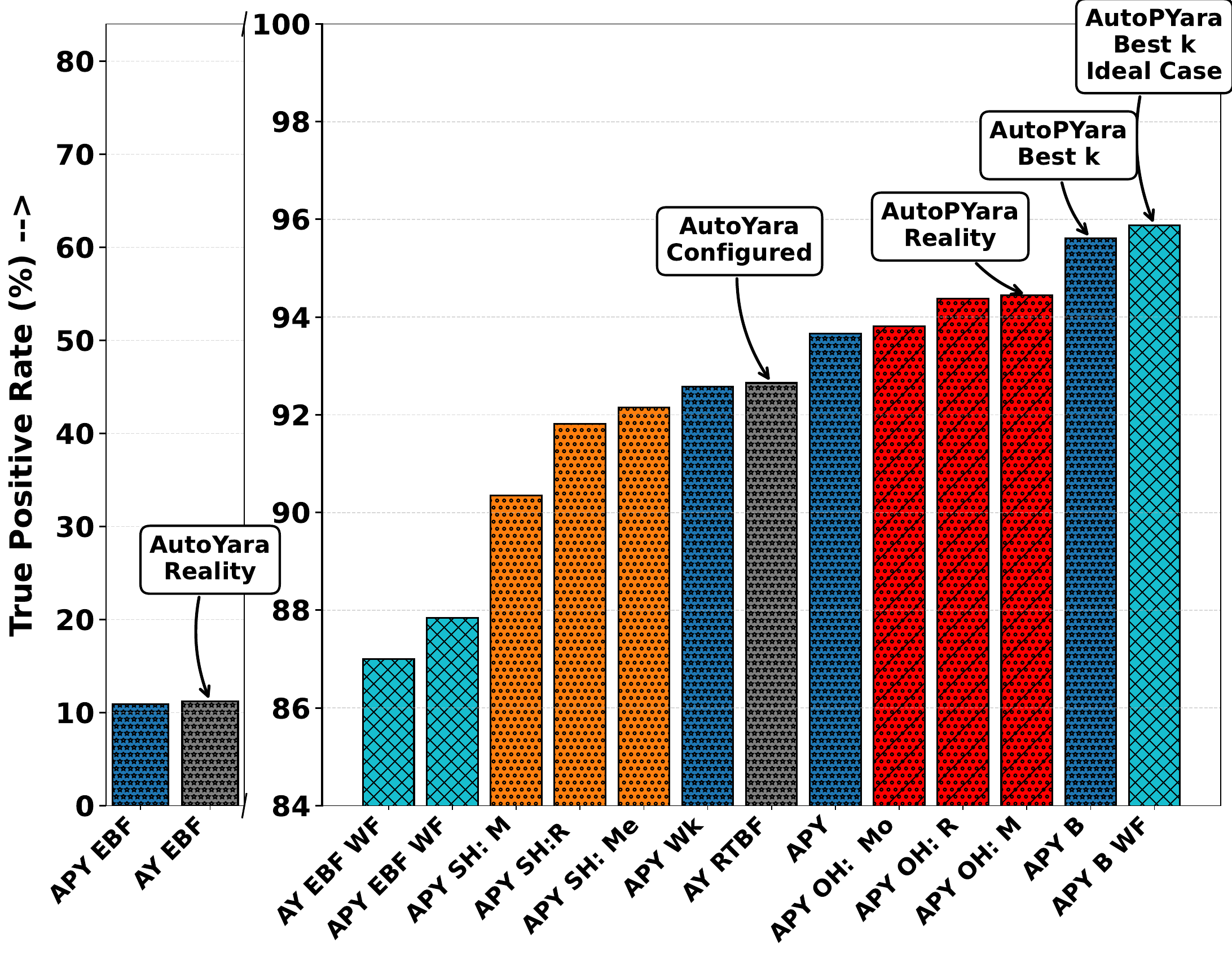}
    }\vspace{-4mm}
    \caption{AutoYara and AutoPYara across different configurations}
    \label{fig:AutoPYaraBigPicture}
    \vspace{-4mm}
\end{figure}

\noindent In this paper, we characterize the AutoYara framework in a real-world setting and highlight pitfalls in its evaluation methodology and how these can skew results, inflating the reported performance. Figure~\ref{fig:AutoPYaraBigPicture} illustrates the overall reconstruction from AutoYara through its theoretical maximum. In practice, we show that our oracle-based approach comes within 2\% of the best outcomes, whereas AutoYara, if not properly characterized, can perform up to  76\% worse than optimal rules in reality. We characterize the performance gains on large clusters in Figure~~\ref{fig:AutoPYaraBigPictureTOP} in Appendix.~\ref{app:bp}, highlighting the gains in larger (stable) clusters.


\section{Conclusion}
\label{sec:conc}
\vspace{-2mm}

\noindent This work introduces AutoPYara, an evolved framework for automatically generating YARA rules for malware family clusters. It advances the SOTA via \textbf{(1)} precise cluster centroid selection based on similarity hashing and \textbf{(2)} better selection of the number of subclusters representing a malware family based on historical VirusTotal data. Our evaluation identified the limits of the automatic rule generation process and demonstrates that AutoPYara's optimal configuration, achieving an 80\% improvement, surpasses existing benchmarks to define the current state-of-the-art. Finally, to facilitate the practical application of our findings, we outline guidelines (Appendix~\ref{app:rec}) for achieving this optimal performance.

\noindent \textbf{Ethical Considerations.} The dataset utilized consists of malware that have been vetted for academic research. Because the dataset contains no active threats, no personally identifiable information, and no sensitive user telemetry, this research poses no risk of harm to users or third parties.

\noindent \textbf{Artifact Availability Statement.} To support open science and full evaluation replication for ACSAC, we provide open-access links to all components:

\begin{itemize}
    \item \textbf{ACSAC Permanent Artifact:} \href{https://zenodo.org/records/23107801}{Zenodo} [\url{https://zenodo.org/records/23107801}]
    
    \item \textbf{ACSAC Artifact Source:} \href{https://github.com/Botacin-s-Lab/AutoPYara/}{GitHub} [\url{https://github.com/Botacin-s-Lab/AutoPYara/}]
    
    \item \textbf{API Documentation:} \href{https://botacin-s-lab.github.io/AutoPYaraPyPI/}{GitHub Pages} [\url{https://botacin-s-lab.github.io/AutoPYaraPyPI/}]
        
    \item \textbf{PyPI Source Repository:} \href{https://github.com/Botacin-s-Lab/AutoPYaraPyPI}{GitHub} [\url{https://github.com/Botacin-s-Lab/AutoPYaraPyPI}]
   
    \item \textbf{Java Backend Engine:} \href{https://github.com/Botacin-s-Lab/AutoPYaraBackend}{GitHub} [\url{https://github.com/Botacin-s-Lab/AutoPYaraBackend}]
    
    \item \textbf{YARA Rules \& Benchmark Data:} \href{https://zenodo.org/records/22665898}{Zenodo} [\url{https://zenodo.org/records/22665898}]
    
    \item \textbf{Python Package (PyPI):} \href{https://pypi.org/project/autopyara/}{PyPI} [\url{https://pypi.org/project/autopyara/}] (\texttt{pip install autopyara})

\end{itemize}
\noindent \textbf{Acknowledgments.} 
We thank the reviewers, shepherd and the artifact evaluators for their helpful insights. We also thank NSF for supporting this work via grant CNS-2327427.

\subsection*{LLM Usage Statement}
\noindent LLMs were used for editorial purposes in this manuscript, and all outputs were inspected by the authors to ensure accuracy and originality. LLMs also assisted in generating the AutoPYara code, which the authors thoroughly tested, validated, and take full responsibility for.
\bibliographystyle{IEEEtran}
%
\bibliography{ref}

\appendices
\vspace{-3mm}

\section{System Architecture and Algorithms}
\vspace{-2mm}

\noindent This appendix provides a deep dive into the system’s architectural components and mathematical logic.

\vspace{-4mm}

\subsection{Frame Work Representation }
\label{app:framework}
\vspace{-4mm}

\vspace{-.5cm}
\begin{figure}[!htpb]
    \centering
    \includegraphics[width=0.95\linewidth]{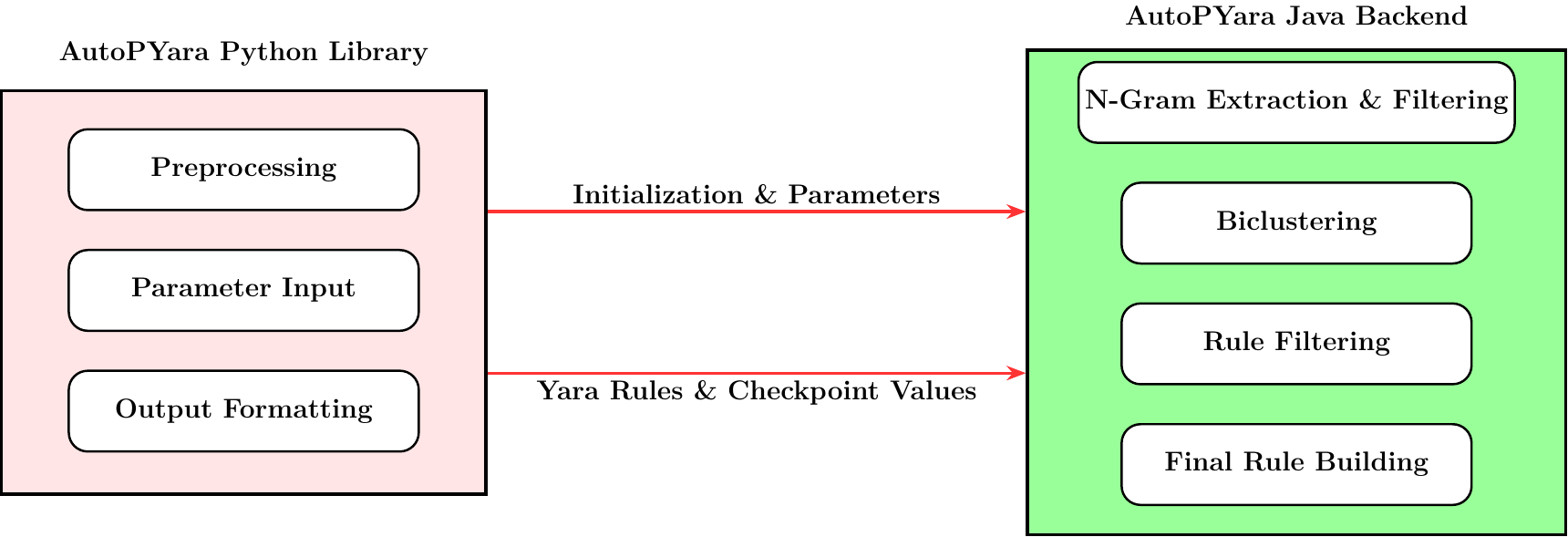}
    \vspace{-3.0mm}
    \caption{The AutoPYara interface and integration pipeline.}
    \label{fig:pyara interface}
\end{figure}

 \noindent This appendix provides a technical breakdown of the design discussed in Section~\ref{sec:Background}. Figure~\ref{fig:yara pipeline} shows the  execution flow of the AutoYara pipeline, highlighting the transition from raw data ingestion to the final generation of signature sets. Figure~\ref{fig:pyara interface} illustrates the specific Python interface logic, detailing how external modules interact with the core engine.

\subsection{Biclustering Algorithms}

\label{app:biclustering}

 \noindent Section~\ref{subsec:OverView} discusses AutoYara and its use of Biclustering, in this appendix we expand its definition. Biclustering is the process in which rows and columns are clustered simultaneously in a matrix. A single bicluster consists of a subset of rows that correlate with a subset of columns or vice versa. In the context of YARA rule generation, we have a list of files $S$ and n-gram candidates (signatures) $F$ within a discrete binary matrix $A_{S \times F}$, where $A_{i,j} = 0$ indicates that sample $i$ does not have n-gram $j$ and $A_{i,j} = 1$ indicates that it does. Biclustering this matrix produces clusters that tell us which n-grams are strongly associated with which types of files. This allows us to process the bicluster and construct it into a conditional query for matching: \textit{if a new sample X has at least 3 of the signatures F1, F3, F10, F58, then it belongs in this bicluster and therefore this malware family.}

 AutoYara performs Biclustering via the SpectralCoClustering algorithm~\cite{dhillon2001_spectralcoclustering}. By computing the normalized graph Laplacian and eigenvectors of the similarity matrix (in our case, matrix $A_{S \times F}$), it can map both features and samples into a shared latent space using the eigenvectors computed as the points in space (denoted by the matrix $Z$). This reduces the dimensionality of a dataset, preserves its distance similarity, and treats both samples and features as data points before running a clustering algorithm on it, allowing simultaneous clustering of features and samples. While it has costs in memory and computation time due to the need for the graph Laplacian and eigenvectors, it can be done in parallel. SpectralCoClustering will transform matrix $A_{S \times F}$ into matrix $Z_{S + F, l}$ where $l = log_2(k)$ and $k$ is the number of expected clusters. This transformation uses the subset of eigenvectors corresponding to indices $2$ through $l+1$ of the graph Laplacian's eigen decomposition. $log_2(k)$ is chosen as the dimension for $Z$ since each additional $k$ cluster can map to a quadrant of space within. Essentially, each new dimension in $l$ doubles the number of possible cluster assignments. After transformation, any traditional clustering algorithm can be adapted to $Z$.

\subsection{Experimental Methodology}
\label{app:methodology}

\begin{figure}[H]
    \centering
    \includegraphics[width=1\linewidth]{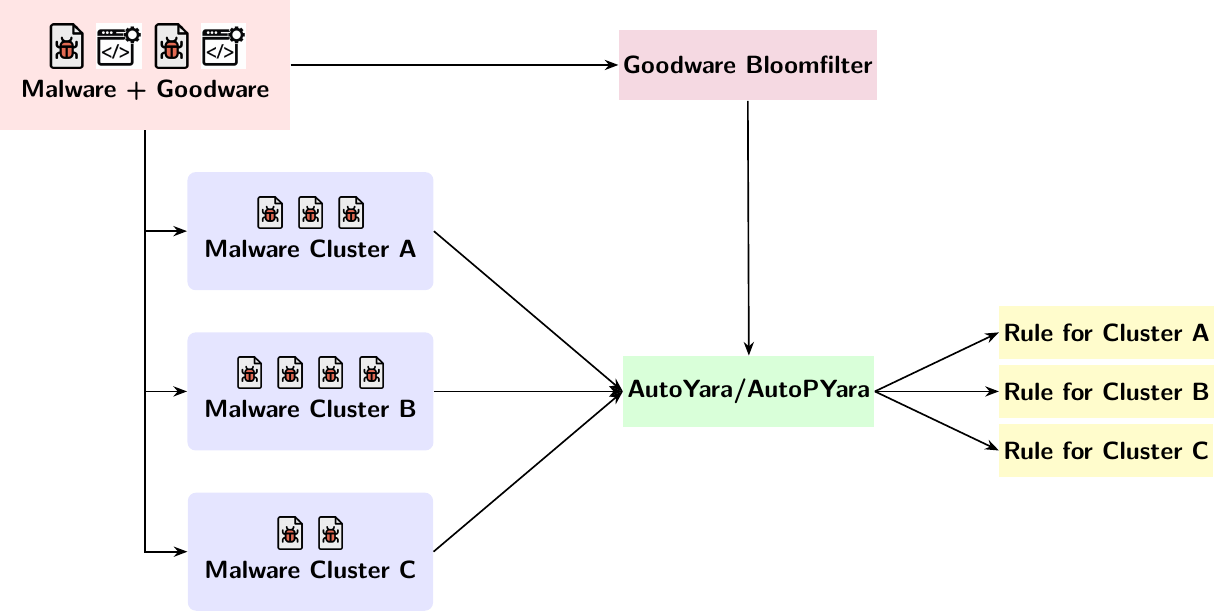}
    \caption{Experimental Methodology}
    \label{fig:meth}
\end{figure}

\noindent Figure~\ref{fig:meth} provides a high-level overview of the AutoYara and AutoPYara frameworks, from labeled data ingestion through cluster-specific YARA rule synthesis. The process begins by using malware and benignware samples to construct Bloom filters that capture salient byte-pattern features. Malware samples are then clustered using fuzzy-hash and labeling signals—SSDEEP, J-SDhash, and VirusTotal-derived family labels—to group closely related variants by shared characteristics. Each resulting cluster, together with the Bloom filters, is passed to the framework backend, which applies the underlying logic to propose candidate indicators while filtering noisy or overly generic substrings. The frameworks finally generates cluster-specific YARA rules.

\begin{algorithm}[H]
\caption{Adaptive Early Stopping Algorithm}
\label{alg:tp-early-stop}
\begin{algorithmic}[1]
\State $T \gets [\,]$, \quad $prs \gets \textsc{None}$
\For{$i = 0$ to $N{-}1$}
    \State $T \gets T \cup \{\textsc{ExtractTP}(\textsc{GenerateRule}(k_i))\}$
    \If{$|T| \geq s$}
        \State $v \gets \sigma(T)$
        \If{$(prs \neq \textsc{None}) \ \textbf{and} \ (|v - prs| < \theta)$}
            \State \textbf{break}
        \Else
            \State $s \gets s + \delta$
            \State $\theta \gets \theta \cdot \alpha$
        \EndIf
        \State $prs \gets v$
    \EndIf
\EndFor
\end{algorithmic}
\end{algorithm}

\noindent The algorithm~\ref{alg:tp-early-stop} incrementally generates candidate rules indexed by \(k_i\), extracts their true positives, and accumulates them in a running set \(T\). Once at least \(s\) items are collected, it computes a stability statistic \(v=\sigma(T)\) (e.g., an estimate of performance). If a previous value \(prs\) exists and the change \(|v-prs|\) is below a tolerance \(\theta\), it stops early, treating the metric as converged. Otherwise, it increases the evidence requirement \(s \leftarrow s+\delta\) and tightens the tolerance \(\theta \leftarrow \theta\cdot\alpha\) (typically \(0<\alpha<1\)), then continues. Thus, it adaptively balances efficiency and reliability: it keeps scanning up to \(N\) candidates but exits once \(\sigma(T)\) stabilizes, using progressively larger samples and stricter criteria to avoid premature stopping. Hyperparameters: \(s\) (initial evidence window), \(\delta\) (window growth), \(\theta\) (stability threshold), \(\alpha\) (threshold shrink factor).

\subsection{Augmented K-Means}
\label{app:aug}

  \noindent Estimating the hyperparameter $\alpha$, which is the upper bound on the predictor label error rate is a complex task. To address this limitation, we use CRDEST to estimate $\alpha$ efficiently. This design rests on four key properties. First, while $\alpha$ is a required hyperparameter bound for the label error rate that is difficult to derive analytically, CRDEST enables its empirical estimation. Second, the algorithm's primary contribution lies in robust initialization and centroid estimation. Third, $k$ estimation and point assignments continue to rely on K-Means as the underlying support engine. Finally, because Algorithm~\ref{alg:crdest} operates in $O(n \log n)$ time due to sorting, its high speed allows for the small-scale brute-force search described in Section~\ref{sec:eval}.

\section{Empirical Evaluation and  Characterization}

In Section~\ref{sec:eval}, we demonstrating how different configurations can yield significantly varied results. This necessitates a critical evaluation of these tools at their extremes across different labeling strategies and Bloom filter configurations. In this appendix, we decompose the evaluation results into their core components to visualize how each framework performs against varying constraints.

\subsection{AutoPYara V/S AutoYara In Practice}
\label{app:yararule}
\noindent In Section~\ref{subsec:QualitativeEval}, we describe how rule complexity plays a vital role in both the interpretability and the practical utility of automatically generated YARA rules. In this appendix, we compare the structural differences between signatures generated by our proposed framework and the baseline AutoYara system. Rules~\ref{lst:autopyara_rule} and~\ref{lst:autoyara_rule} rules generated for the same malware cluster (Cluster 4).

\begin{lstlisting}[language=Python,label={lst:autopyara_rule}, caption={\textbf{AutoPYara Rule}}]
rule Cluster4
{   // Input TP Rate:
    // 609/613
    strings:
    // Benign FP est: -0.0 Malicious FP est: -0.0 Entropy: 3.5, 612 files
    $x0 = { 00 00 C7 45 C4 AC 4E 48 00 E8 95 F8 FF FF 8B 4D } 
    // Benign FP est: -0.0 Malicious FP est: -0.0 Entropy: 4.0, 612 files
    $x1 = { 51 E8 D4 69 FB FF 83 C4 04 5F 5B 33 C0 5E 8B E5 } 
    // Benign FP est: -0.0 Malicious FP est: -0.0 Entropy: 3.75, 612 files
    $x2 = { 15 E4 26 48 00 5F 8B C6 5E 5D C2 2C 00 55 8B EC } 
    // Benign FP est: -0.0 Malicious FP est: -0.0 Entropy: 4.0, 613 files
    $x3 = { 6A 01 8B FA E8 E9 0F FE FF 83 C4 08 85 C0 74 24 } 
    // Benign FP est: -0.0 Malicious FP est: -0.0 Entropy: 2.5, 612 files
    $x4 = { 18 40 00 A7 18 40 00 1A 18 40 00 62 18 40 00 D9 } 
    condition:
    5 of ($x0,$x1,$x2,$x3,$x4)}\end{lstlisting}
\vspace{-2mm}
\begin{lstlisting}[language=Python, label={lst:autoyara_rule},caption={\textbf{AutoYara Rule}}]
rule Cluster4 
{   // TP Rate: 515/613
    strings:
    $x0 = { 53 E8 .. E8 D3 FC 01 } // Entropy: 6.22, Files: 612
    $x1 = { 83 C8 .. 10 53 56 57 } // Entropy: 6.19, Files: 611
    $x2 = { 6A 11 .. 5C EB 10 83 } // Entropy: 6.03, Files: 612
        .
        .   
        .
    $x14 = { FF 75 .. C4 CC CC }  // Entropy: 6.31, Files: 608
    $x15 = { 00 6C .. 00 20 00 }  // Entropy: 3.22, Files: 266
    $x16 = { 40 89 .. 83 79 08 }  // Entropy: 6.15, Files: 612
    condition:
    16 of ($x0,$x1,$x2,$x3,$x4,$x5,$x6, $x7,$x8,$x9,$x10,$x11,$x12,$x13, $x14,$x15,$x16)}\end{lstlisting}

\subsection{False Positive Rates In Practice}
\label{app:fpa}

 \noindent False positive rates are an important metric crucial to the operational capabilities of an automatic YARA rule generator. In practice, YARA rules need to maintain a FP rate as close to zero to make the generated rules practical to deploy as discussed in section~\ref{sec:eval}. Here we characterize the operational precision of AutoPYara by evaluating False Positive Rates (FPR) across diverse clustering strategies.

\begin{table}[ht]
\centering
\caption{FPR Results (\%): Extra and Intra Class}
\label{tab:FPR-Single-Col}
\footnotesize
\setlength{\tabcolsep}{4pt}
\renewcommand{\arraystretch}{0.95}
\begin{tabular}{lcc}
\toprule
\textbf{Method} & \textbf{Ember BF} & \textbf{Retrained BF} \\
\midrule
\multicolumn{3}{l}{\textbf{Part A: Extra-Class FPR (Benign Samples)}} \\
\textit{SSdeep} & & \\
\hspace{1em}Auto(P)Yara & $0.9 \pm 0.8$ & $1.1 \pm 0.8$ \\
\hspace{1em}Best/Worst K & -- & $\approx 0.21$ \\
\textit{J-SDhash} & & \\
\hspace{1em}Auto(P)Yara & $1.1 \pm 0.8$ & $1.0 \pm 0.9$ \\
\hspace{1em}Best/Worst K & -- & $\approx 0.22$ \\
\textit{VirusTotal} & & \\
\hspace{1em}AutoYara & $48.7 \pm 33.3$ & $20.4 \pm 12.8$ \\
\hspace{1em}AutoPYara & $44.4 \pm 19.4$ & $54.3 \pm 22.1$ \\
\hspace{1em}Best/Worst K & $9.9 \pm 4.2$ & $12.5 \pm 5.0$ \\
\midrule
\multicolumn{3}{l}{\textbf{Part B: Intra-Class FPR (Cross-Cluster)}} \\
\textit{All Methods} & & \\
\hspace{1em}AutoYara (SSDeep) & $0.15 \pm 0.33$ & $0.00$ \\
\hspace{1em}All Other Configs & $0.00$ & $0.00$ \\
\bottomrule
\end{tabular}
\end{table}

\subsection{Extra-Class Detection}
\vspace{-.3cm}

\noindent Table \ref{tab:FPR-Single-Col} presents extra-class FPR (\%) across SSDeep, J‑SDhash, and VirusTotal for AutoYara and AutoPYara, evaluated against three baselines: SSDeep similarity matches, J‑SDhash similarity matches, and VirusTotal family-labeling signals. Lower values indicate fewer benign files matched by rules or baseline heuristics, which is desirable for operational deployments. Our experimental results show that byte-based clustering yields a substantially lower (often near-zero) extra class FPR due to its homogeneous nature, compared with heterogeneous, antivirus family label–based clustering. This is critical for proactive deployment of such systems.

\subsection{Intra-Class Detection}

 Table \ref{tab:FPR-Single-Col} reports intra-class FPR (\%), which are uniformly zero in nearly all cases, indicating minimal spurious cross-cluster matches and strong cluster isolation. The only non-zero entry is a negligible \(0.15 \pm 0.33\) for AutoYara under SSDeep with the Ember BF, consistent with measurement noise. These results confirm tight cluster boundaries and a low risk of rule bleed-over in deployment.

\subsection{Note on {k} search}
\label{app:note}

 \noindent
In Section~\ref{subsec:AutoPYaraKEval}, we characterize the performance of AutoPYara and demonstrate how it outperforms the original framework. However, as we speculate that AutoPYara may occasionally under-estimate the optimal value of $k$, we utilize a brute-force search across a range of $k$ values to identify the local minima that yield the highest-performing YARA rules. In this appendix, we provide additional information regarding the resulting search space and its impact on rule efficacy.  In practice, we test the ±10 neighborhood—i.e., $k-10$ through  $k+10$— because empirical observations show that the optimal $k$ consistently lies near AutoYara’s estimate, making an exhaustive search unnecessary. Although we could brute-force all possibilities to find the optimal $k$, we observed that the best value consistently lies near AutoYara’s estimate.

\subsection{Empirical Evaluation of Rule Complexity}
\label{app:rme}

 \noindent To supplement the high-level qualitative findings in Section~\ref{subsec:QualitativeEval}, we here provide a granular breakdown of our rule complexity metrics.

\noindent \textbf{1)} \textit{Number of Strings} ($N$): The \textbf{number of strings} reflects the breadth of features used and is independent of the logical operators in the condition block.\newline
\noindent \textbf{2)} \textit{Hex Length} ($H$):  
AutoPYara generates byte n-grams from kilobyte-sized chunks of binaries and encodes them as hexadecimal strings. Given $N$ hex strings $s_1, s_2,..s_N$, we define Hex Length ($H$) as average string length.\newline
\noindent \textbf{3)} \textit{Conditions Count} ($C$): Counts logical operators (AND, OR) in the rule's condition section. Higher $C$ indicates complex rules capturing greater malware variability and heterogeneous sub-clusters, where different variants require different string combinations. Complex conditions ($C > 5$) are harder to interpret and maintain, but may be necessary for diverse malware families. Rules with simpler conditions are easier to maintain, while complex families require higher operator counts.
Given two rules with similar TP and FP rates, compactness is preferable. We measure compactness using two separate metrics: the number of strings ($N$) and their average length ($H$). Fewer and shorter strings simplify rule interpretation and validation. The conditions count ($C$) is independent and lower, especially when a single logical statement captures many strings. Malware with a simple structure yields concise rules with low $C$ and $N$, while complex malware families require more strings and potentially more conditions for robust detection, increasing the validation burden.  Table~\ref{tab:ToolAlgoStats} provides detailed metrics for all generated rules across the configurations tested across thresholds (Th50–Th90). SSDeep and J-SDhash configurations produce shorter, sparser rules (OR typically near 1; AND roughly 2–5), while VT-driven configurations yield the longest hex strings and highest AND counts, reflecting more complex conjunctions.

\begin{table}
\centering

\caption{Rule Characteristics: Ember vs. Retrained Bloom Filters}
\label{tab:ToolAlgoStats}
\footnotesize
\setlength{\tabcolsep}{1.8pt}
\renewcommand{\arraystretch}{0.85}
\resizebox{\columnwidth}{!}{%
\begin{tabular}{lcccccccccccc}
\toprule
 & \multicolumn{6}{c}{\textbf{Ember Bloom Filters}} & \multicolumn{6}{c}{\textbf{Retained Bloom Filters}} \\
\cmidrule(lr){2-7} \cmidrule(lr){8-13}
\textbf{Method} & 
\makecell{\textbf{Avg}\\\textbf{HexLen}} & 
\makecell{\textbf{Avg}\\\textbf{OR}} & 
\makecell{\textbf{Avg}\\\textbf{AND}} & 
\makecell{\textbf{Med}\\\textbf{HexLen}} & 
\makecell{\textbf{Med}\\\textbf{OR}} & 
\makecell{\textbf{Med}\\\textbf{AND}} &
\makecell{\textbf{Avg}\\\textbf{HexLen}} & 
\makecell{\textbf{Avg}\\\textbf{OR}} & 
\makecell{\textbf{Avg}\\\textbf{AND}} & 
\makecell{\textbf{Med}\\\textbf{HexLen}} & 
\makecell{\textbf{Med}\\\textbf{OR}} & 
\makecell{\textbf{Med}\\\textbf{AND}} \\
\midrule

\multicolumn{13}{l}{\textbf{AutoYara}} \\
SSdeep-Th50 & 471.9 & 1.1 & 3.8 & 32 & 1 & 3 & 362.9 & 1.1 & 2.6 & 8 & 1 & 1 \\
SSdeep-Th60 & 446.1 & 1.1 & 3.8 & 24 & 1 & 3 & 315.5 & 1.0 & 2.4 & 8 & 1 & 1 \\
SSdeep-Th70 & 391.1 & 1.1 & 3.2 & 24 & 1 & 3 & 277.4 & 1.0 & 2.2 & 8 & 1 & 1 \\
SSdeep-Th80 & 306.7 & 1.1 & 3.3 & 24 & 1 & 3 & 239.0 & 1.0 & 2.0 & 8 & 1 & 1 \\
SSdeep-Th90 & 296.6 & 1.1 & 3.3 & 24 & 1 & 3 & 180.4 & 1.0 & 1.7 & 8 & 1 & 1 \\
J-SDhash-Th50 & 341.9 & 1.1 & 3.1 & 8 & 1 & 0 & 403.5 & 1.0 & 2.7 & 8 & 1 & 1 \\
J-SDhash-Th60 & 394.8 & 1.1 & 2.8 & 8 & 1 & 0 & 377.0 & 1.1 & 2.4 & 8 & 1 & 1 \\
J-SDhash-Th70 & 410.0 & 1.1 & 3.0 & 8 & 1 & 0 & 337.7 & 1.1 & 2.4 & 8 & 1 & 1 \\
J-SDhash-Th80 & 449.5 & 1.1 & 3.2 & 8 & 1 & 0 & 310.6 & 1.0 & 2.2 & 8 & 1 & 1 \\
J-SDhash-Th90 & 362.3 & 1.1 & 2.4 & 8 & 1 & 1 & 249.4 & 1.0 & 1.9 & 8 & 1 & 1 \\
VT & 966.1 & 1.3 & 26.7 & 128 & 1 & 5 & 1041.7 & 1.3 & 37.5 & 160 & 1 & 5 \\
\midrule

\multicolumn{13}{l}{\textbf{AutoPYara}} \\
SSdeep-Th50 & 318.3 & 1.1 & 4.2 & 24 & 1 & 3 & 259.1 & 1.1 & 2.8 & 8 & 1 & 1 \\
SSdeep-Th60 & 296.8 & 1.1 & 3.9 & 24 & 1 & 3 & 243.5 & 1.1 & 2.6 & 8 & 1 & 1 \\
SSdeep-Th70 & 258.5 & 1.1 & 3.1 & 16.0 & 1 & 2 & 201.7 & 1.0 & 2.2 & 8 & 1 & 1 \\
SSdeep-Th80 & 222.0 & 1.1 & 3.1 & 24 & 1 & 3 & 179.9 & 1.0 & 1.9 & 8 & 1 & 1 \\
SSdeep-Th90 & 244.9 & 1.1 & 3.0 & 24 & 1 & 3 & 138.1 & 1.0 & 1.7 & 8 & 1 & 1 \\
J-SDhash-Th50 & 235.3 & 1.1 & 3.2 & 8 & 1 & 1 & 325.1 & 1.1 & 2.9 & 8 & 1 & 1 \\
J-SDhash-Th60 & 208.3 & 1.1 & 2.7 & 8 & 1 & 1 & 287.0 & 1.1 & 2.6 & 8 & 1 & 1 \\
J-SDhash-Th70 & 193.0 & 1.1 & 2.7 & 8 & 1 & 1 & 246.9 & 1.1 & 2.3 & 8 & 1 & 1 \\
J-SDhash-Th80 & 174.8 & 1.1 & 2.8 & 8 & 1 & 1 & 225.6 & 1.0 & 2.2 & 8 & 1 & 1 \\
J-SDhash-Th90 & 138.5 & 1.1 & 2.3 & 8 & 1 & 1 & 181.9 & 1.0 & 1.9 & 8 & 1 & 1 \\
VT & 671.5 & 2.7 & 34.7 & 80 & 1 & 5 & 678.3 & 3.2 & 36.7 & 96 & 1 & 5 \\
\midrule

\multicolumn{13}{l}{\textbf{AutoPyara BEST K}} \\
SSdeep-Th50 & 417.2 & 1.3 & 4.8 & 24 & 1 & 3 & 302.3 & 1.1 & 3.1 & 8 & 1 & 1 \\
SSdeep-Th60 & 325.9 & 1.2 & 4.1 & 24 & 1 & 3 & 275.5 & 1.1 & 2.6 & 8 & 1 & 1 \\
SSdeep-Th70 & 316.3 & 1.2 & 3.1 & 16 & 1 & 2 & 237.3 & 1.1 & 2.3 & 8 & 1 & 1 \\
SSdeep-Th80 & 248.7 & 1.2 & 3.0 & 24 & 1 & 3 & 201.2 & 1.1 & 2.1 & 8 & 1 & 1 \\
SSdeep-Th90 & 223.2 & 1.2 & 2.9 & 24 & 1 & 3 & 154.5 & 1.1 & 1.7 & 8 & 1 & 1 \\
J-SDhash-Th50 & 321.4 & 1.2 & 3.5 & 8 & 1 & 1 & 366.8 & 1.1 & 3.1 & 8 & 1 & 1 \\
J-SDhash-Th60 & 278.2 & 1.2 & 3.0 & 8 & 1 & 1 & 314.9 & 1.1 & 2.8 & 8 & 1 & 1 \\
J-SDhash-Th70 & 244.8 & 1.1 & 2.8 & 8 & 1 & 1 & 279.2 & 1.1 & 2.5 & 8 & 1 & 1 \\
J-SDhash-Th80 & 208.8 & 1.1 & 2.7 & 8 & 1 & 1 & 256.2 & 1.1 & 2.3 & 8 & 1 & 1 \\
J-SDhash-Th90 & 180.3 & 1.1 & 2.5 & 8 & 1 & 1 & 208.8 & 1.1 & 1.9 & 8 & 1 & 1 \\
VT & 727.2 & 2.7 & 31.8 & 88 & 1 & 4 & 721.6 & 3.2 & 35.7 & 96 & 1 & 5 \\
\midrule

\multicolumn{13}{l}{\textbf{AutoPyara Worst K}} \\
SSdeep-Th50   & 587.8 & 1.1 & 4.3 & 24 & 1 & 3 & 382.3 & 1.1 & 2.5 & 8 & 1 & 1 \\
SSdeep-Th60   & 584.0 & 1.1 & 4.1 & 24 & 1 & 3 & 342.8 & 1.0 & 2.3 & 8 & 1 & 1 \\
SSdeep-Th70   & 459.1 & 1.1 & 3.1 & 16 & 1 & 2 & 300.4 & 1.0 & 2.1 & 8 & 1 & 1 \\
SSdeep-Th80   & 508.6 & 1.1 & 3.3 & 24 & 1 & 3 & 252.4 & 1.0 & 1.9 & 8 & 1 & 1 \\
SSdeep-Th90   & 477.9 & 1.1 & 3.3 & 24 & 1 & 3 & 187.1 & 1.0 & 1.7 & 8 & 1 & 1 \\
J-SDhash-Th50 & 386.5 & 1.1 & 3.1 & 8  & 1 & 1 & 432.1 & 1.1 & 2.6 & 8 & 1 & 1 \\
J-SDhash-Th60 & 357.9 & 1.1 & 2.7 & 8  & 1 & 1 & 388.3 & 1.0 & 2.4 & 8 & 1 & 1 \\
J-SDhash-Th70 & 328.4 & 1.1 & 2.7 & 8  & 1 & 1 & 344.1 & 1.0 & 2.4 & 8 & 1 & 1 \\
J-SDhash-Th80 & 295.8 & 1.1 & 3.1 & 8  & 1 & 1 & 301.2 & 1.0 & 2.2 & 8 & 1 & 1 \\
J-SDhash-Th90 & 261.6 & 1.1 & 2.3 & 8  & 1 & 1 & 265.6 & 1.0 & 1.9 & 8 & 1 & 1 \\
VT            & 1068.2 & 1.3 & 33.6 & 128 & 1 & 5 & 1157.2 & 1.3 & 39.6 & 144 & 1 & 5 \\

\midrule

\multicolumn{13}{l}{\textbf{VT-Heuristics - Retrained Bloom Filters}} \\
\multicolumn{13}{l}{\hspace{1em}MODE} \\
SSdeep-Th50 & - & - & - & - & - & - & 929.1 & 2.0 & 16.7 & 128 & 1 & 4 \\
SSdeep-Th60 & - & - & - & - & - & - & 889.7 & 1.7 & 14.8 & 101 & 1 & 4 \\
SSdeep-Th70 & - & - & - & - & - & - & 663.7 & 1.5 & 7.5 & 64 & 1 & 3 \\
SSdeep-Th80 & - & - & - & - & - & - & 481.7 & 1.3 & 4.8 & 32 & 1 & 3 \\
SSdeep-Th90 & - & - & - & - & - & - & 504.0 & 1.3 & 4.3 & 24 & 1 & 2 \\
J-SDhash-Th50 & - & - & - & - & - & - & 770.5 & 1.3 & 5.7 & 64 & 1 & 3 \\
J-SDhash-Th60 & - & - & - & - & - & - & 629.6 & 1.2 & 4.7 & 48 & 1 & 3 \\
J-SDhash-Th70 & - & - & - & - & - & - & 570.2 & 1.2 & 4.8 & 32 & 1 & 3 \\
J-SDhash-Th80 & - & - & - & - & - & - & 625.7 & 1.2 & 5.8 & 32 & 1 & 2 \\
J-SDhash-Th90 & - & - & - & - & - & - & 591.8 & 1.1 & 3.4 & 16 & 1 & 2 \\

\multicolumn{13}{l}{\hspace{1em}MAX} \\
SSdeep-Th50 & - & - & - & - & - & - & 341.6 & 1.1 & 3.7 & 8 & 1 & 2 \\
SSdeep-Th60 & - & - & - & - & - & - & 301.2 & 1.1 & 4.1 & 8 & 1 & 1 \\
SSdeep-Th70 & - & - & - & - & - & - & 256.0 & 1.1 & 2.8 & 8 & 1 & 1 \\
SSdeep-Th80 & - & - & - & - & - & - & 190.5 & 1.0 & 2.3 & 8 & 1 & 1 \\
SSdeep-Th90 & - & - & - & - & - & - & 190.2 & 1.0 & 2.2 & 8 & 1 & 1 \\
J-SDhash-Th50 & - & - & - & - & - & - & 358.2 & 1.1 & 3.5 & 8 & 1 & 2 \\
J-SDhash-Th60 & - & - & - & - & - & - & 323.0 & 1.1 & 2.8 & 8 & 1 & 1 \\
J-SDhash-Th70 & - & - & - & - & - & - & 324.7 & 1.1 & 2.8 & 8 & 1 & 1 \\
J-SDhash-Th80 & - & - & - & - & - & - & 306.5 & 1.1 & 2.9 & 8 & 1 & 1 \\
J-SDhash-Th90 & - & - & - & - & - & - & 244.2 & 1.0 & 2.6 & 8 & 1 & 1 \\

\multicolumn{13}{l}{\hspace{1em}RANDOM} \\
SSdeep-Th50 & - & - & - & - & - & - & 515.0 & 1.3 & 7.2 & 8 & 1 & 2 \\
SSdeep-Th60 & - & - & - & - & - & - & 453.4 & 1.2 & 4.8 & 8 & 1 & 1 \\
SSdeep-Th70 & - & - & - & - & - & - & 389.5 & 1.2 & 3.9 & 8 & 1 & 1 \\
SSdeep-Th80 & - & - & - & - & - & - & 306.4 & 1.2 & 2.8 & 8 & 1 & 1 \\
SSdeep-Th90 & - & - & - & - & - & - & 290.4 & 1.1 & 2.5 & 8 & 1 & 1 \\
J-SDhash-Th50 & - & - & - & - & - & - & 750.7 & 1.5 & 6.0 & 96 & 1 & 3 \\
J-SDhash-Th60 & - & - & - & - & - & - & 720.2 & 1.3 & 5.3 & 48 & 1 & 3 \\
J-SDhash-Th70 & - & - & - & - & - & - & 641.6 & 1.4 & 4.8 & 64 & 1 & 3 \\
J-SDhash-Th80 & - & - & - & - & - & - & 776.6 & 1.4 & 5.5 & 40 & 1 & 3 \\
J-SDhash-Th90 & - & - & - & - & - & - & 514.9 & 1.2 & 3.8 & 24 & 1 & 2 \\

\bottomrule
\end{tabular}
}
\end{table}

\subsection{Evaluation of YARA Rules on  Large Cluster}
\label{app:bp}

\noindent In Section~\ref{sec:benchmarkfinal}, we provide a comprehensive characterization of both the AutoYara and AutoPYara frameworks, demonstrating how performance observations can vary significantly if not correctly contextualized. We explore the operational extremes of these tools, including their theoretical performance limits, and offer guidance on achieving optimal results in real-world deployments. However, the initial results encompassed the entire dataset, which includes a high volume of small clusters that may skew the overall performance metrics.

\begin{figure}[H]
\vspace{-3mm}

    \centering
    {%
        \includegraphics[width=0.6\linewidth]{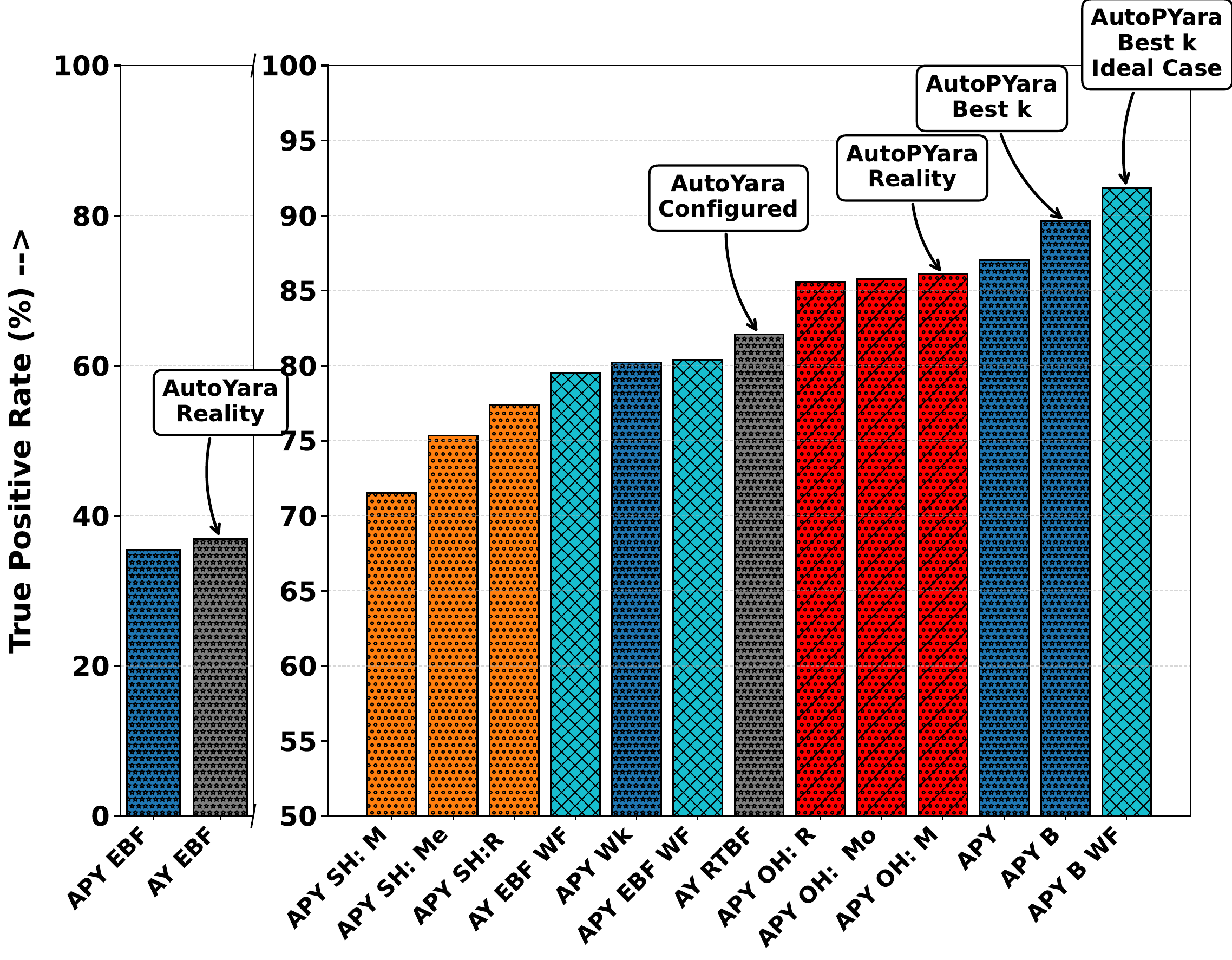}
    }\vspace{-4mm}
    \caption{AutoYara v/s AutoPYara using large clusters}
    \label{fig:AutoPYaraBigPictureTOP}

\end{figure}

In this appendix, we isolate and demonstrate the impact of larger clusters on tool efficacy. Figure~\ref{fig:AutoPYaraBigPictureTOP} illustrates the performance delta between the two frameworks across top-performing configurations. Our findings show that AutoPYara maintains a consistently higher detection rate compared to the baseline AutoYara, even when small clusters are excluded from the analysis. This distinction is critical, as small clusters often introduce noise into evaluation metrics and tend to display inflated performance trends due to a lack of structural variability.

\section{Empirical Evaluation and  Characterization}

In Section~\ref{sec:eval}, we demonstrate how different configurations can yield significantly varied results. This necessitates a critical evaluation of these tools at their extremes across different labeling strategies and Bloom filter configurations. In this appendix, we decompose the evaluation results into their core components to visualize how each framework performs against varying constraints.

\subsection{Extra-Class Detection}
\vspace{-.3cm}

\subsection{Evaluation of YARA Rules on  Large Cluster}

\noindent In Section~\ref{sec:benchmarkfinal}, we provide a comprehensive characterization of both the AutoYara and AutoPYara frameworks, demonstrating how performance observations can vary significantly if not correctly contextualized. We explore the operational extremes of these tools, including their theoretical performance limits, and offer guidance on achieving optimal results in real-world deployments. However, the initial results encompassed the entire dataset, which includes a high volume of small clusters that may skew the overall performance metrics.

In this appendix, we isolate and demonstrate the impact of larger clusters on tool efficacy. Figure~\ref{fig:AutoPYaraBigPictureTOP} illustrates the performance delta between the two frameworks across top-performing configurations. Our findings show that AutoPYara maintains a consistently higher detection rate compared to the baseline AutoYara, even when small clusters are excluded from the analysis. This distinction is critical, as small clusters often introduce noise into evaluation metrics and tend to display inflated performance trends due to a lack of structural variability.

\section{Recommendations}
\label{app:rec}

\noindent Throughout the evaluation in Section~\ref{sec:eval}, we demonstrated how different technical specifications and strategies can result in varied performance, emphasizing that these results should not be taken at face value. To transition our scientific investigation into practical application, we have compiled our findings into the following best-practice guidelines:

\noindent \textbf{1)} \textbf{\textit{Retrain Bloom Filters}} [C2]
\newline
Retrain with up-to-date malware and goodware samples. Outdated filters miss new byte patterns, reduce rule quality, and increase false positives.

\noindent \textbf{2)} \textbf{\textit{Compare Average not Single Shot Rules}} [C3,C6]
\newline
Clustering results can change with random seeds. Run small ensembles with different seeds for robustness. Track stability with cumulative metrics rather than single-run averages. Generating a single rule might mislead the user given the non-deterministic nature of the framework.

\noindent \textbf{3)} \textbf{\textit{Accuracy is Important But Ensure Low FP Rate}} [C4]
\newline
It is important that generated rules are validated against both malicious and critically goodware samples (FP rates matter). Characterized into two categories (Goodware FP rate and Cross Cluster FP rate), Goodware FP is more critical as it impacts usability at endpoint devices.

\noindent \textbf{4)}  \textbf{\textit{Favor Byte-Based Similarity Clustering Over AntiVirus Family Clustering}} [C1,C4]
\newline
Given the heterogeneous nature of the task, favor similarity-based clustering over family-based clustering, as intrinsic correlations in the data are better captured and result in higher quality rules. 

\noindent \textbf{5)} \textbf{\textit{Calibrate Threshold for Coverage not Precision}} [C5]
\newline
Thresholds impact the formation of the clusters. Higher thresholds create tighter clusters and more specific rules; lower thresholds broaden coverage, but the performance gain is not due to the changes in threshold rather just an illusion due to the fewer number (higher the threshold fewer the samples per cluster) of samples. 

\noindent \textbf{6)}  \textbf{\textit{Calibrate Rule Generation with Centroid Selection to Improve Precision}} [F1]
\newline
We demonstrate how Centroid selection can be adapted into the AutoYara framework, and show that it results in up to 13 \% absolute gain in performance. We also highlight how the gain of centroid selection is greater in larger, more realistic clusters.  

\noindent \textbf{7)} \textbf{\textit{Calibrate $k$ Values When Possible}} [F2]
\newline
We demonstrate that the values of $k$ (in other words, the number of clusters) have a profound impact on the performance of the framework. The absolute difference in performance between the best and worst $k$ can be up to 35\% in large clusters, highlighting its importance.

\noindent \textbf{8)}  \textbf{\textit{Use AntiVirus Family to Build an Oracle for $k$}}  [F3] \newline
Build an oracle to suggest $k$ from historical results. Use family labels as a coarse filter to narrow the $k$ range, and use similarity hashes for centroids in actual clustering. Keep the oracle updated as new data arrives.

\section{AutoPYara Interface}
\label{app:autopyara:interface}

\noindent We develop and publish AutoPYara on PyPI as a package that is easy to install and use natively in Python. The user interface is as follows.

\begin{lstlisting}[language=Python, caption={\textbf{Typical use case of AutoPYara.}}, label=pyara3]
print("AutoYara Cluster: (AugmentedKMeansDBSCANSoft)")

yara = AutoPYara()
pyara_obj = yara.generate(
    get_project_path("input_testing", "malicious", 
                     "output_preprocessed", cluster_folder),
    get_project_path("intermediate", "bloom_filters", 
                     "malicious-bytes"),
    get_project_path("intermediate", "bloom_filters", 
                     "benign-bytes"),
    output_dir=get_project_path("output", cluster_folder),
    bicluster_alg="SpectralCoCluster",
    cluster_alg="AugmentedKMeansDBSCANSoft",
    output_format="string",
    similarity_threshold=80 )
print(f"TP: {pyara_obj['TP']} out of {pyara_obj['file_count']} files!")
print("string output", pyara_obj['output'])
print("full output object", pyara_obj)
\end{lstlisting}

\section{Visualizing AutoPYara's Centroid Selection}
\label{app:visual}

\begin{figure}[ht]
\centering
\includegraphics[width=\linewidth]{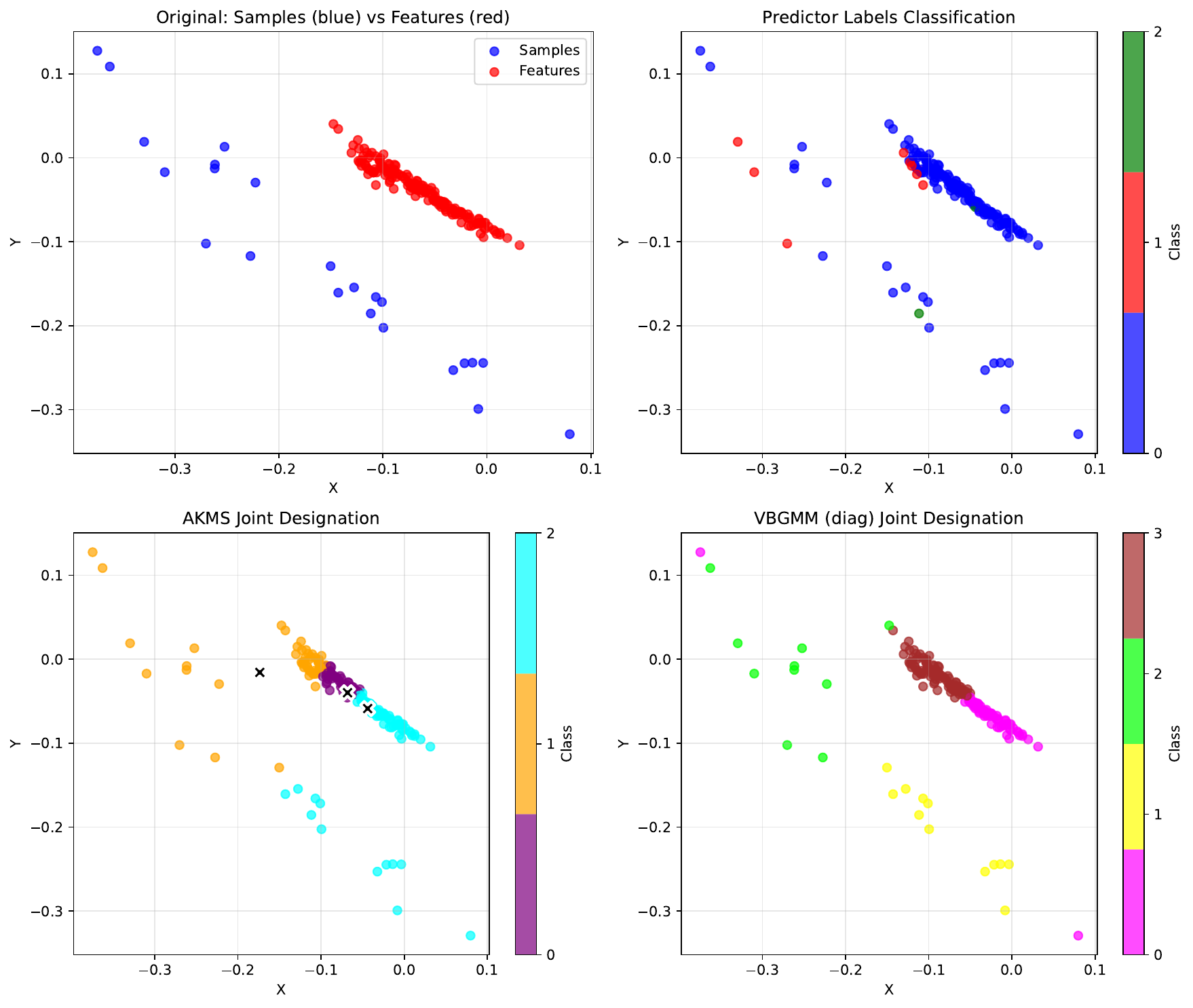}
\caption{\textbf{Algorithm classification performance.} A comparison between the latent Z space classification for AKMS ($k=3$) VBGMM ($k=4$). This Z space is derived from the Cluster 5 matrix with 23 samples and 165 features (gram size 32). Features and samples are distinguished on the top left graph, and predictor labels with mode extension are given in the top right. The X and Y axis represent the 2nd and 3rd singular vector values for each sample/feature data point, which is the clustering space used to create biclusters~\cite{dhillon2001_spectralcoclustering}. Ideally, clusters should include a combination of features and samples.}\label{fig:algorithm_classifications}
\end{figure}

\noindent To better understand how AutoPYara improves classification via cluster centroid selection, consider Figure~\ref{fig:algorithm_classifications}, which exemplifies one clustering scenario in a comparative way between VBGMM and AKMS (our algorithm). This scenario was picked as a representative example of the multiple clusters generated during our experimentation process. Figure~\ref{fig:algorithm_classifications} presents four subgraphs that represent: \textit{Original: Samples (blue) vs Features (red)}, showing the distinction between features and samples (extracted from

 \noindent Spectral Co-Clustering), \textit{Predictor Labels Classification} showing predictor labels and the extension labels, \textit{AKMS Joint Designation} showing AKMS clustering, and \textit{VBGMM (diag) Joint Designation} for VBGMM.

\noindent VBGMM is observed to create globally unoptimal and locally optimal clusters. This is because it is a model that assumes a Gaussian distribution~\cite{corduneanu2001_vbgmm}. As such, when Spectral Co-Clustering produces a latent space where features and samples are far apart, VBGMM will cluster feature points and samples separately. In consequence, VBGMM can produce biclusters with empty samples or empty features, which are removed by the biclustering algorithm, wasting its biclusters.

\noindent AKMS is provided with predictor labels as global context along with sample/feature distinction. Since the extended algorithm maps all features to the same shared labels as samples, this coerces AKMS to place centroids between features and samples. This works even when features and samples are far apart in the latent space (like Figure~\ref{fig:algorithm_classifications}), creating ``bridges'' (such as class 1-orange of \textit{AKMS Joint Designation}) to tie both groups together. This creates locally unoptimal but globally optimal clusters where all biclusters contain samples and features and have local distance similarity in the latent space. This produces biclusters with nonzero rows and columns or larger clusters, as no bicluster was wasted from clustering feature/sample ``islands'' of points. Another observation is that Figure~\ref{fig:algorithm_classifications}'s \textit{Predictor Labels Classification} graph produces a heavily imbalanced prediction. This is disadvantageous as this produces one large bicluster and many small biclusters. The AutoPYara filtering algorithm will remove small biclusters, leading to an overall waste of biclusters, which is a similar problem as VBGMM. AKMS addresses this by using the predictor labels only as guides for placing the centroids, which the algorithm then classifies points based on centroid proximity similar to K-Means/Fuzzy C-Means (local context), thereby not relying on only the predictor labels (or global context). \noindent Figure~\ref{fig:algorithm_classifications} demonstrates that integrating domain knowledge improves performance. Algorithms utilizing local knowledge (data points and distances) may get stuck in suboptimal/greedy clustering solutions. Our algorithm, AKMS, considers both local and global knowledge. It avoids the disadvantage of greedy clustering, avoids relying on only noisy/unfavorable global knowledge from our security domain, and considers the statistical relevance that Spectral Co-Clustering has embedded in the latent space. This leads to a hybridization that can achieve advantages from both local (latent space information) and global solutions (domain context). As such, the visualization clearly demonstrates where the approach is superior.

\section{Clusters Distribution}
\label{app:dataset}
\vspace{-2mm}

\begin{figure}
    \centering
    \subfloat[SSdeep Clusters Avg.\label{fig:ssdeepCluster}]{%
        \includegraphics[width=0.48\linewidth]{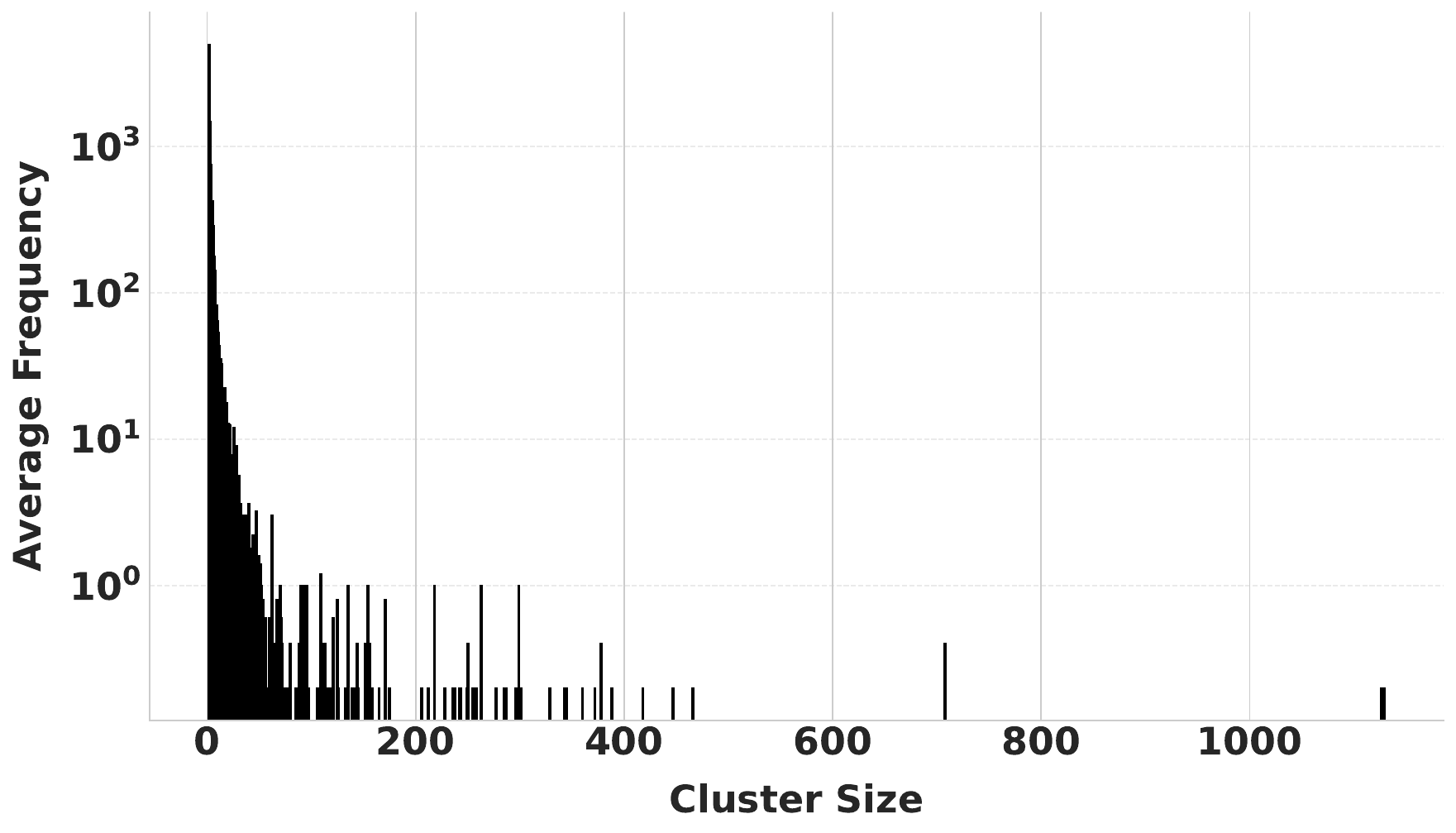}
    }\hfill
    
    \subfloat[J-SDhash Clusters Avg.\label{fig:sdhashCluster}]{%
        \includegraphics[width=0.48\linewidth]{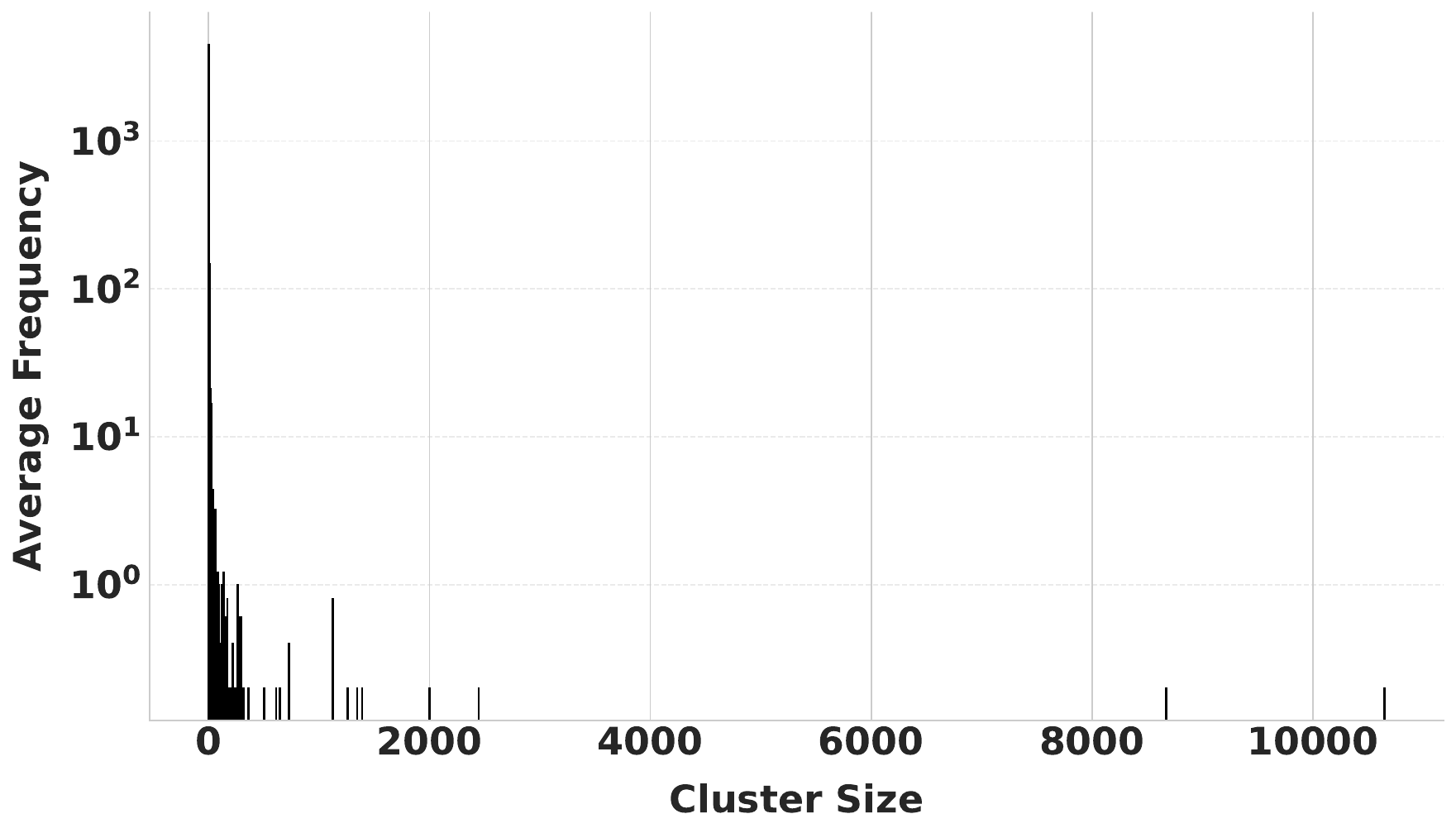}
    }\hfill
    
    \subfloat[VT Clusters Avg.\label{fig:vtCluster}]{%
        \includegraphics[width=0.48\linewidth]{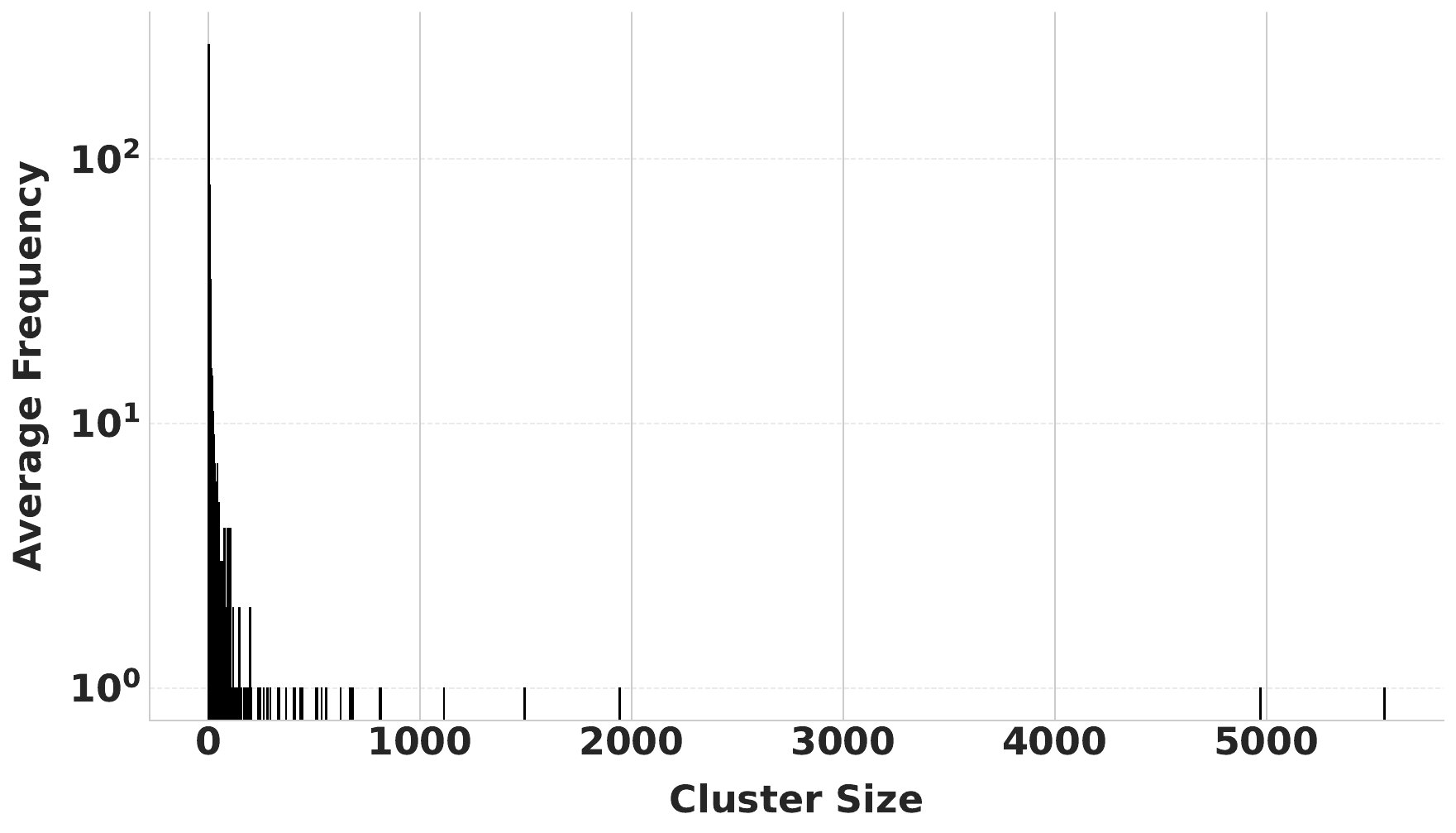}
    }
    \caption{Comparision of Cluster Distribution}
    \label{fig:clusterDistributions}
\end{figure}

 \noindent Figure~\ref{fig:clusterDistributions} summarizes how cluster sizes are distributed across the datasets for three clustering strategies. It reports the average frequency at each cluster size for SSdeep~\ref{fig:ssdeepCluster}, J‑SDhash~\ref{fig:sdhashCluster}, and VirusTotal family label groupings~\ref{fig:vtCluster}, averaged over the evaluated thresholds where applicable. Together, these plots show that algorithm choice changes how samples aggregate: the same corpus yields different cluster‑size profiles under content‑similarity digests~\cite{191669} versus AV label–based families. For SSdeep, the threshold parameter filters matches by minimum similarity score; lower thresholds typically merge more samples into larger clusters, while higher thresholds fragment clusters into smaller groups and singletons. For J‑SDhash, varying the acceptance criterion on digest similarity produces analogous trends. For VirusTotal family clusters, differences arise from AV label normalization and alias resolution rather than digest similarity, so distribution shifts reflect labeling practices and any normalization tooling, not content‑similarity thresholds. We emphasize both the differences among the clustering algorithms and their impact on the resulting cluster‑size distributions.

\vspace{-2mm}

\section{Evaluation of Rules in Threat Hunting}
\label{app:threat}

\begin{figure}
    \centering
    \includegraphics[width=1\linewidth]{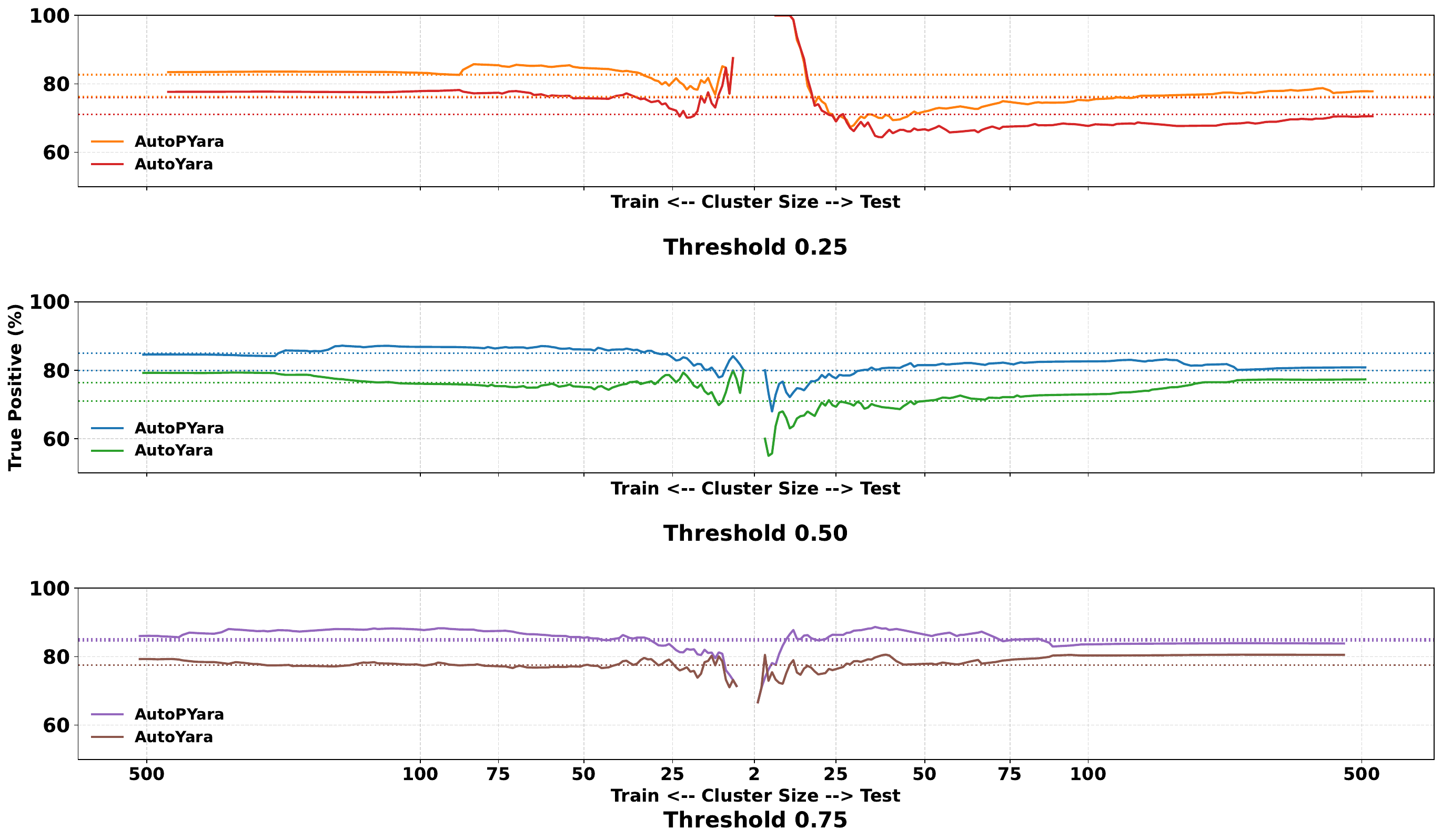}

    \caption{Threat Hunting SSdeep}
    \label{fig:ThreatHuntingSSdeep}
\end{figure}

\noindent We evaluate generalization in a threat-hunting setting by partitioning each cluster into training and test subsets at three ratios—75/25, 50/50, and 25/75—to emulate limited analyst visibility during rule authoring and unseen samples in operations. For each split, we randomly select training samples within every cluster to generate rules and score the remaining samples as test data. Figures~\ref{fig:ThreatHuntingSSdeep}summarize performance across SSdeep- and SDhash-derived clusters, respectively, showing how detection degrades as training data decreases, while cross-cluster detections remain near zero and goodware matches stay low. These results quantify the impact of data availability on rule effectiveness and robustness in realistic threat-hunting workflows. The plots clear highlight and demonstrate and how AutoPYara outperforms AutoYara across all different configurations. 
\label{app:tha}

\section{Ablation Study and Sensitivity Analysis}
\label{app:plots}

\noindent This section presents a detailed characterization of AutoYara and AutoPYara, providing an in-depth visualization of the tools' performance across various configurations and thresholds.

\clearpage

\begin{figure*}
    \centering

    \vspace{2mm}
    \begin{minipage}{0.95\linewidth}
    \centering
    \subfloat[SSdeep Avg. EBF]{\includegraphics[width=0.23\linewidth]{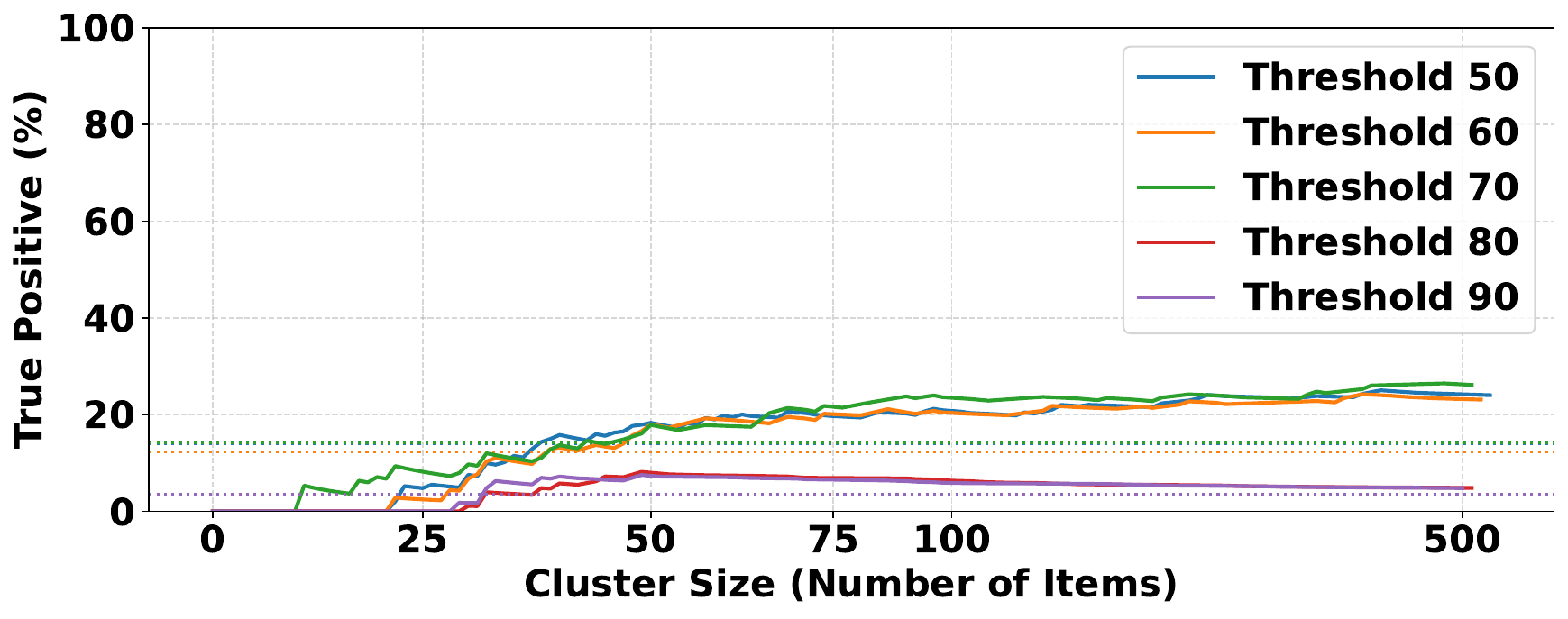}}
    \hfill
    \subfloat[SSdeep Avg. RTBF]{\includegraphics[width=0.23\linewidth]{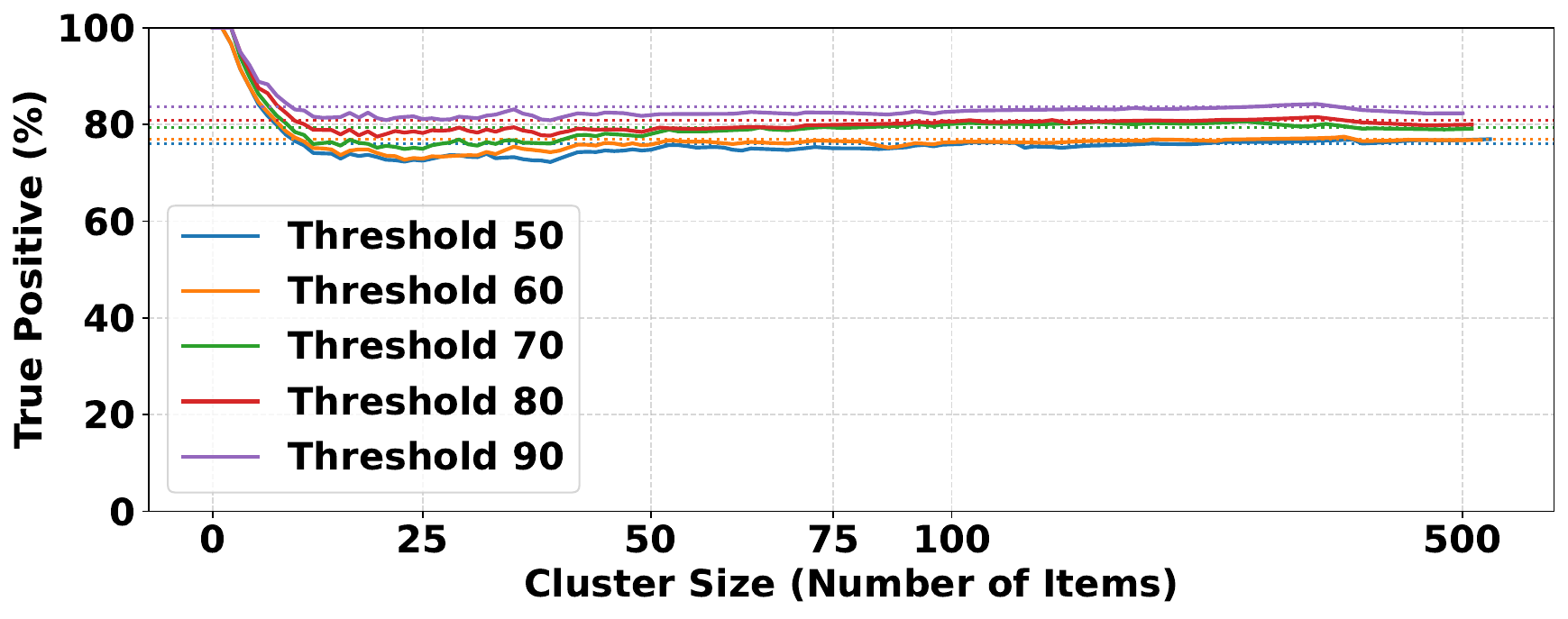}}
    \hfill
    \subfloat[SSdeep True EBF]{\includegraphics[width=0.23\linewidth]{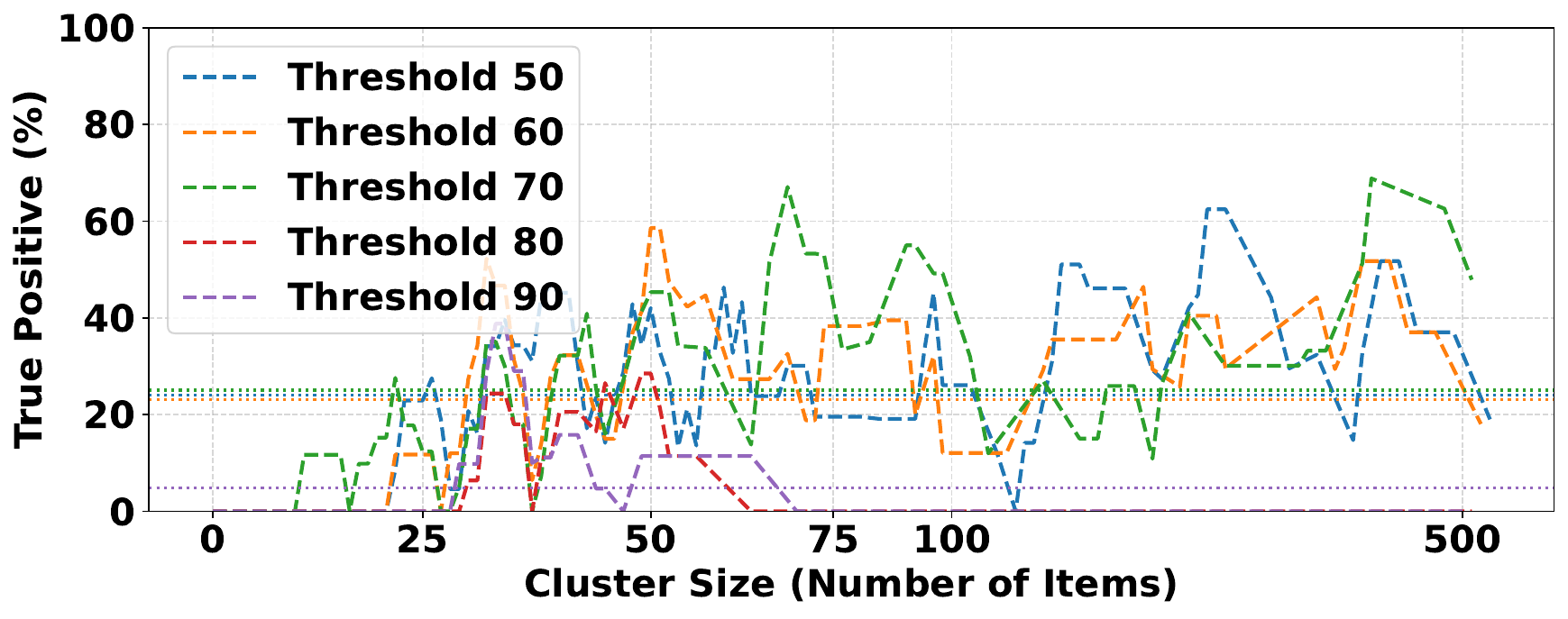}}
    \hfill
    \subfloat[SSdeep True RTBF]{\includegraphics[width=0.23\linewidth]{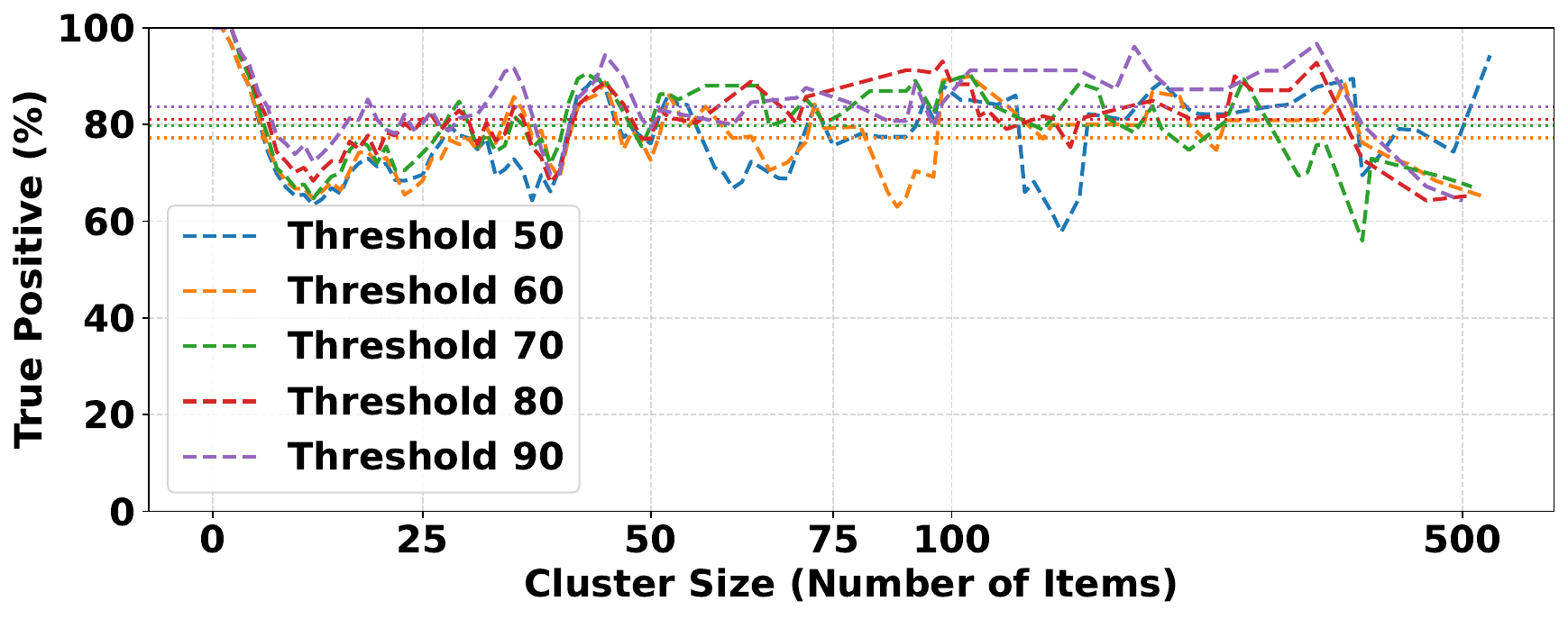}} \\[2mm]
    \subfloat[J-SDhash Avg. EBF]{\includegraphics[width=0.23\linewidth]{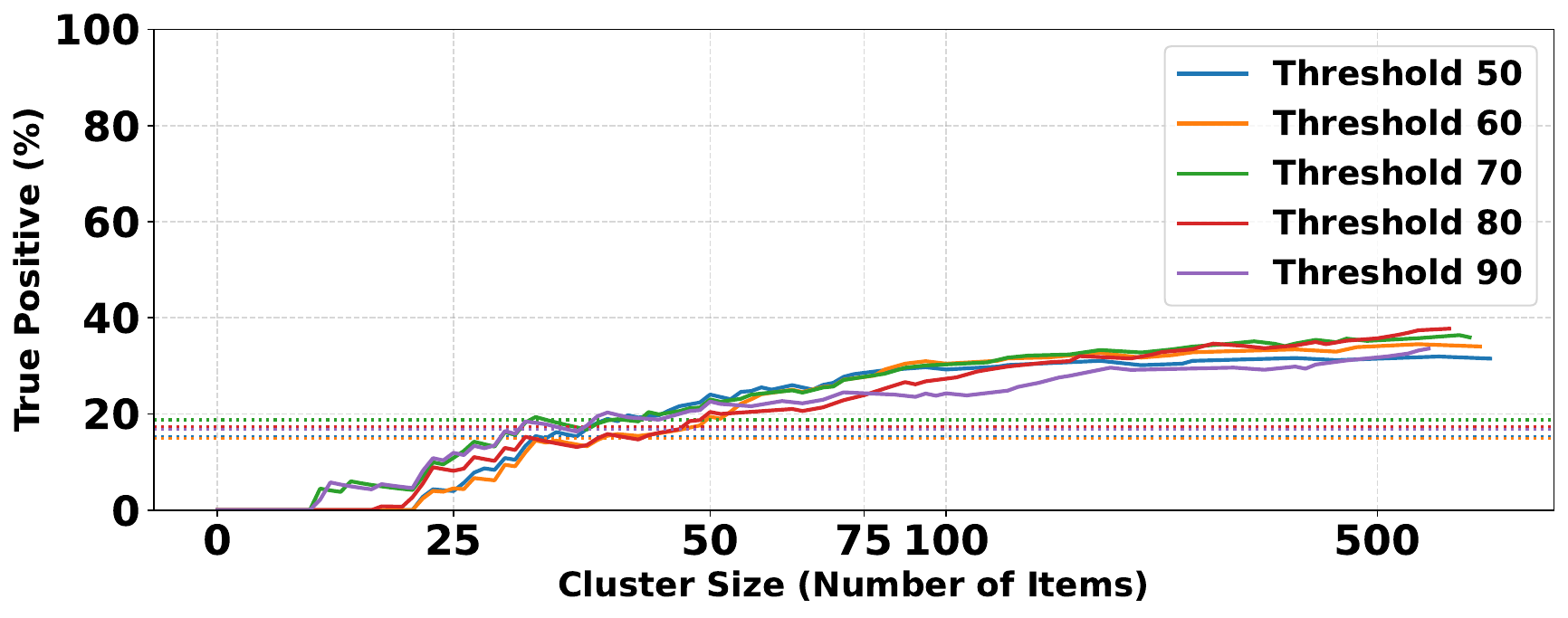}}
    \hfill
    \subfloat[J-SDhash Avg. RTBF]{\includegraphics[width=0.23\linewidth]{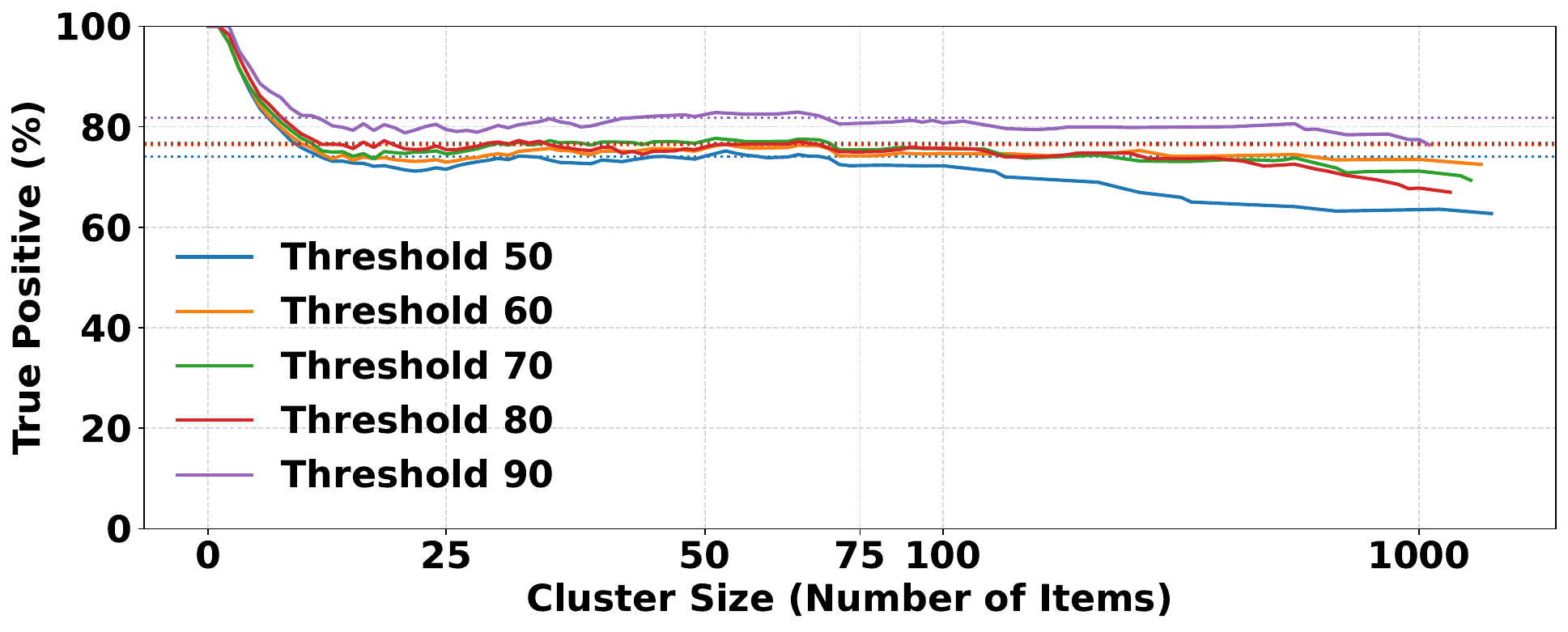}}
    \hfill
    \subfloat[J-SDhash True EBF]{\includegraphics[width=0.23\linewidth]{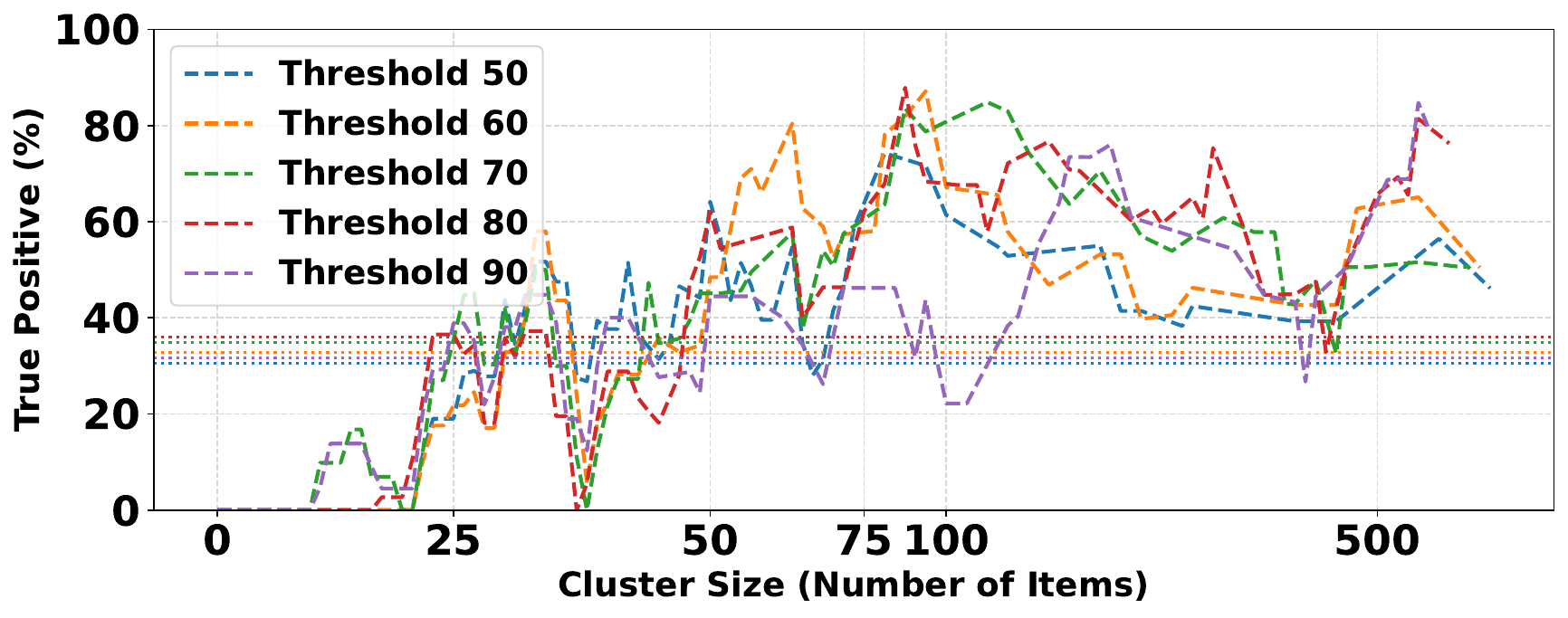}}
    \hfill
    \subfloat[J-SDhash True RTBF]{\includegraphics[width=0.23\linewidth]{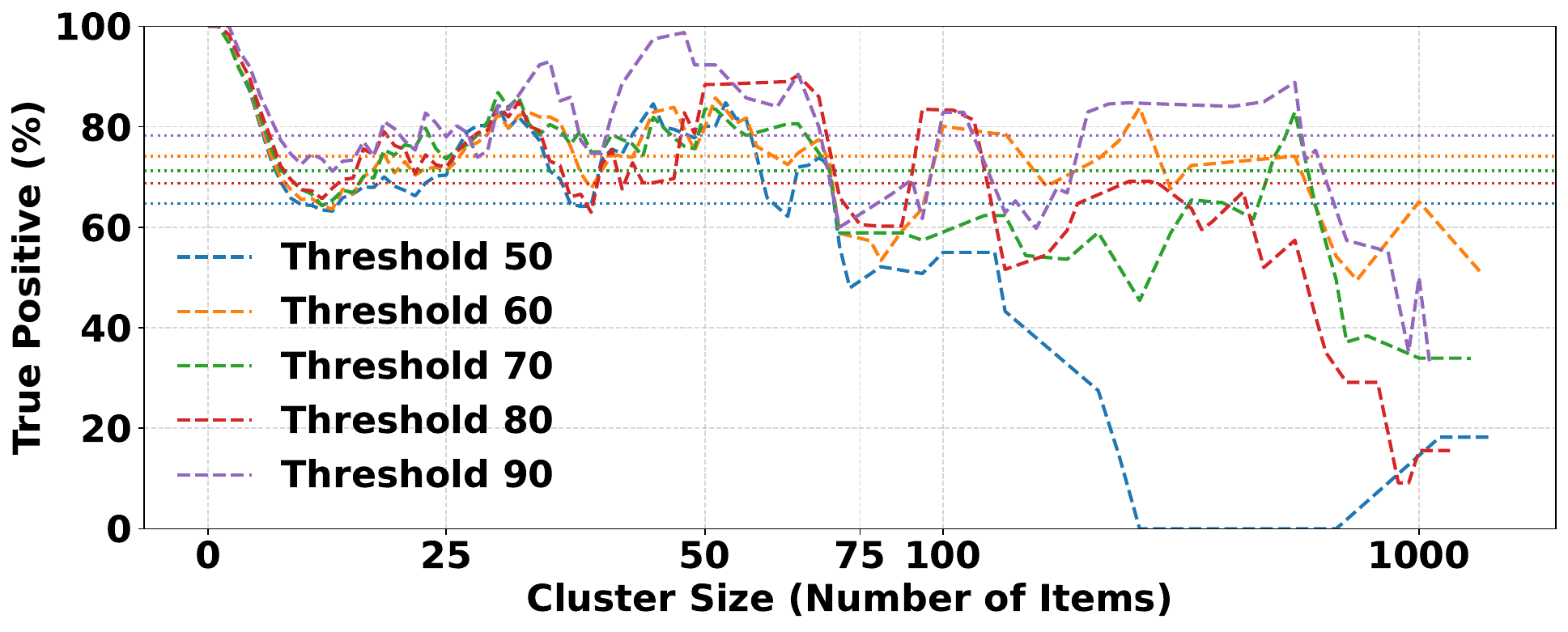}} \\[2mm]
    \subfloat[VT Avg. EBF]{\includegraphics[width=0.23\linewidth]{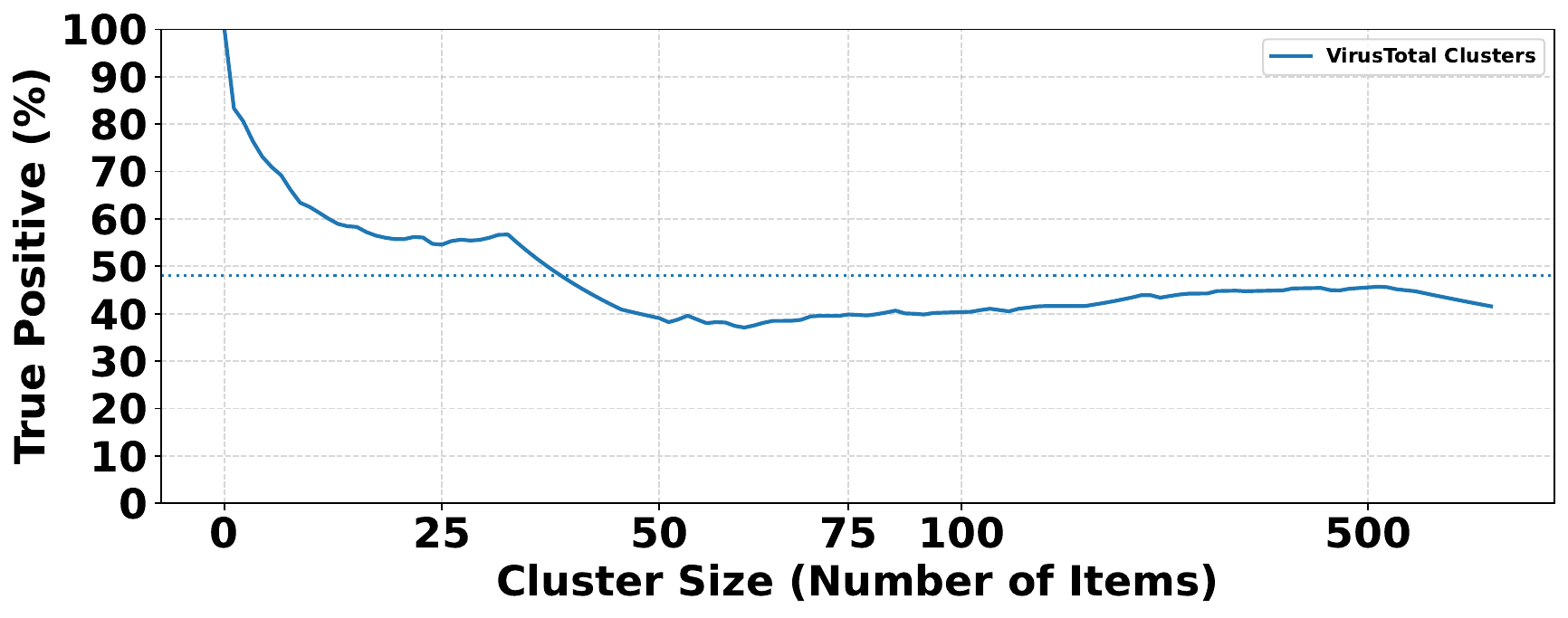}}
    \hfill
    \subfloat[VT Avg. RTBF]{\includegraphics[width=0.23\linewidth]{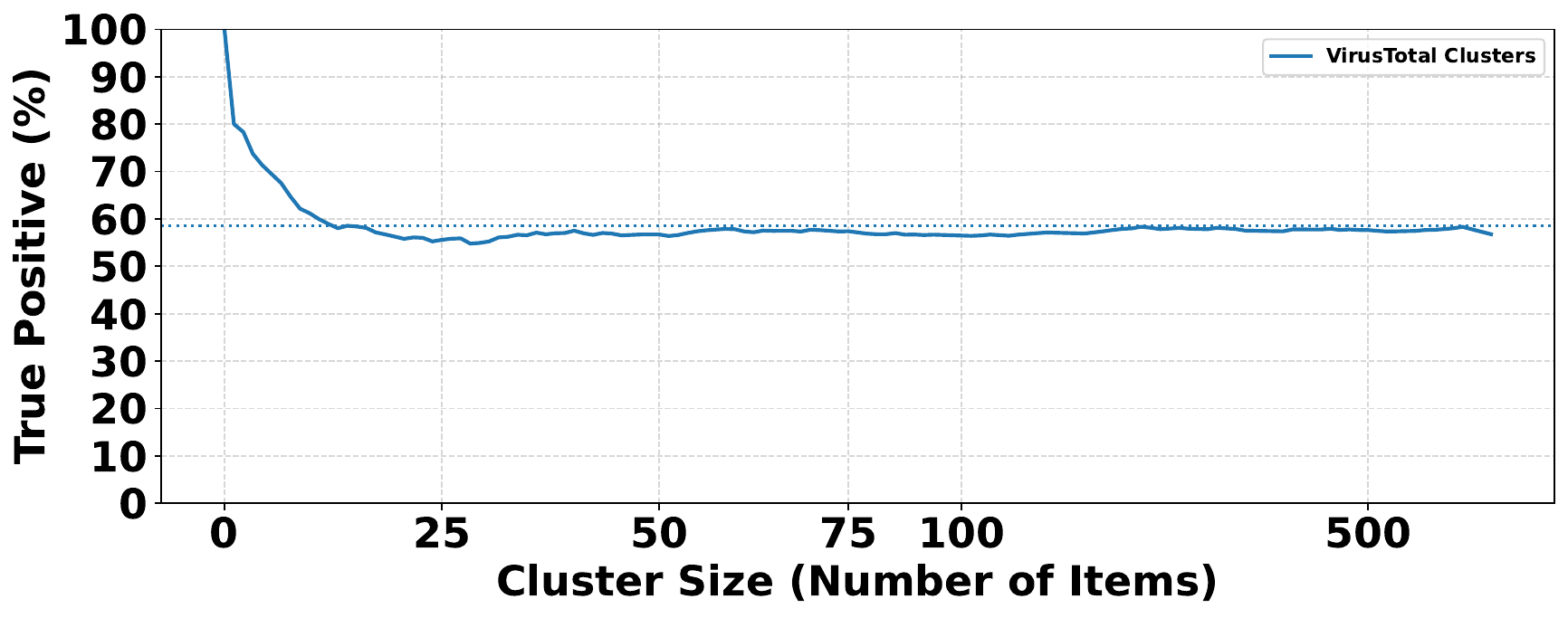}}
    \hfill
    \subfloat[VT True EBF]{\includegraphics[width=0.23\linewidth]{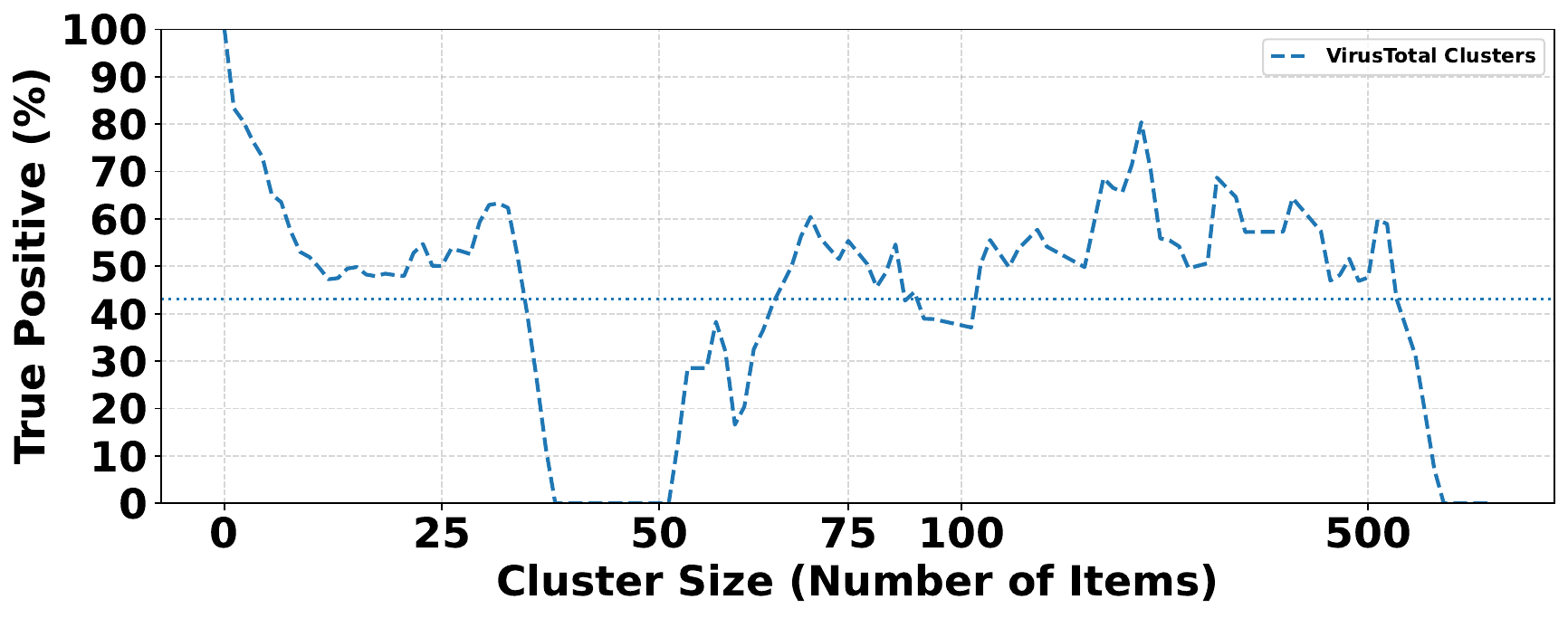}}
    \hfill
    \subfloat[VT True RTBF]{\includegraphics[width=0.23\linewidth]{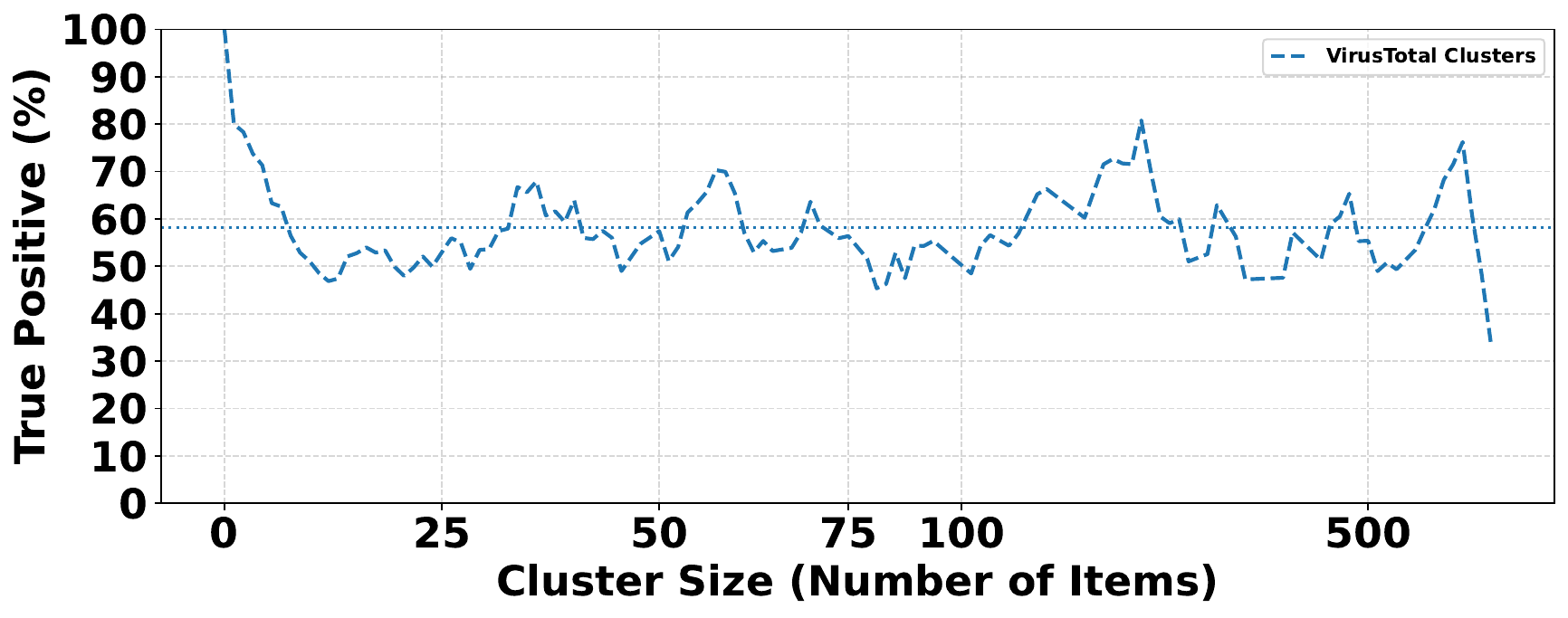}}
    \caption{AutoYara Rules by Thresholds (EBF and RTBF)}
    \label{fig:AutoYara_Thresholds_Comparison_TwoCol}
    \end{minipage}

    \vspace{2mm}
    \begin{minipage}{0.95\linewidth}
    \centering
    \subfloat[SSdeep Avg. EBF]{\includegraphics[width=0.23\linewidth]{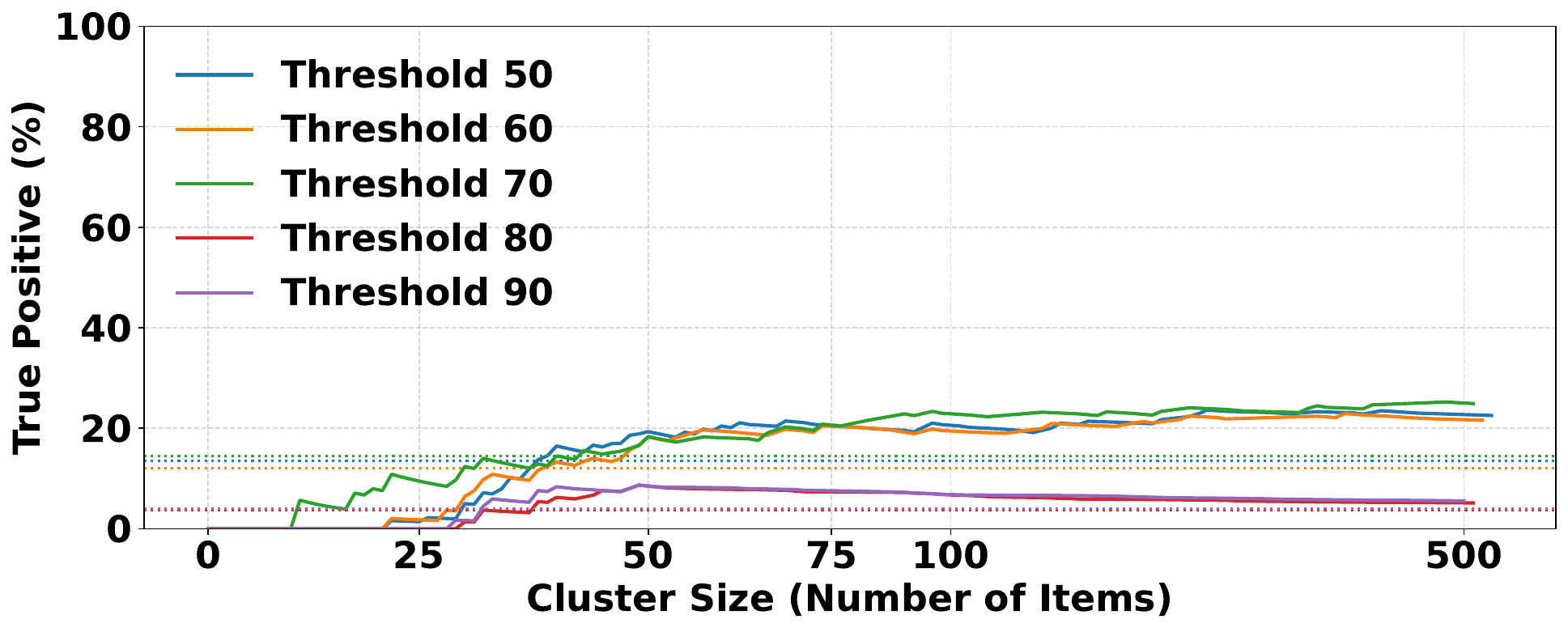}}
    \hfill
    \subfloat[SSdeep Avg. RTBF]{\includegraphics[width=0.23\linewidth]{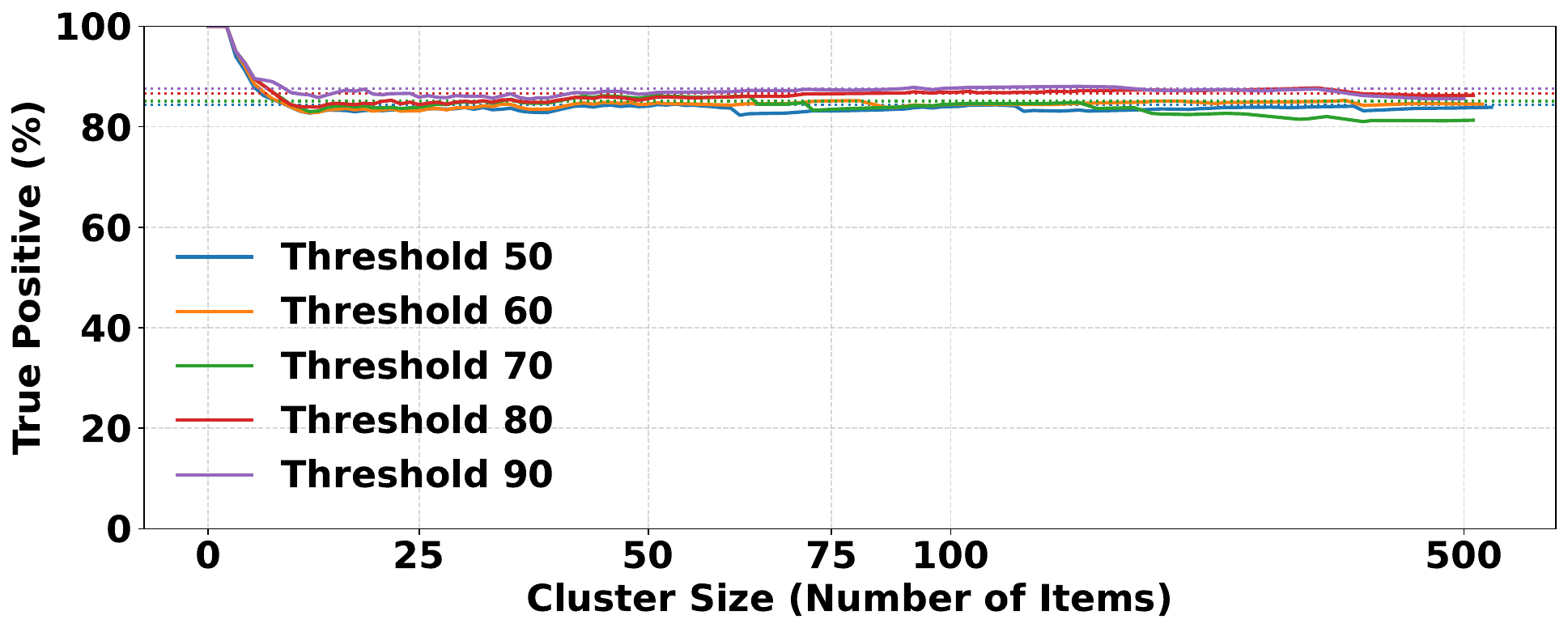}}
    \hfill
    \subfloat[SSdeep True EBF]{\includegraphics[width=0.23\linewidth]{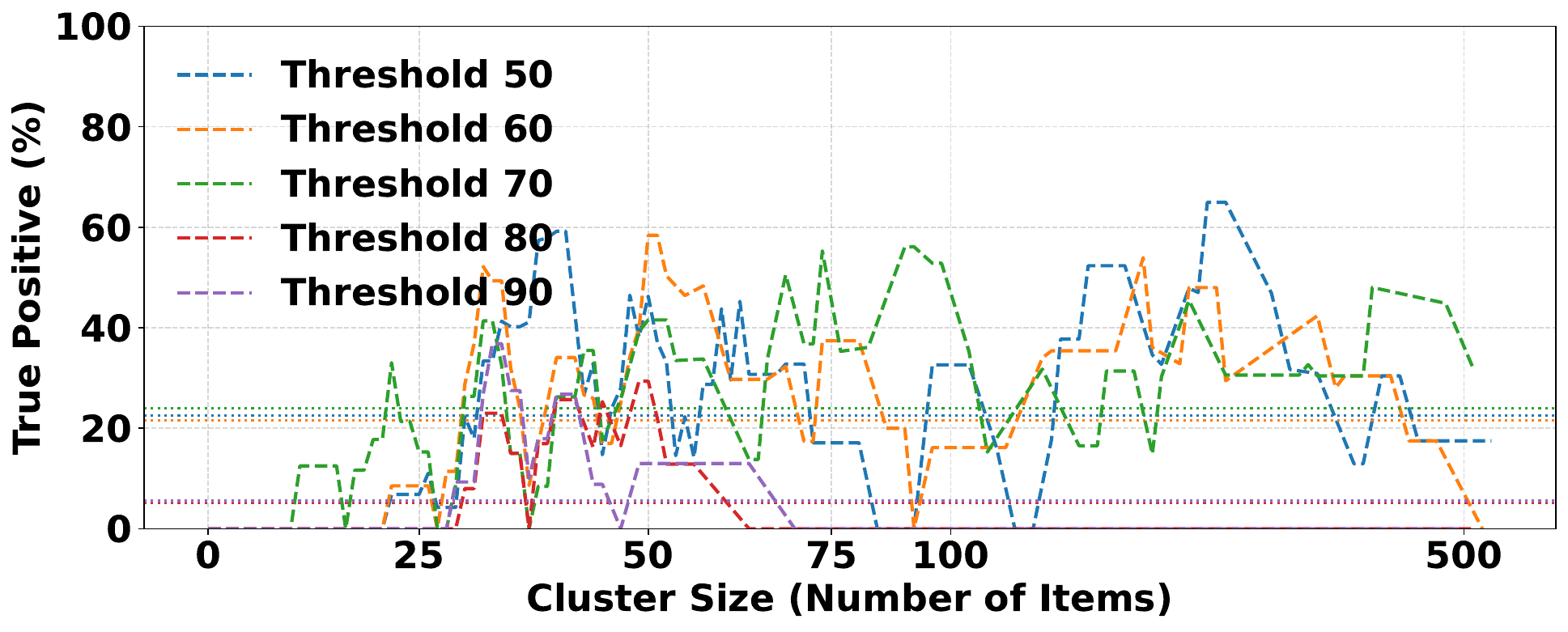}}
    \hfill
    \subfloat[SSdeep True RTBF]{\includegraphics[width=0.23\linewidth]{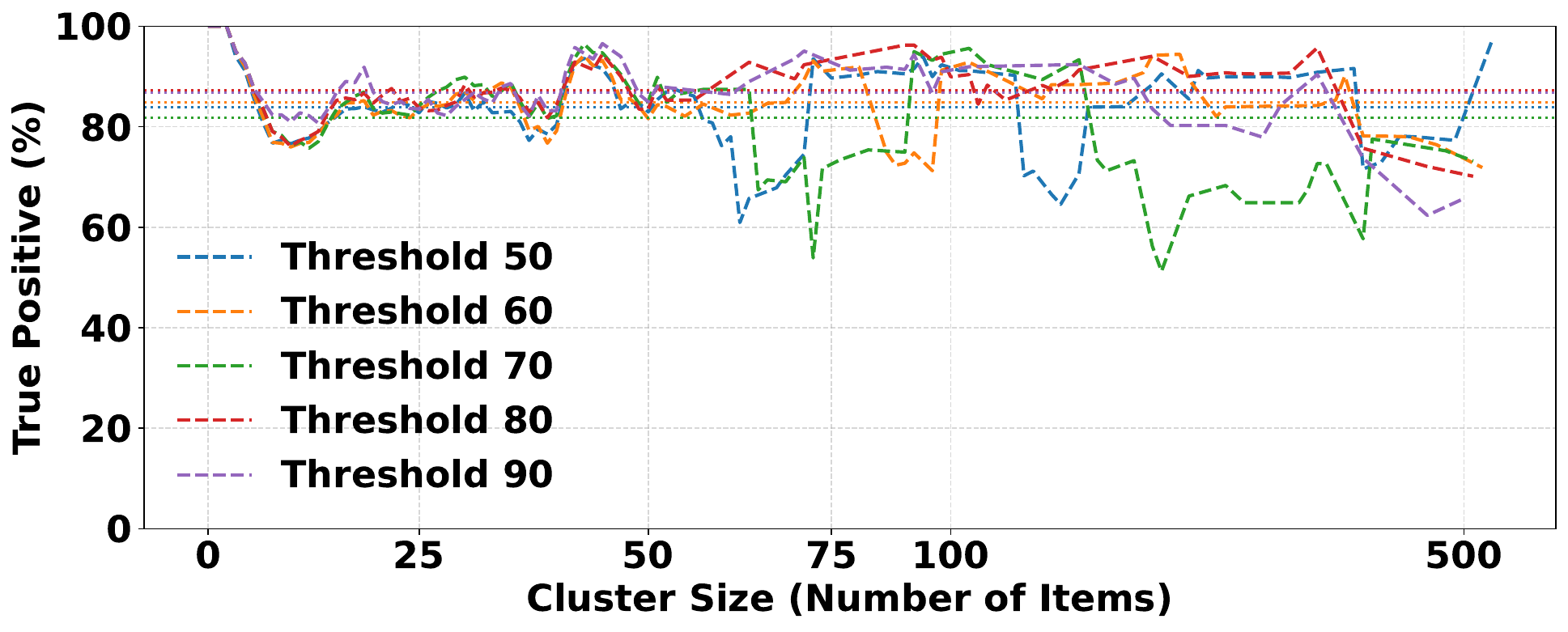}} \\[2mm]
    \subfloat[J-SDhash Avg. EBF]{\includegraphics[width=0.23\linewidth]{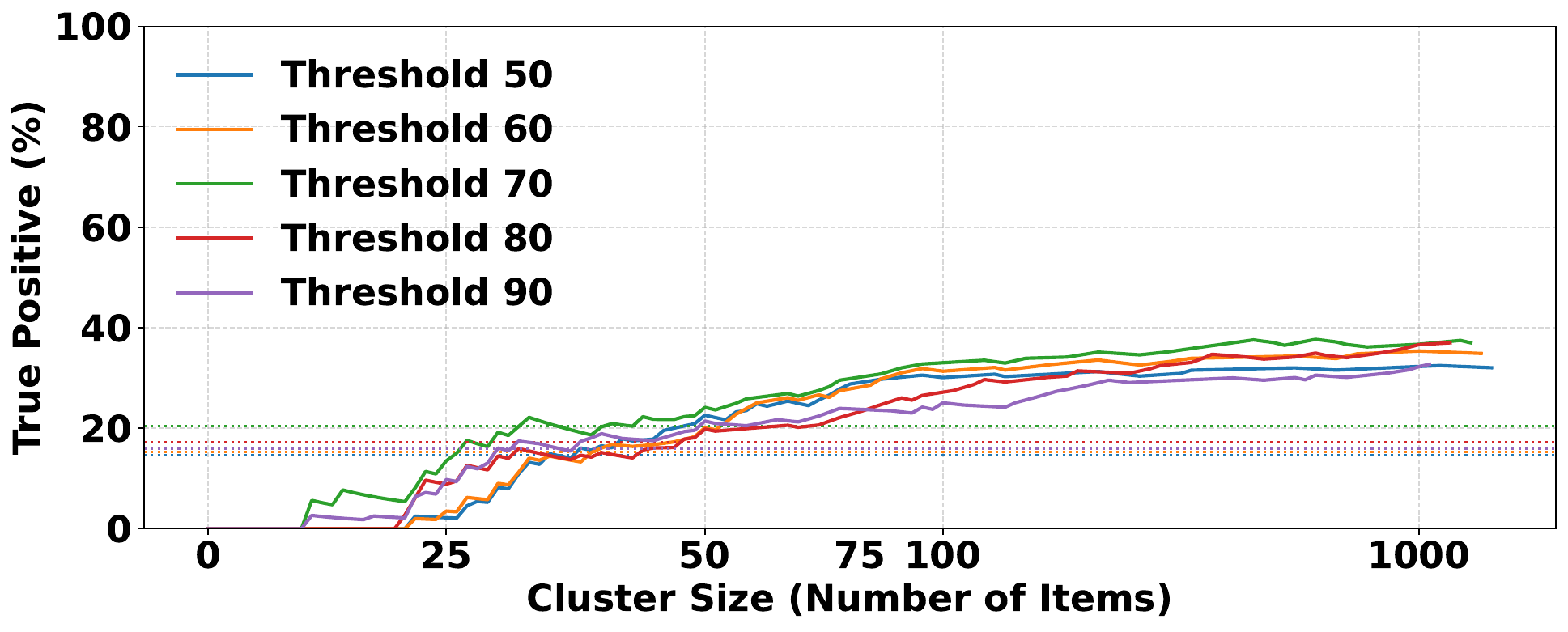}}
    \hfill
    \subfloat[J-SDhash Avg. RTBF]{\includegraphics[width=0.23\linewidth]{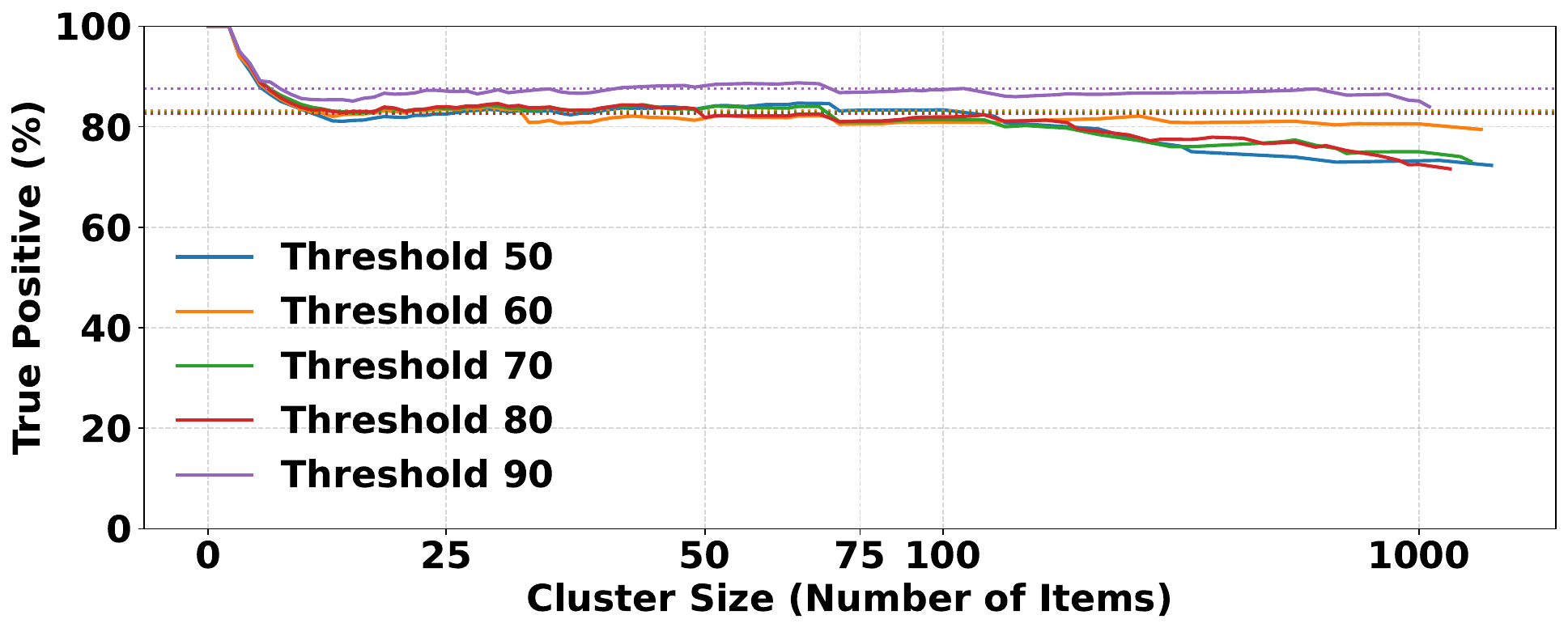}}
    \hfill
    \subfloat[J-SDhash True EBF]{\includegraphics[width=0.23\linewidth]{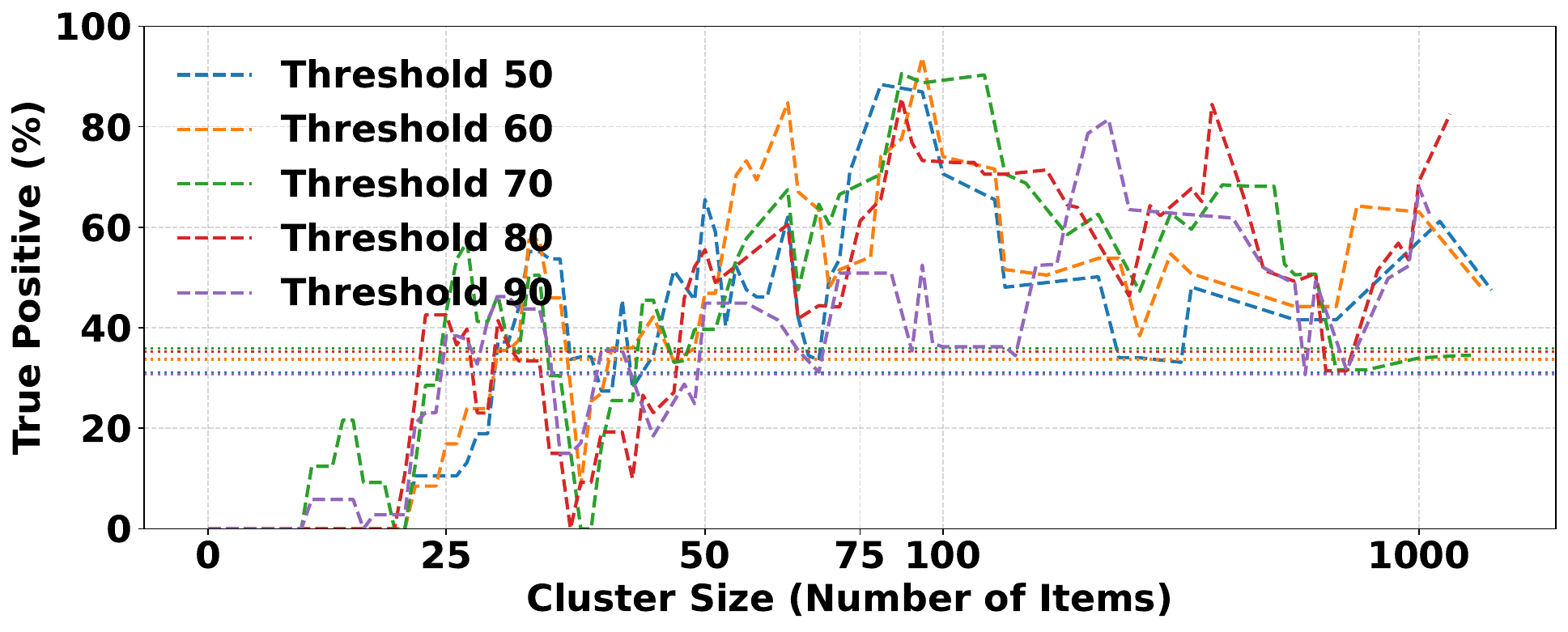}}
    \hfill
    \subfloat[J-SDhash True RTBF]{\includegraphics[width=0.23\linewidth]{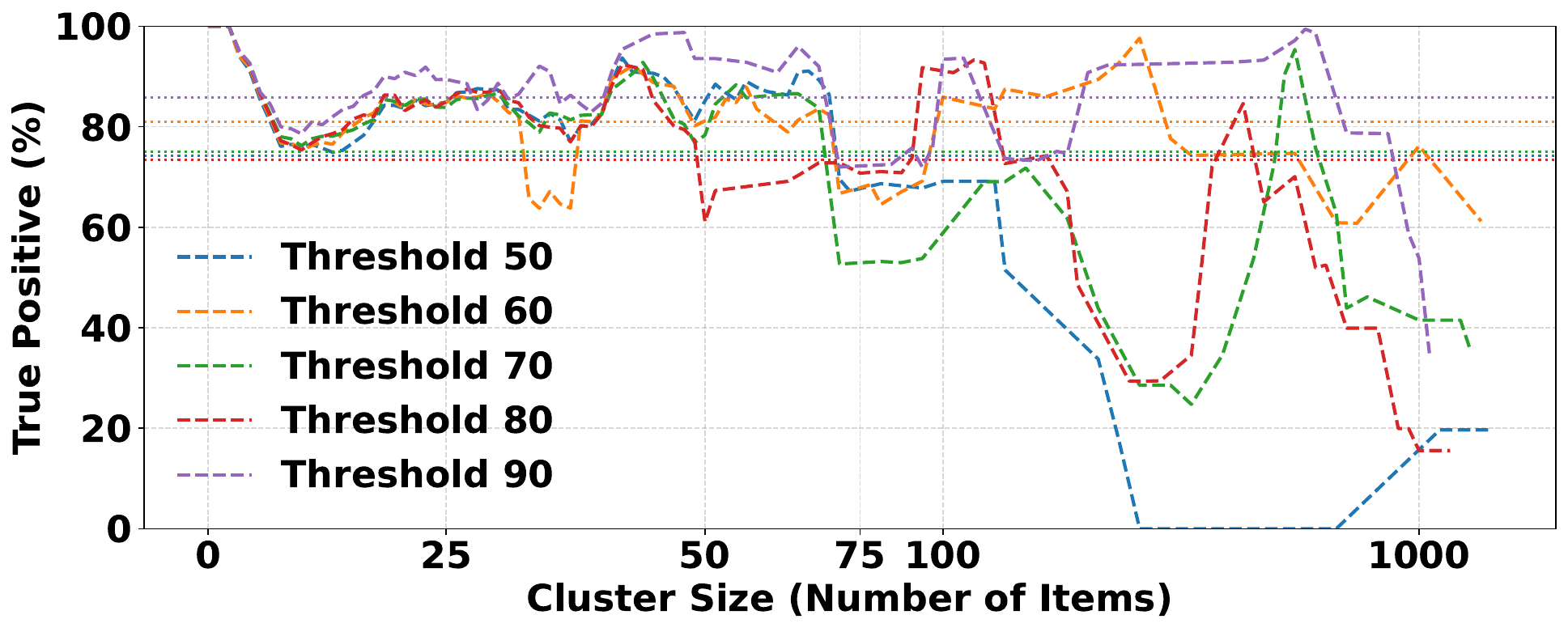}} \\[2mm]
    \subfloat[VT Avg. EBF]{\includegraphics[width=0.23\linewidth]{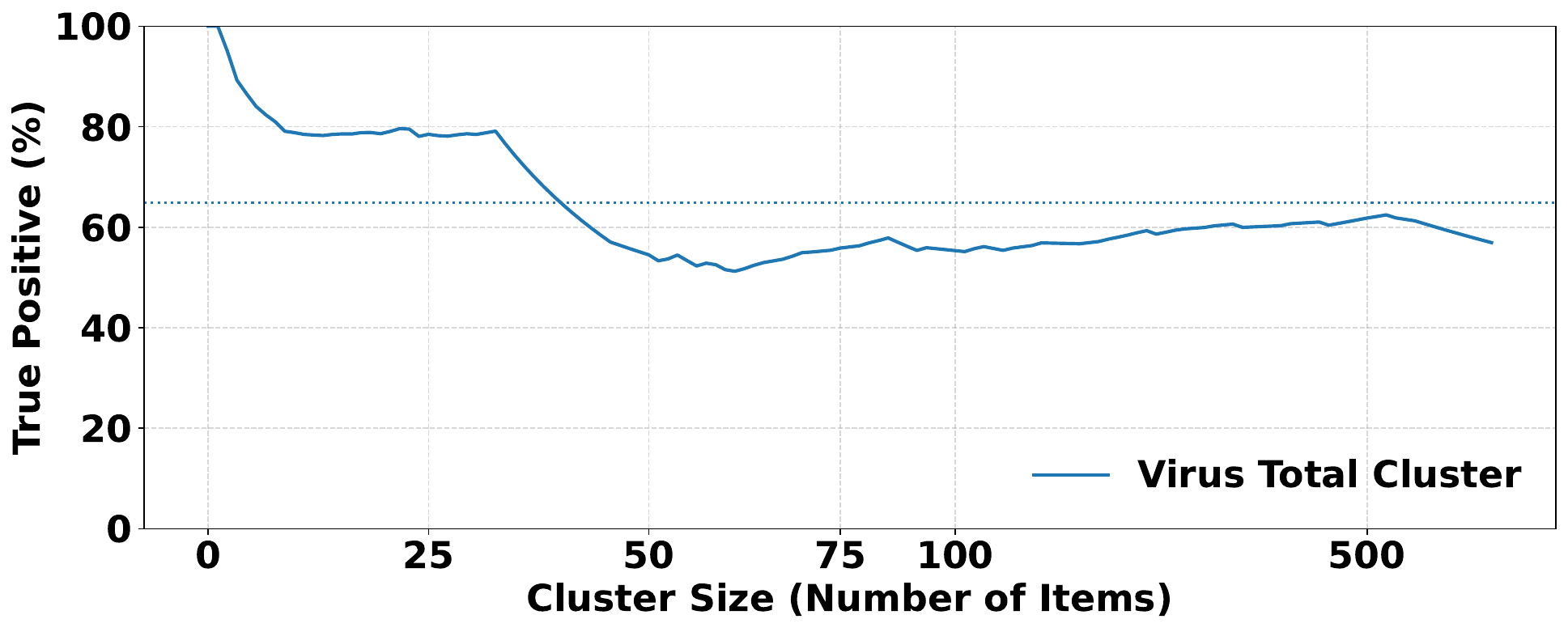}}
    \hfill
    \subfloat[VT Avg. RTBF]{\includegraphics[width=0.23\linewidth]{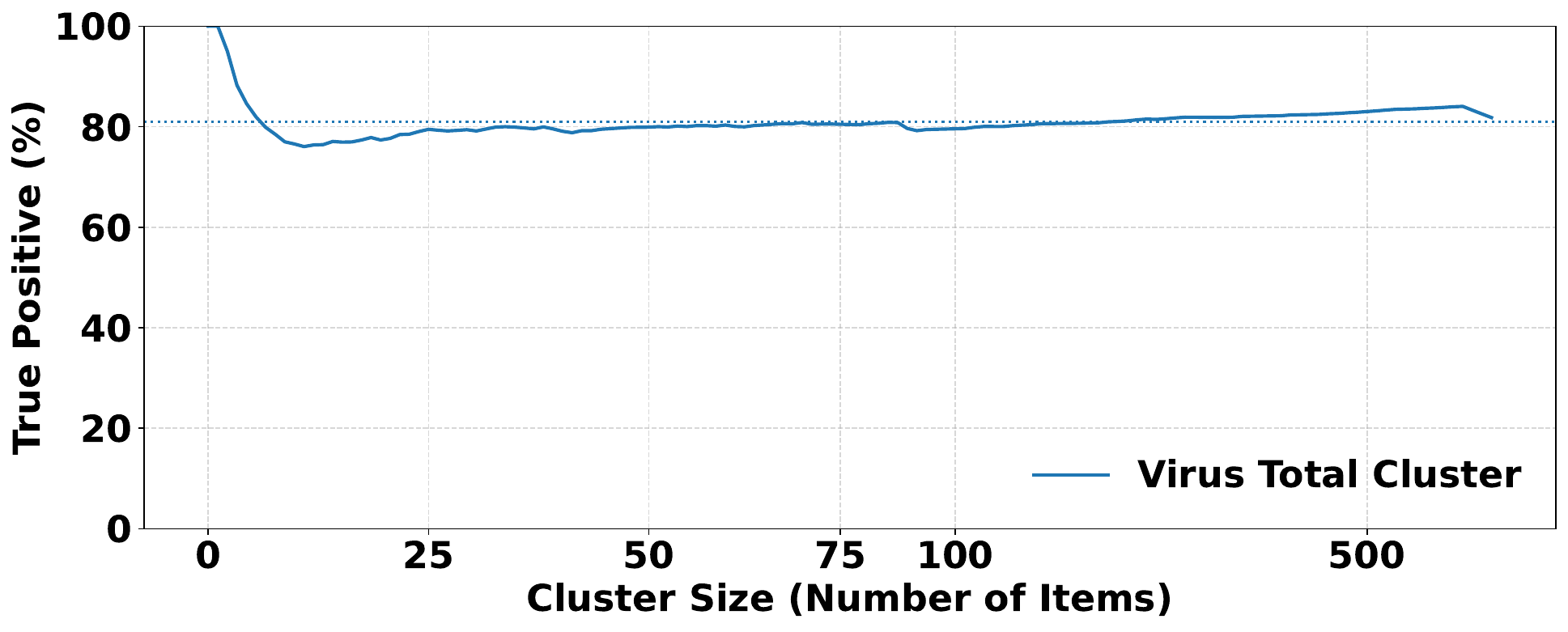}}
    \hfill
    \subfloat[VT True EBF]{\includegraphics[width=0.23\linewidth]{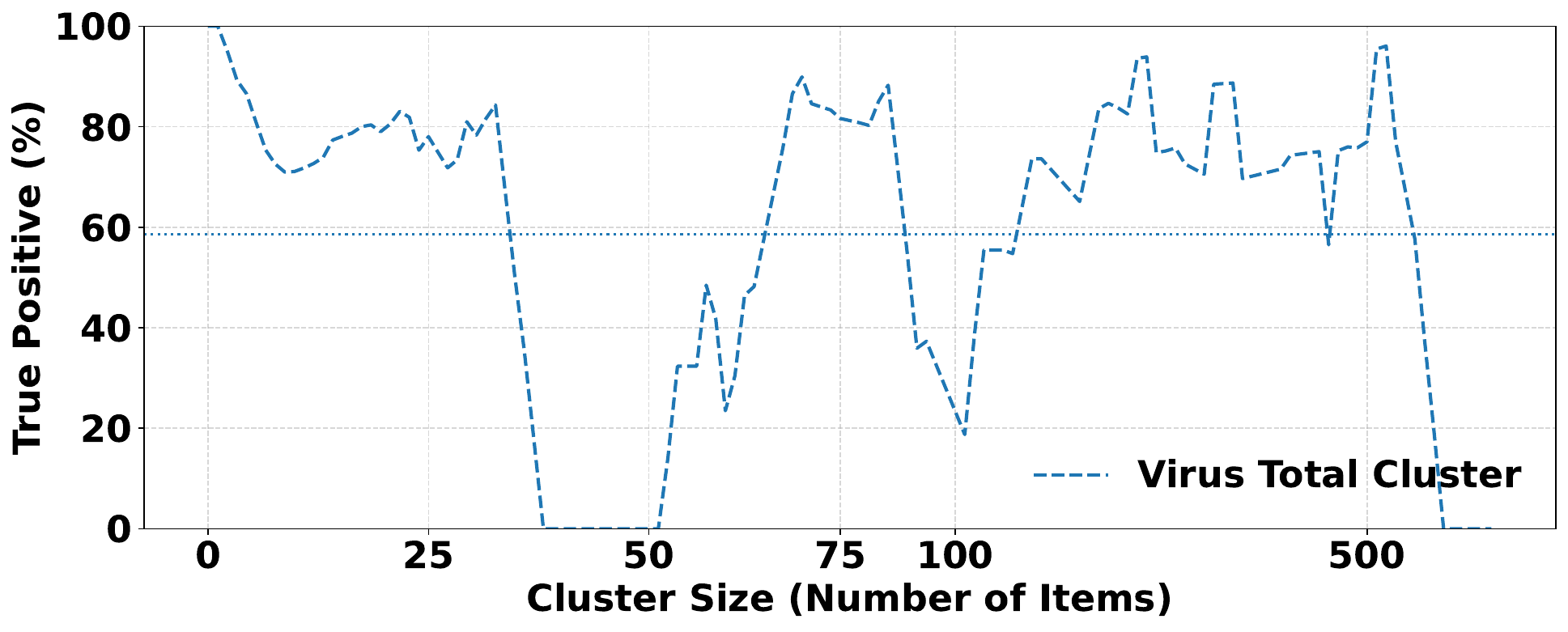}}
    \hfill
    \subfloat[VT True RTBF]{\includegraphics[width=0.23\linewidth]{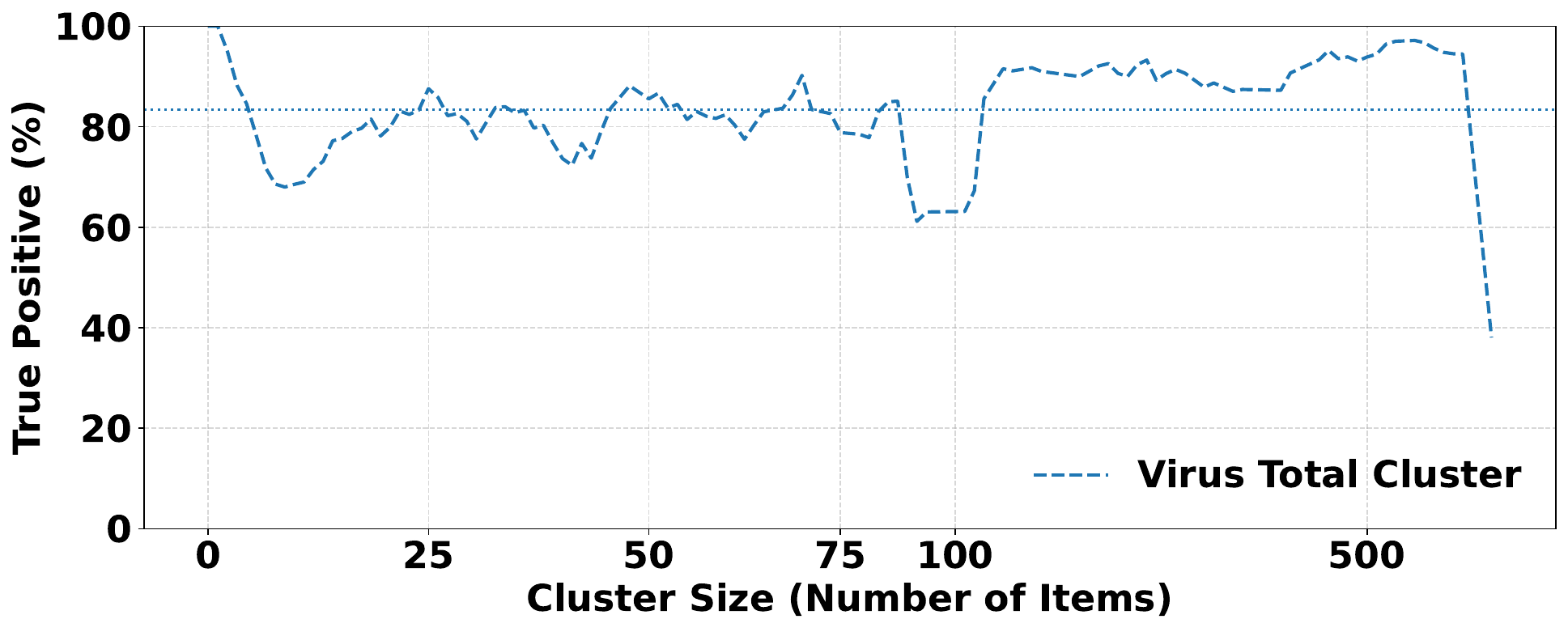}}
    \caption{AutoPYara Rules by Thresholds (EBF and RTBF)}
    \label{fig:AutoPYara_Thresholds_Comparison_TwoCol}
    \end{minipage}

    \vspace{2mm}
    \begin{minipage}{0.95\linewidth}
    \centering
    \scalebox{0.8}{
    \begin{tabular}{ccc}
    \includegraphics[width=0.32\linewidth]{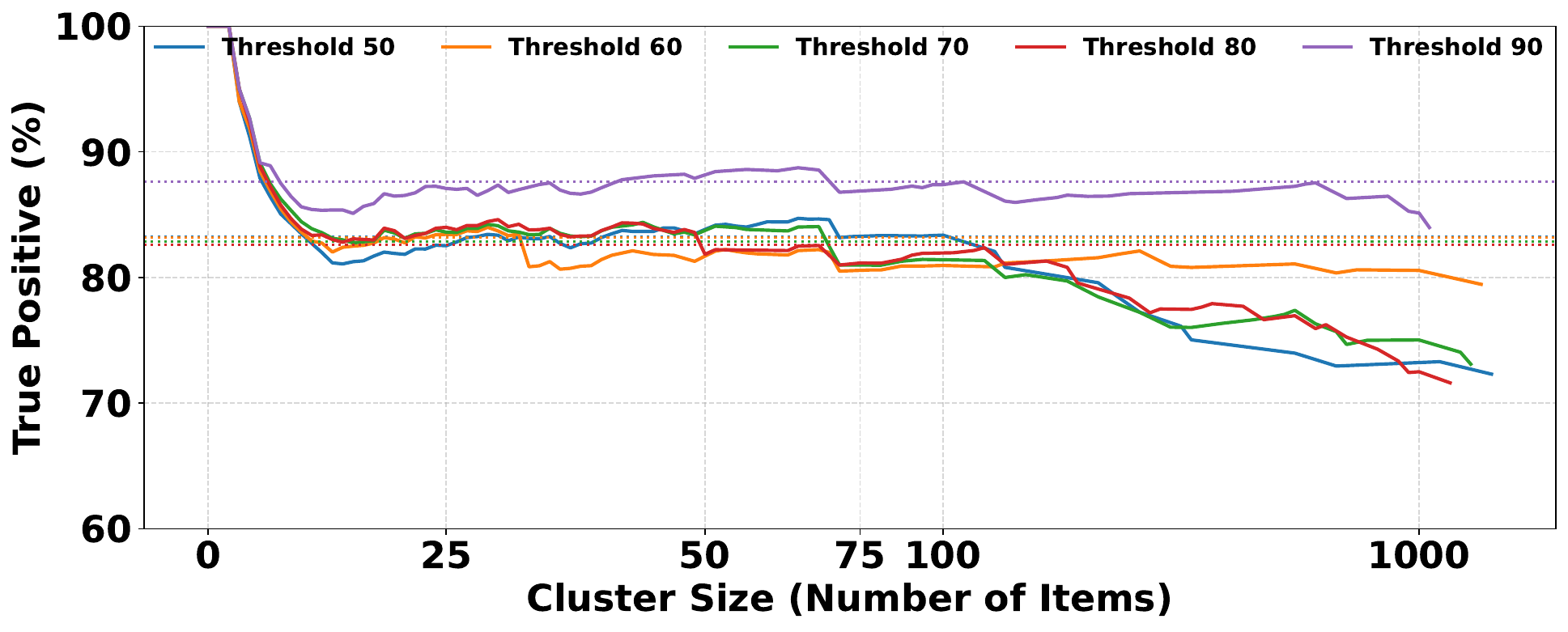} &
    \includegraphics[width=0.32\linewidth]{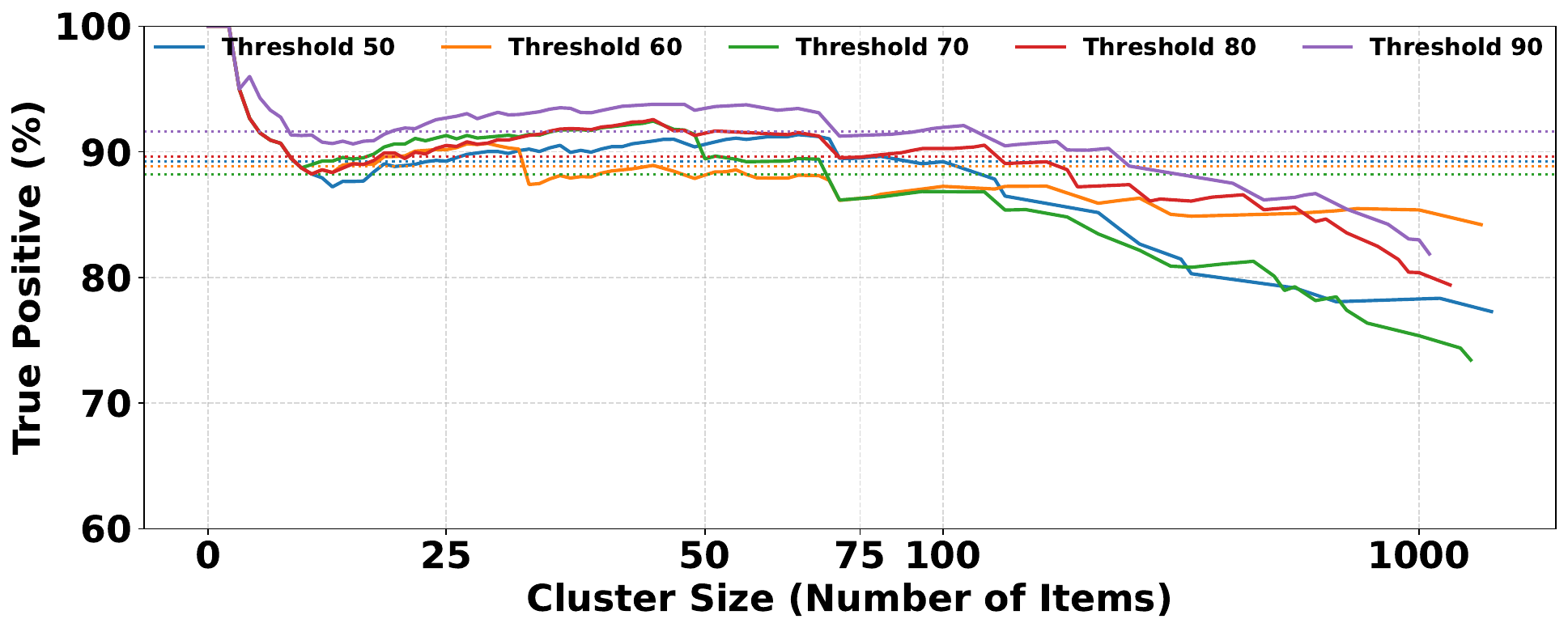} &
    \includegraphics[width=0.32\linewidth]{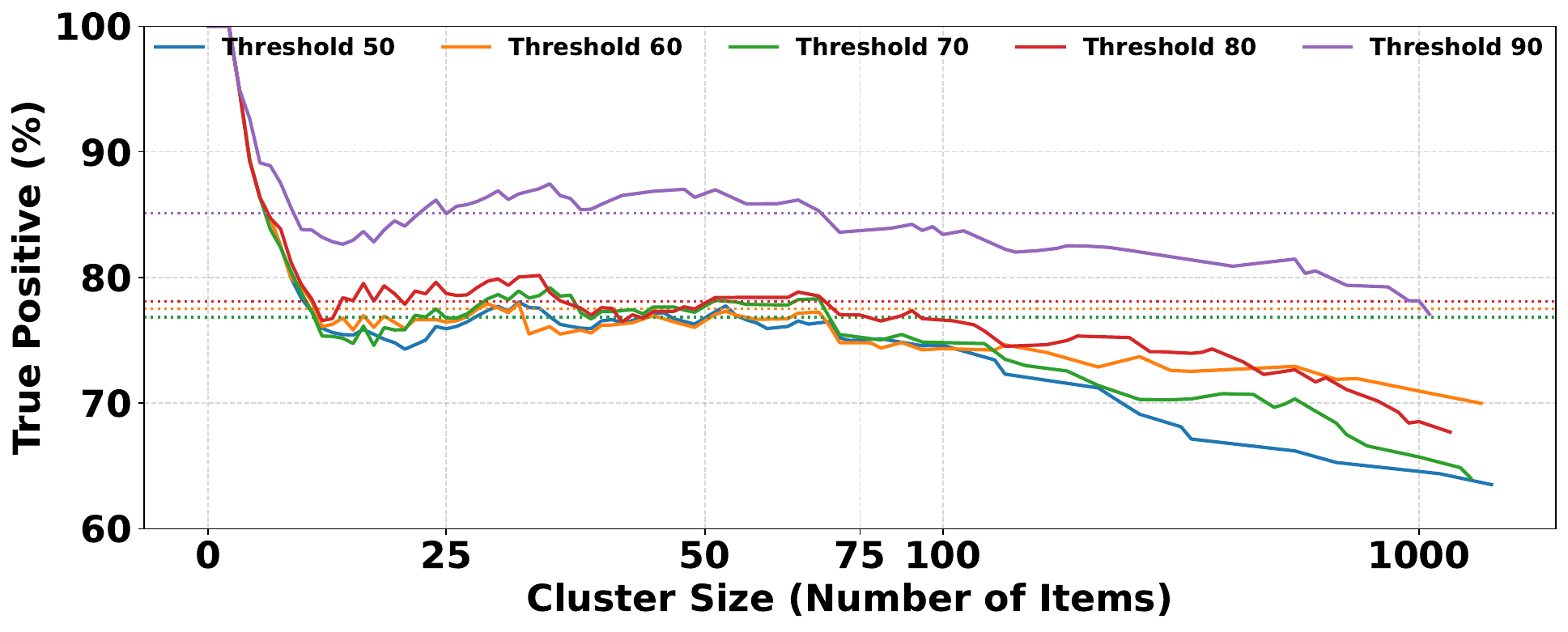} \\
    \includegraphics[width=0.32\linewidth]{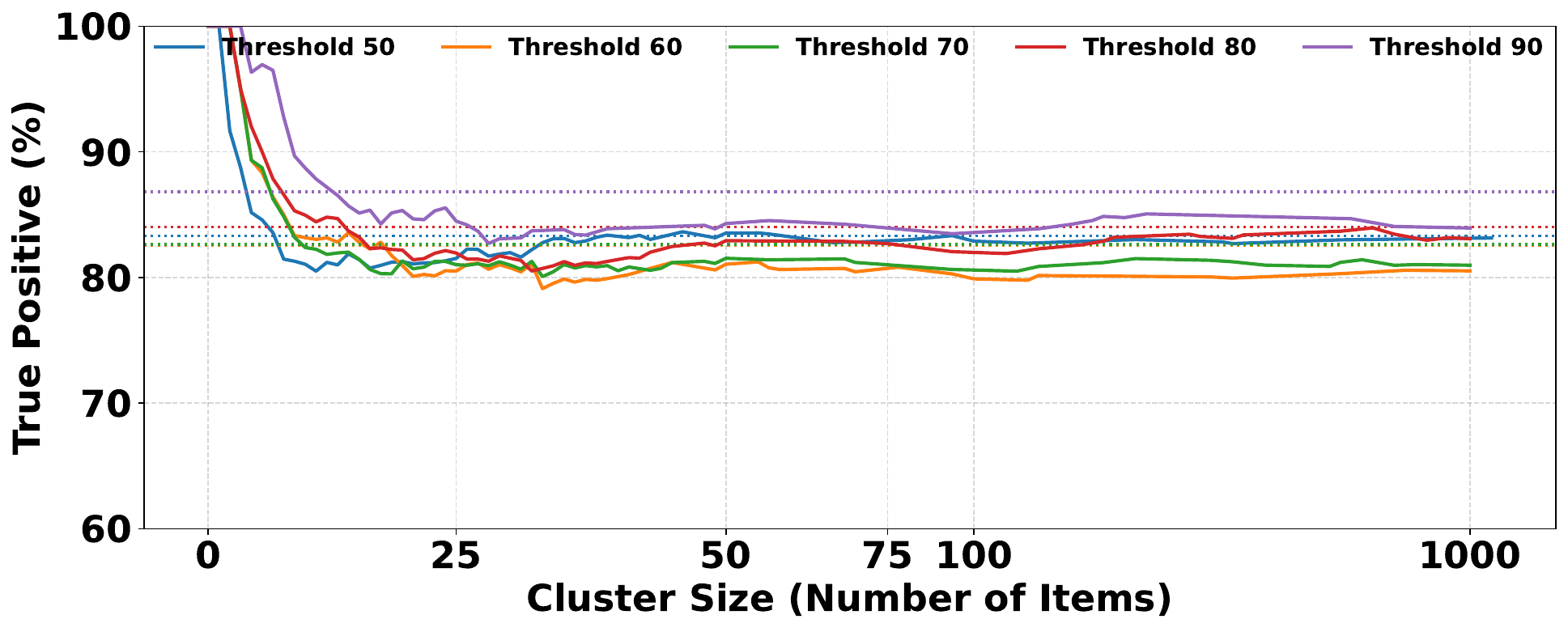} &
    \includegraphics[width=0.32\linewidth]{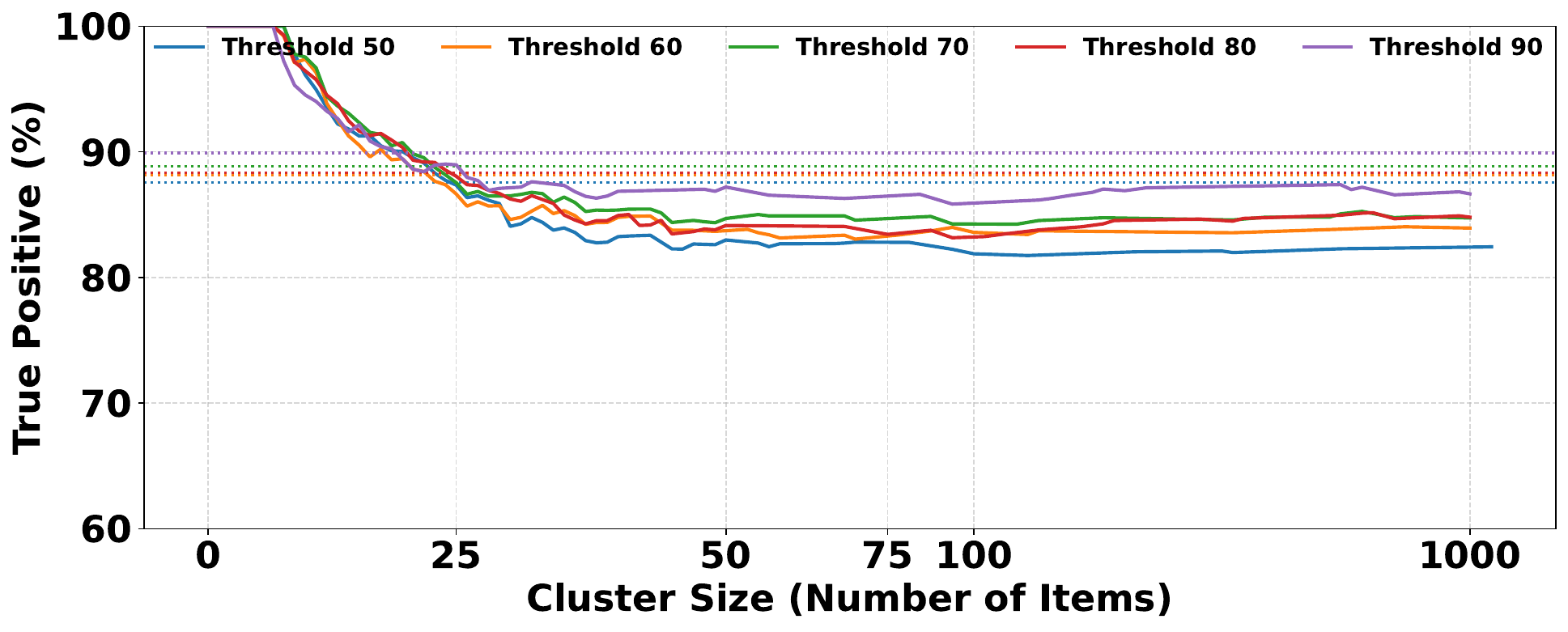} &
    \includegraphics[width=0.32\linewidth]{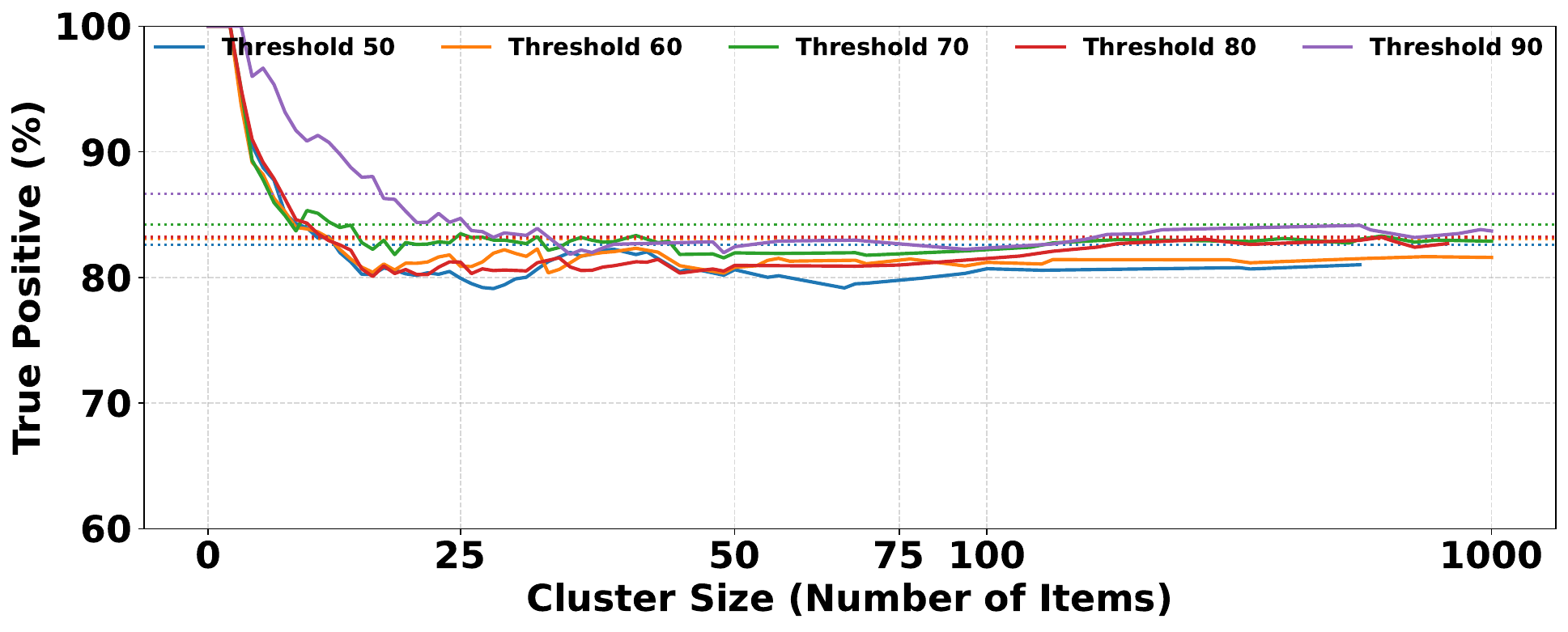} \\
    \includegraphics[width=0.32\linewidth]{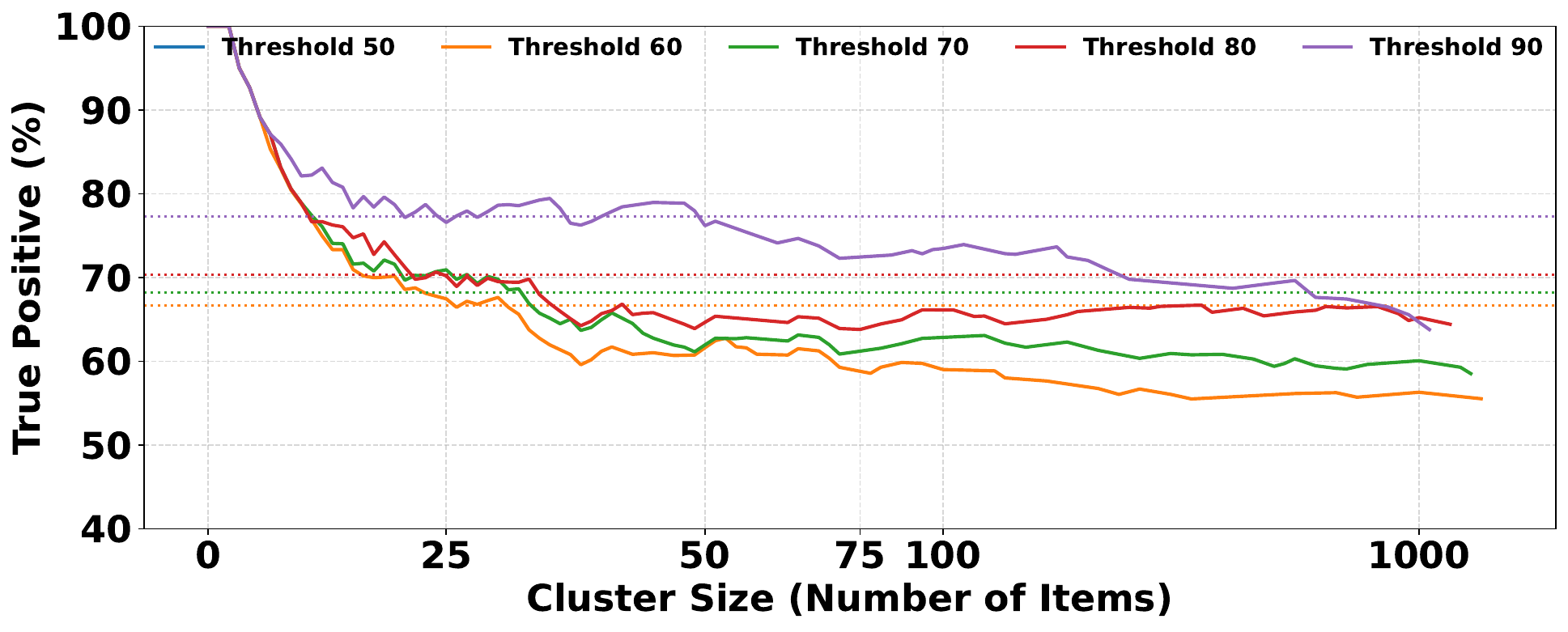} &
    \includegraphics[width=0.32\linewidth]{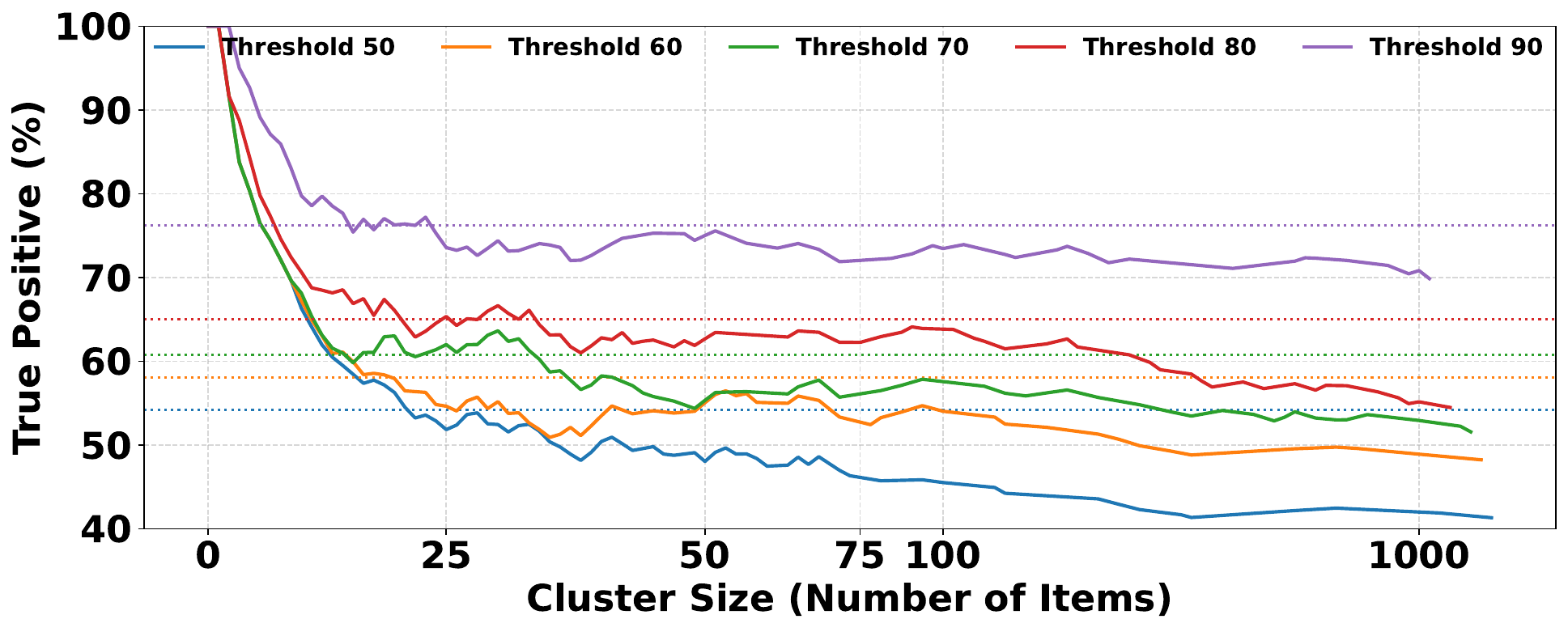} &
    \includegraphics[width=0.32\linewidth]{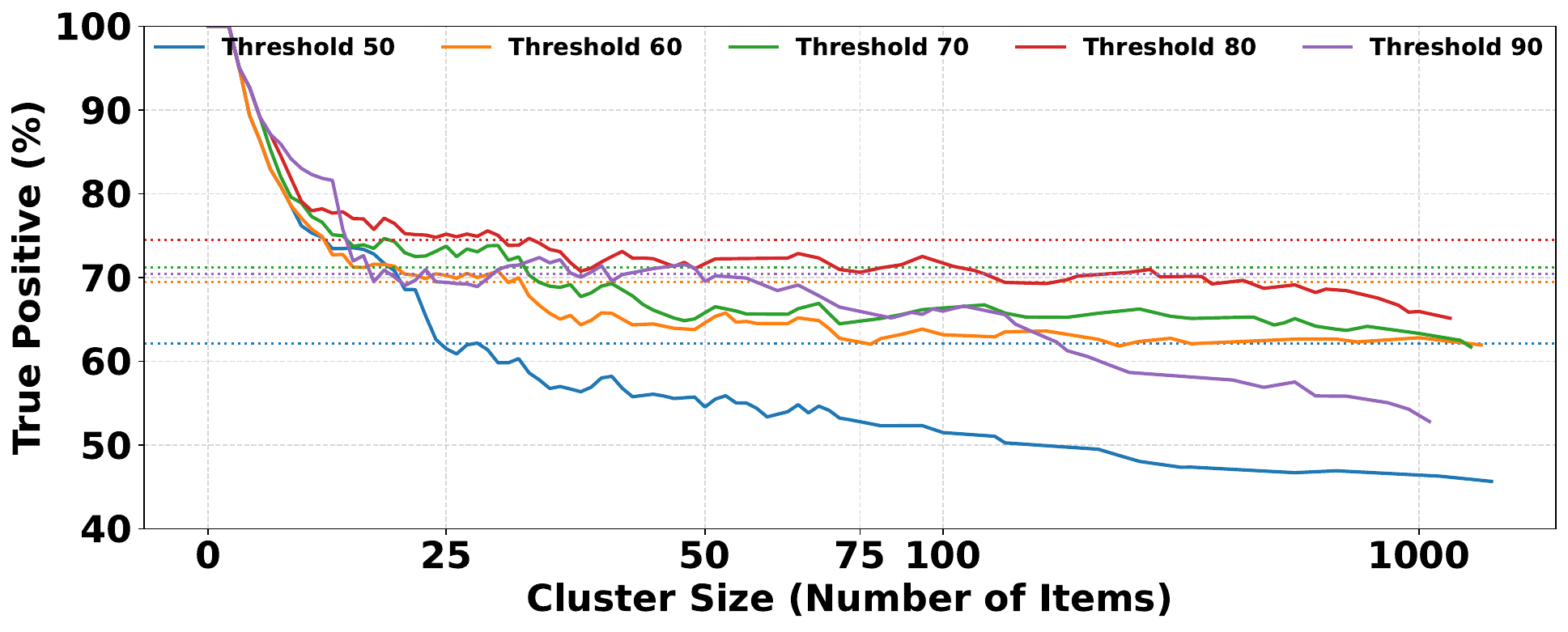}
    \end{tabular}}
    \caption{Comparing AutoPYara in Different Configurations for SDhash}
    \label{fig:autopyara_sdhash_Kimpact}
    \end{minipage}
\end{figure*}

\clearpage

\setlength{\tabcolsep}{1pt} 
\renewcommand{\arraystretch}{1} 

\begin{table*}
    \centering
    \makebox[\textwidth][c]{%
    \begin{tabularx}{\textwidth}{*{5}{>{\centering\arraybackslash}X}}
        \textbf{TH 50} & \textbf{TH 60} & \textbf{TH 70} & \textbf{TH 80} & \textbf{TH 90} \\
                \hline

        \multicolumn{5}{c}{\textbf{AutoYara}} \\
        \includegraphics[width=\linewidth]{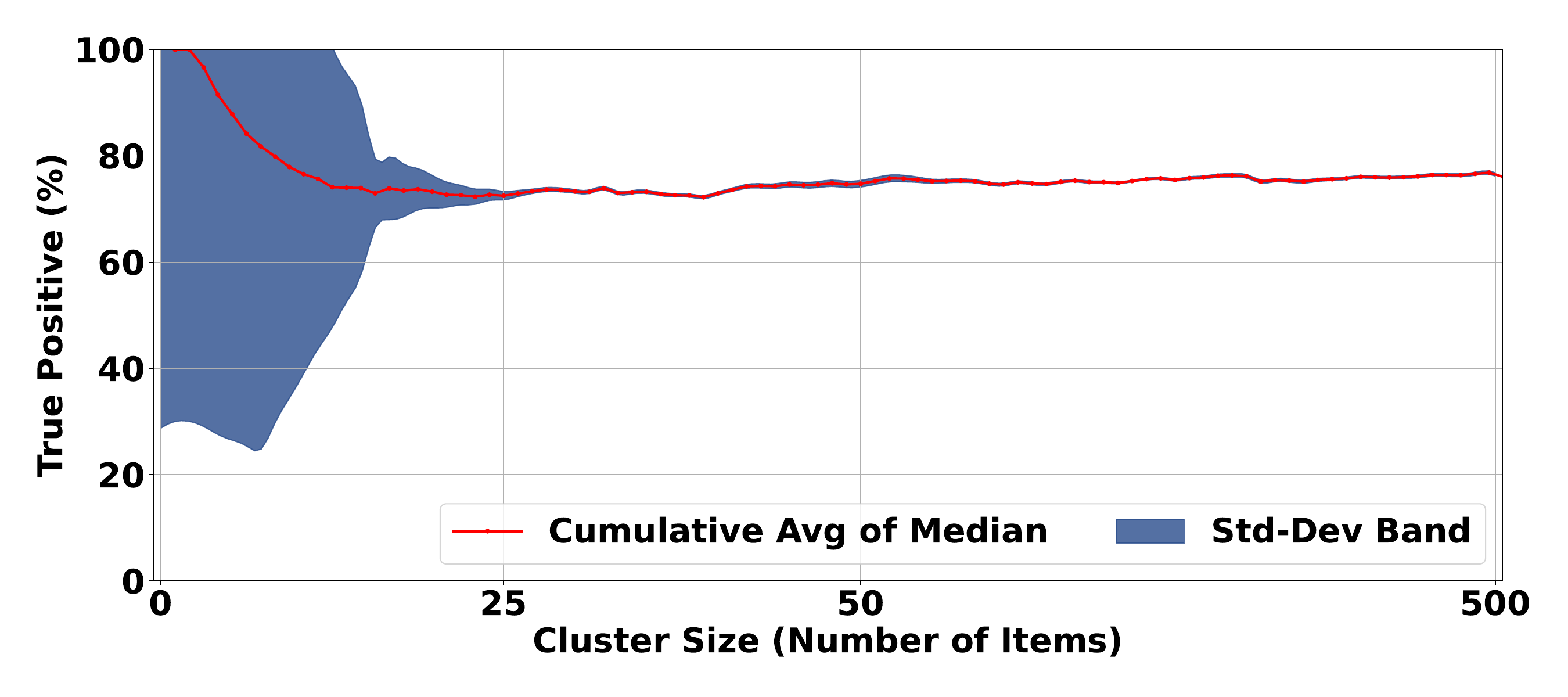} &
        \includegraphics[width=\linewidth]{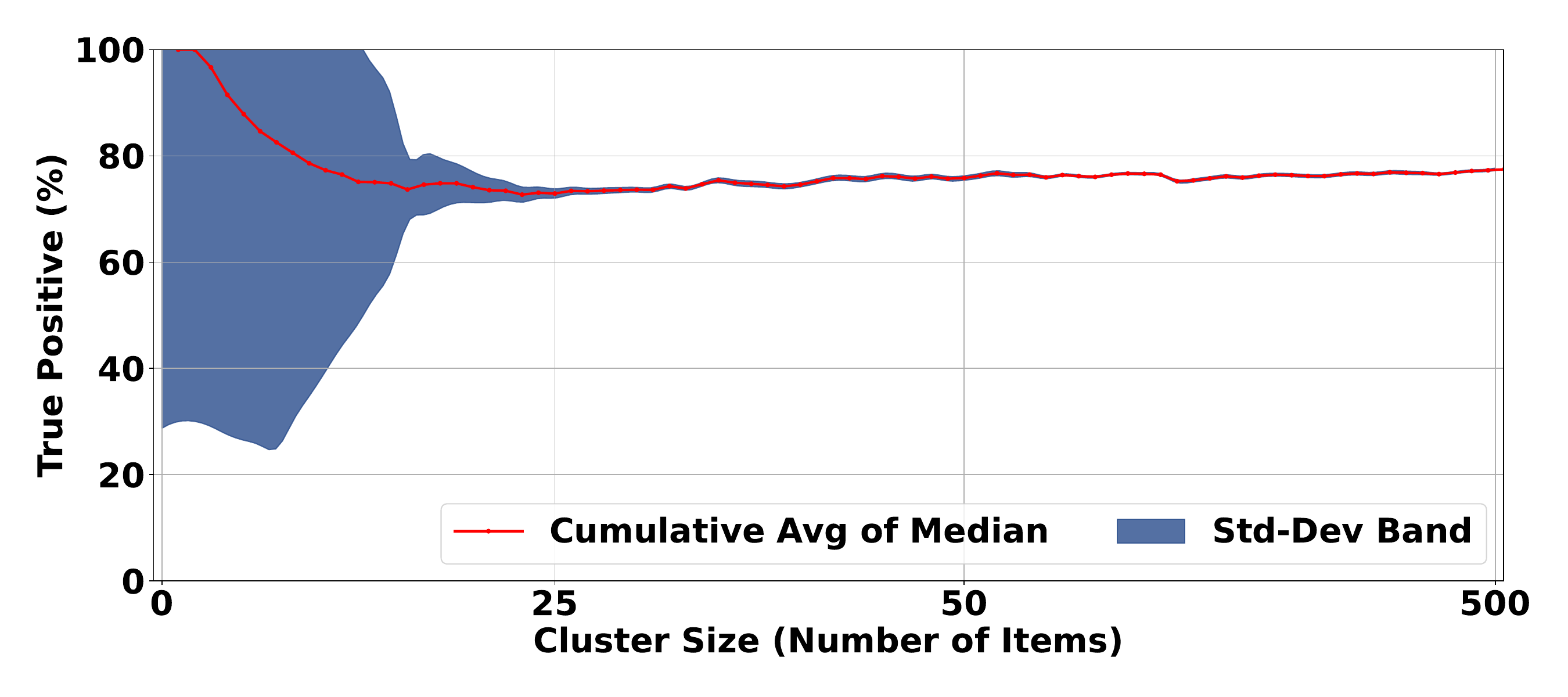} &
        \includegraphics[width=\linewidth]{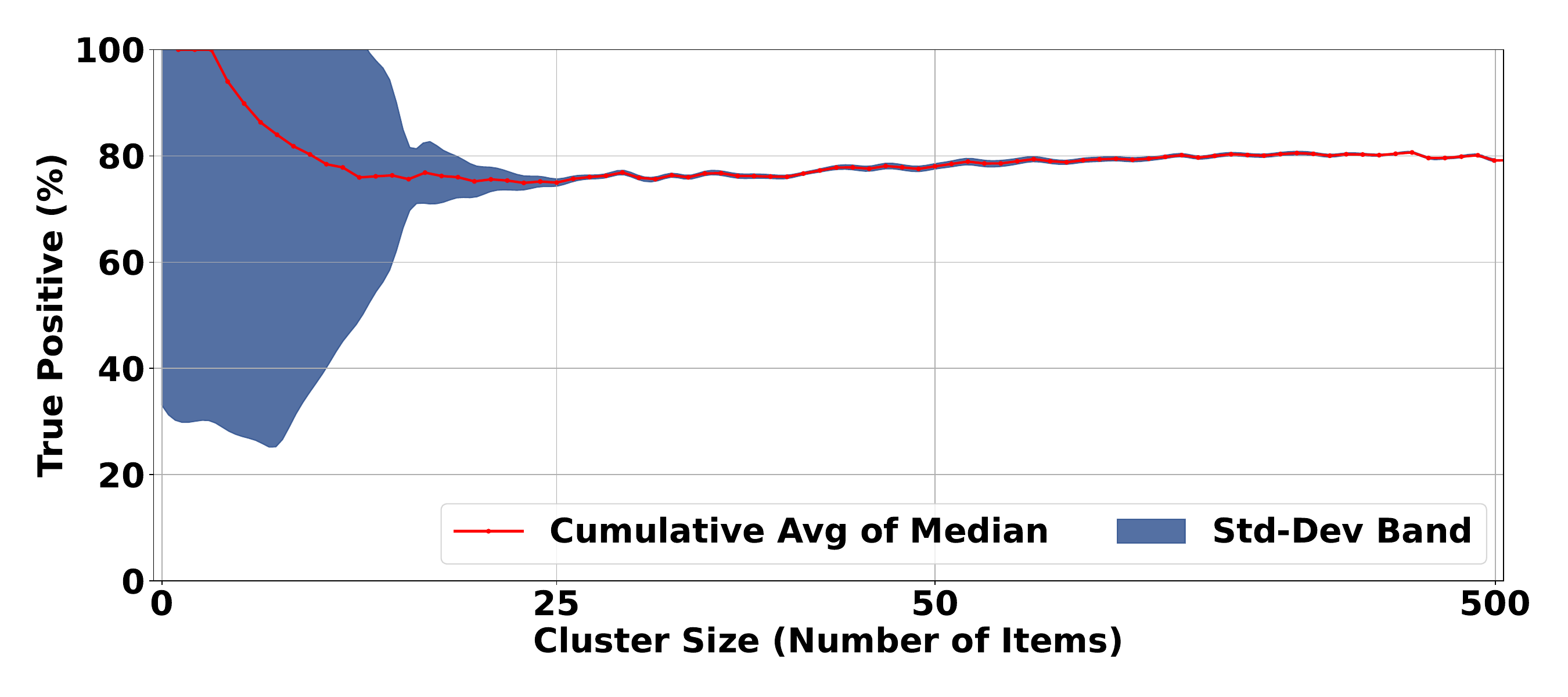} &
        \includegraphics[width=\linewidth]{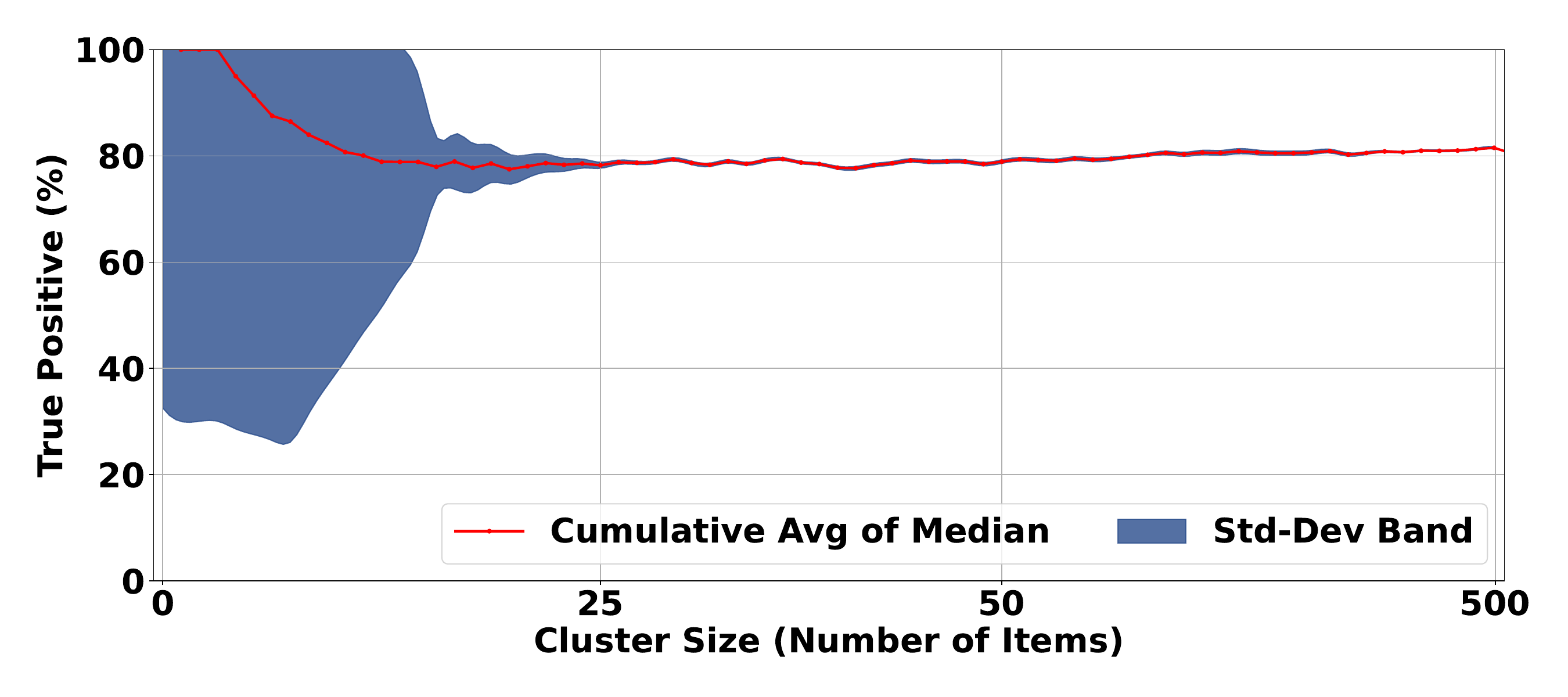} &
        \includegraphics[width=\linewidth]{Figures/AutoYara/RTBF/VairiationPlots/Threshold/SSdeep_Th90.pdf} \\
        \multicolumn{5}{c}{\textbf{AutoPYara}} \\
        \includegraphics[width=\linewidth]{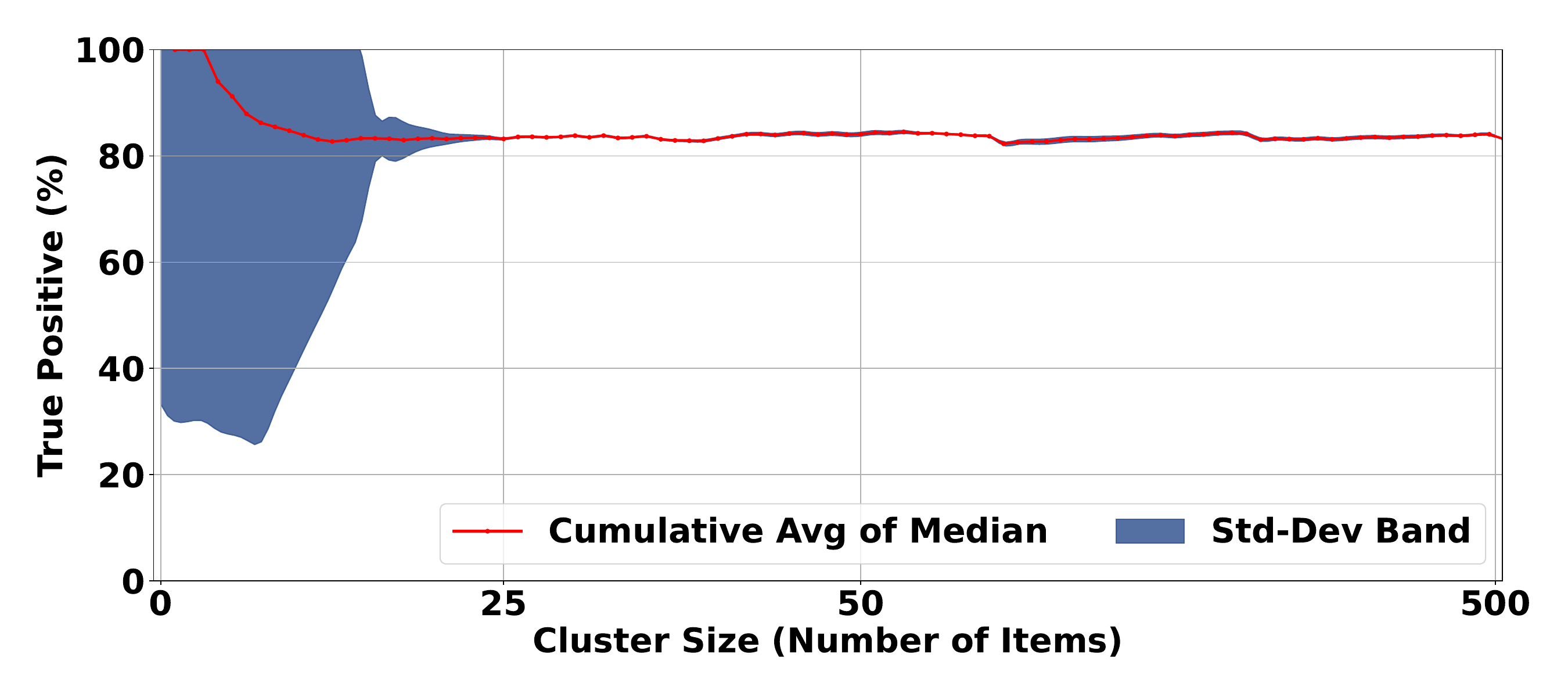} &
        \includegraphics[width=\linewidth]{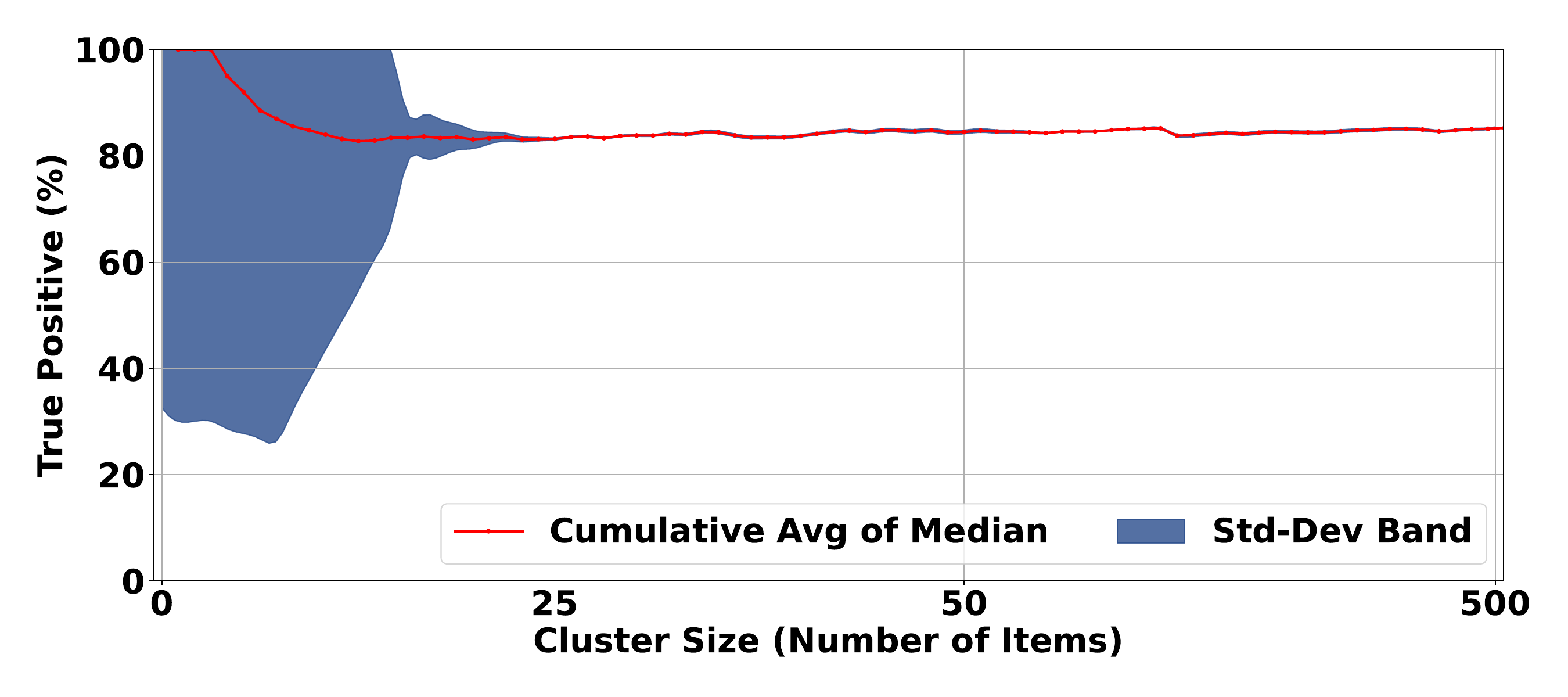} &
        \includegraphics[width=\linewidth]{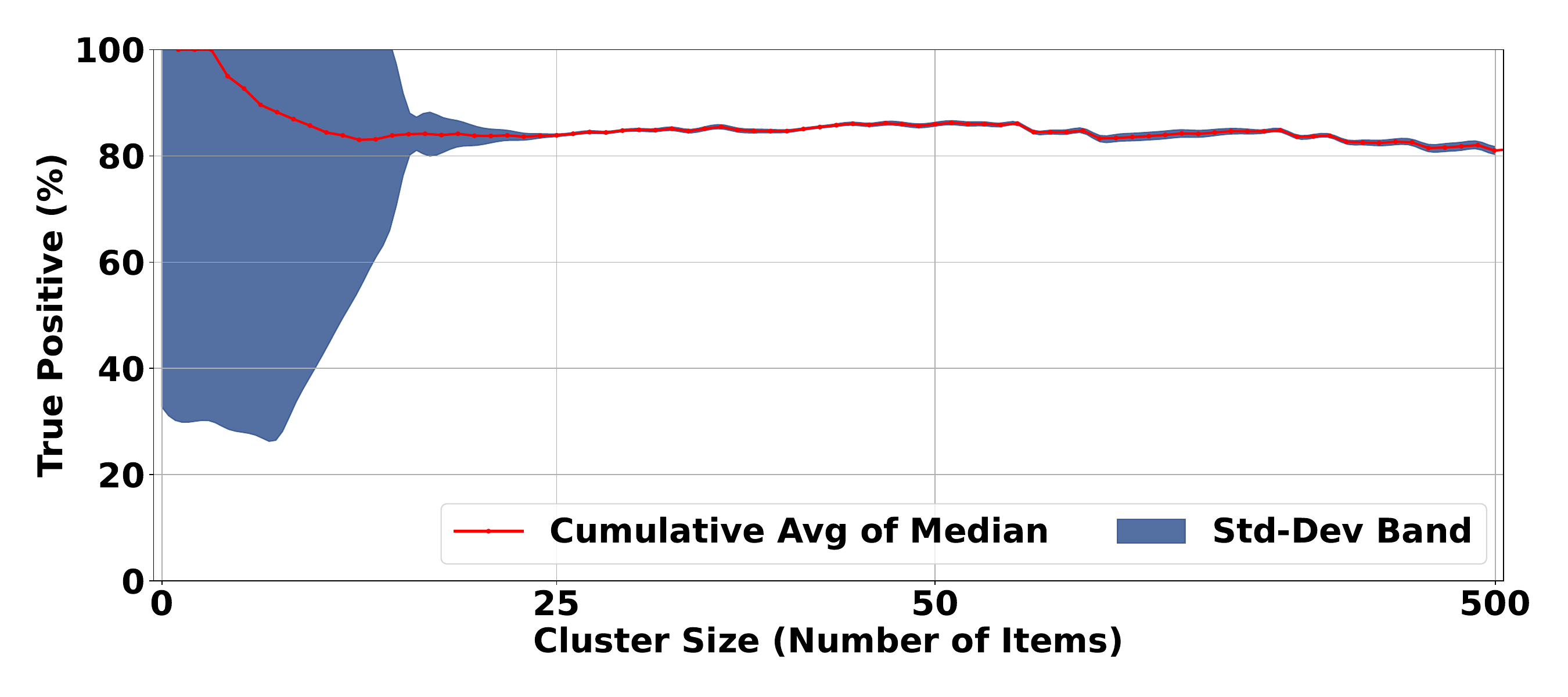} &
        \includegraphics[width=\linewidth]{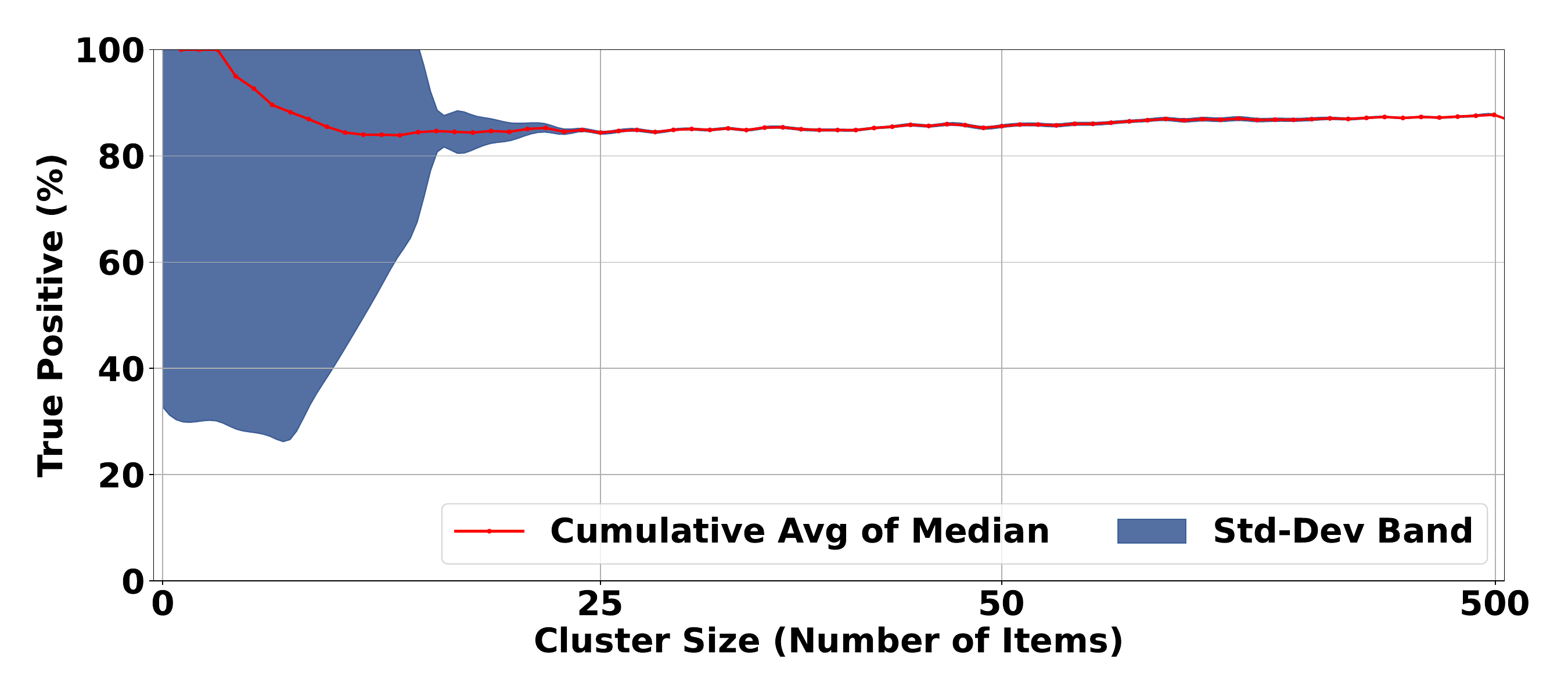} &
        \includegraphics[width=\linewidth]{Figures/AutoPYara/RTBF/VairiationPlots/Threshold/SSdeep_Th90.pdf} \\
        \multicolumn{5}{c}{\textbf{AutoPYara Ideal $K$ Configuration}} \\
        \includegraphics[width=\linewidth]{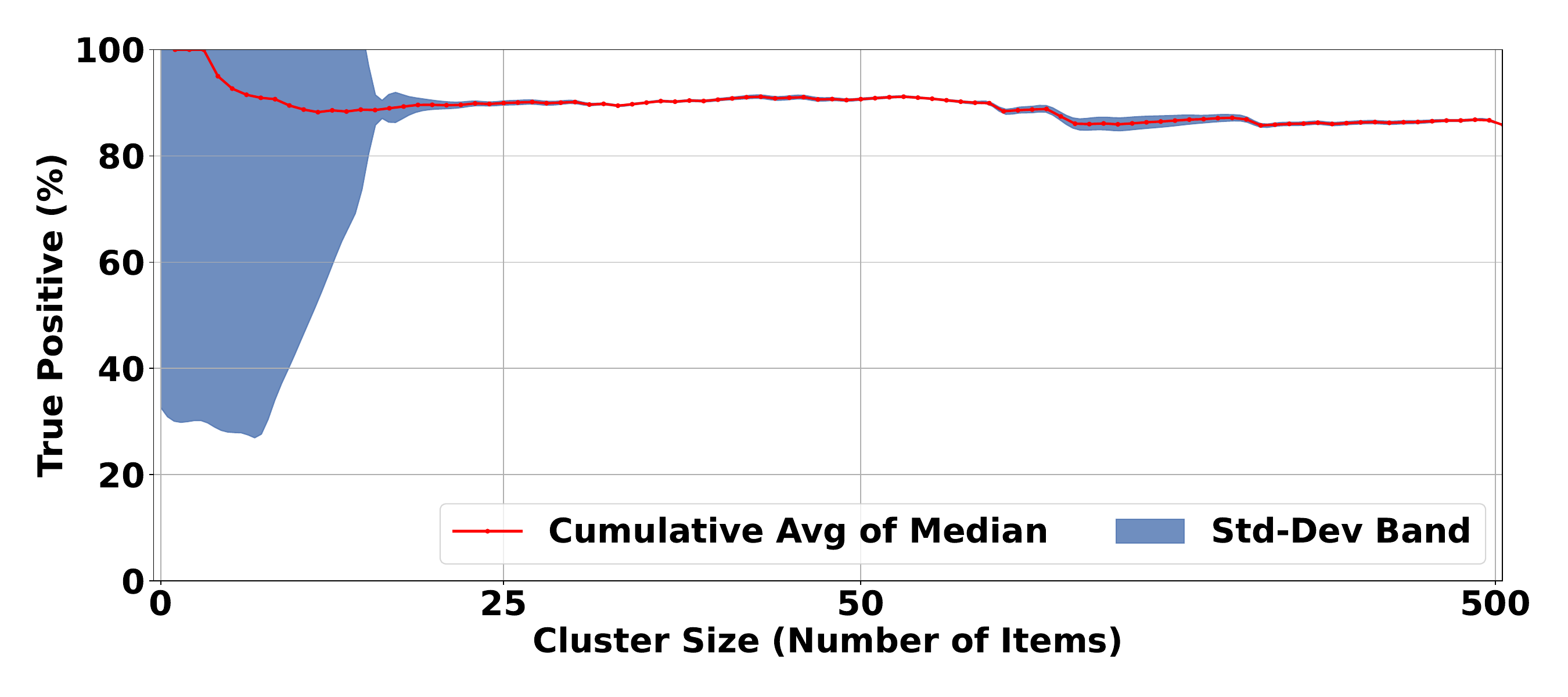} &
        \includegraphics[width=\linewidth]{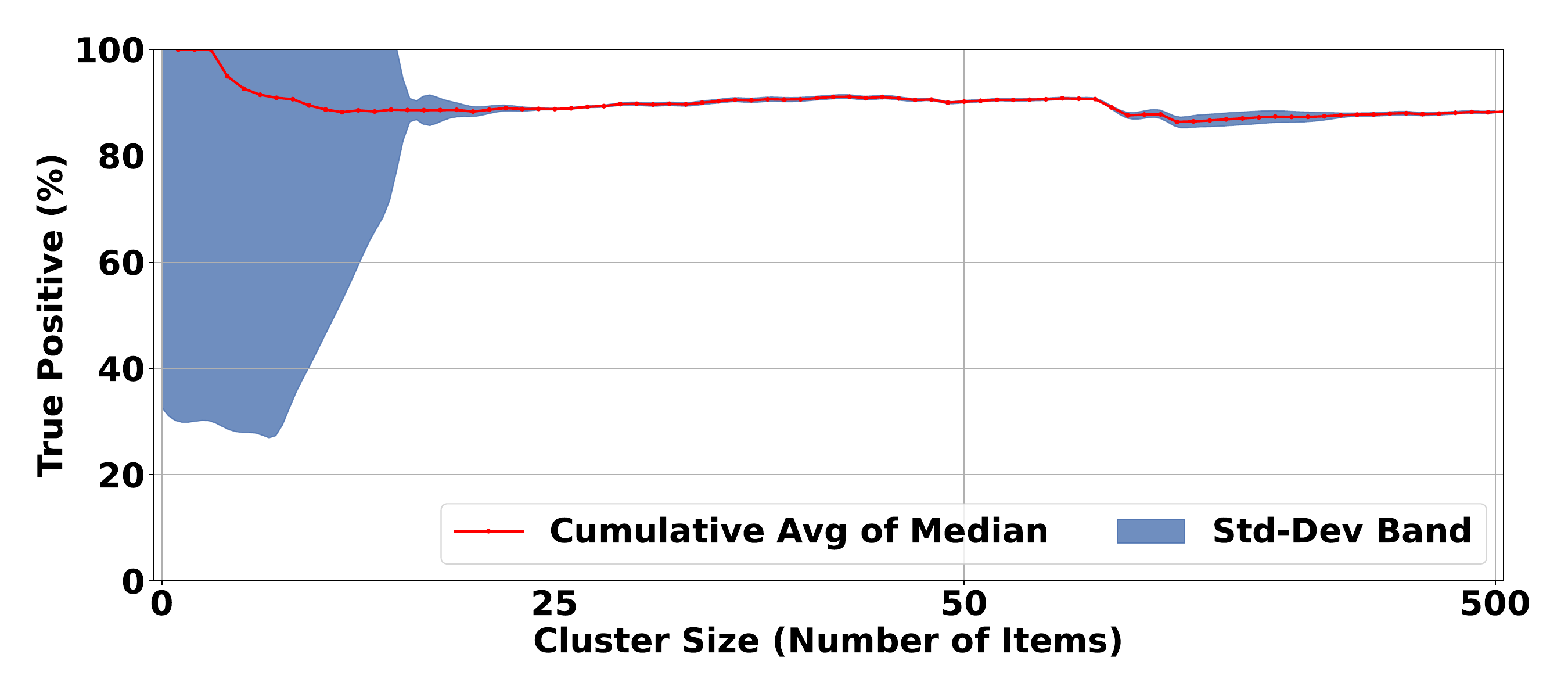} &
        \includegraphics[width=\linewidth]{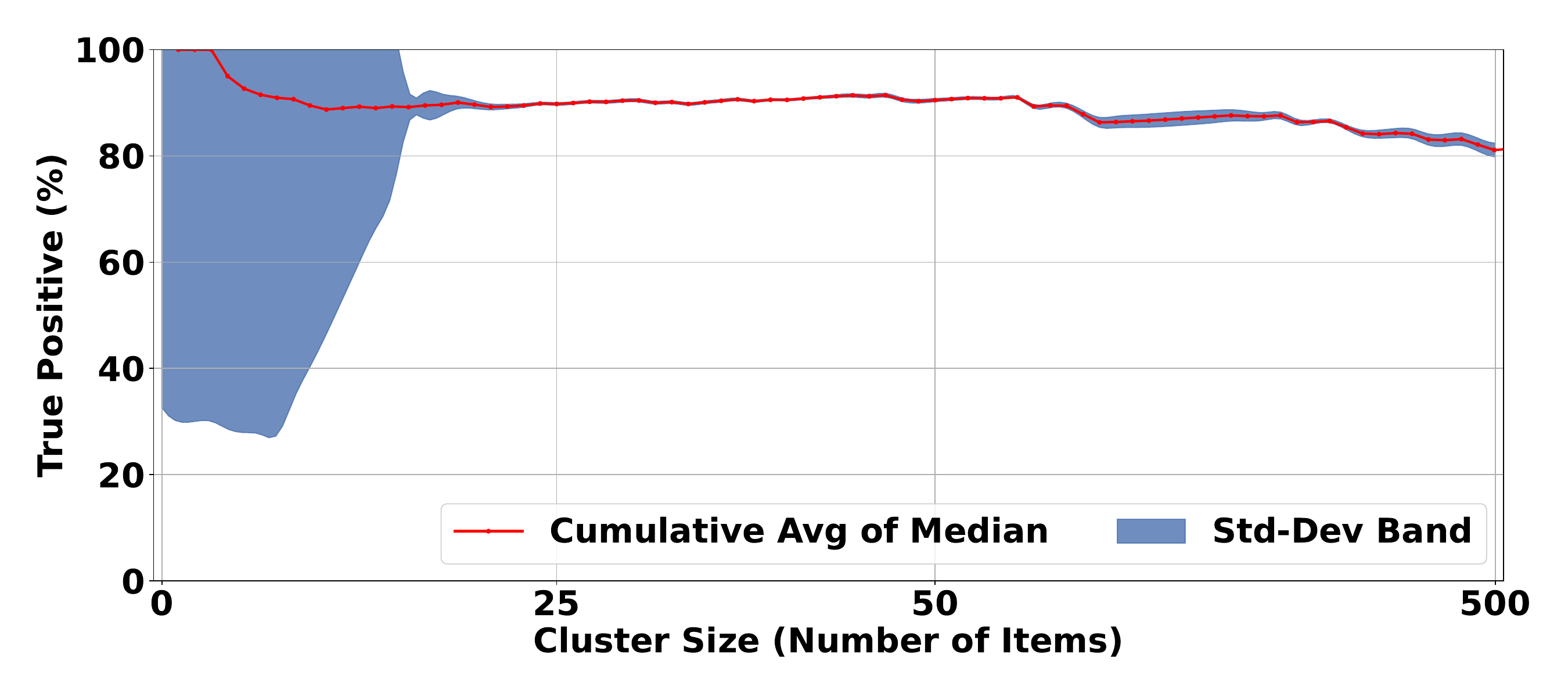} &
        \includegraphics[width=\linewidth]{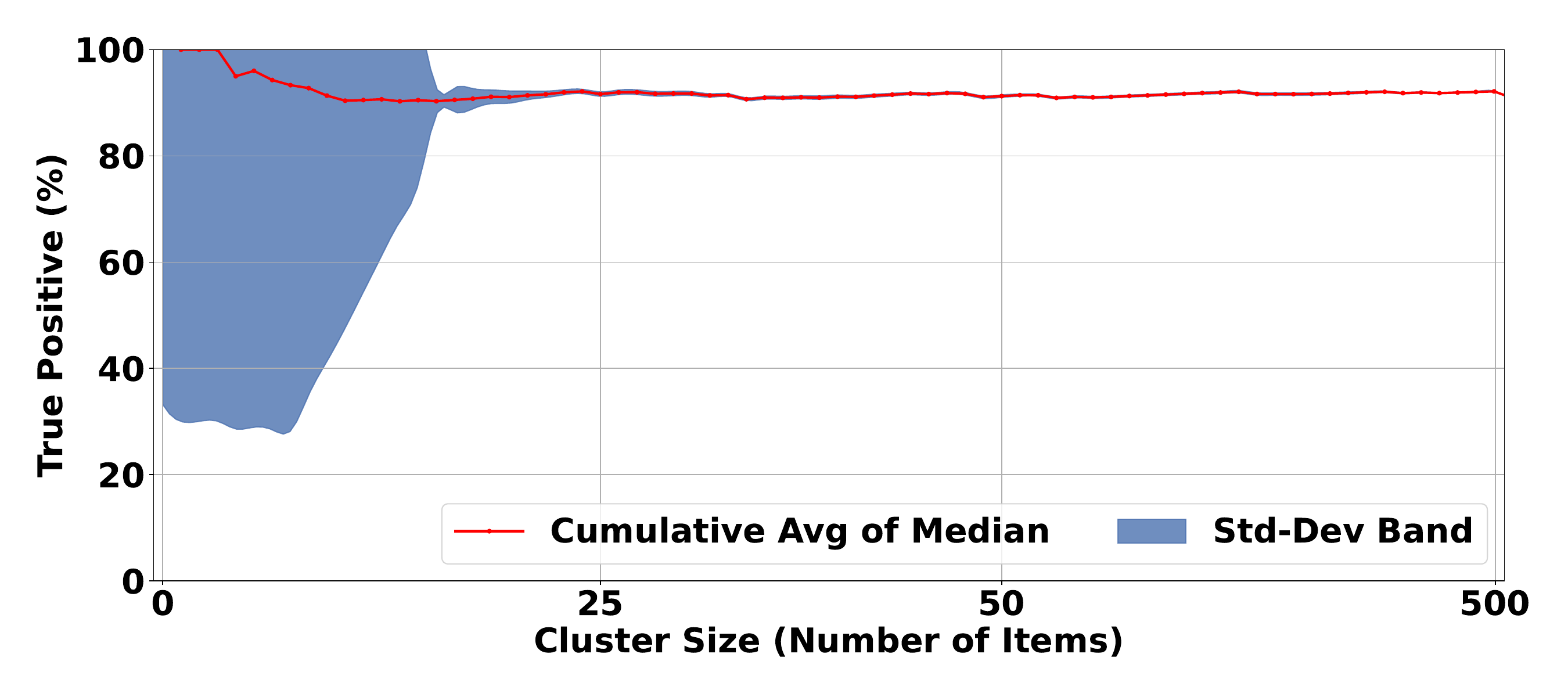} &
        \includegraphics[width=\linewidth]{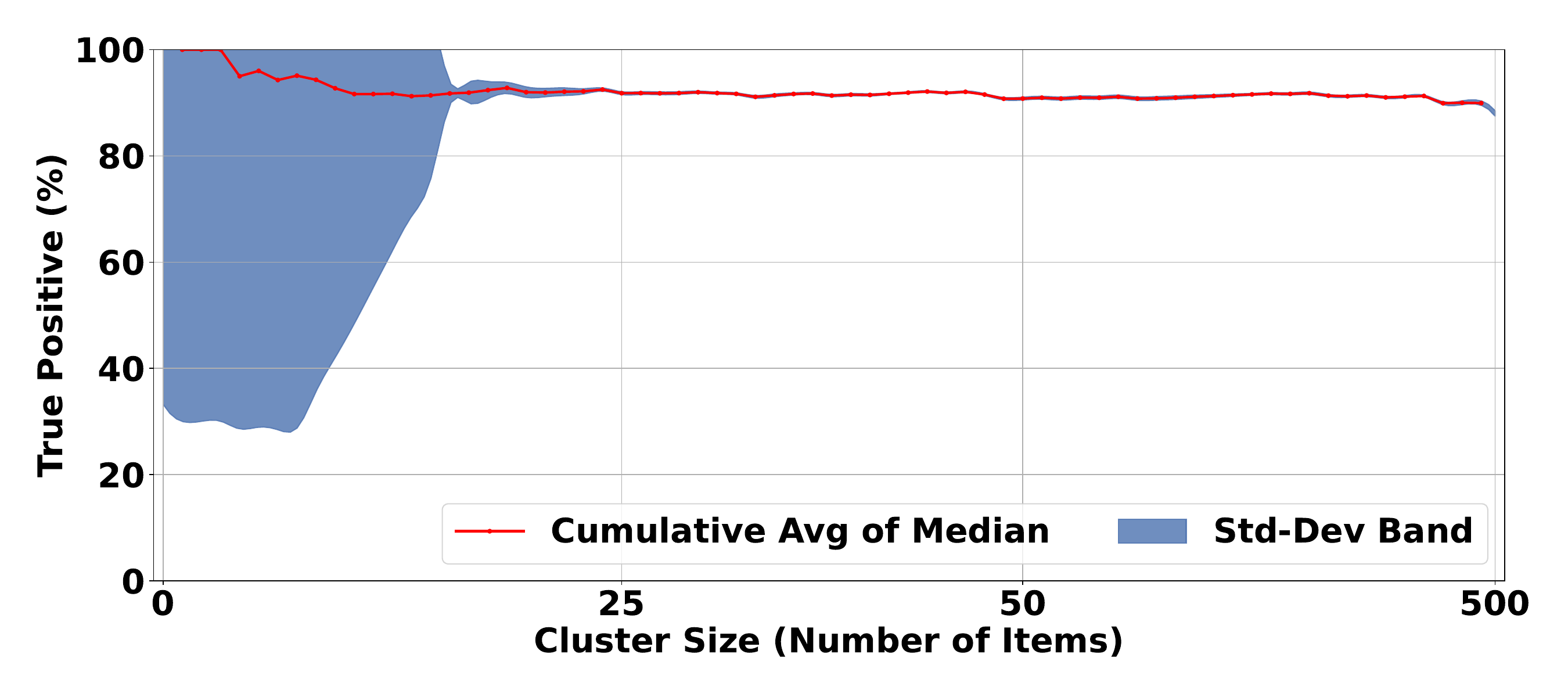} \\        
        \multicolumn{5}{c}{\textbf{AutoPYara Worst $K$ Configuration}} \\
        \includegraphics[width=\linewidth]{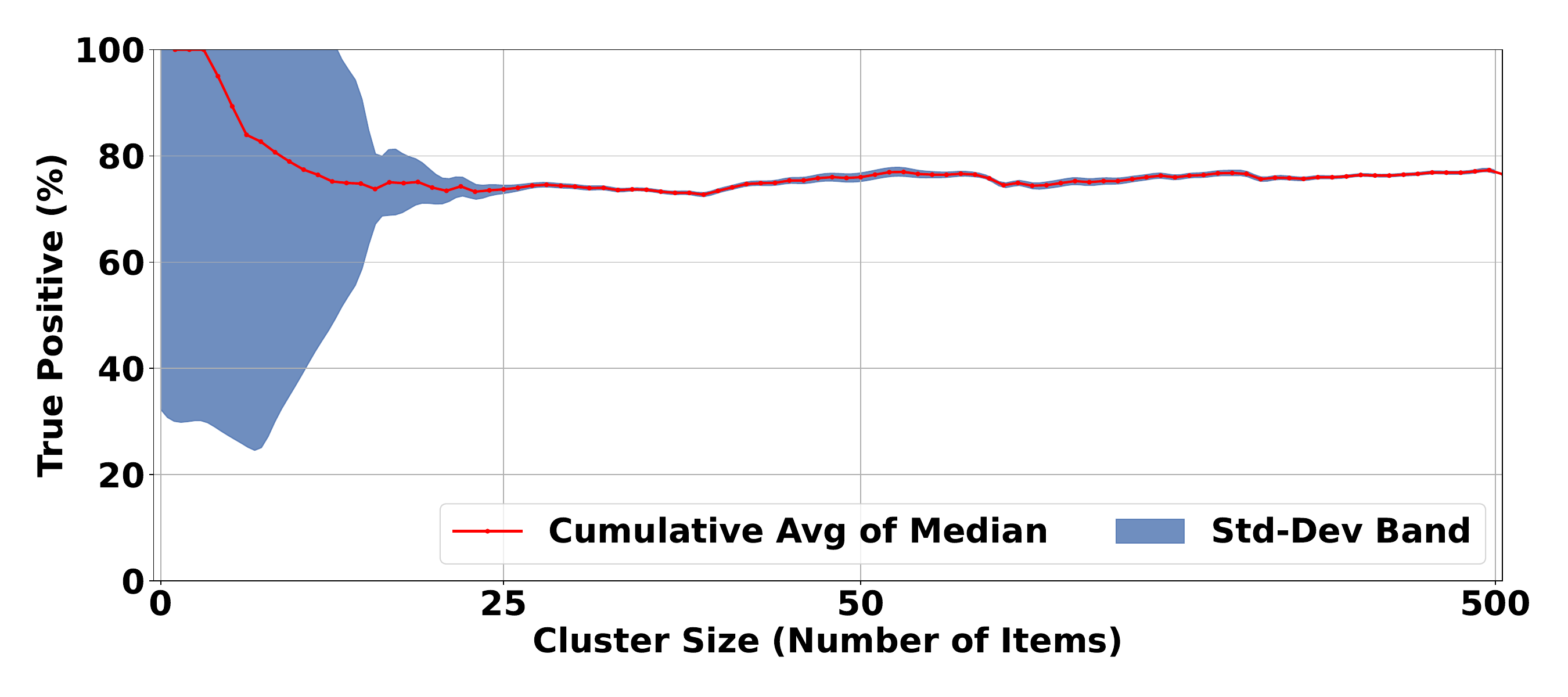} &
        \includegraphics[width=\linewidth]{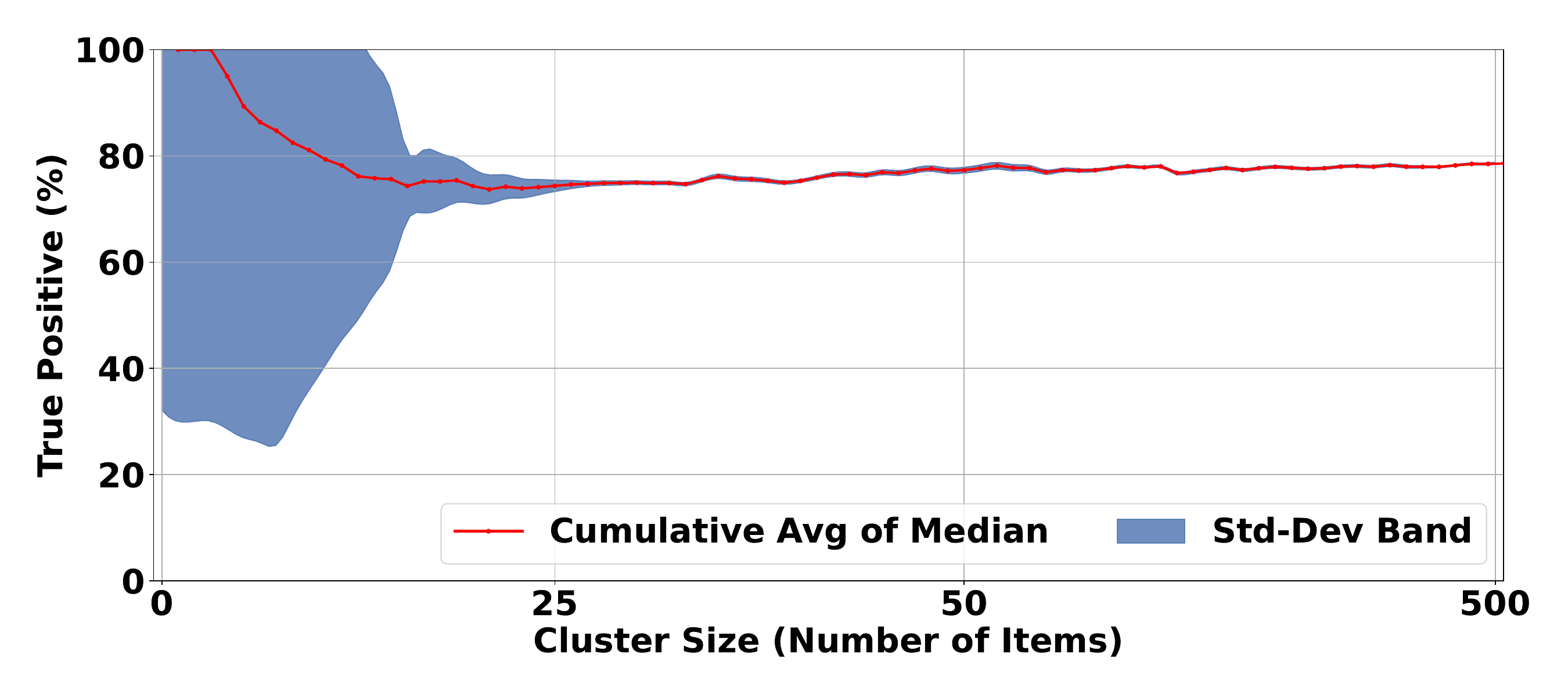} &
        \includegraphics[width=\linewidth]{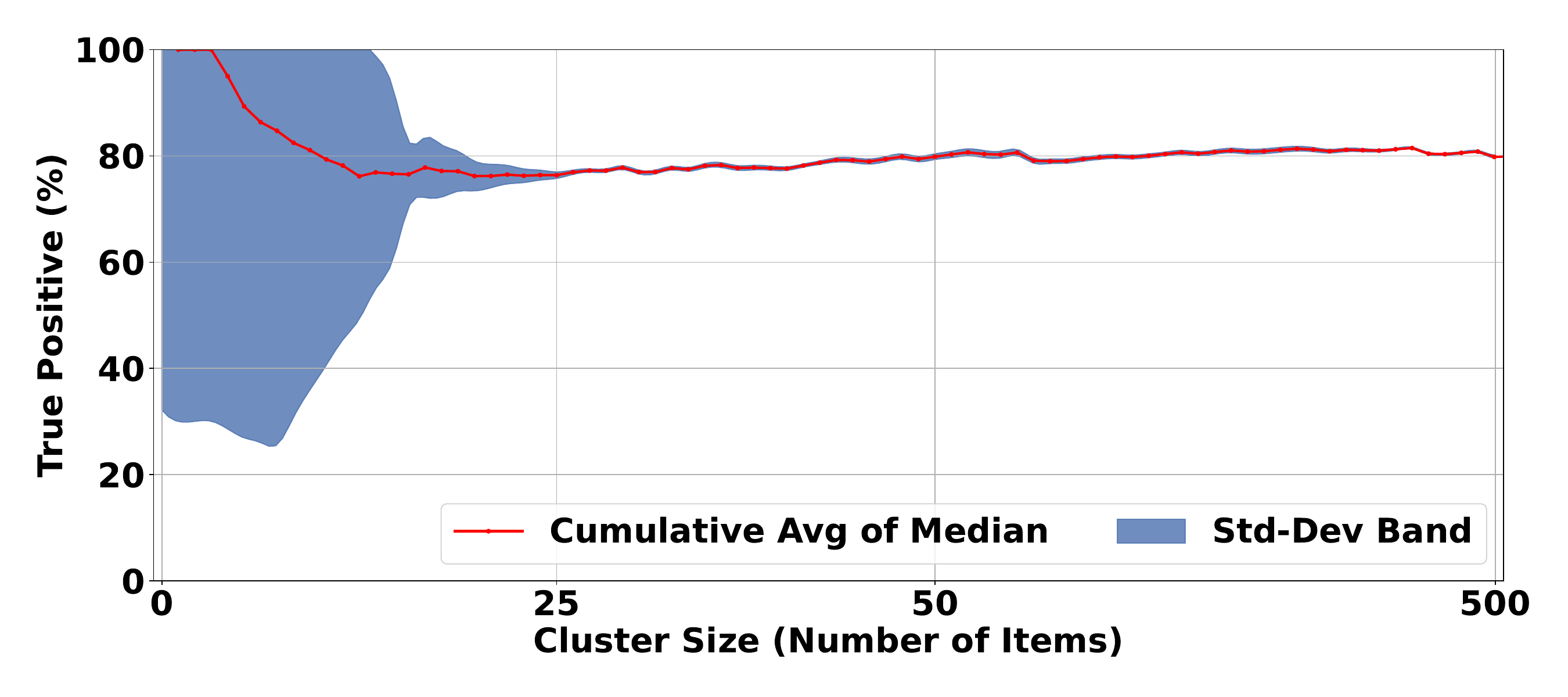} &
        \includegraphics[width=\linewidth]{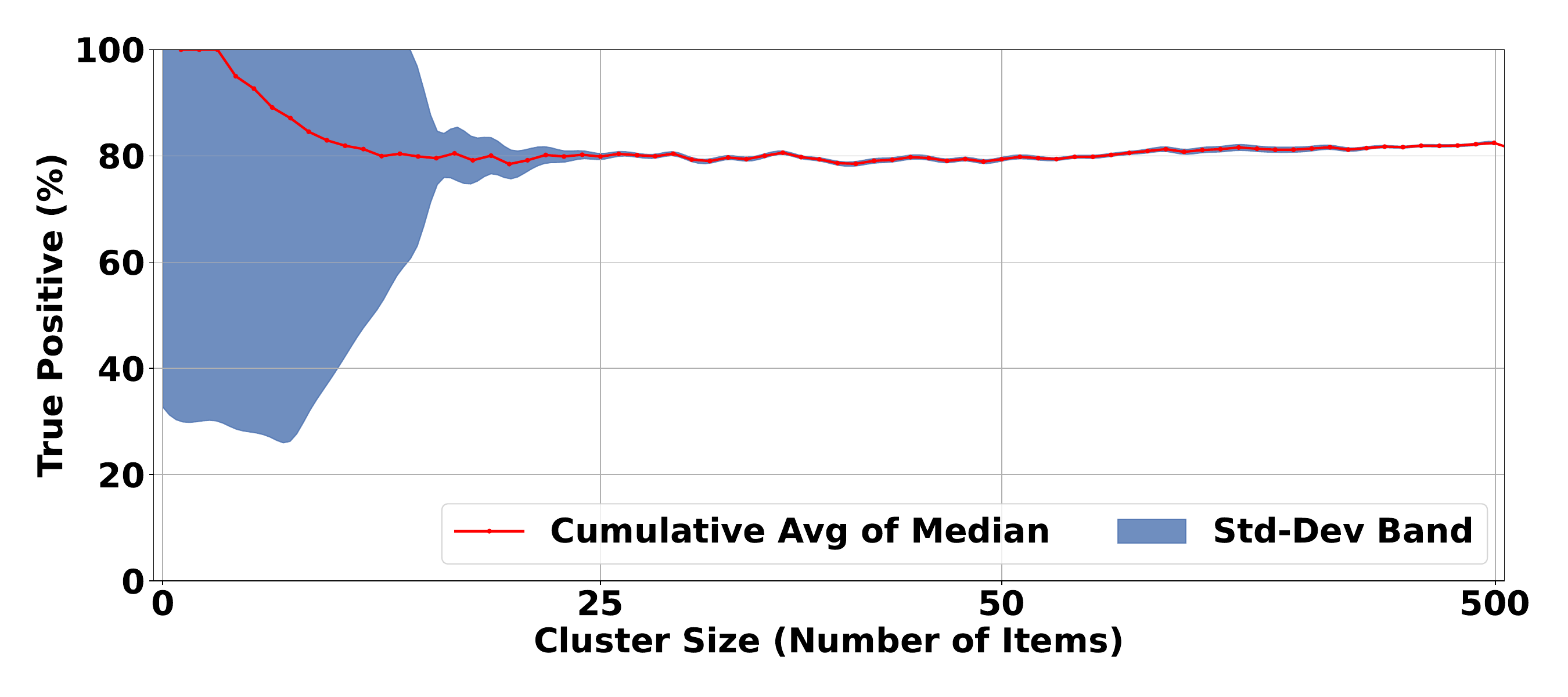} &
        \includegraphics[width=\linewidth]{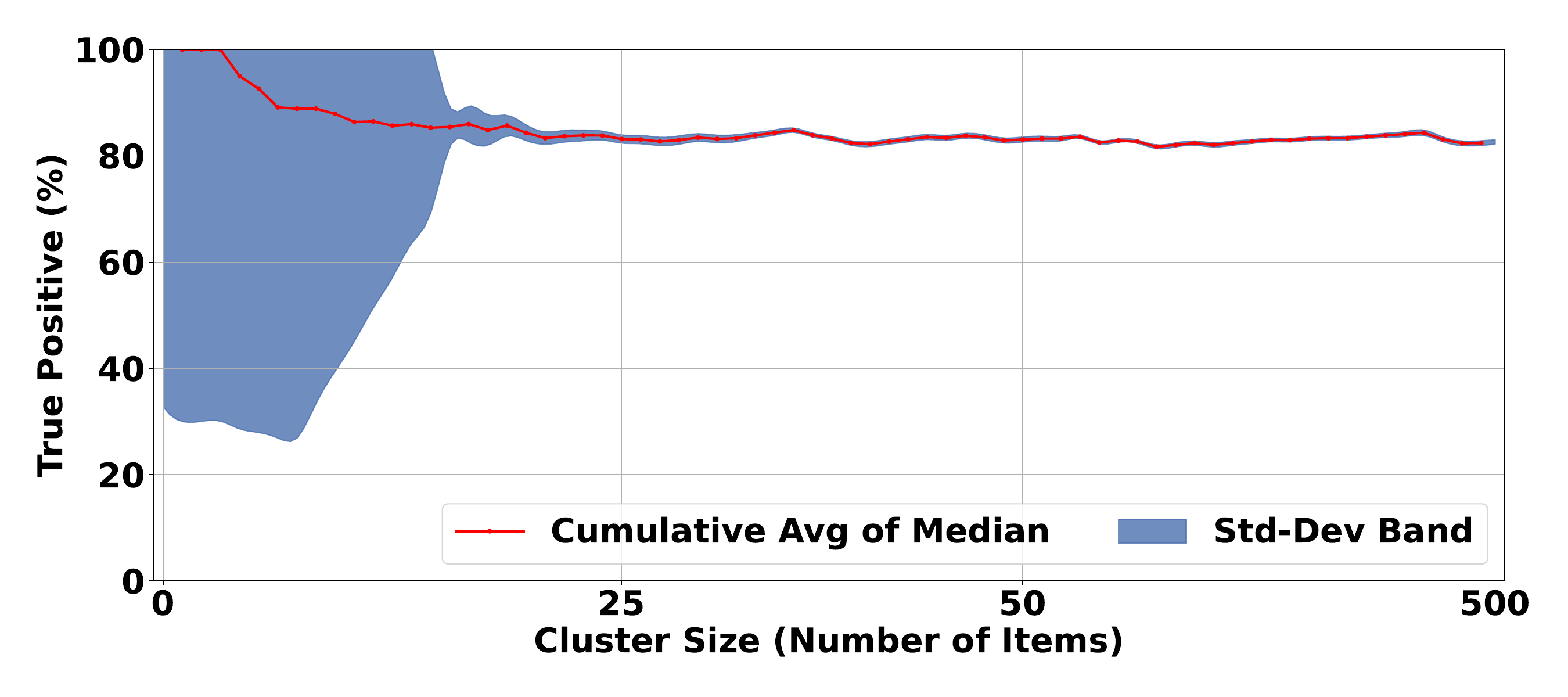} \\
        \multicolumn{5}{c}{\underline{\textbf{Oracle Based Heuristic}}} \\
        \multicolumn{5}{c}{\textbf{Mode $K$ From Oracle}} \\
        \includegraphics[width=\linewidth]{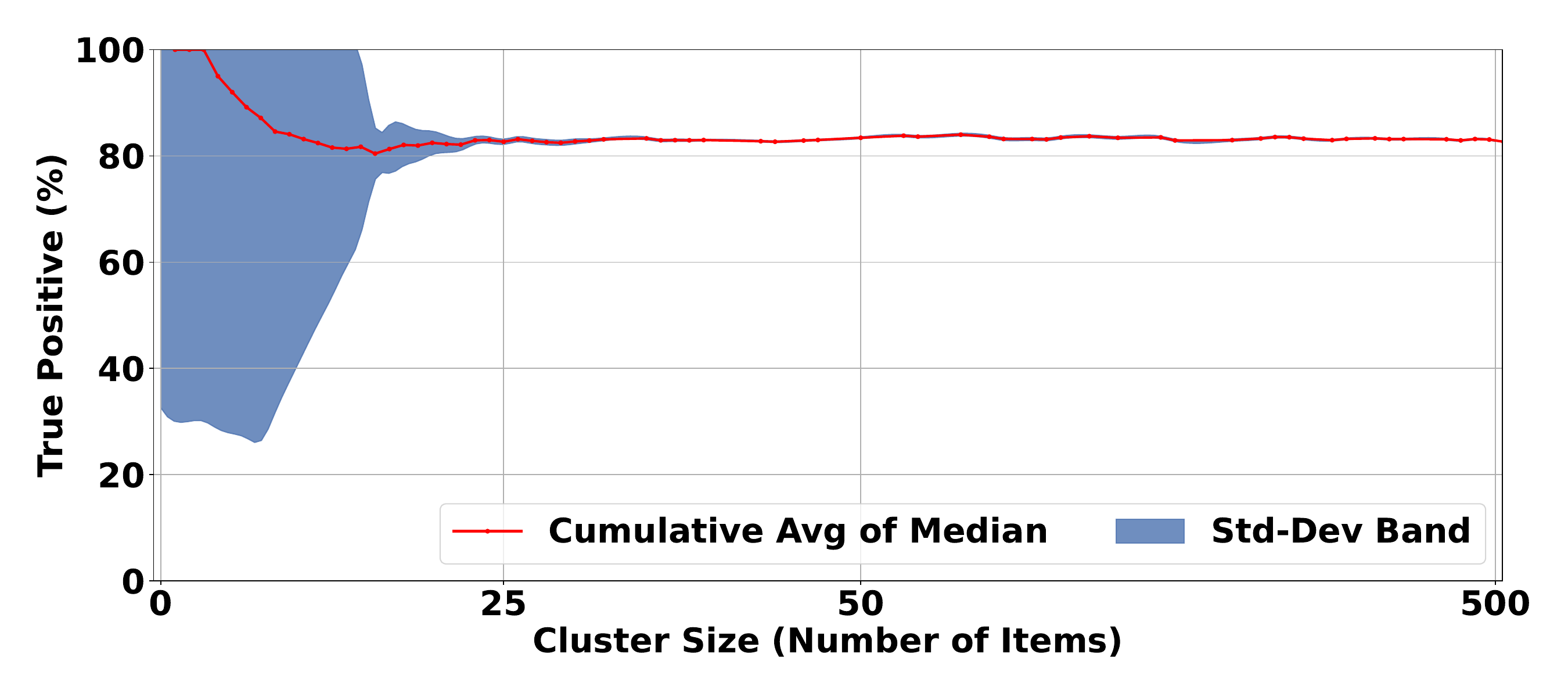} &
        \includegraphics[width=\linewidth]{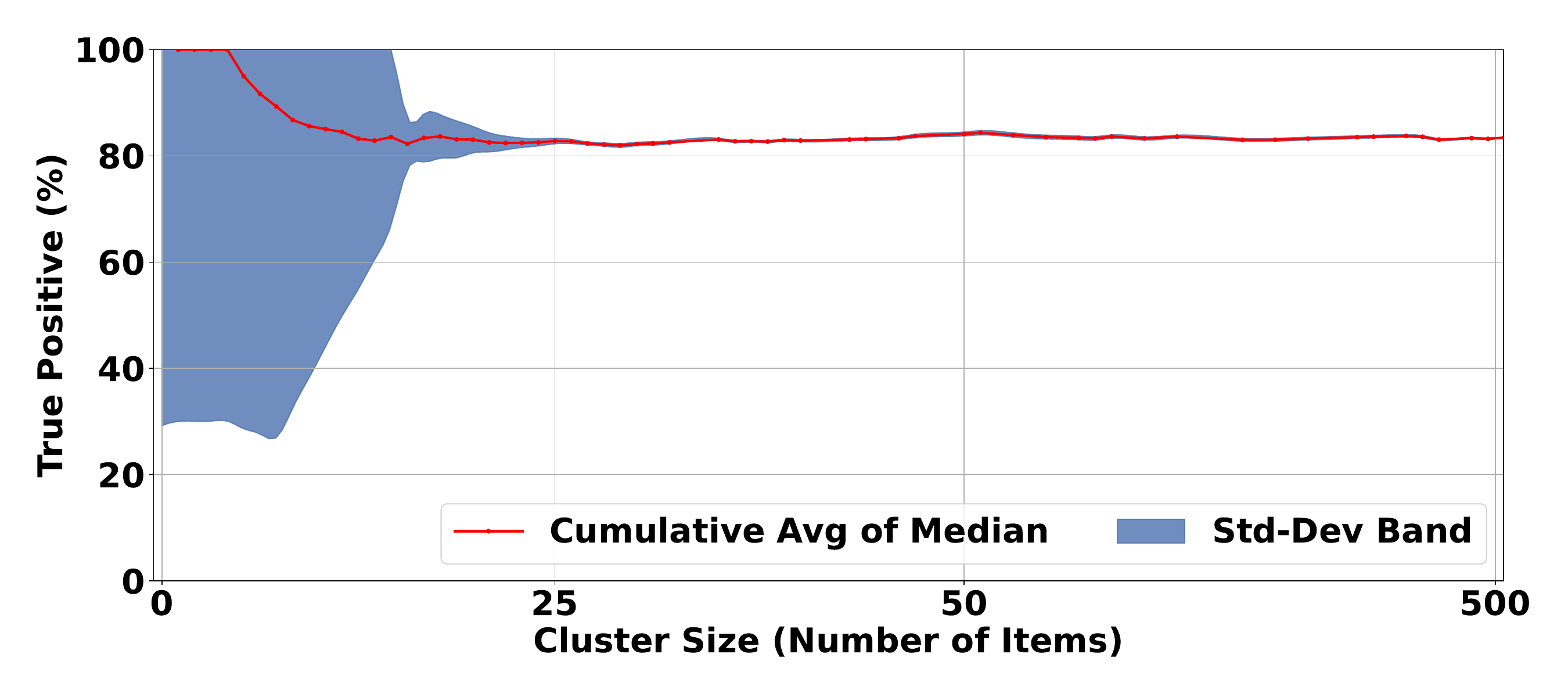} &
        \includegraphics[width=\linewidth]{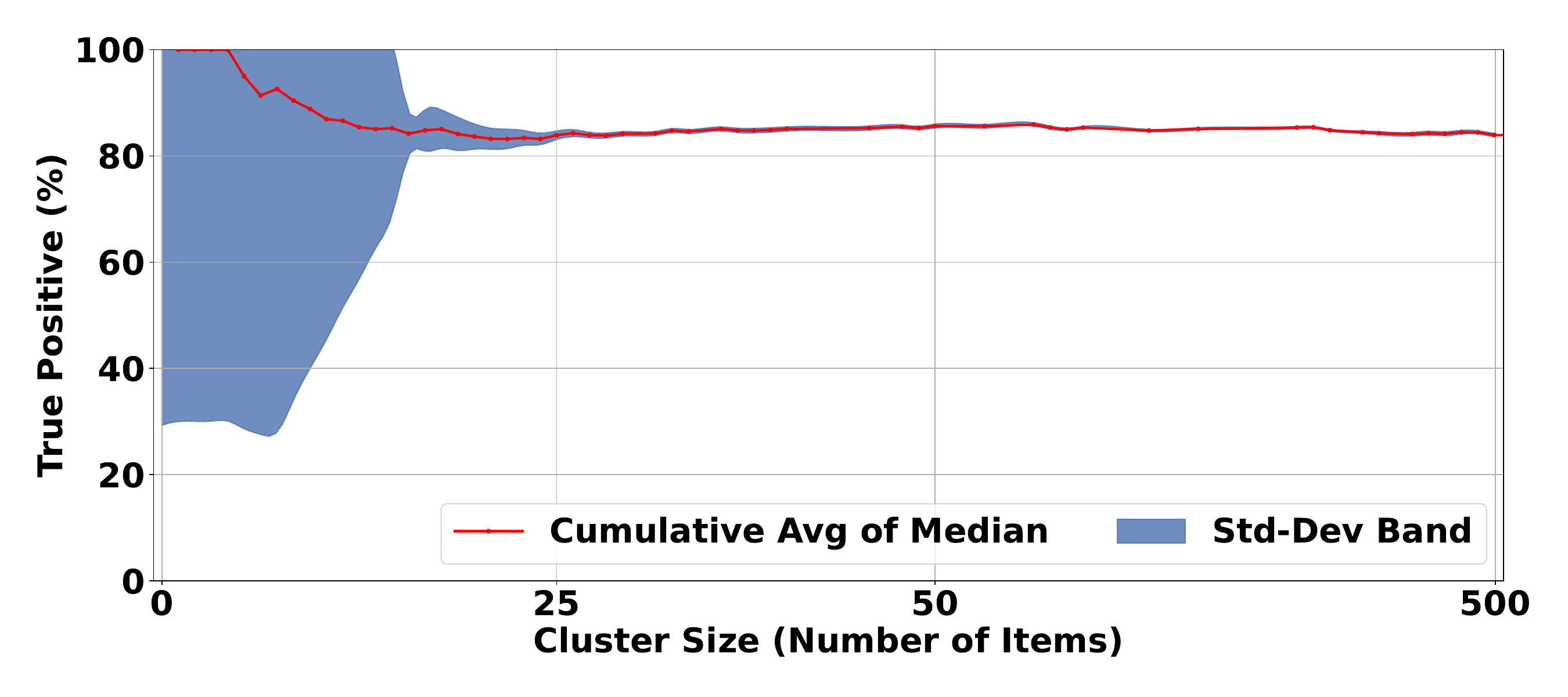} &
        \includegraphics[width=\linewidth]{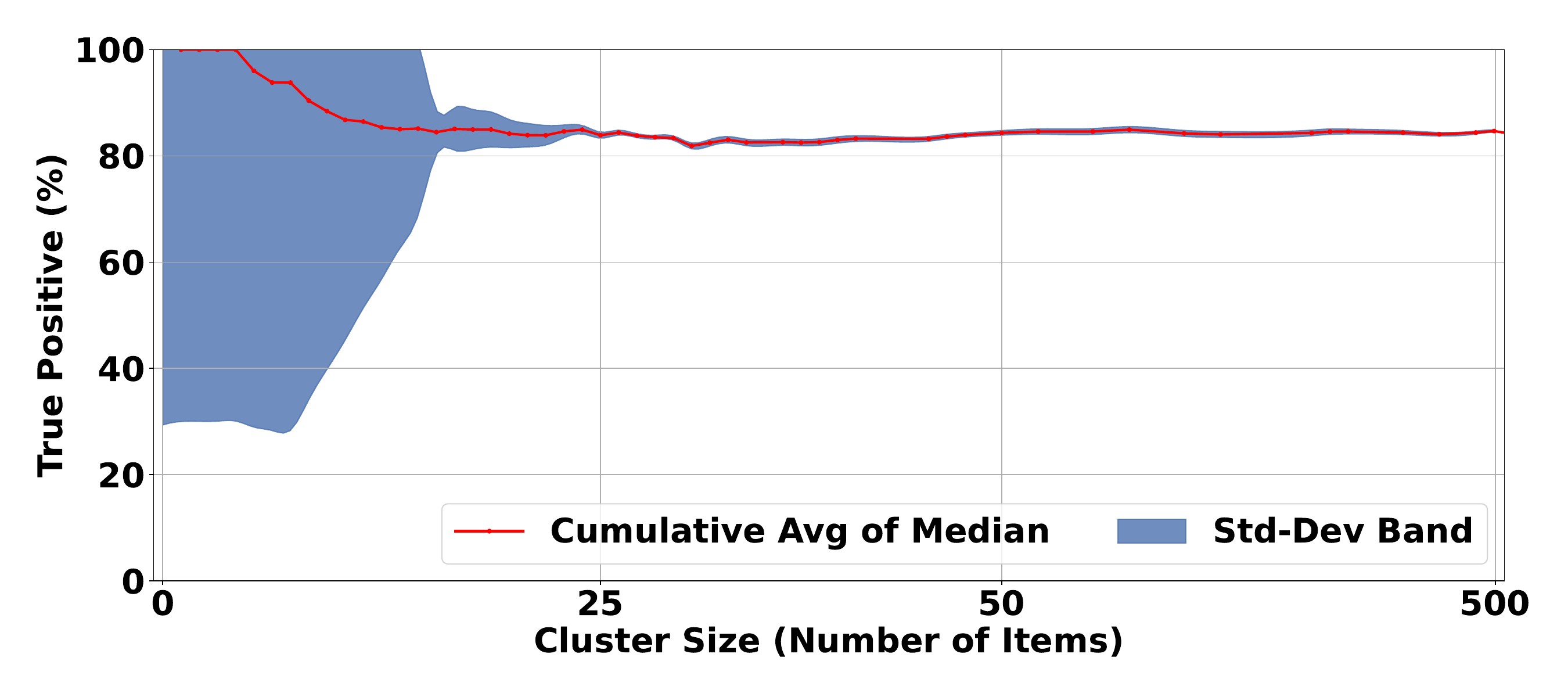} &
        \includegraphics[width=\linewidth]{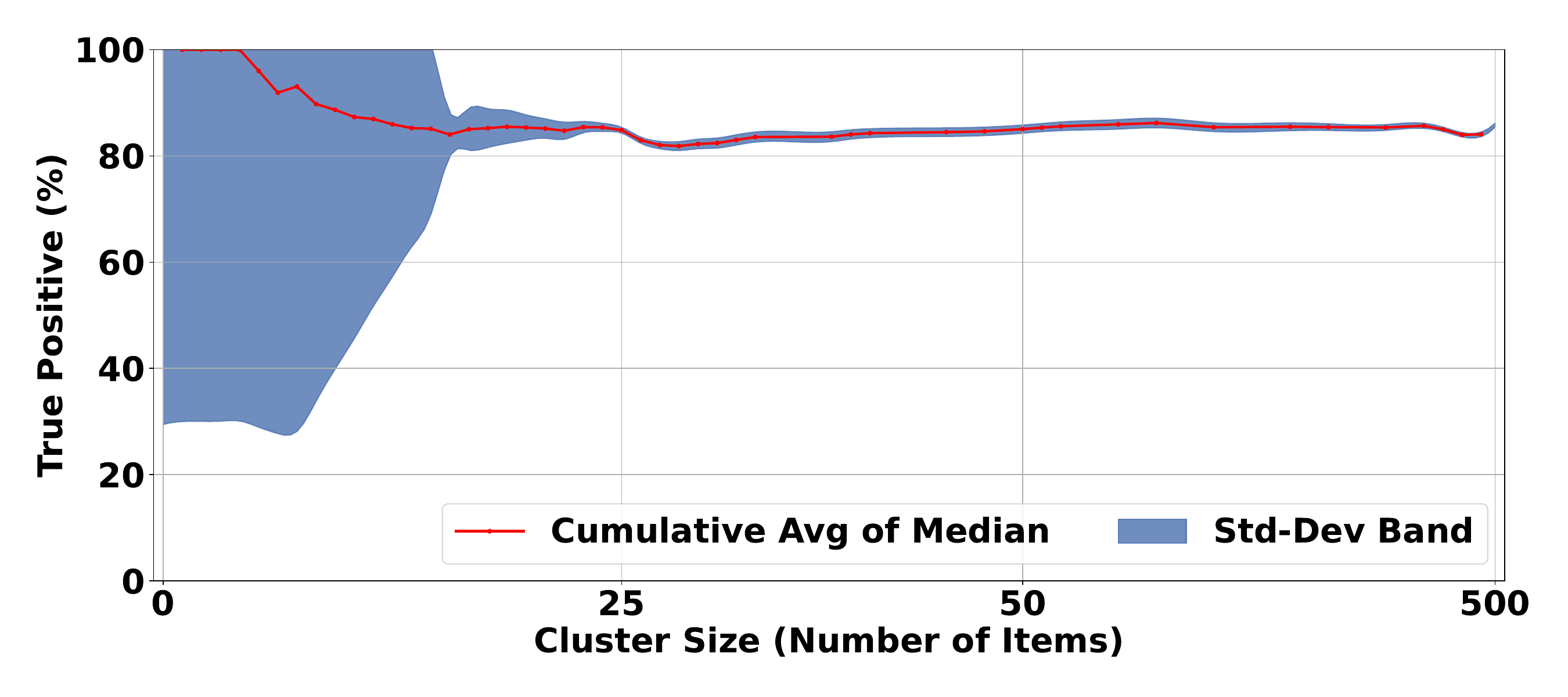} \\    
        \multicolumn{5}{c}{\textbf{Max $K$ From Oracle}} \\
        \includegraphics[width=\linewidth]{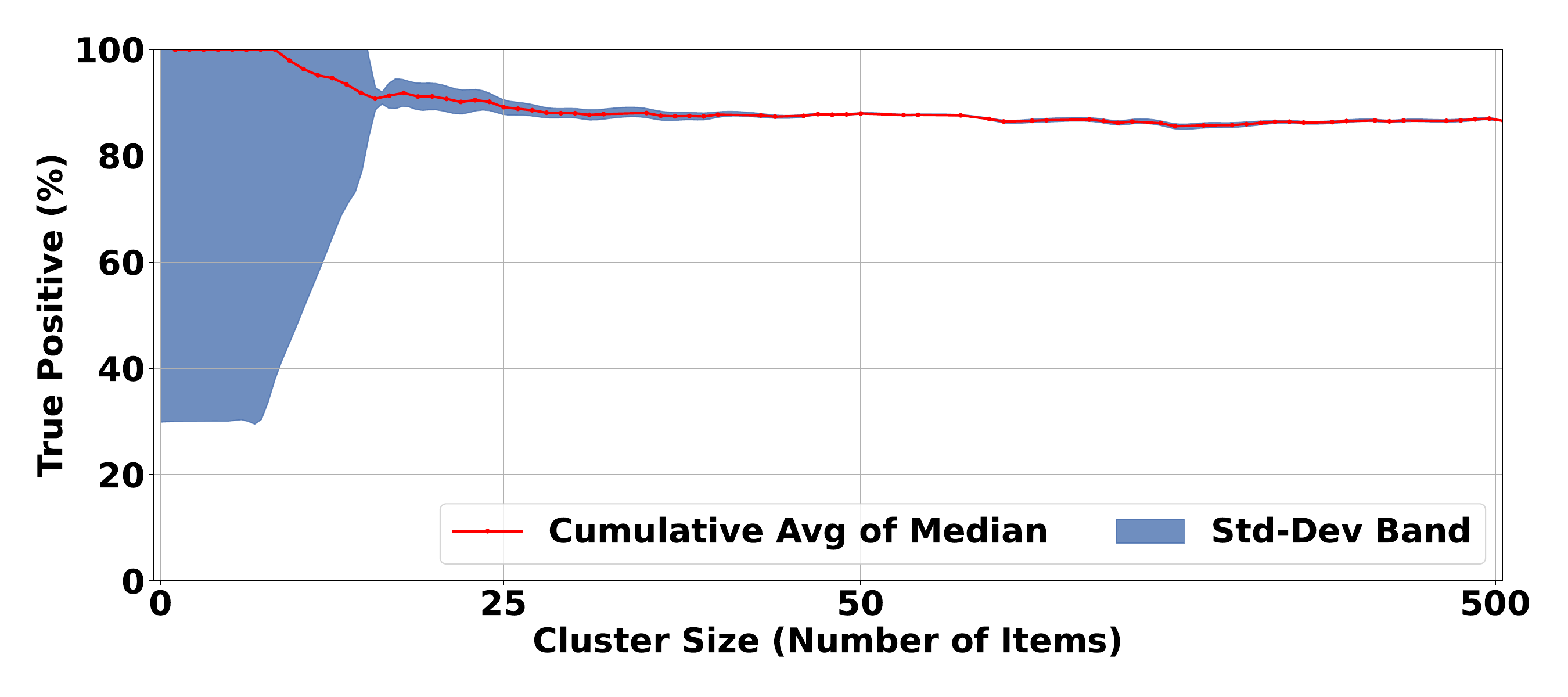} &
        \includegraphics[width=\linewidth]{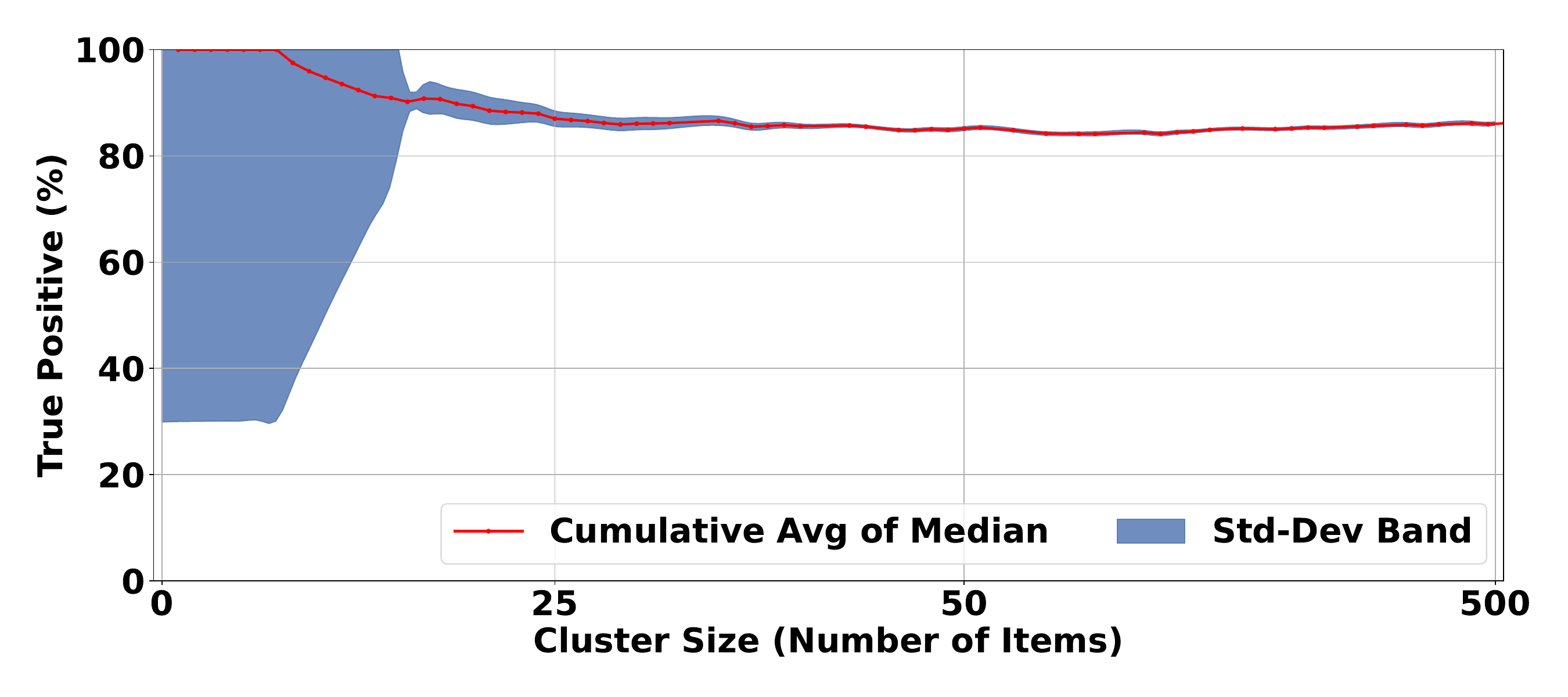} &
        \includegraphics[width=\linewidth]{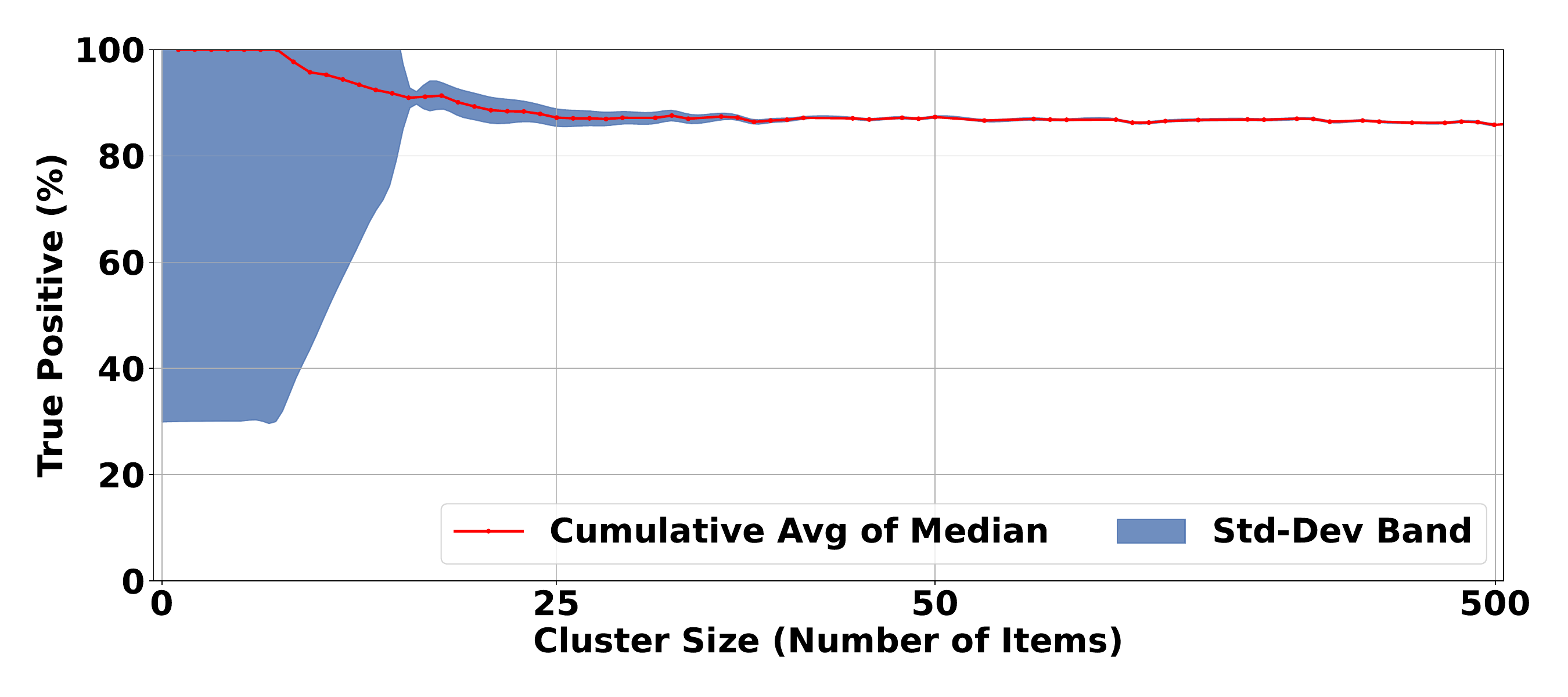} &
        \includegraphics[width=\linewidth]{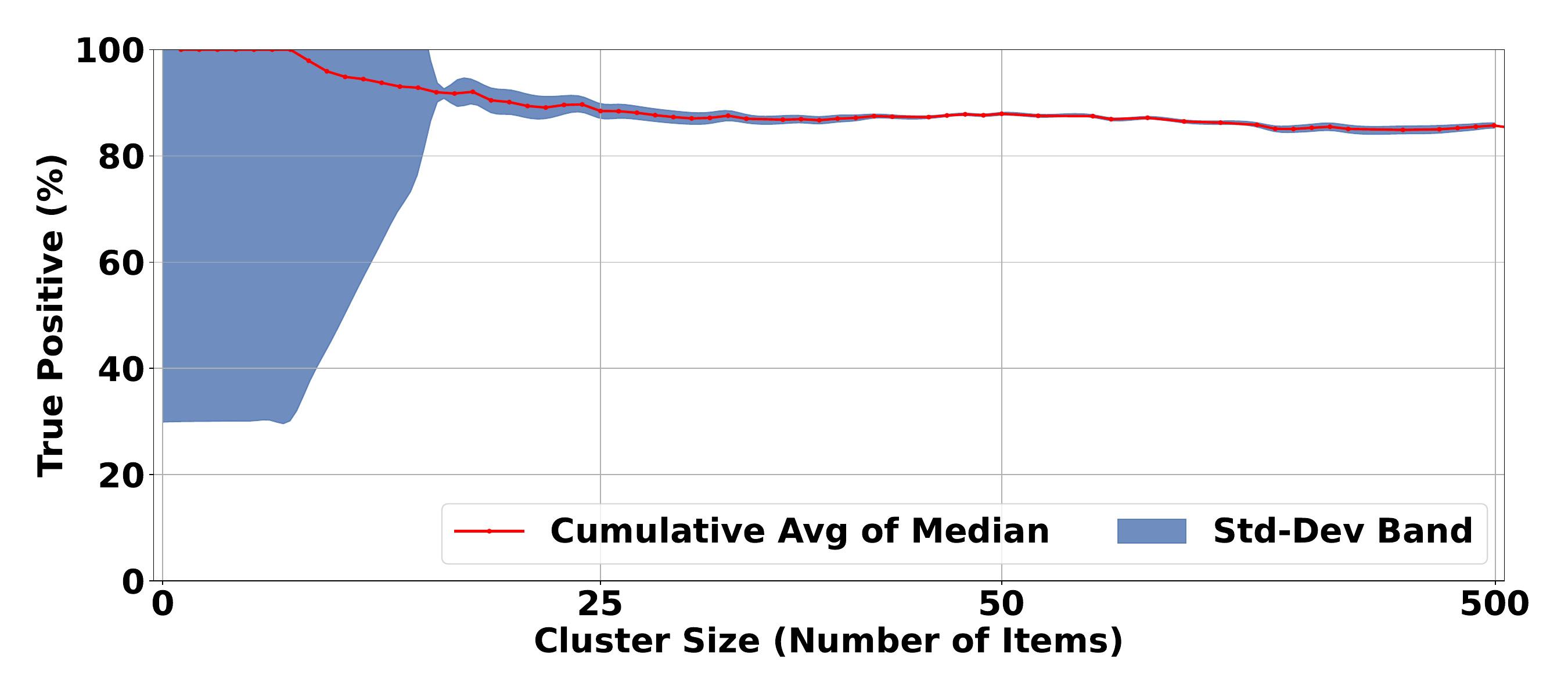} &
        \includegraphics[width=\linewidth]{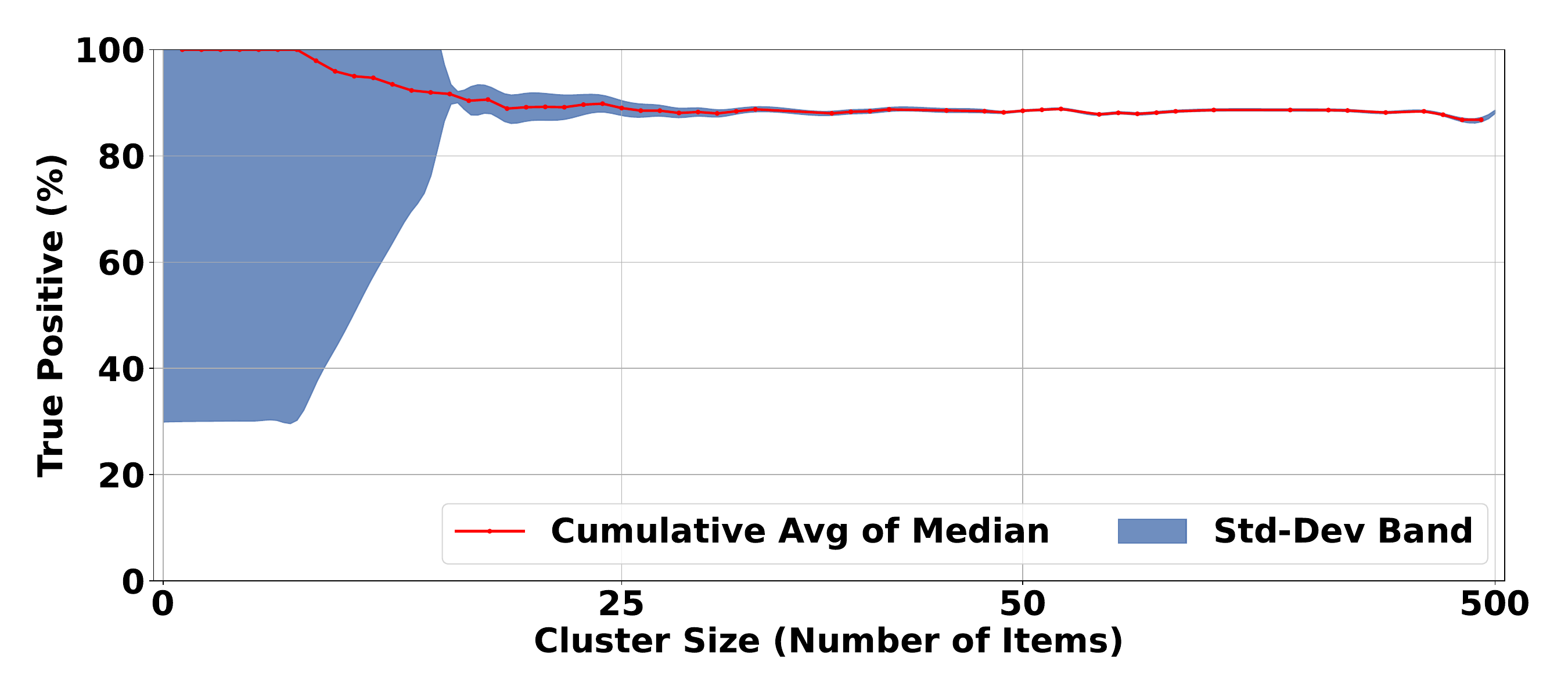} \\  
        \multicolumn{5}{c}{\textbf{Random $K$ From Oracle}} \\
        \includegraphics[width=\linewidth]{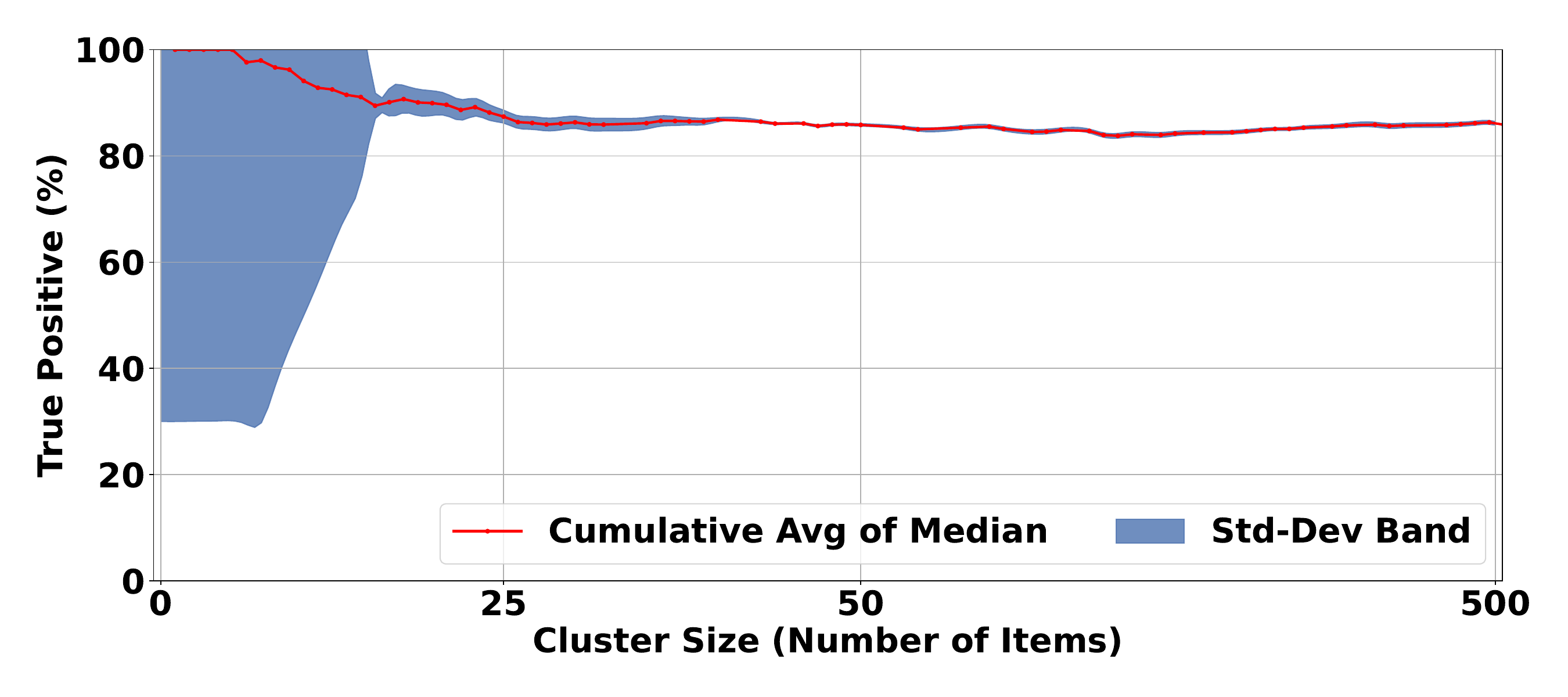} &
        \includegraphics[width=\linewidth]{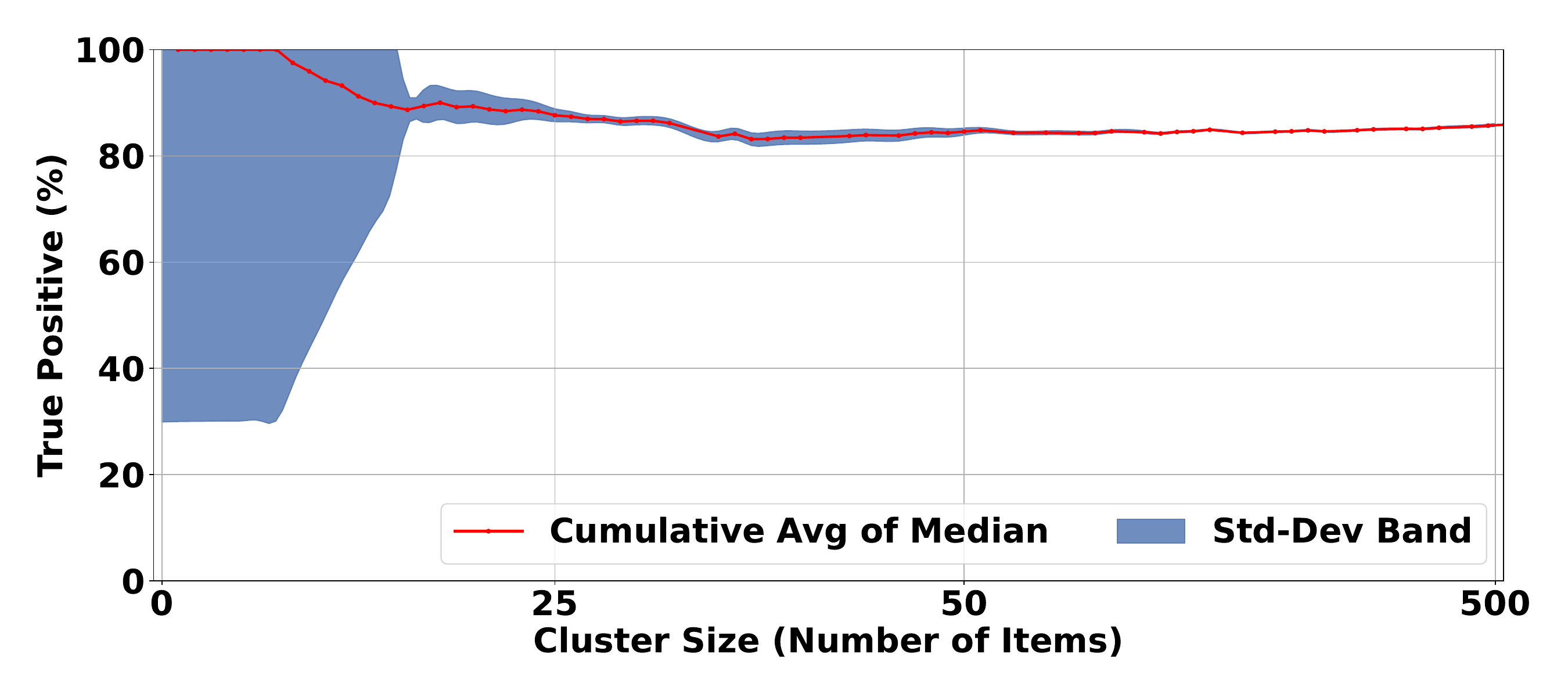} &
        \includegraphics[width=\linewidth]{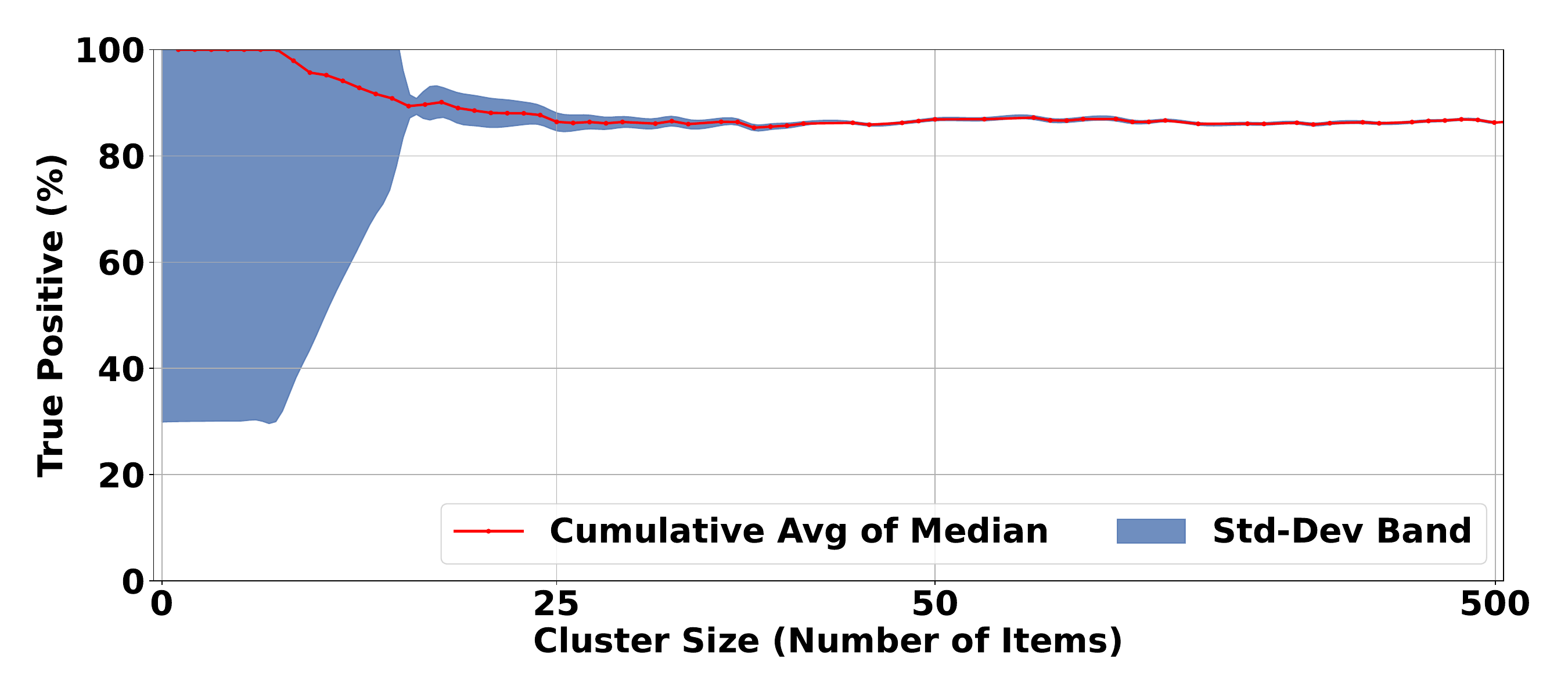} &
        \includegraphics[width=\linewidth]{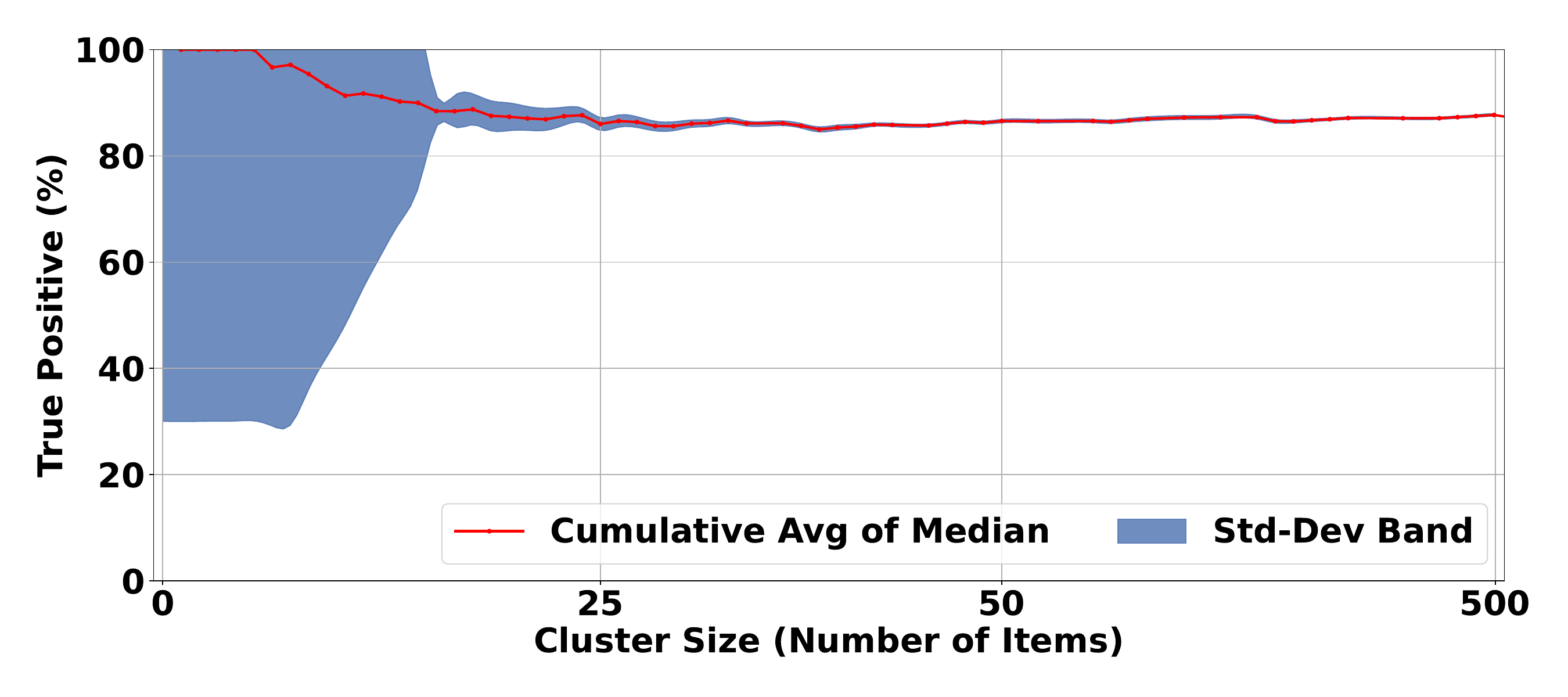} &
        \includegraphics[width=\linewidth]{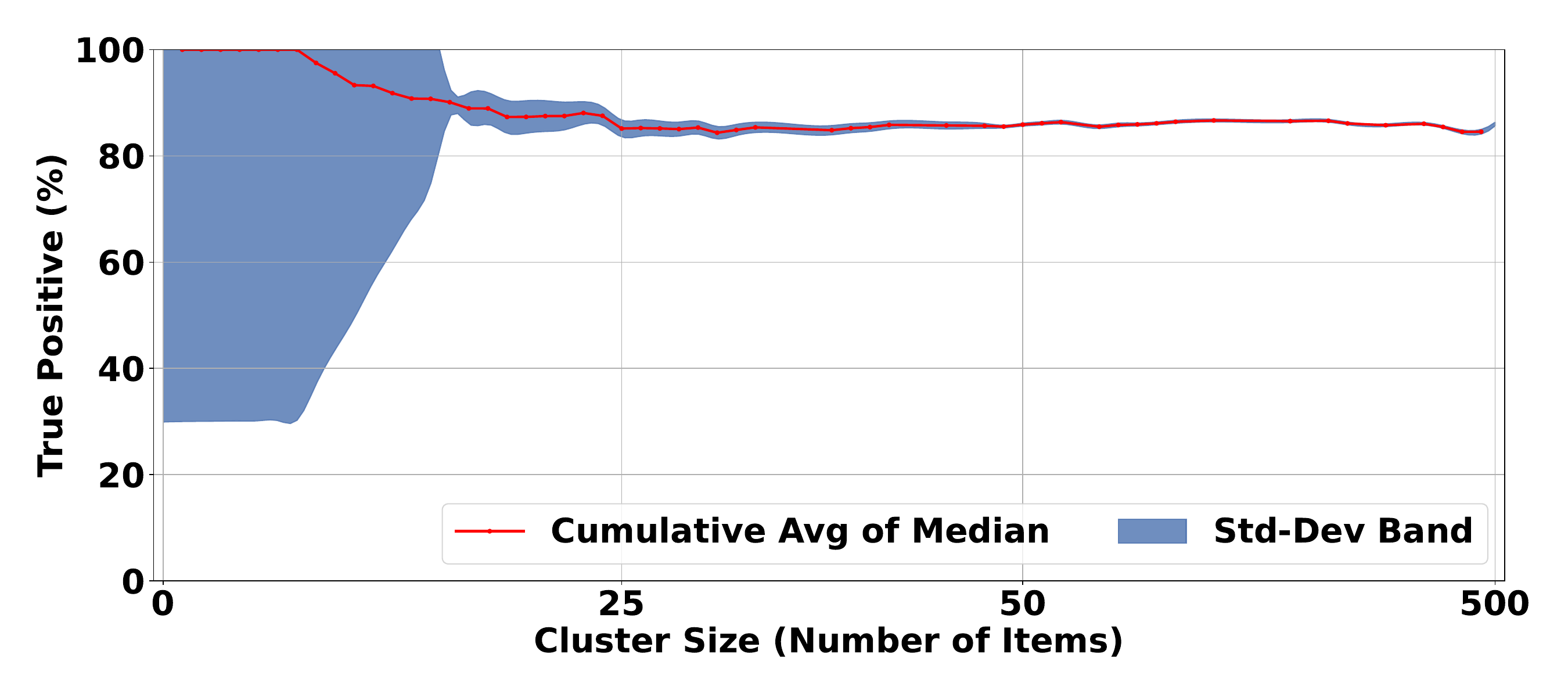} \\ 
        \multicolumn{5}{c}{\underline{\textbf{Sub-Optimal Heuristic}}} \\
        \multicolumn{5}{c}{\textbf{Average $K$}} \\    
        \includegraphics[width=\linewidth]{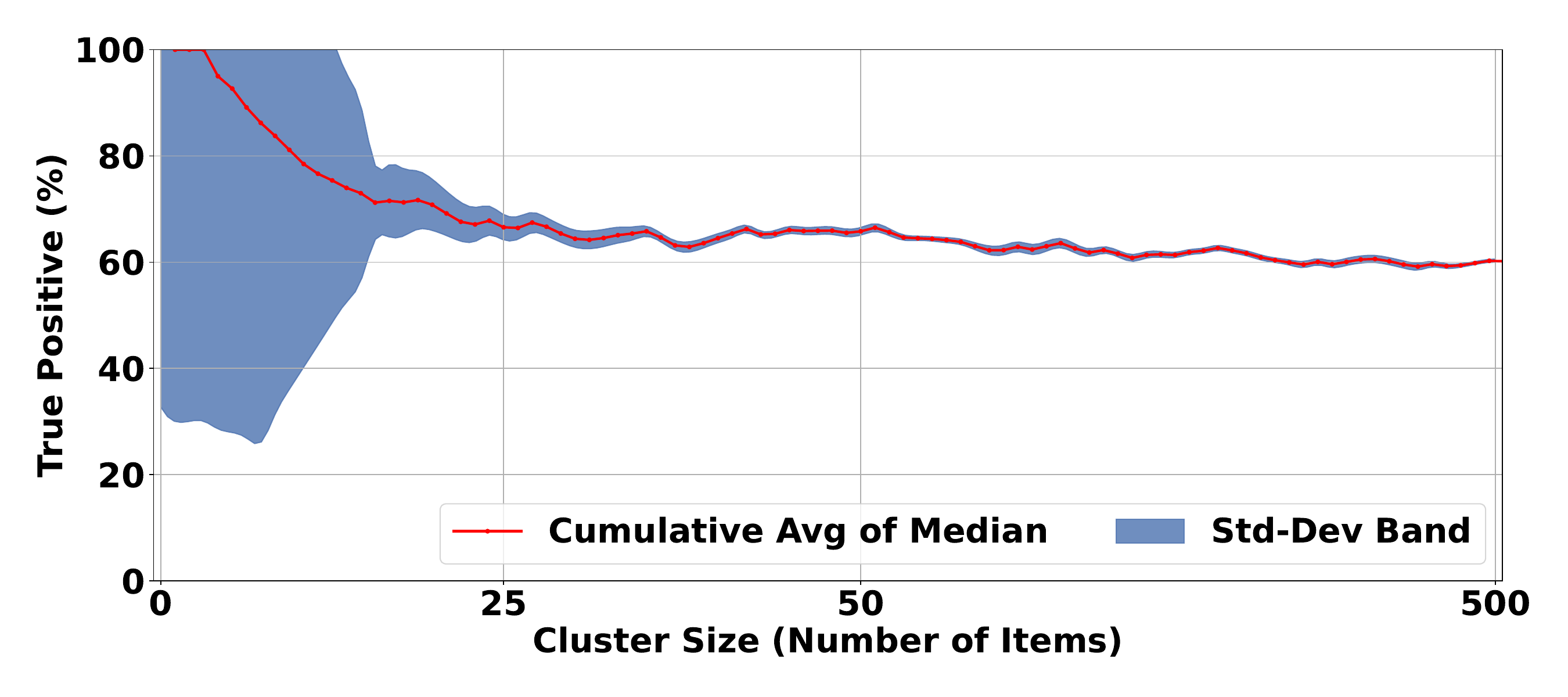} &
        \includegraphics[width=\linewidth]{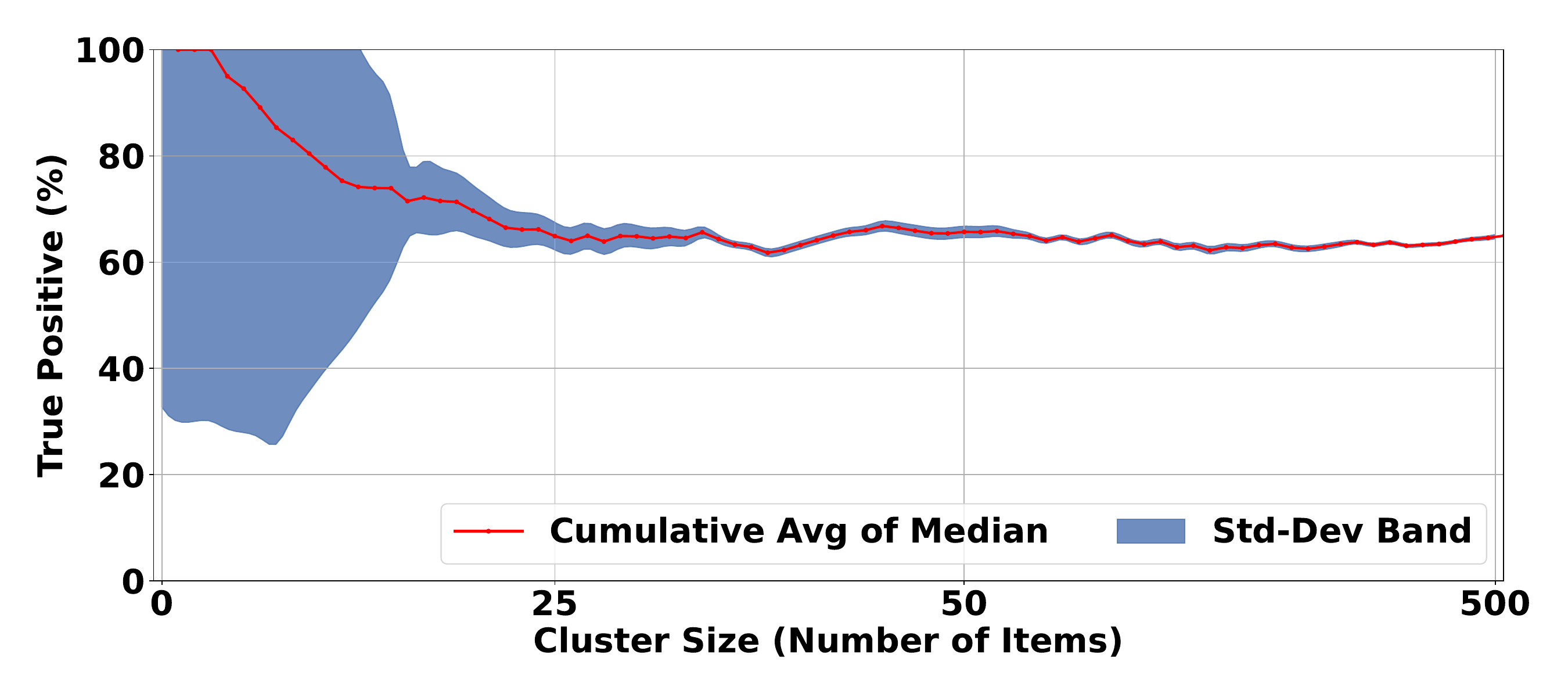} &
        \includegraphics[width=\linewidth]{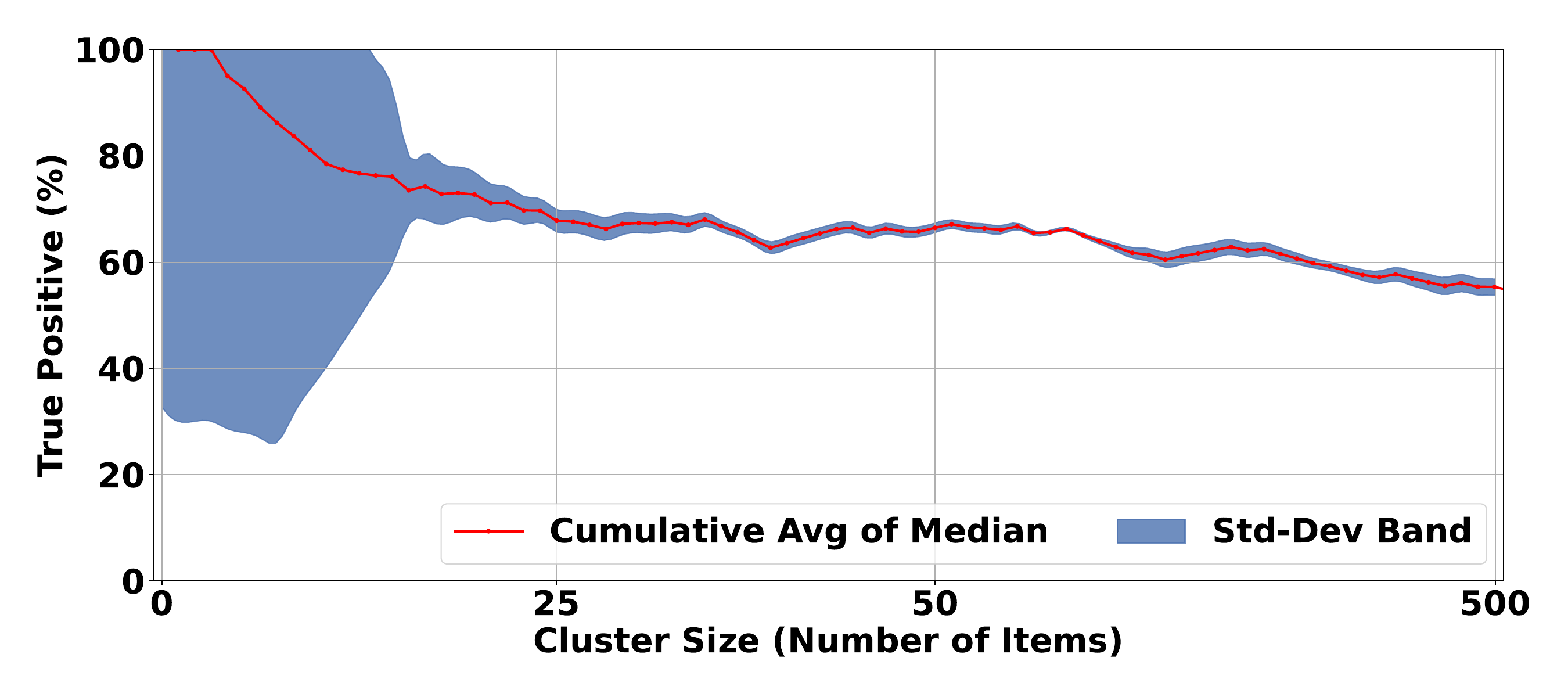} &
        \includegraphics[width=\linewidth]{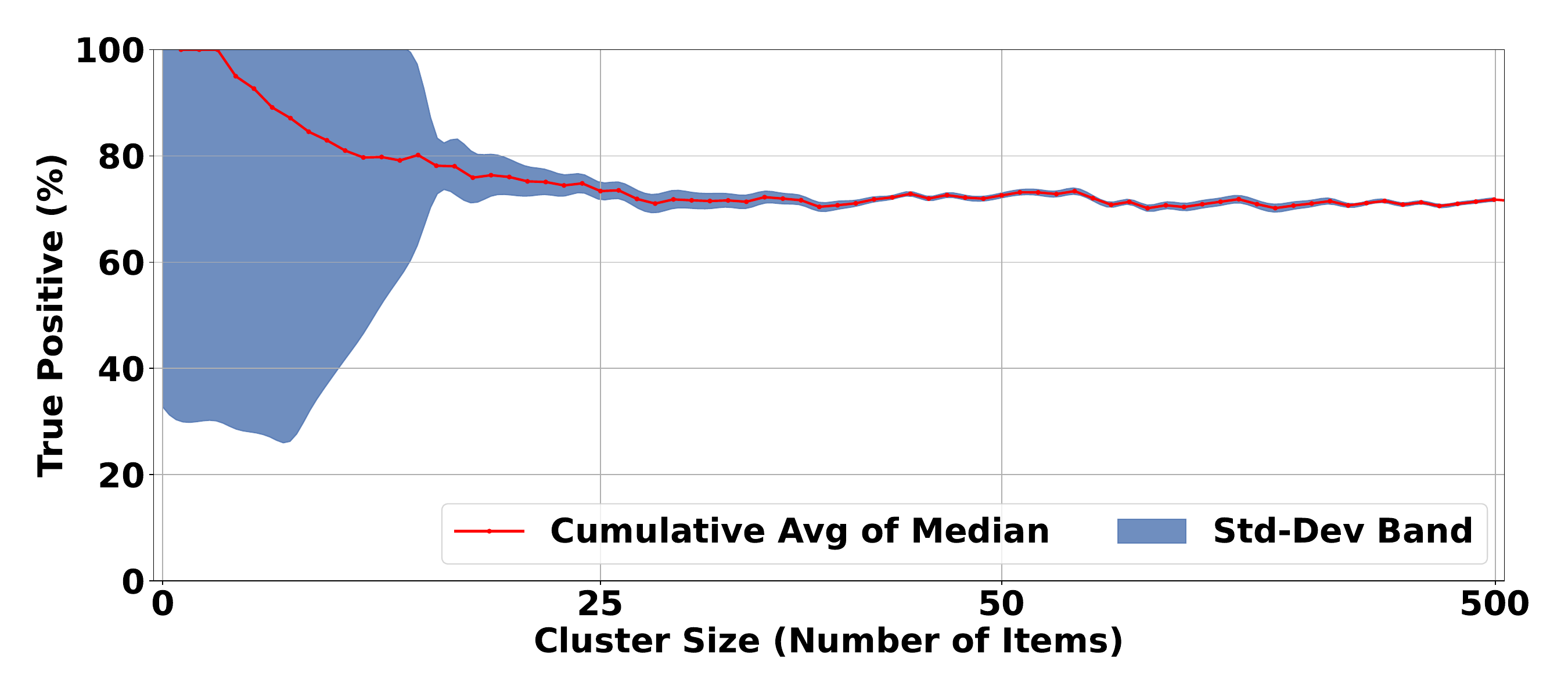} &
        \includegraphics[width=\linewidth]{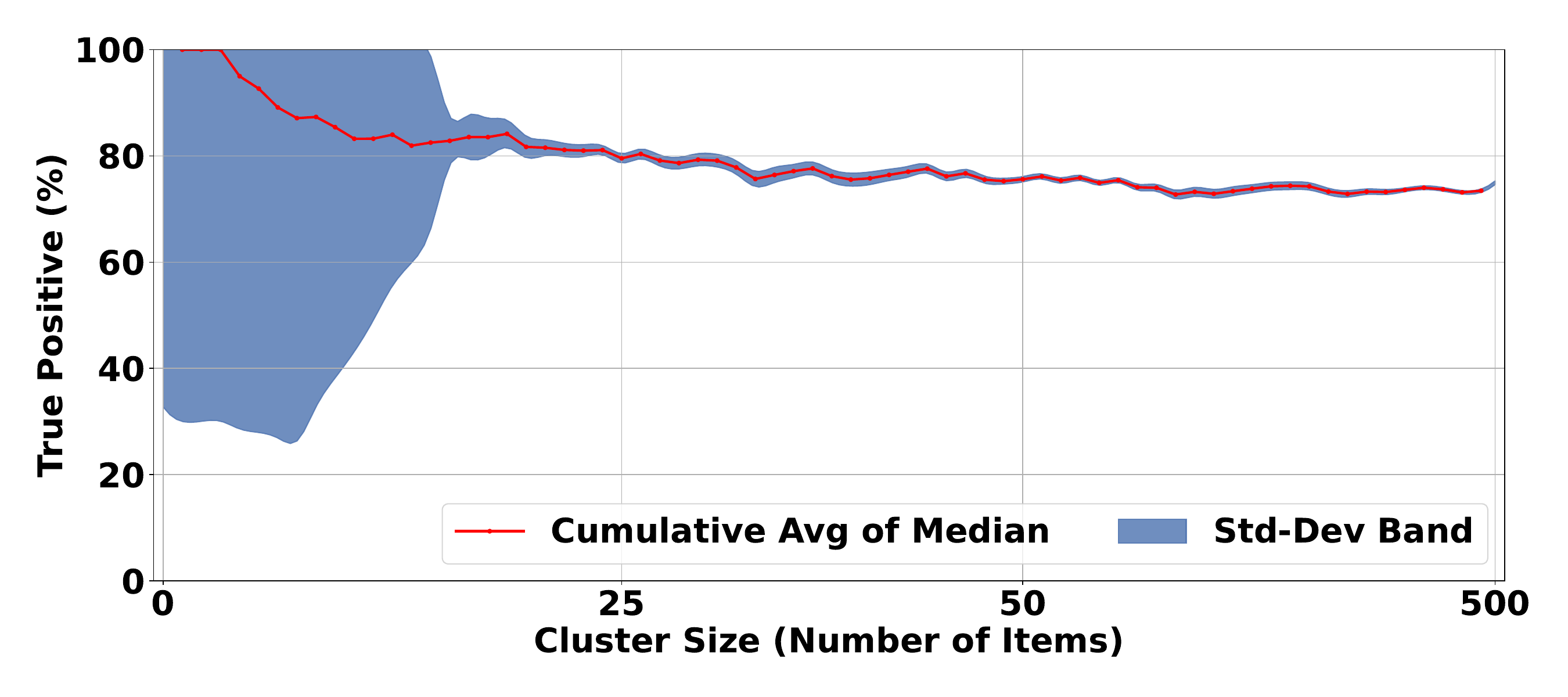} \\         
        \multicolumn{5}{c}{\textbf{Maximum $K$}} \\       
        \includegraphics[width=\linewidth]{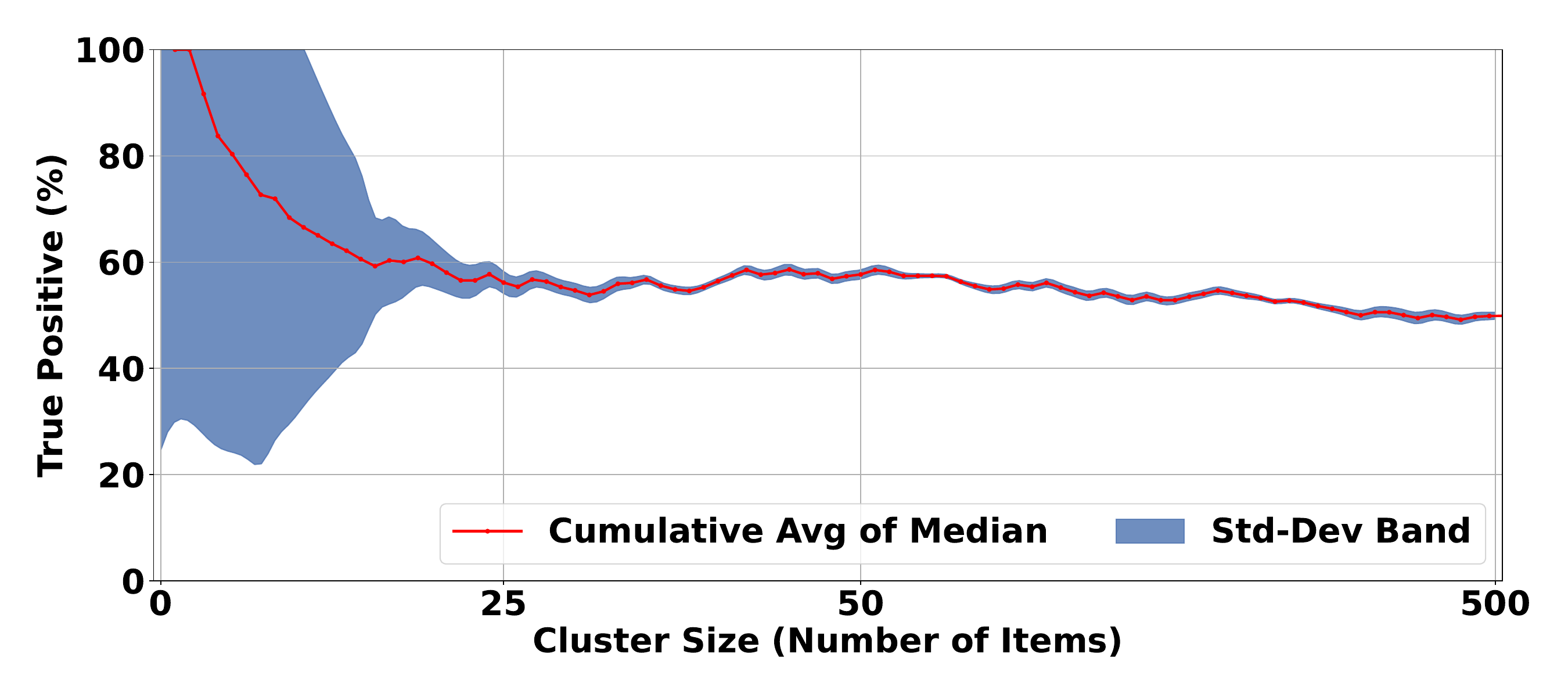} &
        \includegraphics[width=\linewidth]{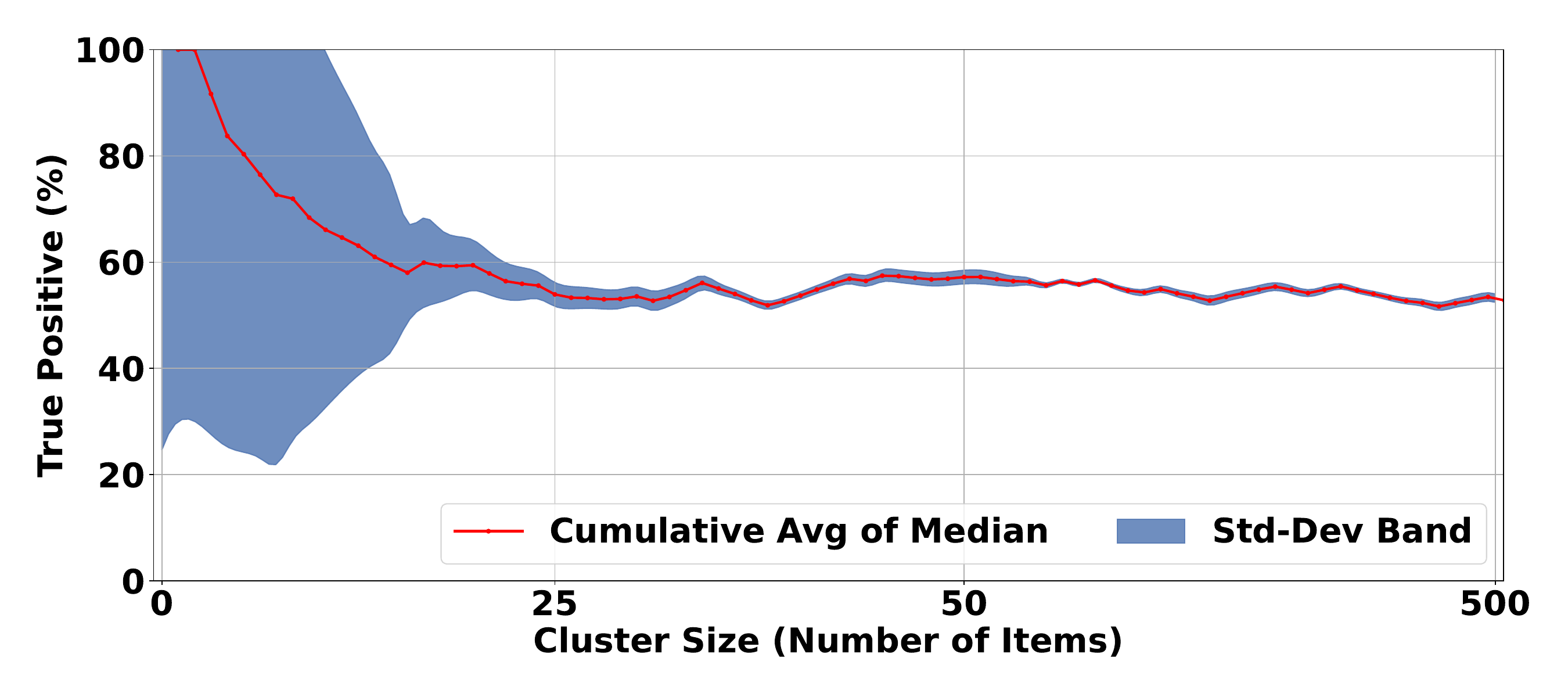} &
        \includegraphics[width=\linewidth]{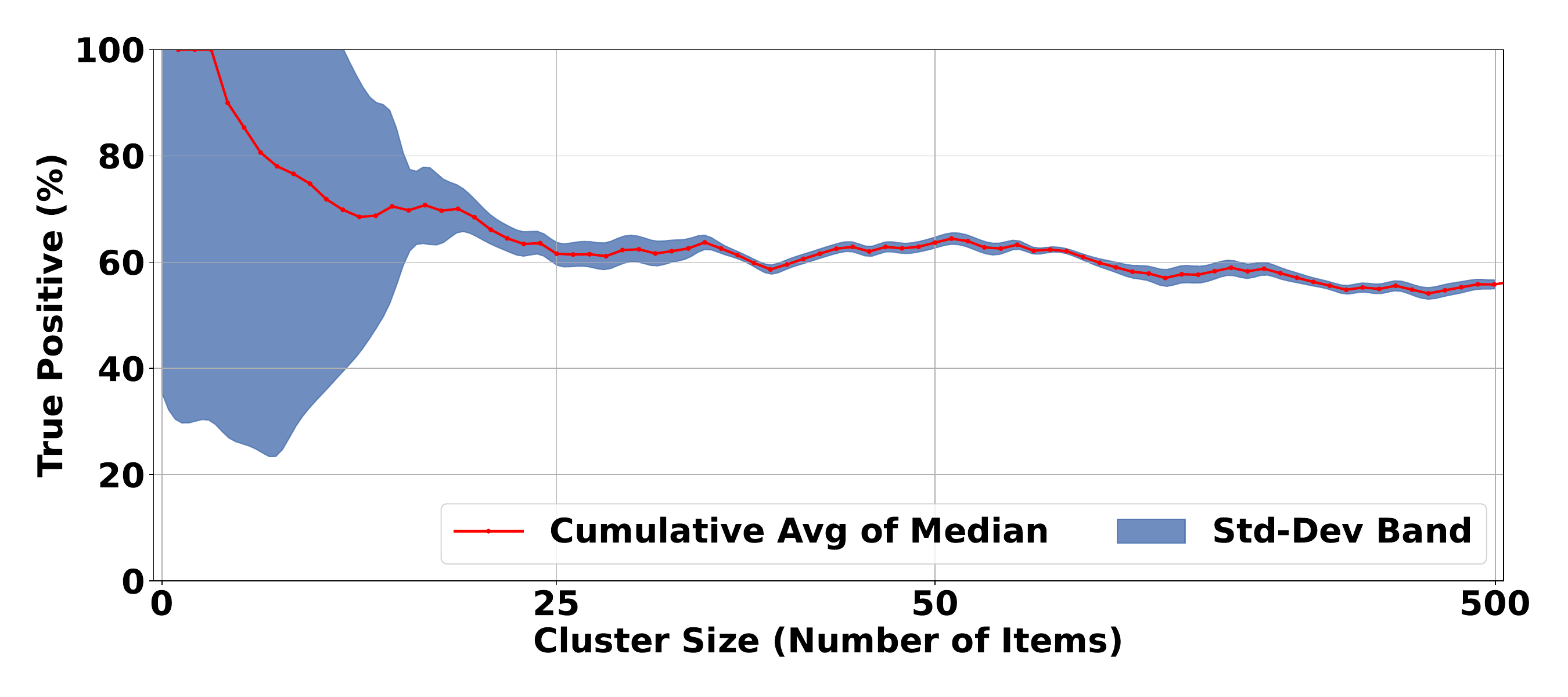} &
        \includegraphics[width=\linewidth]{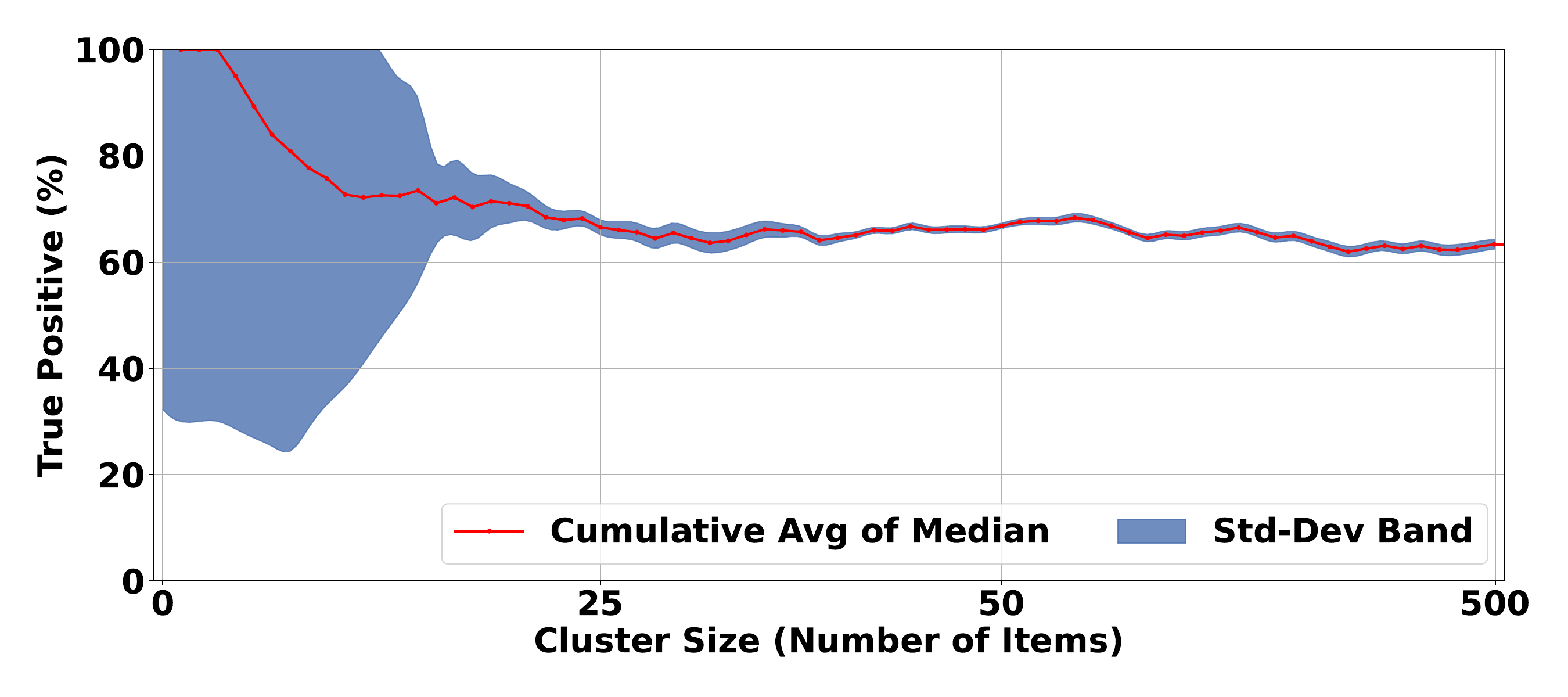} &
        \includegraphics[width=\linewidth]{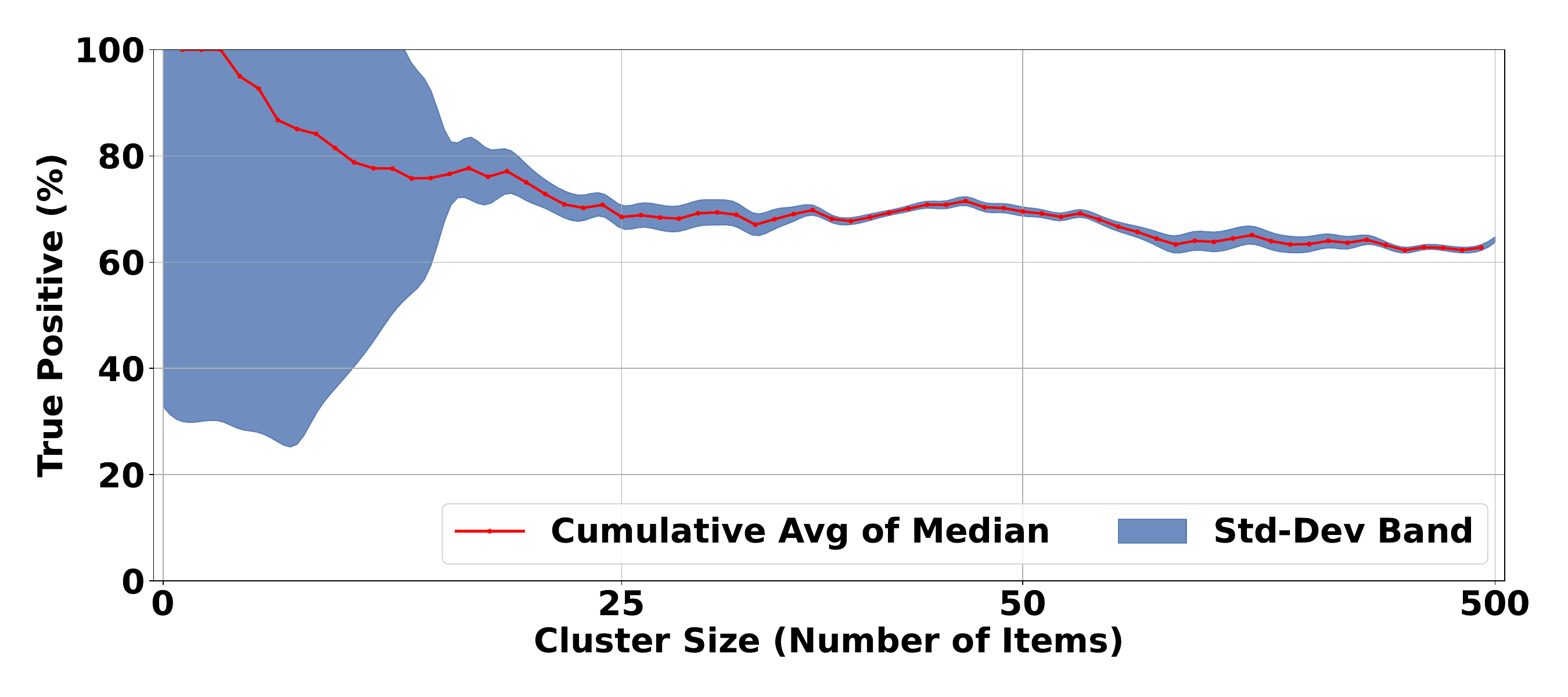} \\          
        \multicolumn{5}{c}{\textbf{Random $K$}} \\ 
        \includegraphics[width=\linewidth]{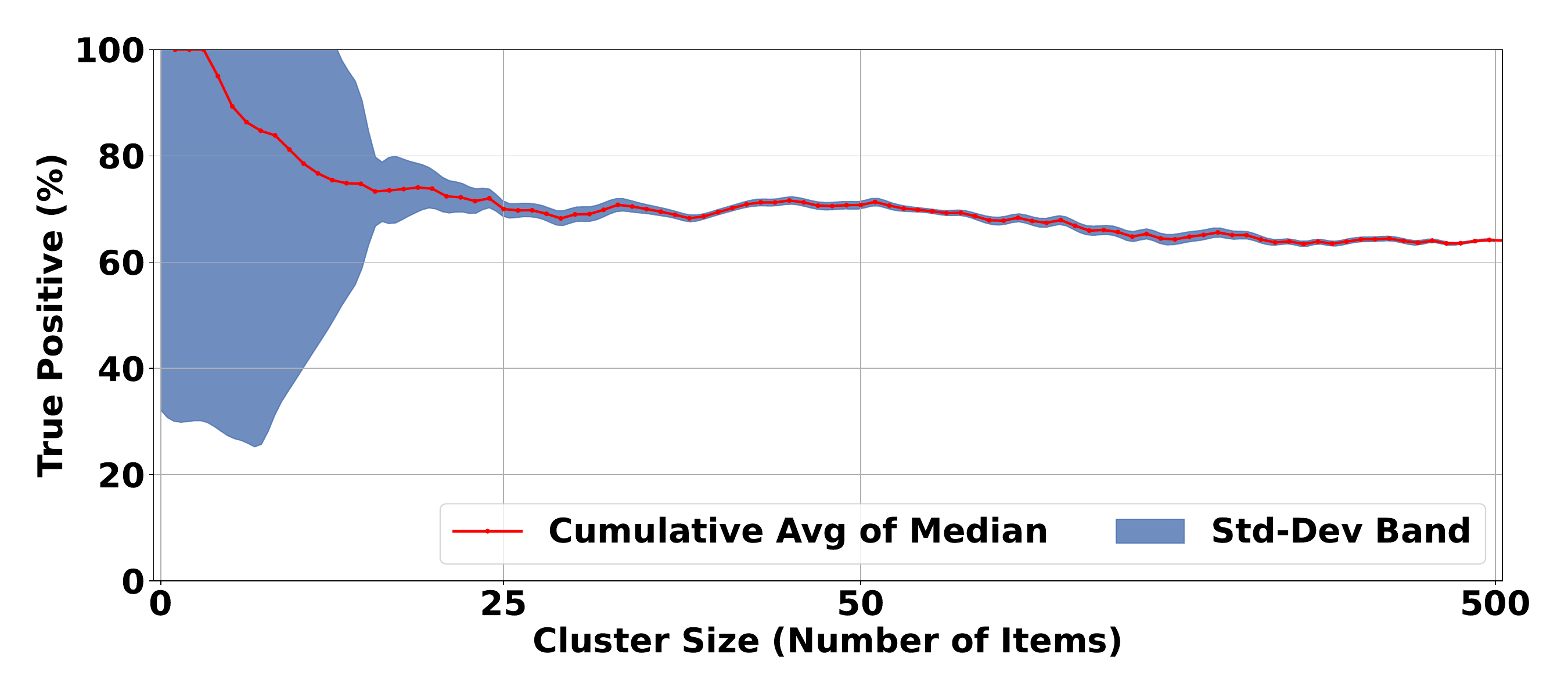} &
        \includegraphics[width=\linewidth]{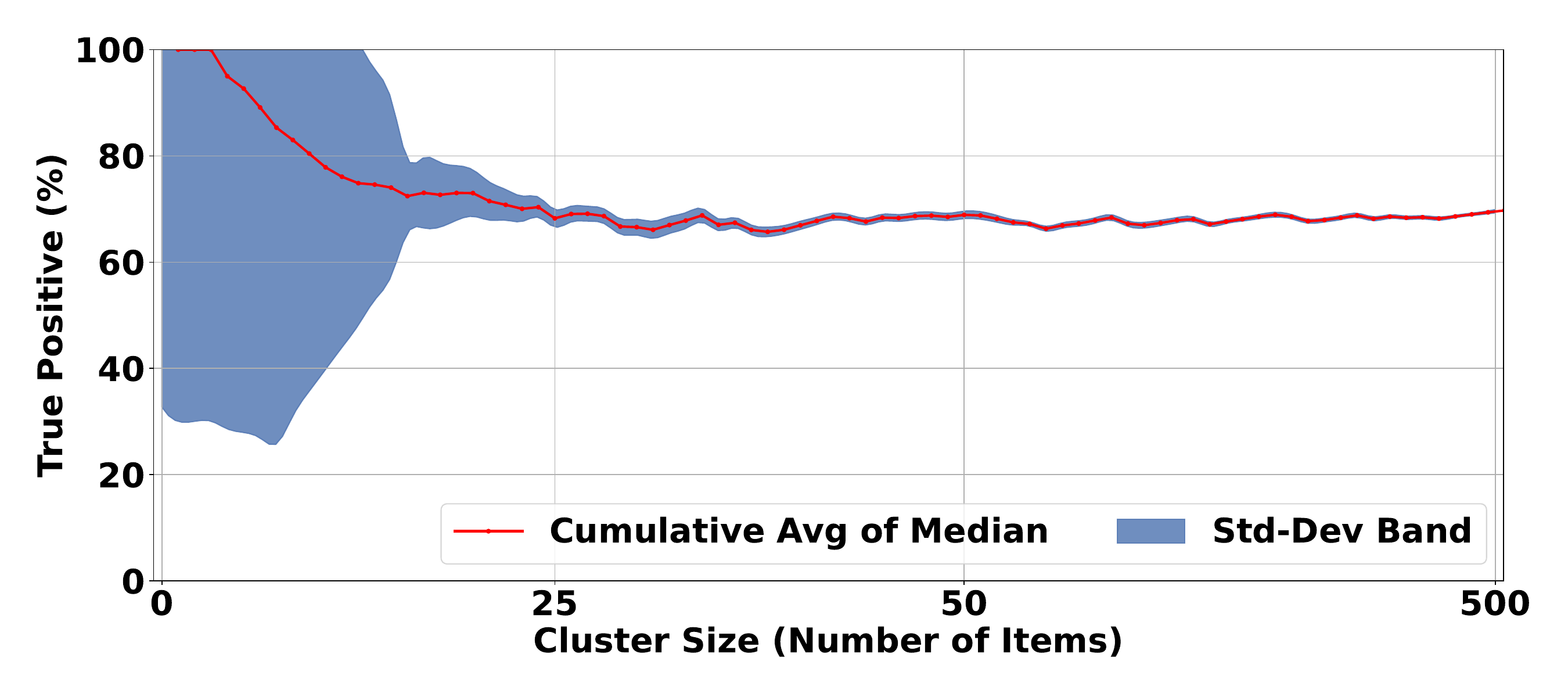} &
        \includegraphics[width=\linewidth]{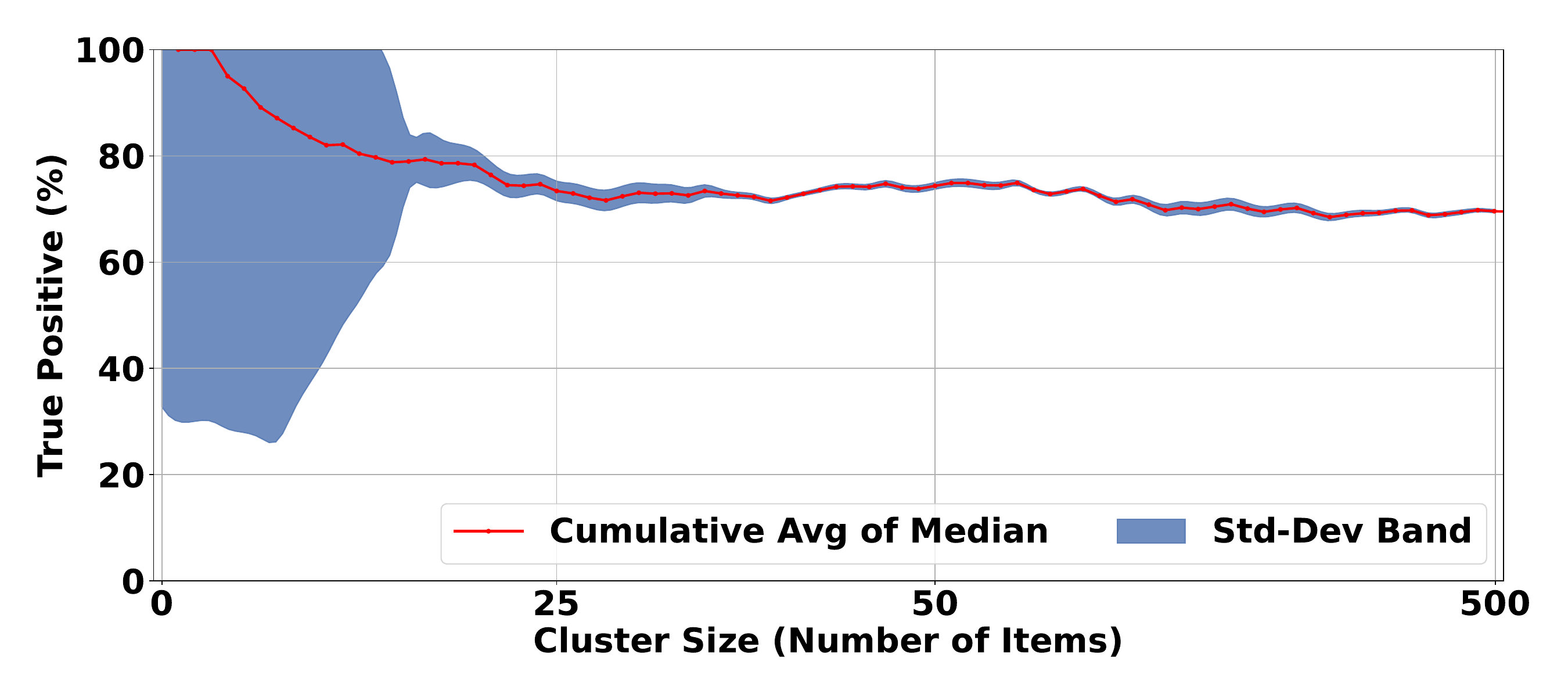} &
        \includegraphics[width=\linewidth]{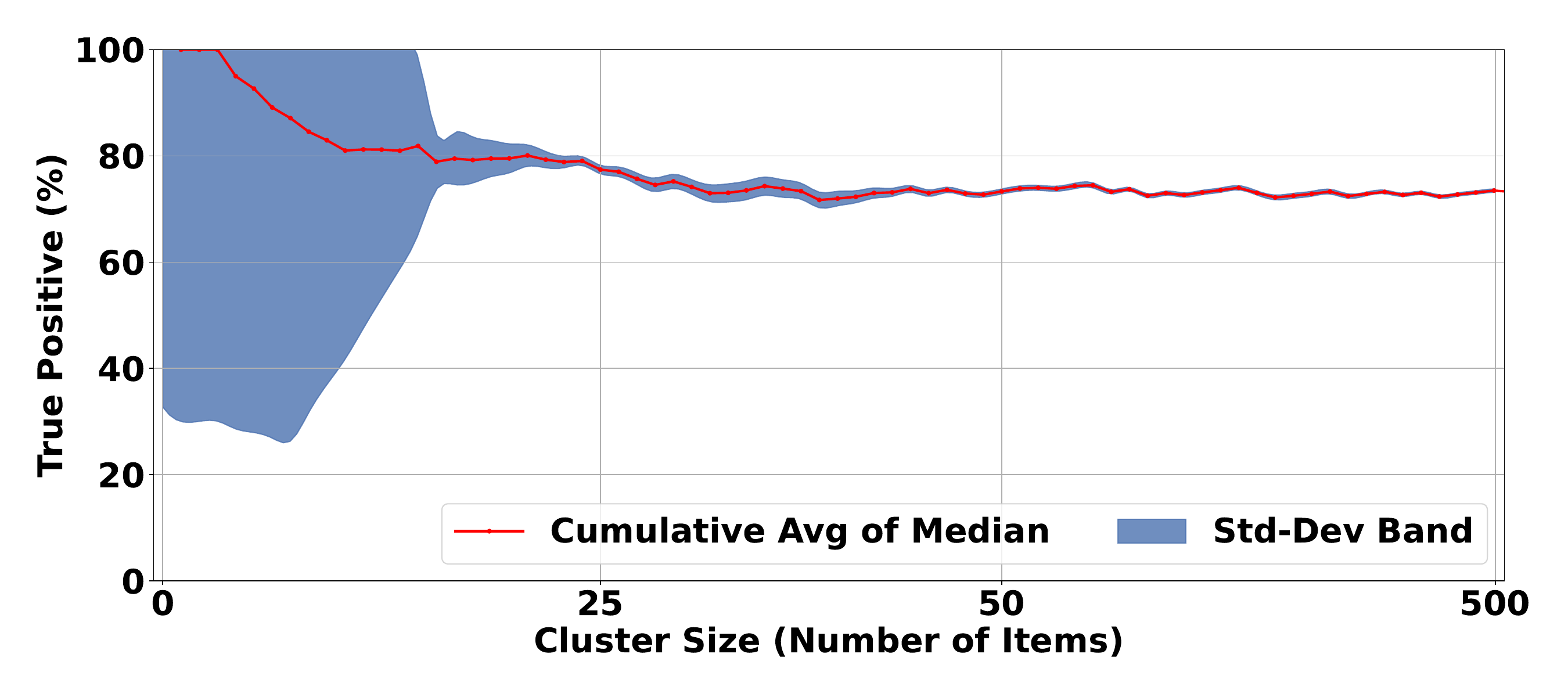} &
        \includegraphics[width=\linewidth]{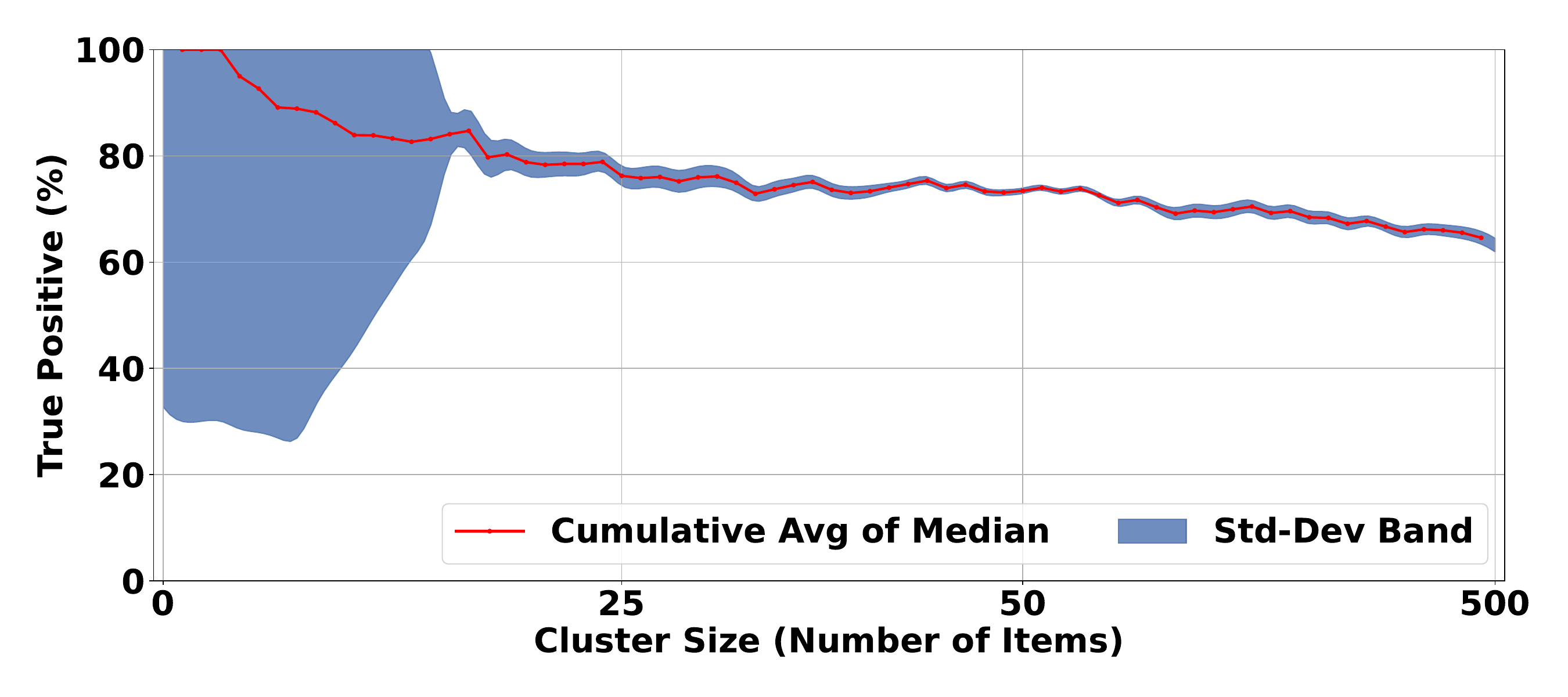} \\         
        
    \end{tabularx}%
    }
    \caption{SSdeep Threshold Variation with Retrained Bloom Filters}
    \label{tab:autoyara_autopyara_ssdeep_variation}
\end{table*}
\clearpage

\begin{table*}
    \centering
    \makebox[\textwidth][c]{%
    \begin{tabularx}{\textwidth}{*{5}{>{\centering\arraybackslash}X}}
        \textbf{TH 50} & \textbf{TH 60} & \textbf{TH 70} & \textbf{TH 80} & \textbf{TH 90} \\
                \hline

        \multicolumn{5}{c}{\textbf{AutoYara}} \\
        \includegraphics[width=\linewidth]{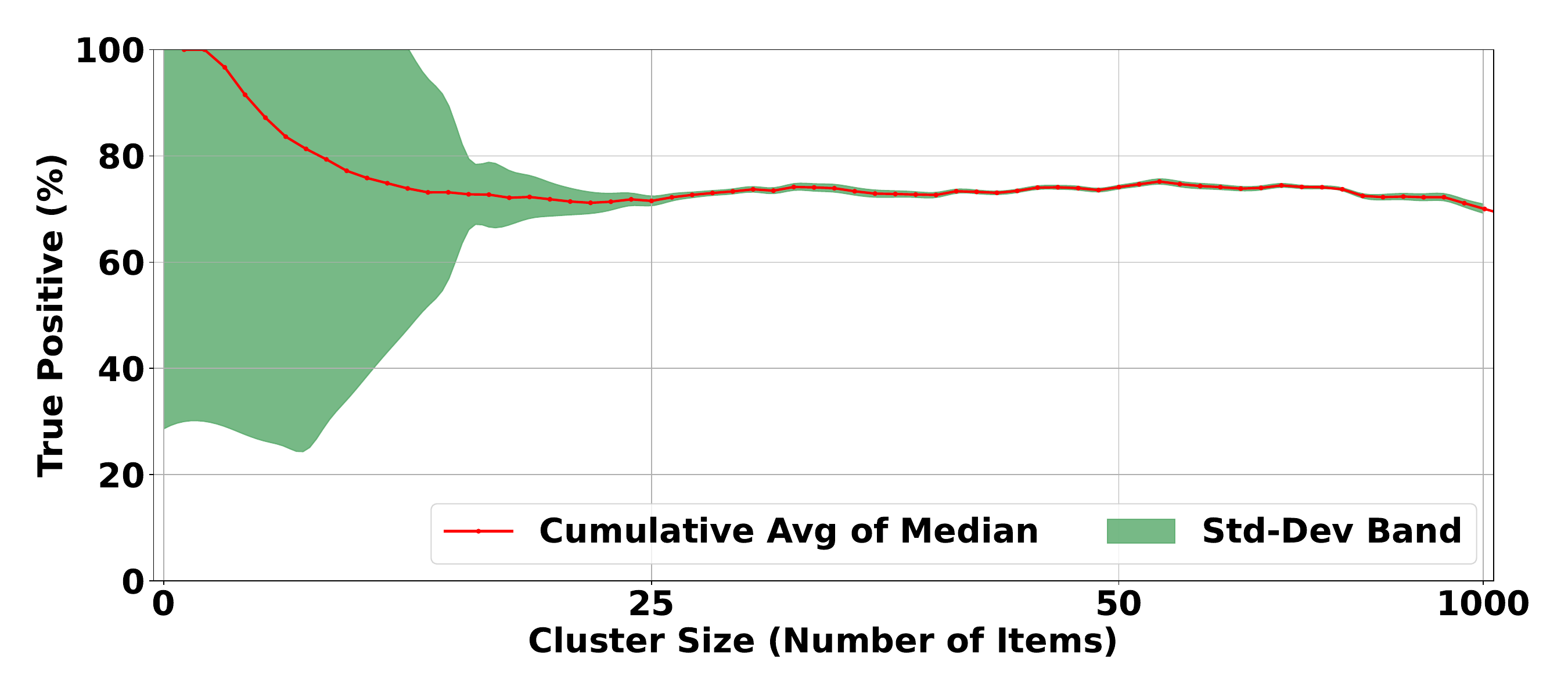} &
        \includegraphics[width=\linewidth]{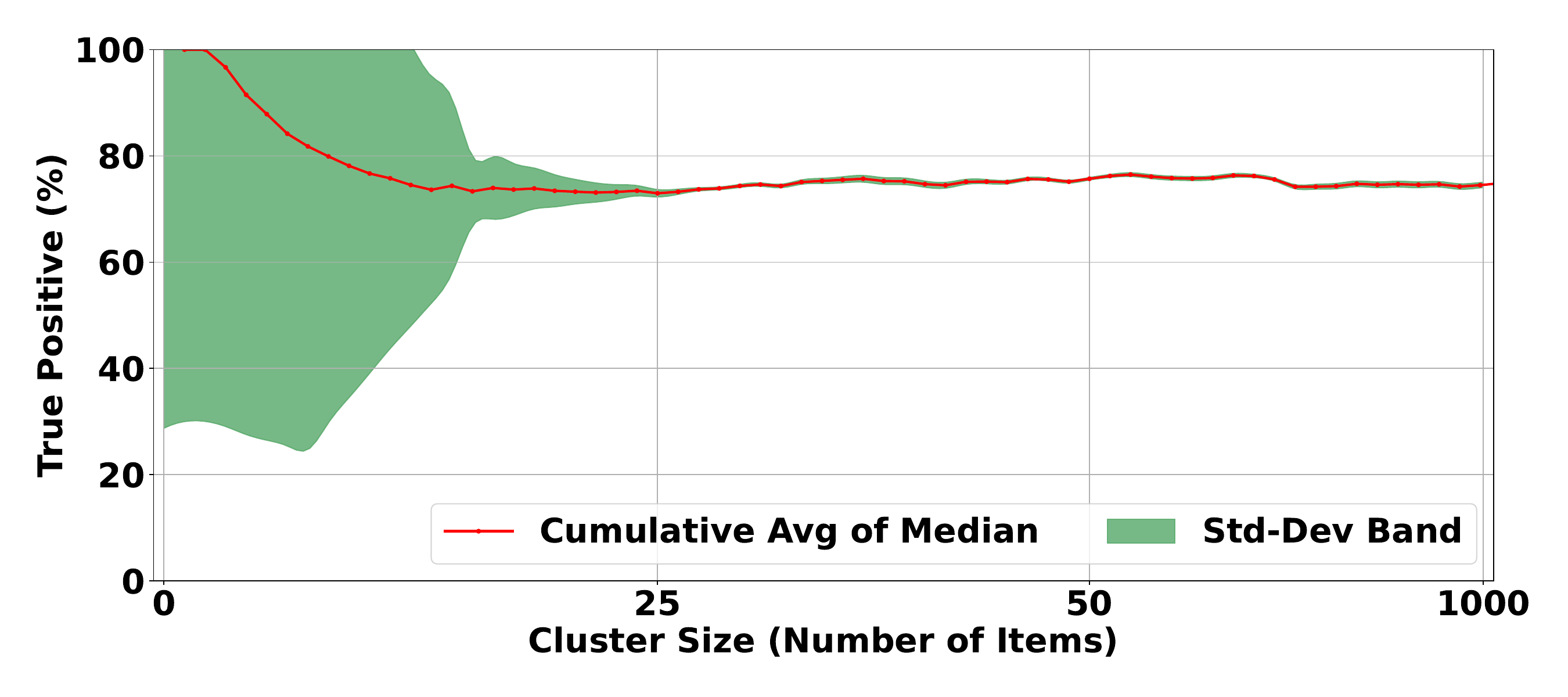} &
        \includegraphics[width=\linewidth]{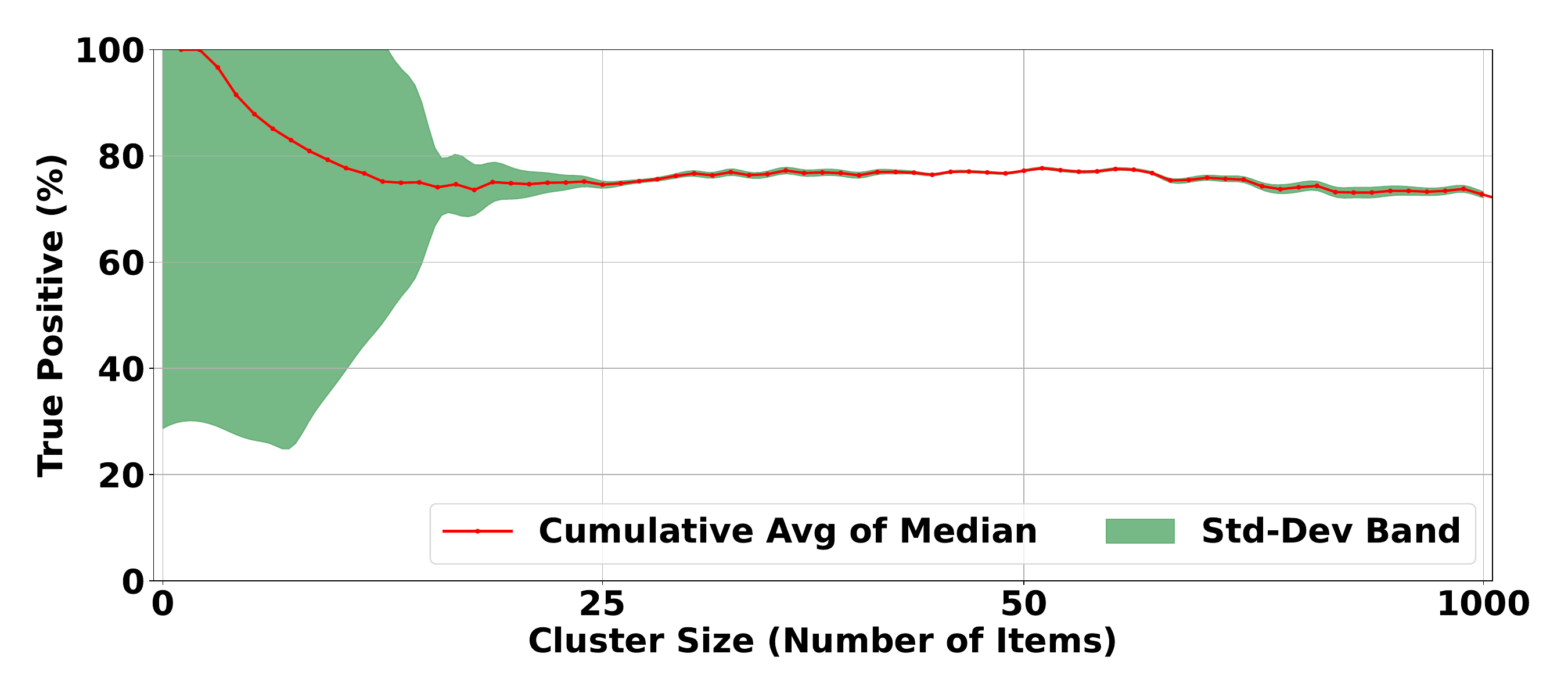} &
        \includegraphics[width=\linewidth]{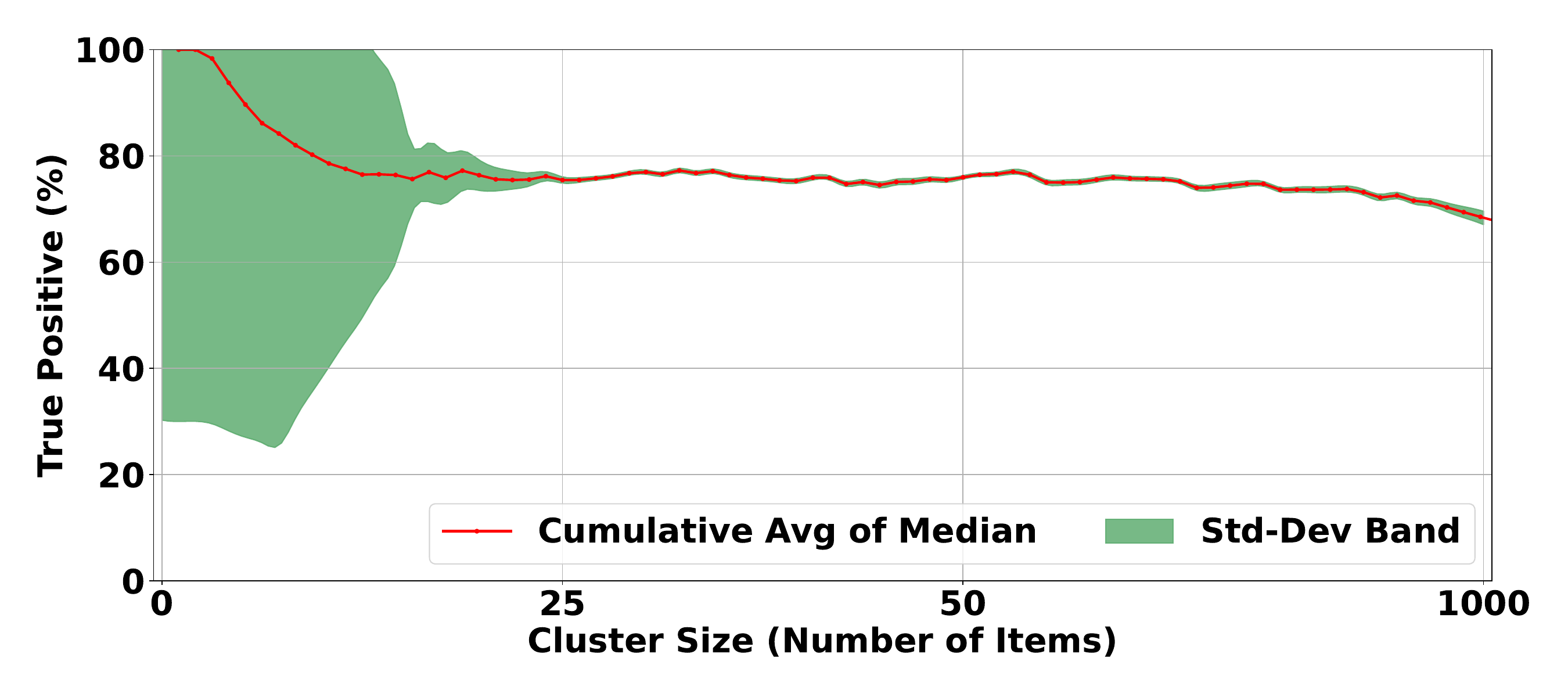} &
        \includegraphics[width=\linewidth]{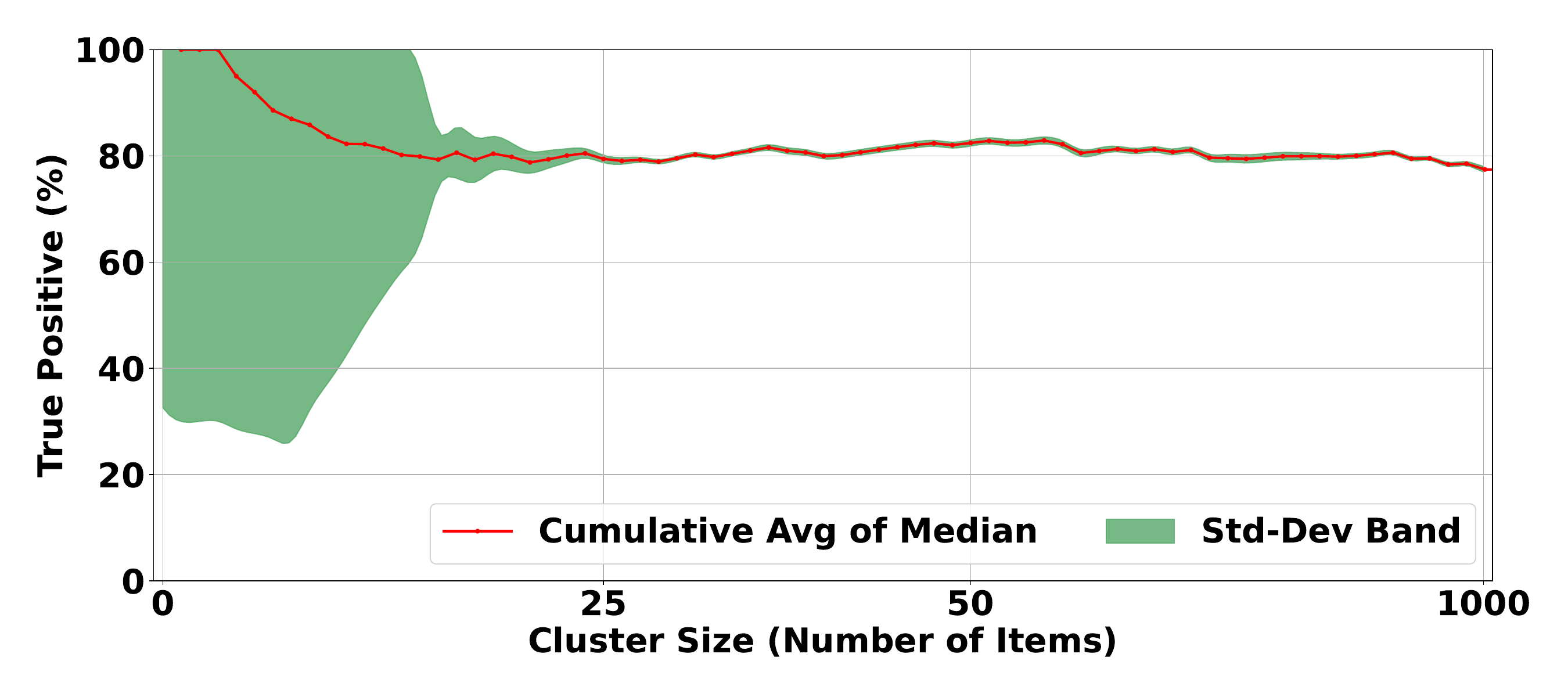} \\
        \multicolumn{5}{c}{\textbf{AutoPYara}} \\
        \includegraphics[width=\linewidth]{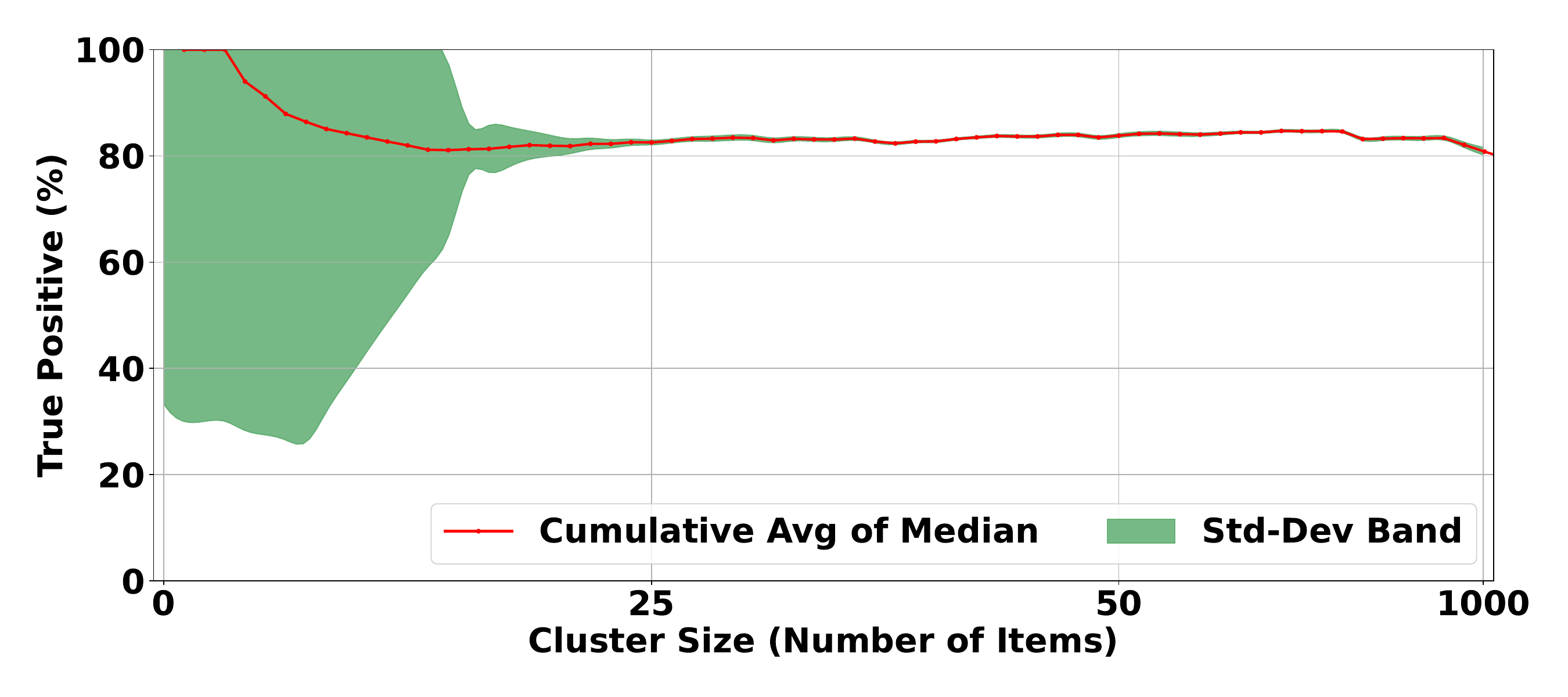} &
        \includegraphics[width=\linewidth]{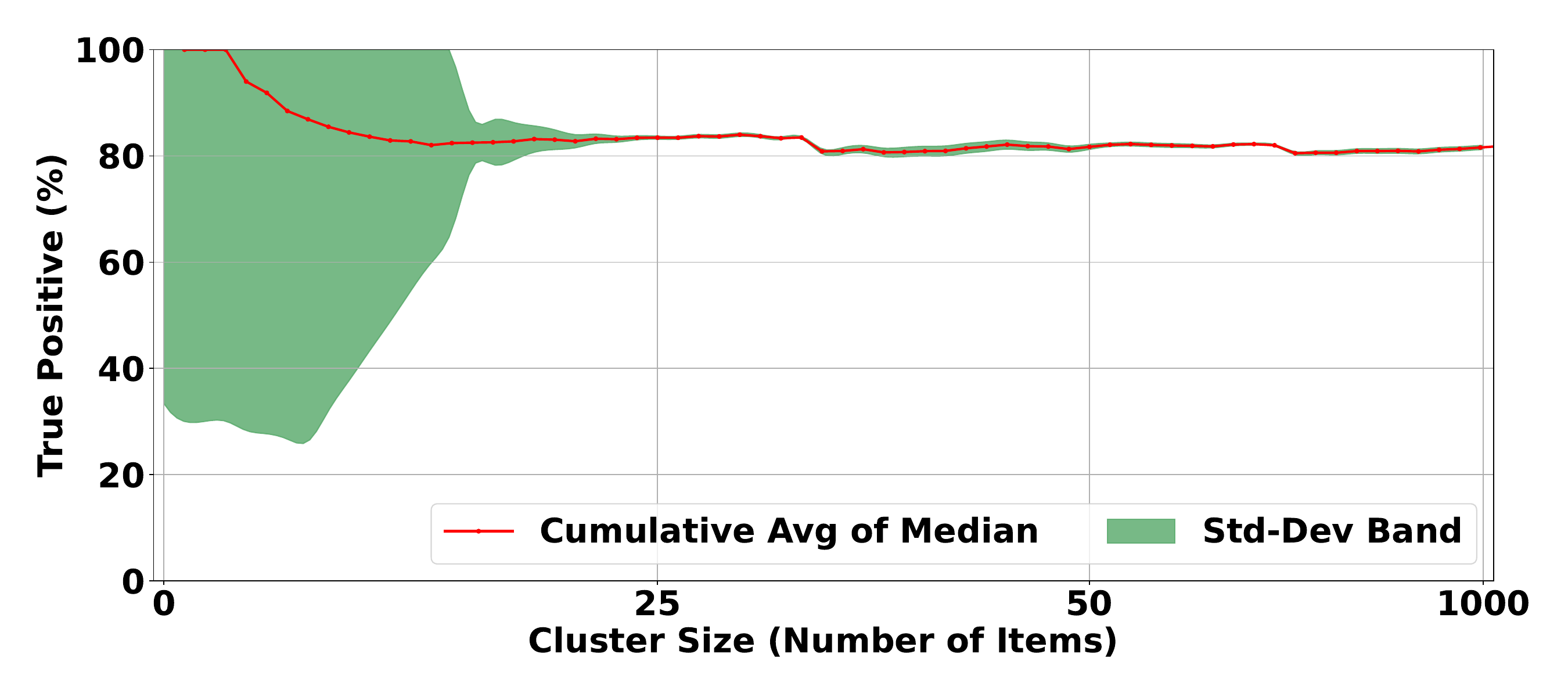} &
        \includegraphics[width=\linewidth]{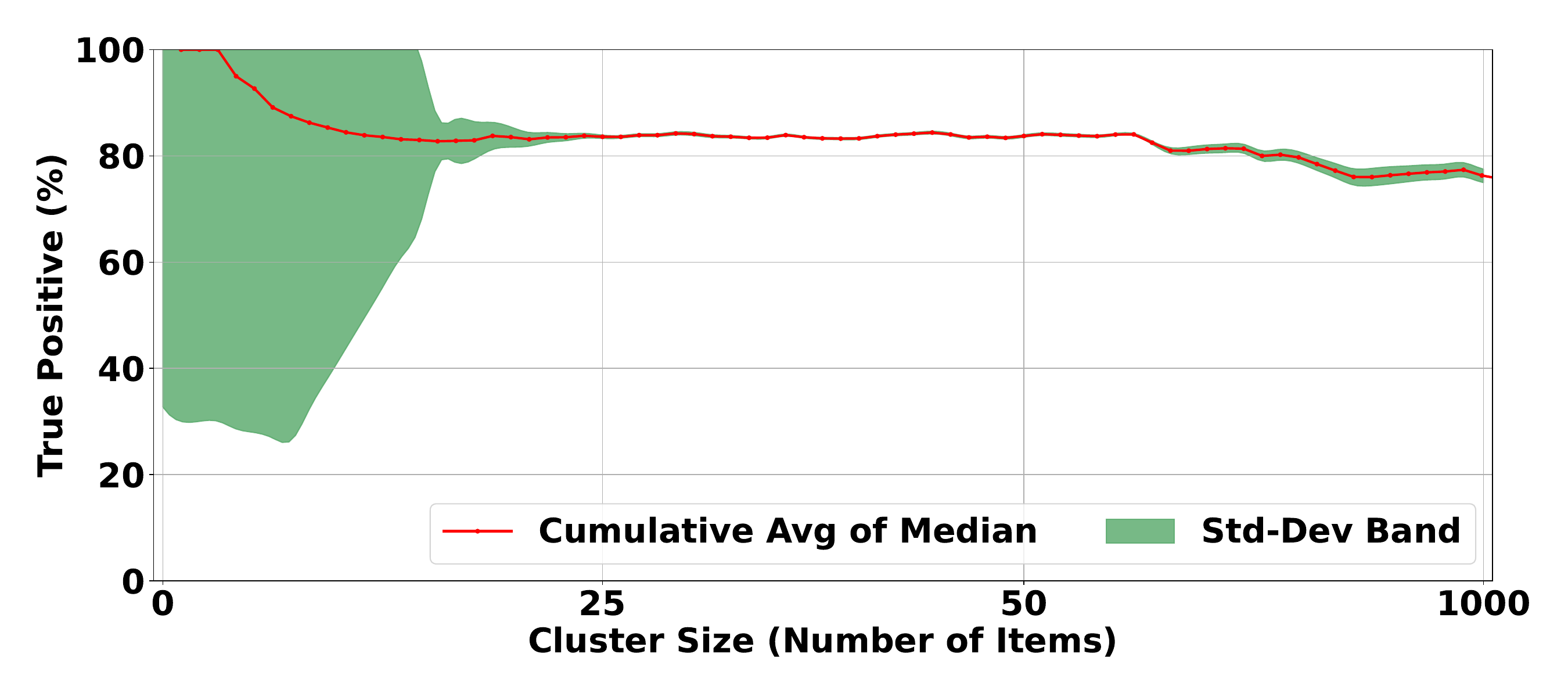} &
        \includegraphics[width=\linewidth]{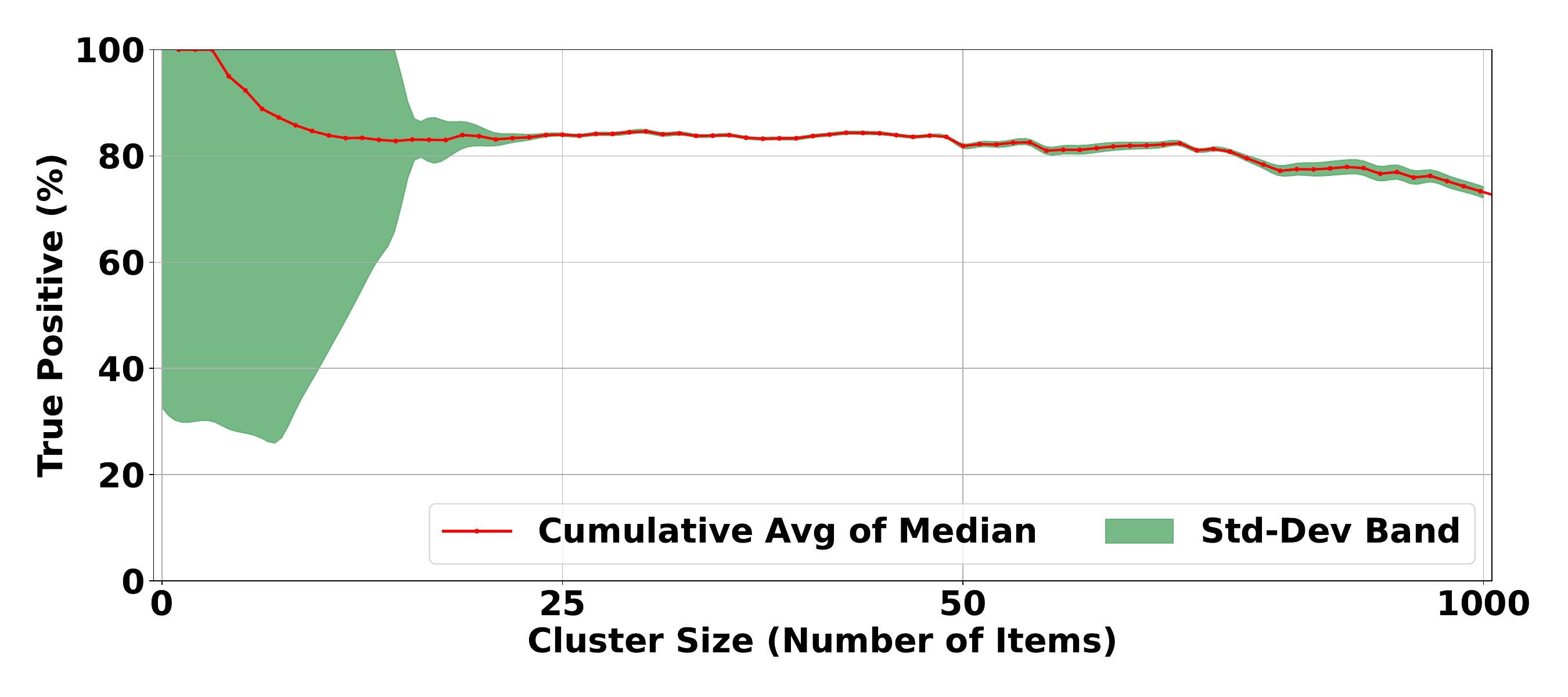} &
        \includegraphics[width=\linewidth]{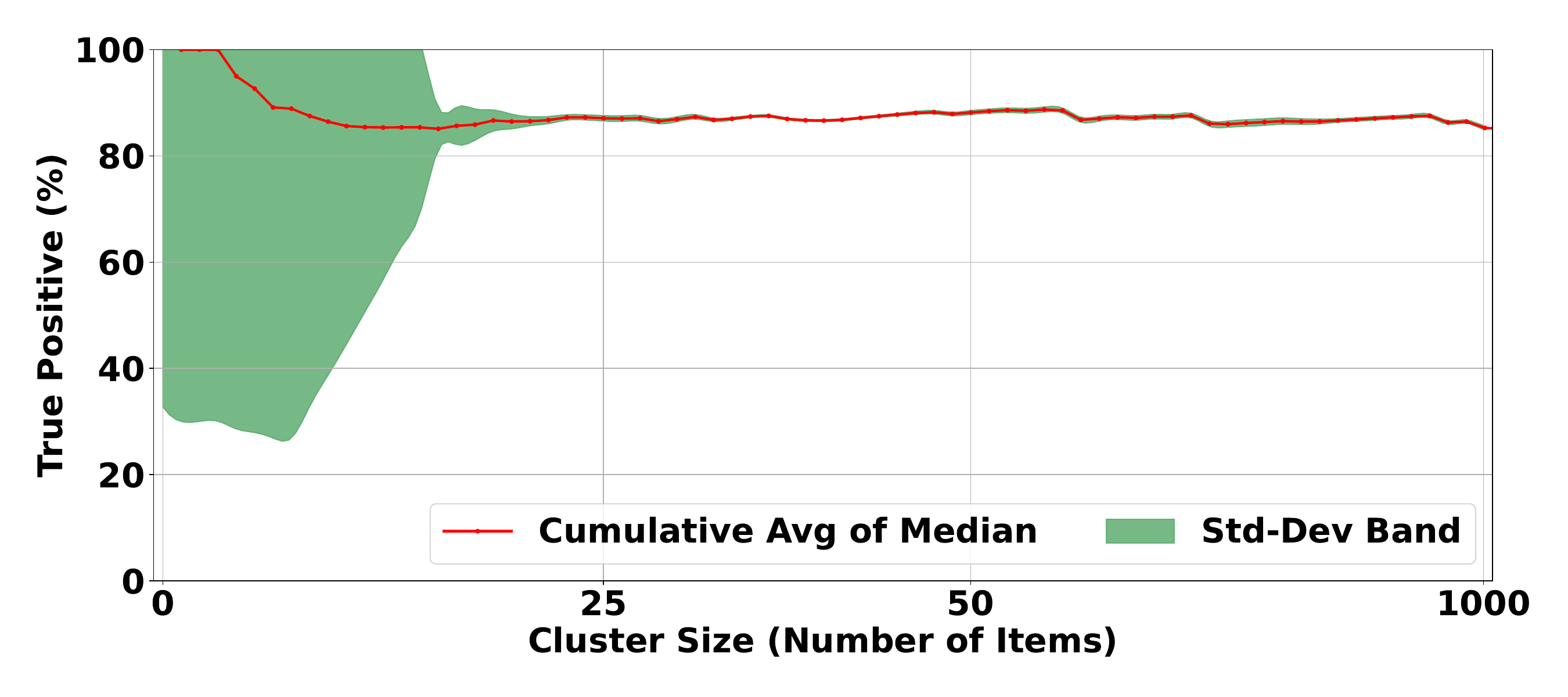} \\
        
        \multicolumn{5}{c}{\textbf{AutoPYara Ideal $K$ Configuration}} \\
        \includegraphics[width=\linewidth]{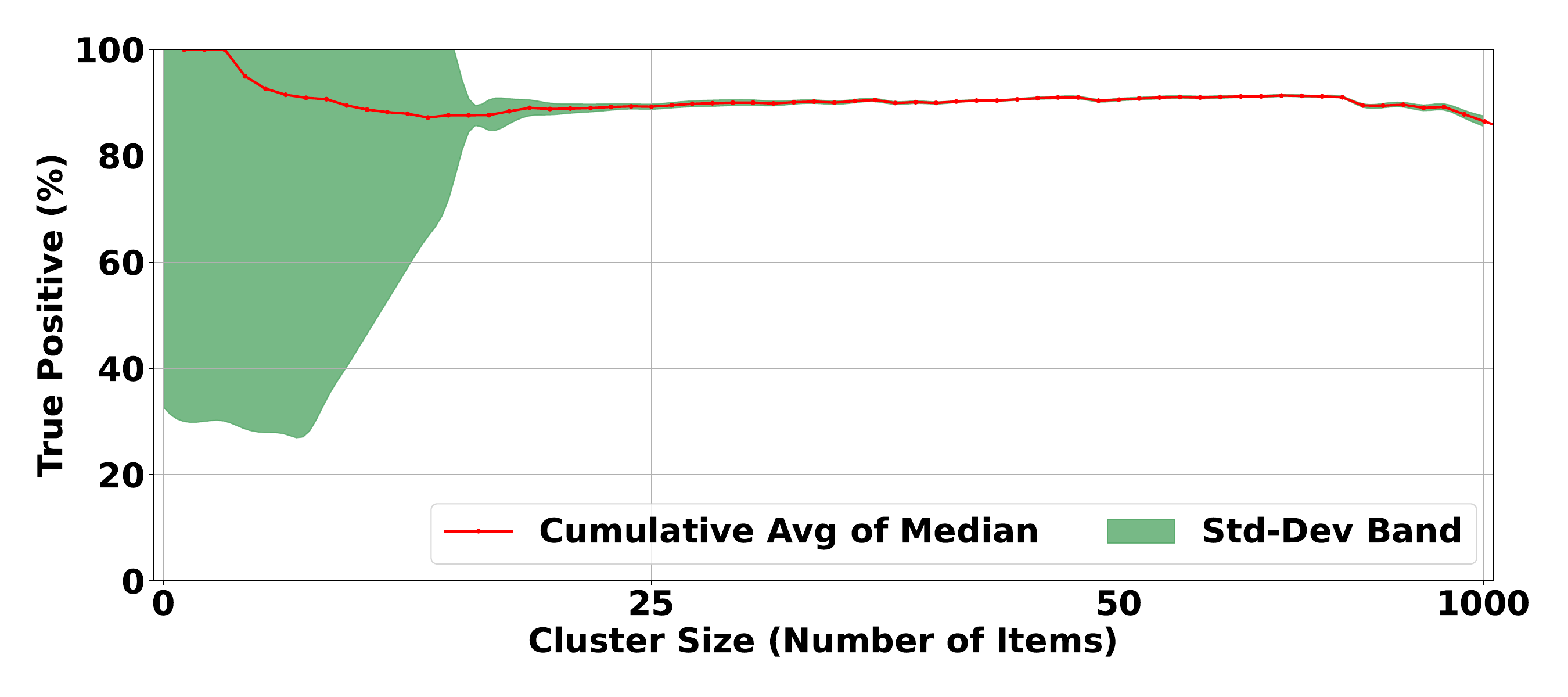} &
        \includegraphics[width=\linewidth]{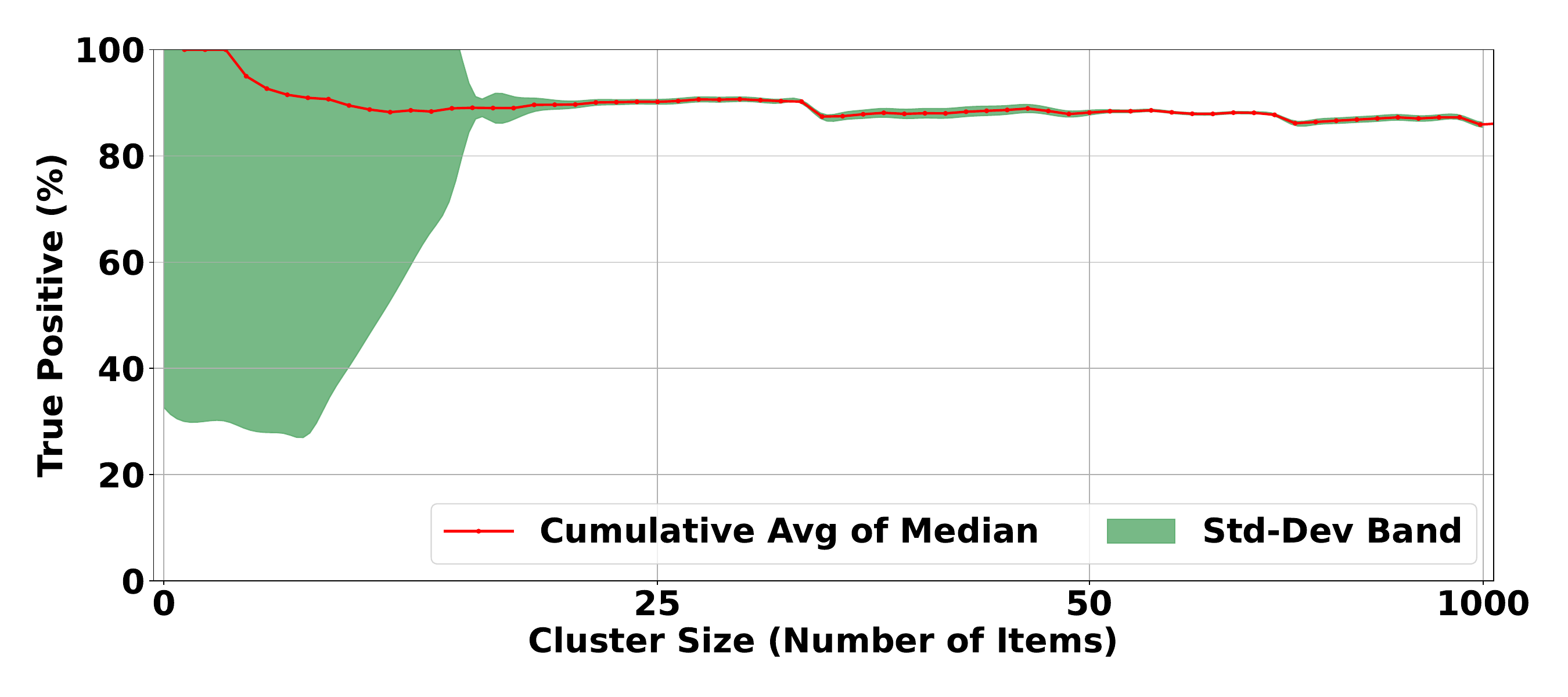} &
        \includegraphics[width=\linewidth]{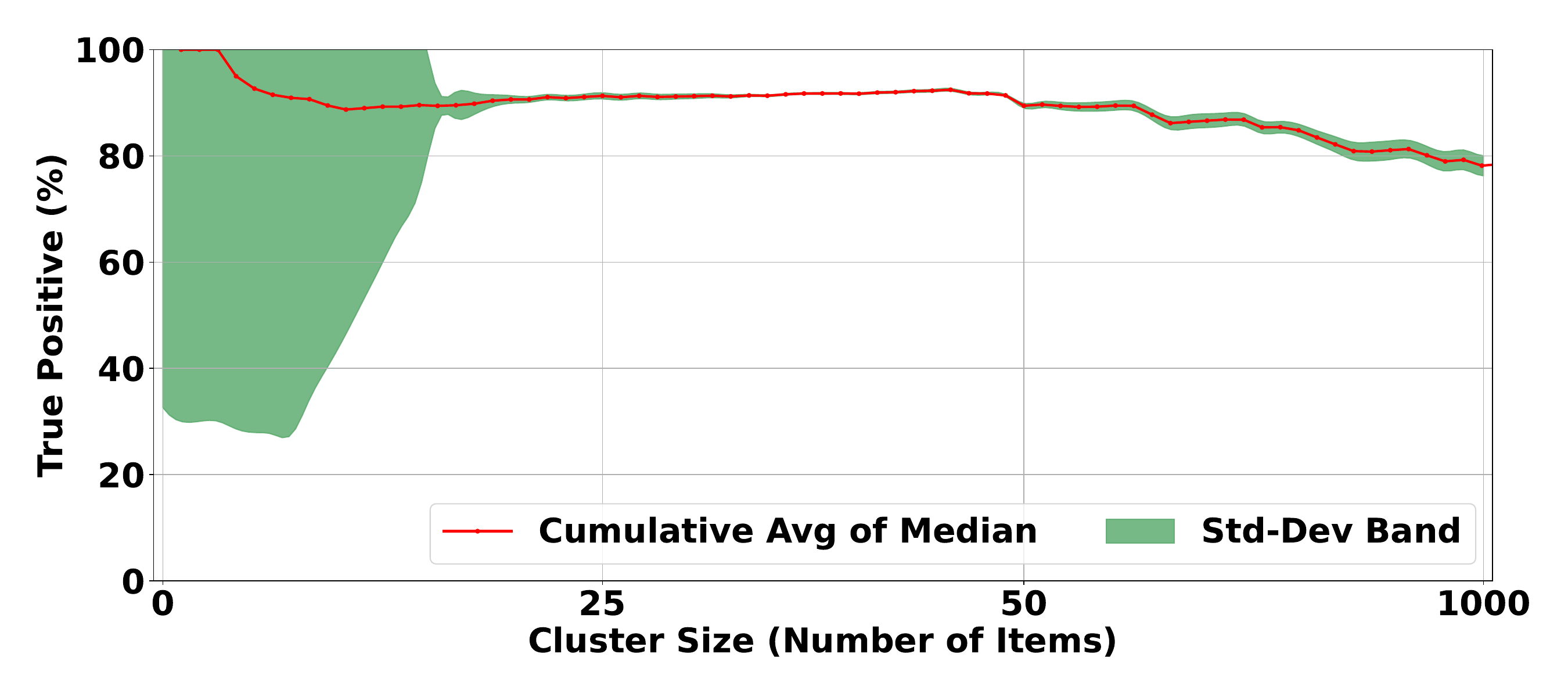} &
        \includegraphics[width=\linewidth]{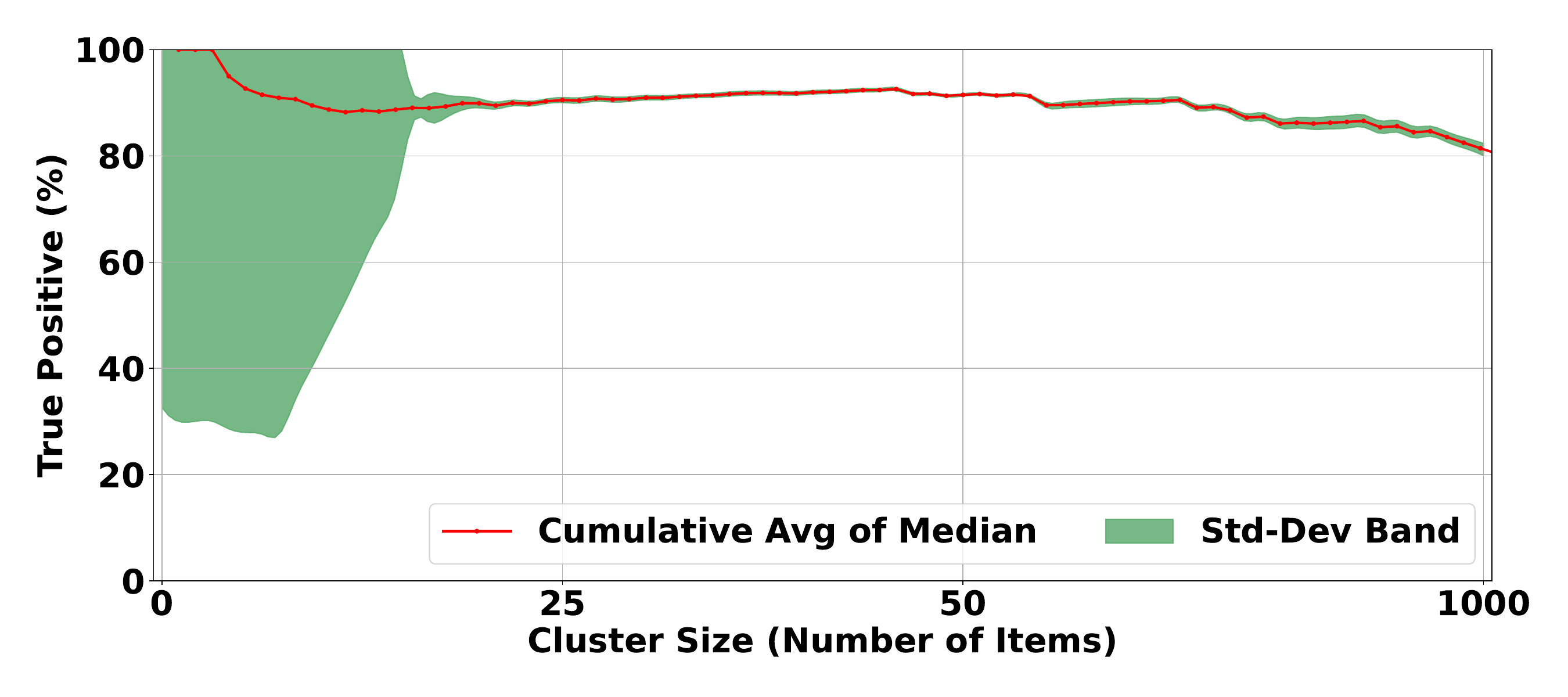} &
        \includegraphics[width=\linewidth]{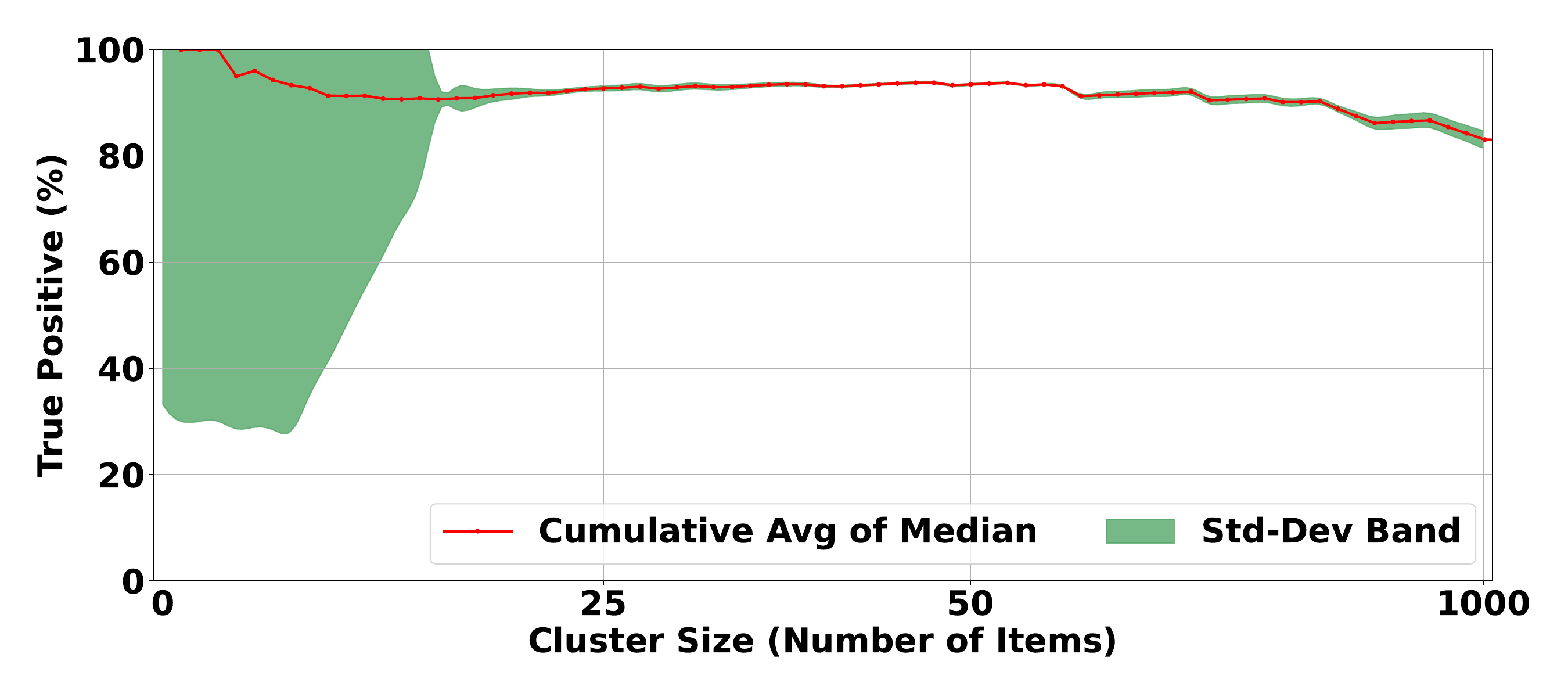} \\        
        \multicolumn{5}{c}{\textbf{AutoPYara Worst $K$ Configuration}} \\
        \includegraphics[width=\linewidth]{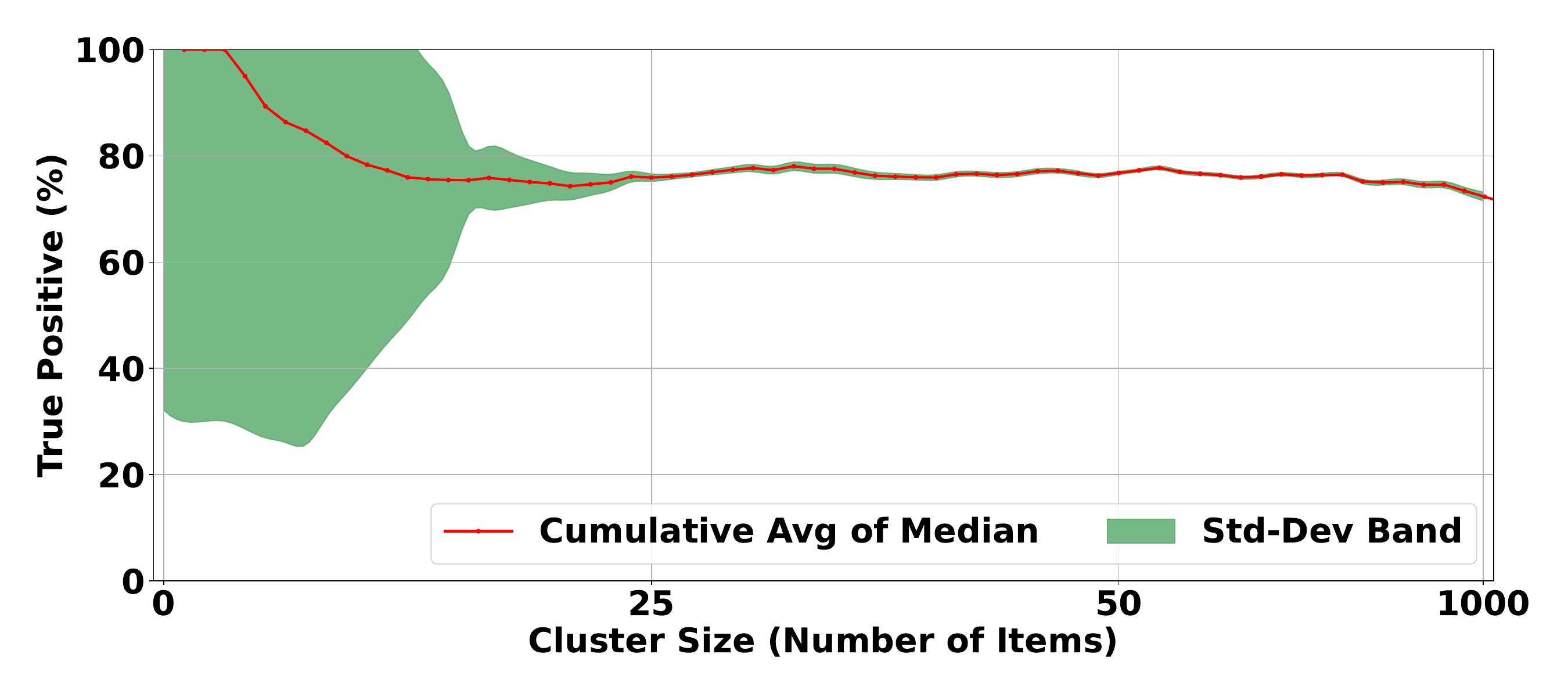} &
        \includegraphics[width=\linewidth]{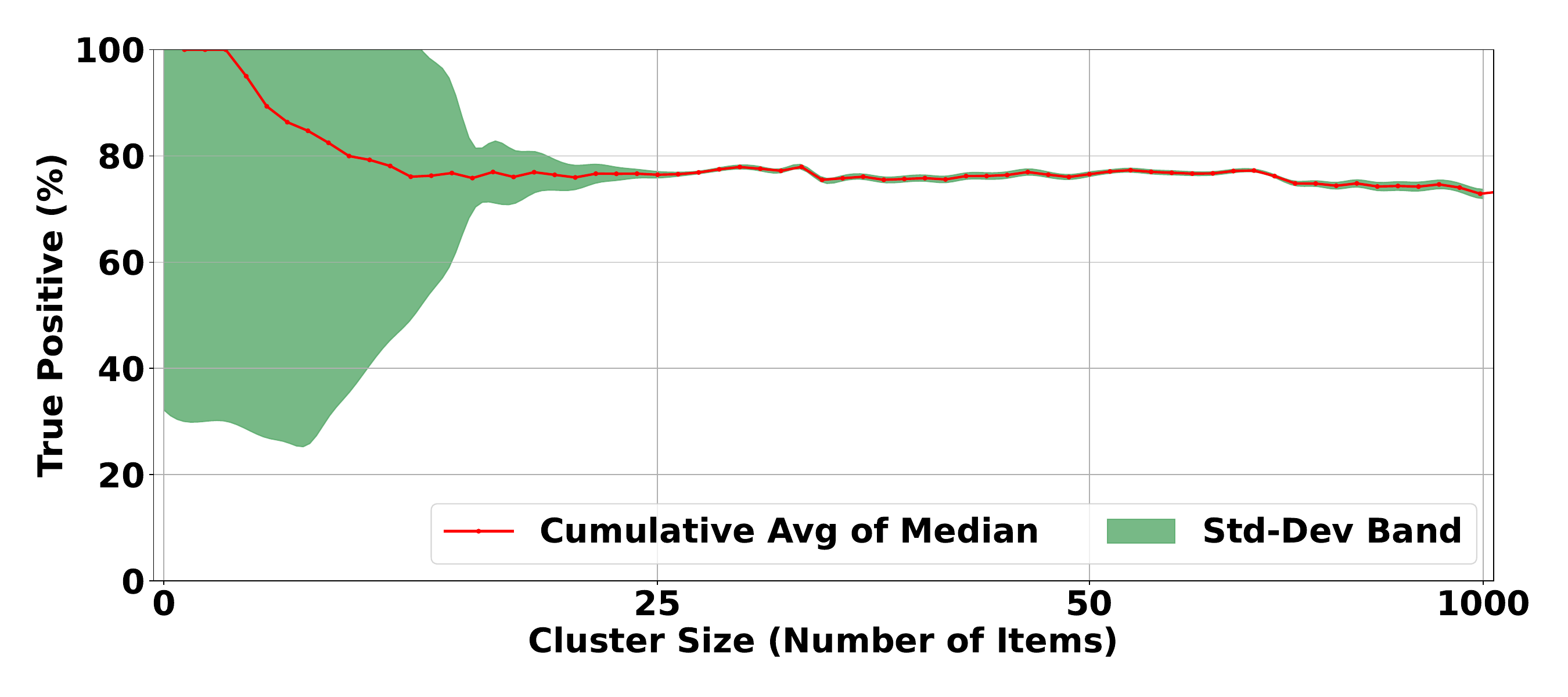} &
        \includegraphics[width=\linewidth]{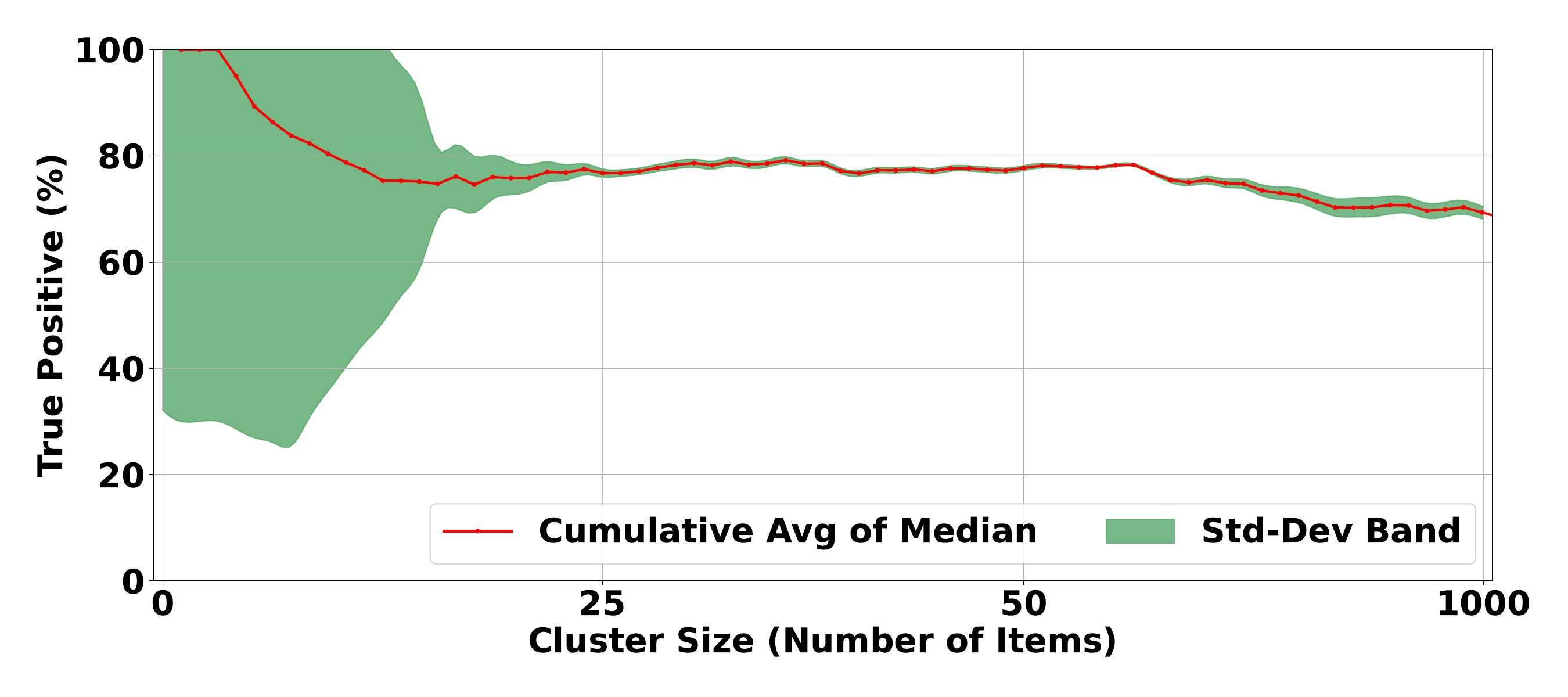} &
        \includegraphics[width=\linewidth]{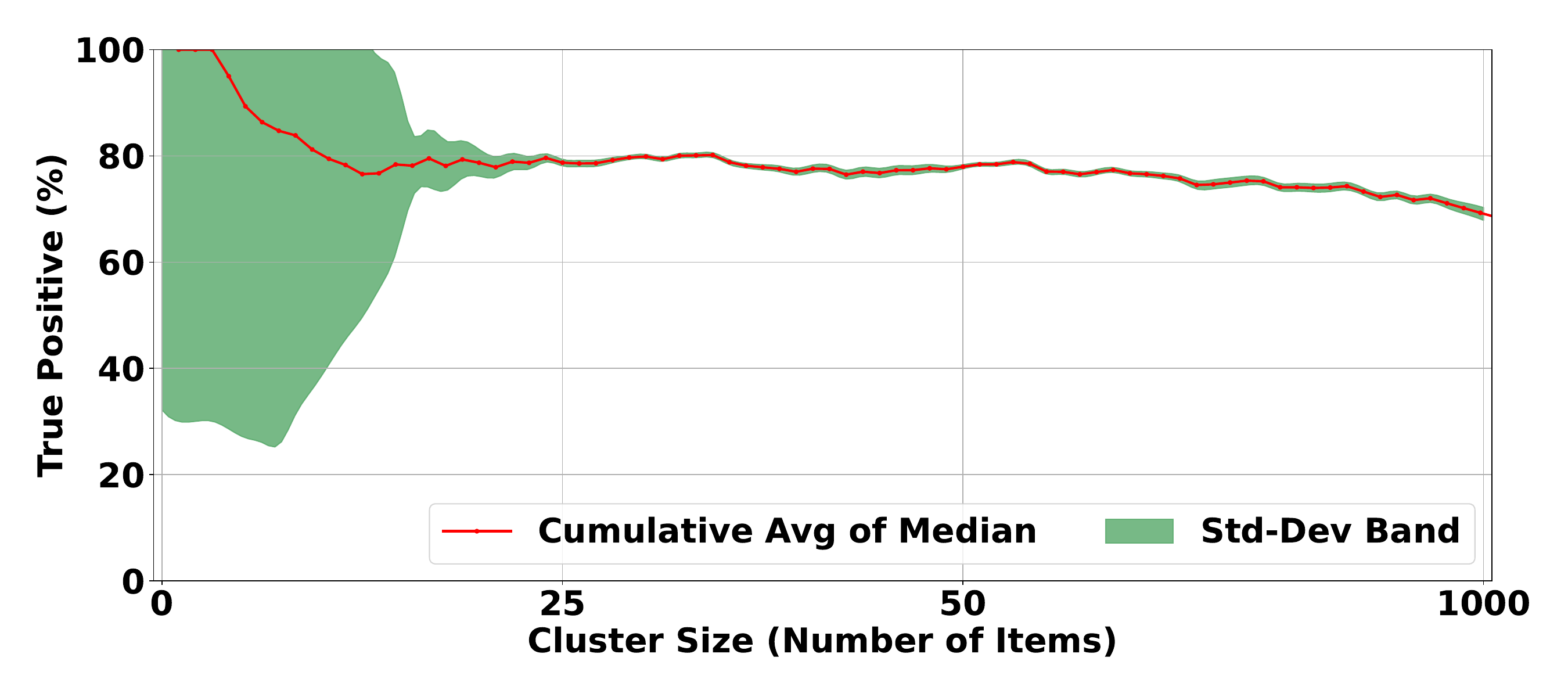} &
        \includegraphics[width=\linewidth]{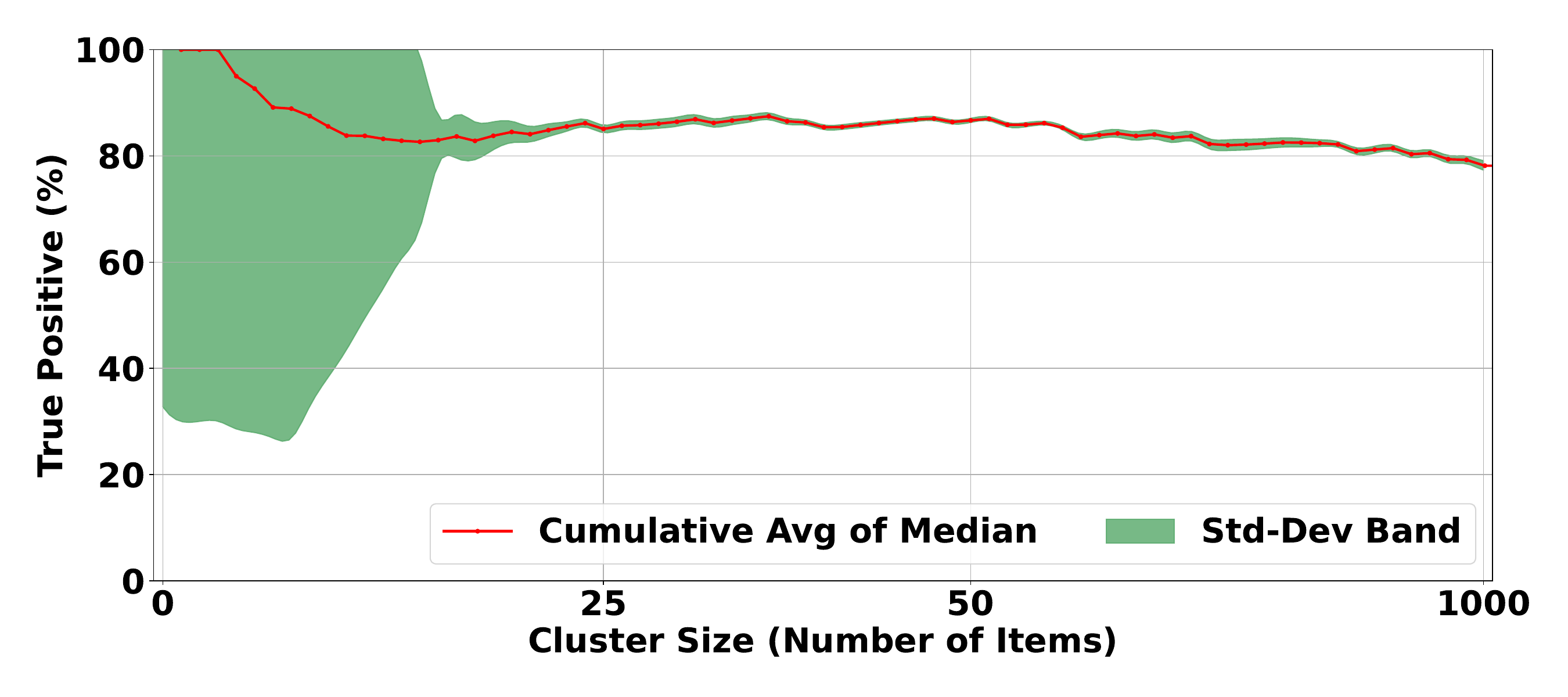} \\

        \multicolumn{5}{c}{\underline{\textbf{Oracle Based Heuristic}}} \\
        \multicolumn{5}{c}{\textbf{Mode $K$ From Oracle}} \\
        \includegraphics[width=\linewidth]{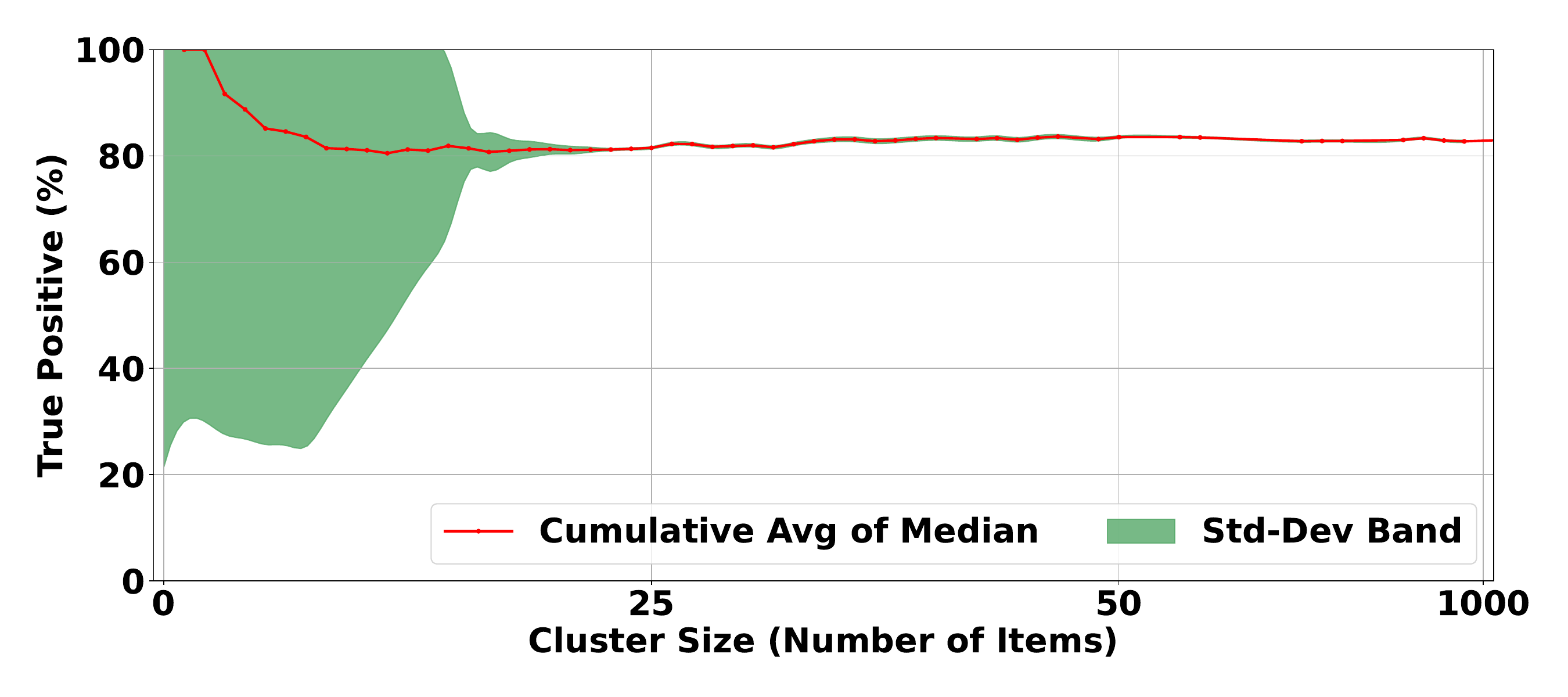} &
        \includegraphics[width=\linewidth]{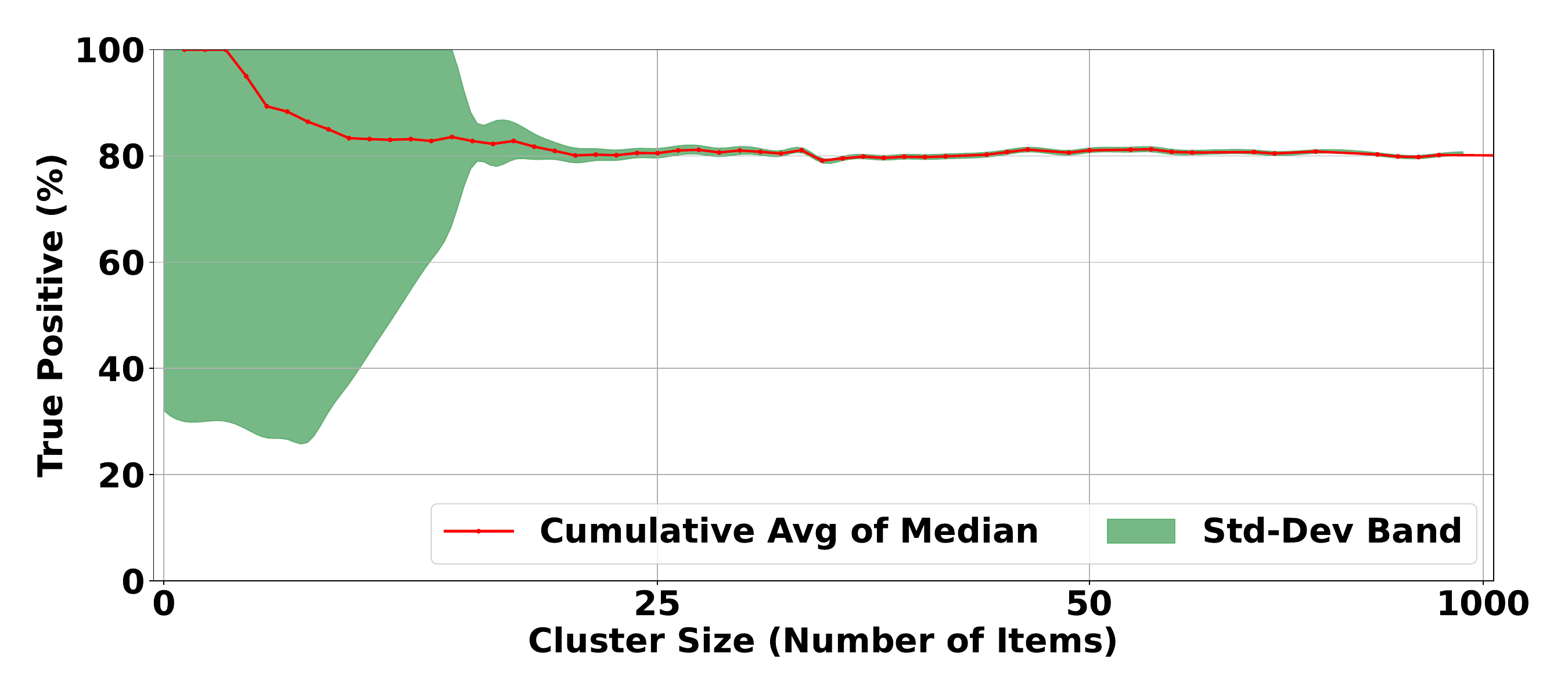} &
        \includegraphics[width=\linewidth]{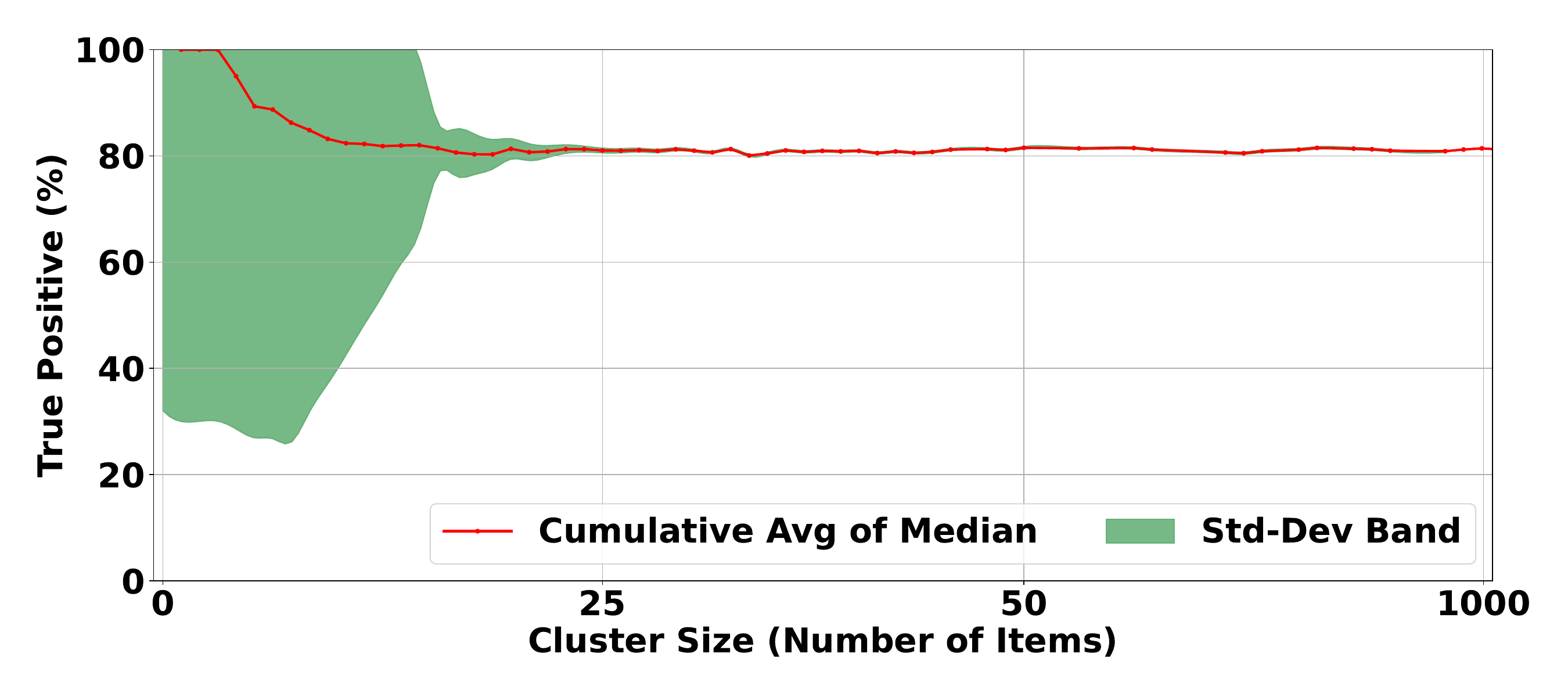} &
        \includegraphics[width=\linewidth]{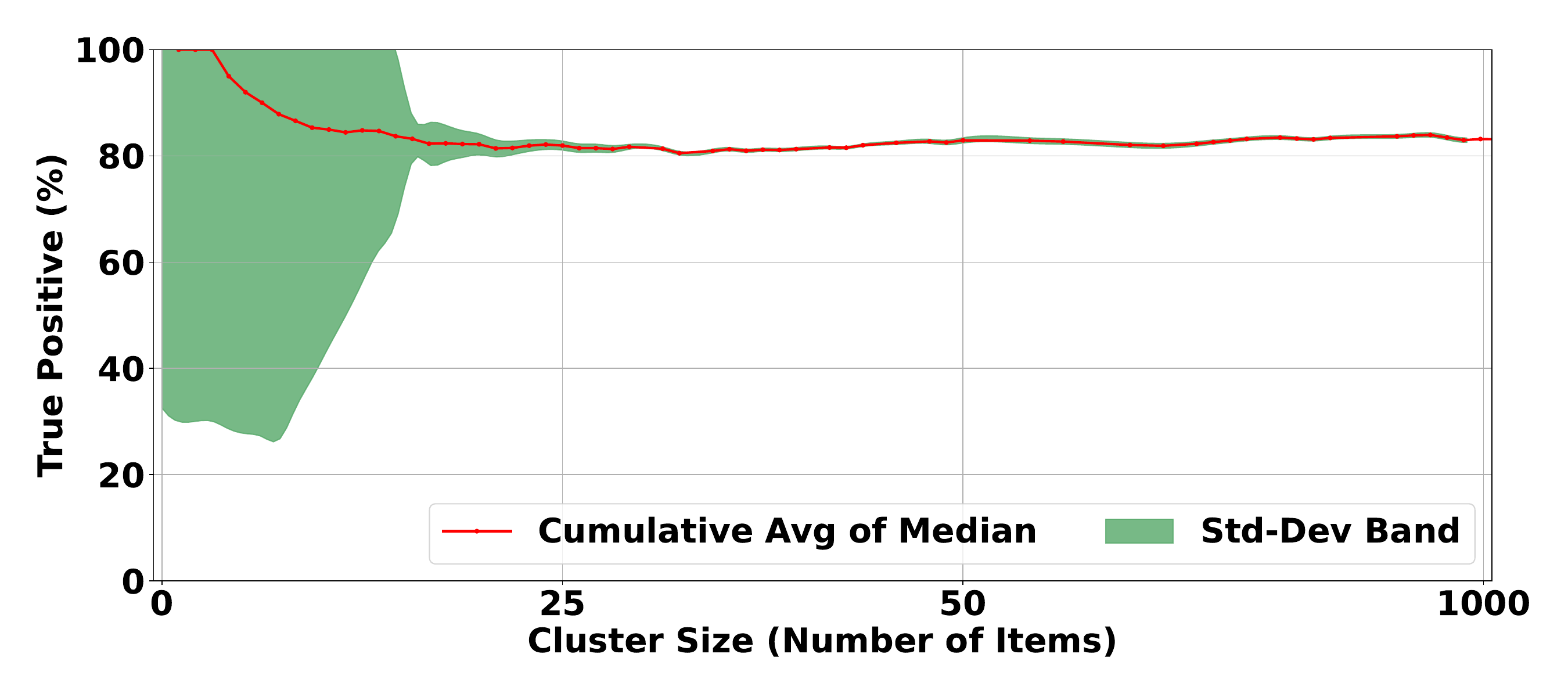} &
        \includegraphics[width=\linewidth]{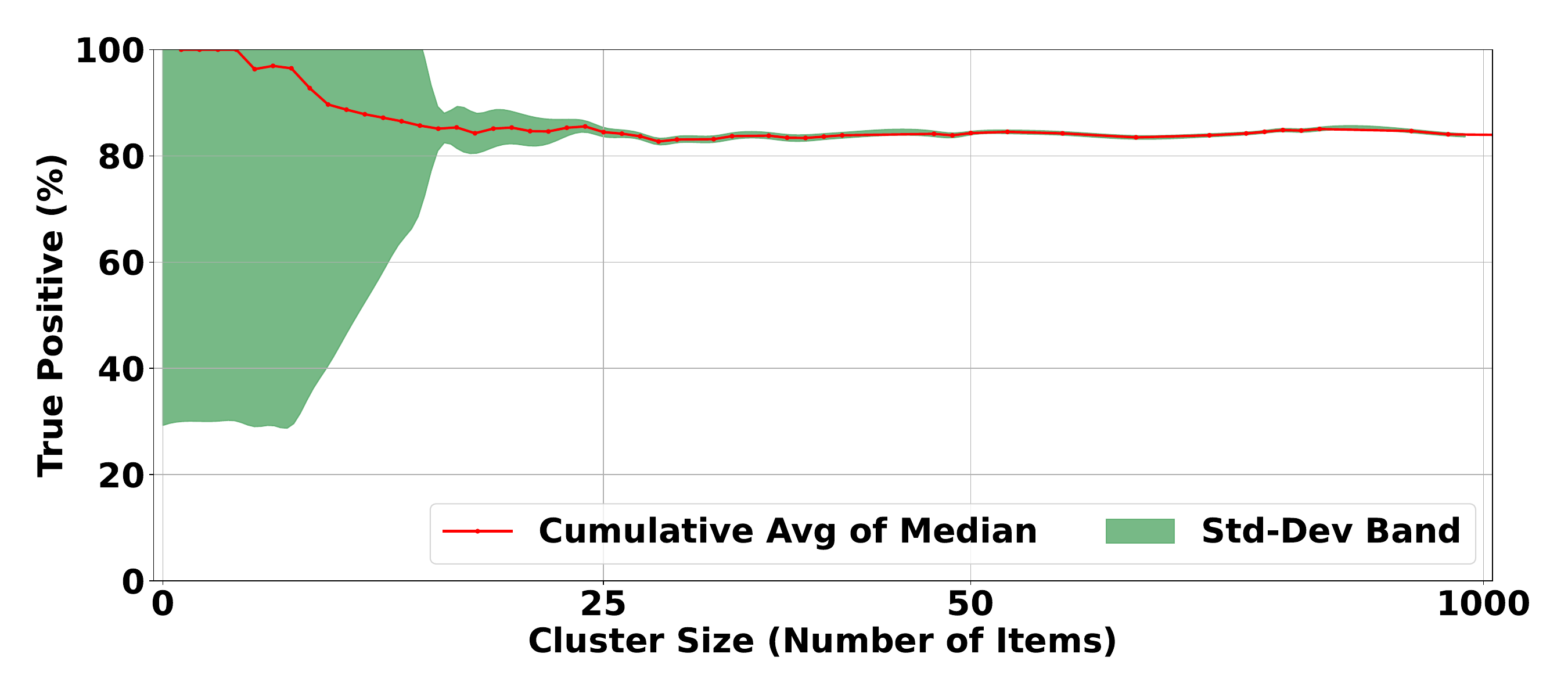} \\    
        \multicolumn{5}{c}{\textbf{Max $K$ From Oracle}} \\
        \includegraphics[width=\linewidth]{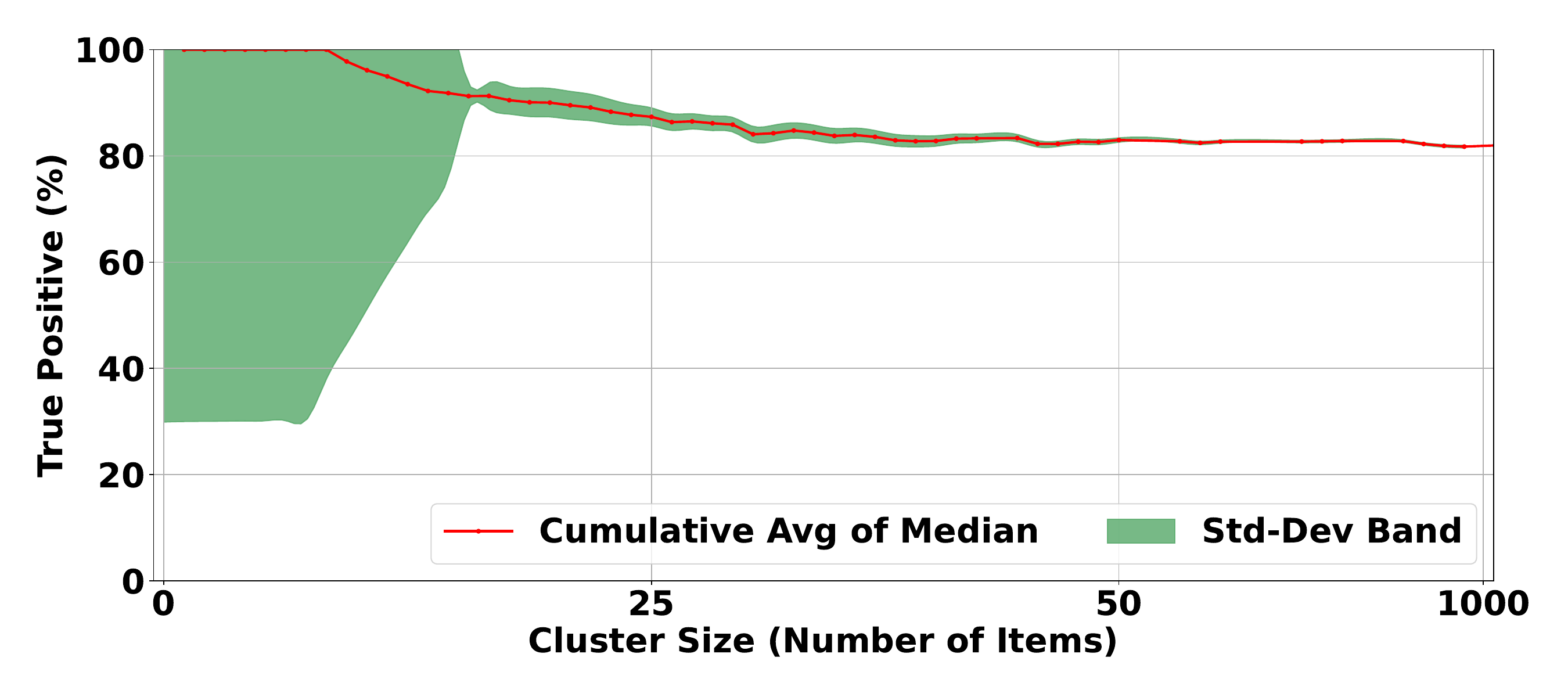} &
        \includegraphics[width=\linewidth]{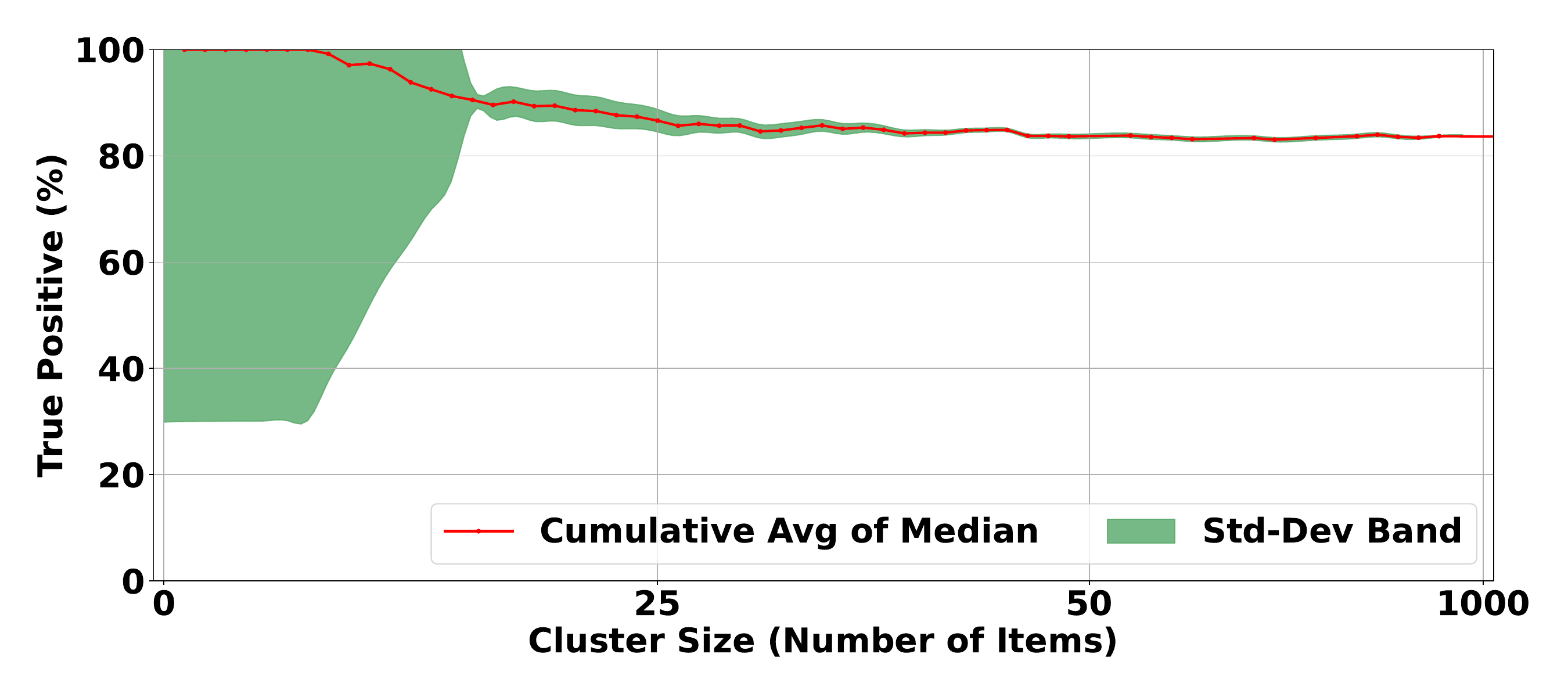} &
        \includegraphics[width=\linewidth]{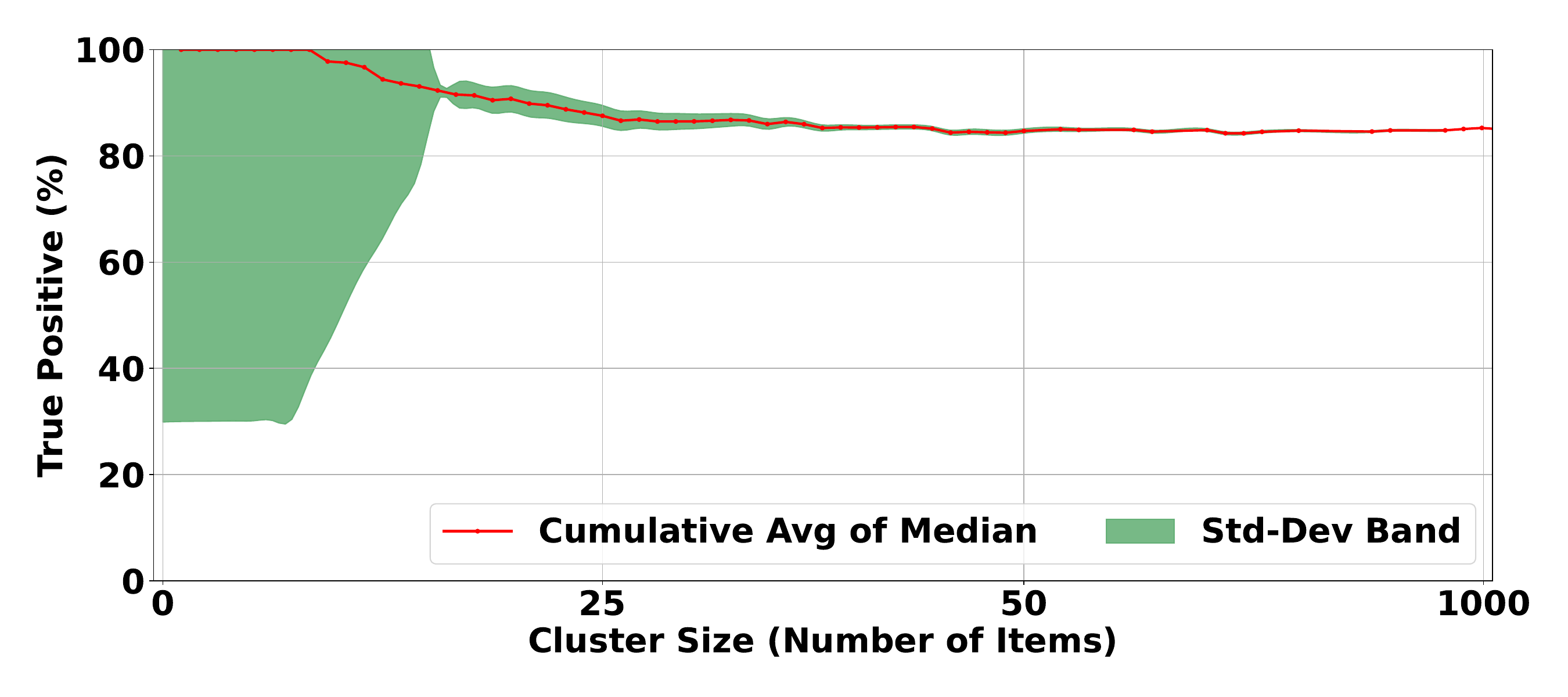} &
        \includegraphics[width=\linewidth]{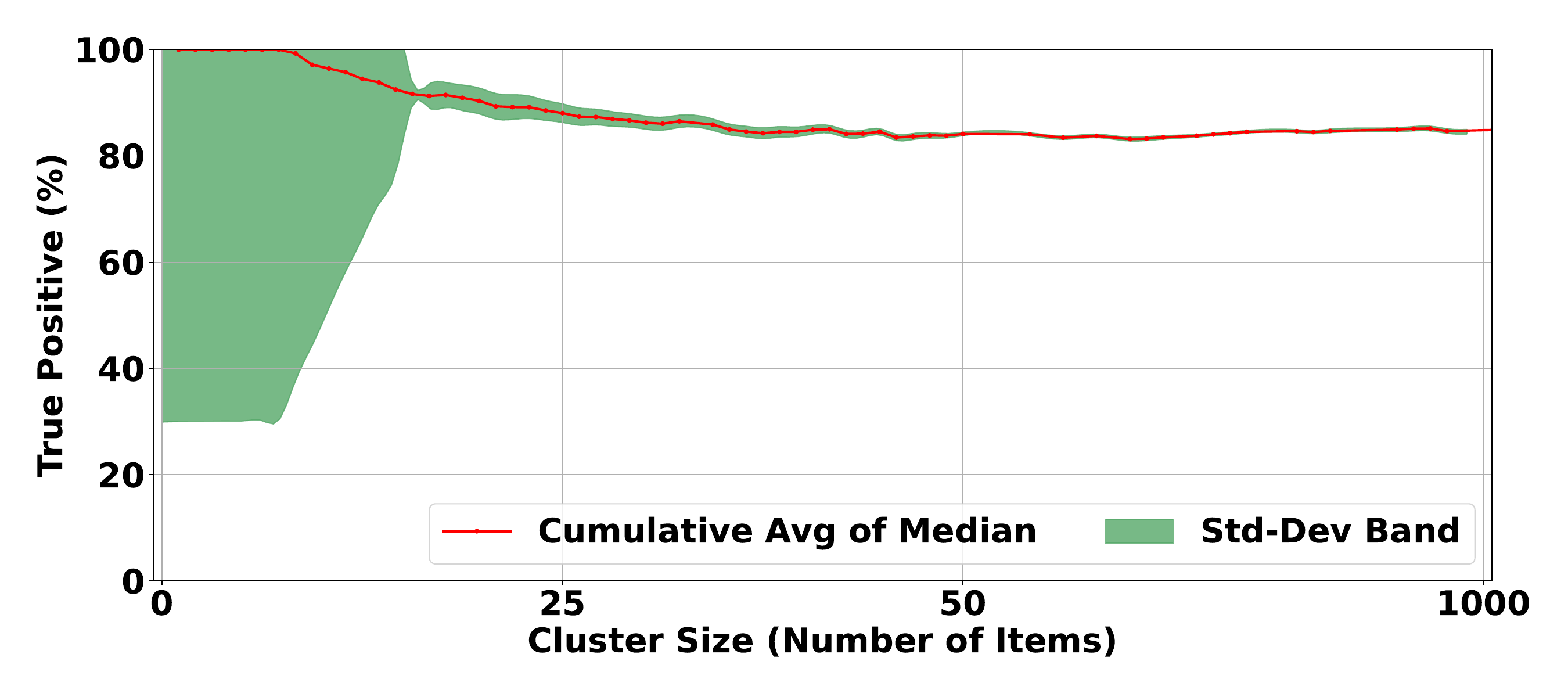} &
        \includegraphics[width=\linewidth]{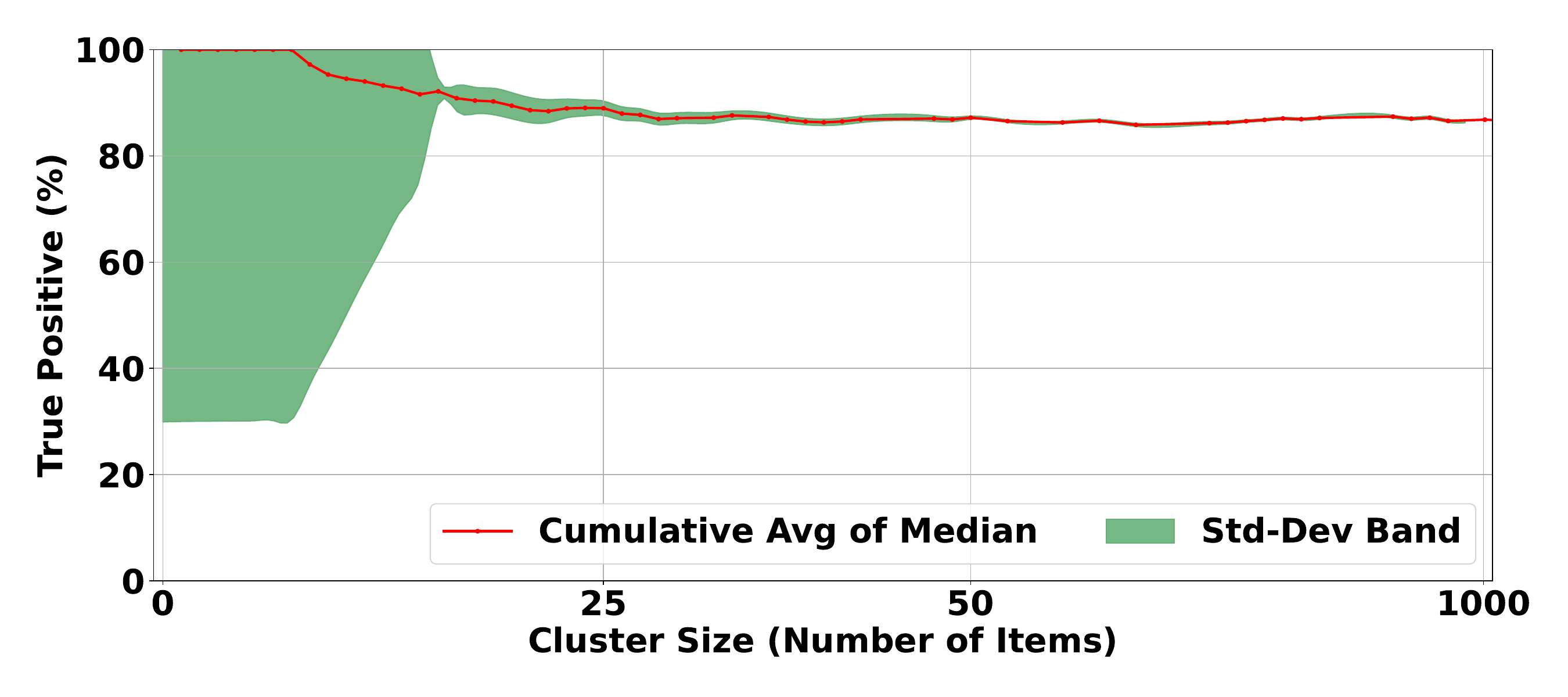} \\  
        \multicolumn{5}{c}{\textbf{Random $K$ From Oracle}} \\
        \includegraphics[width=\linewidth]{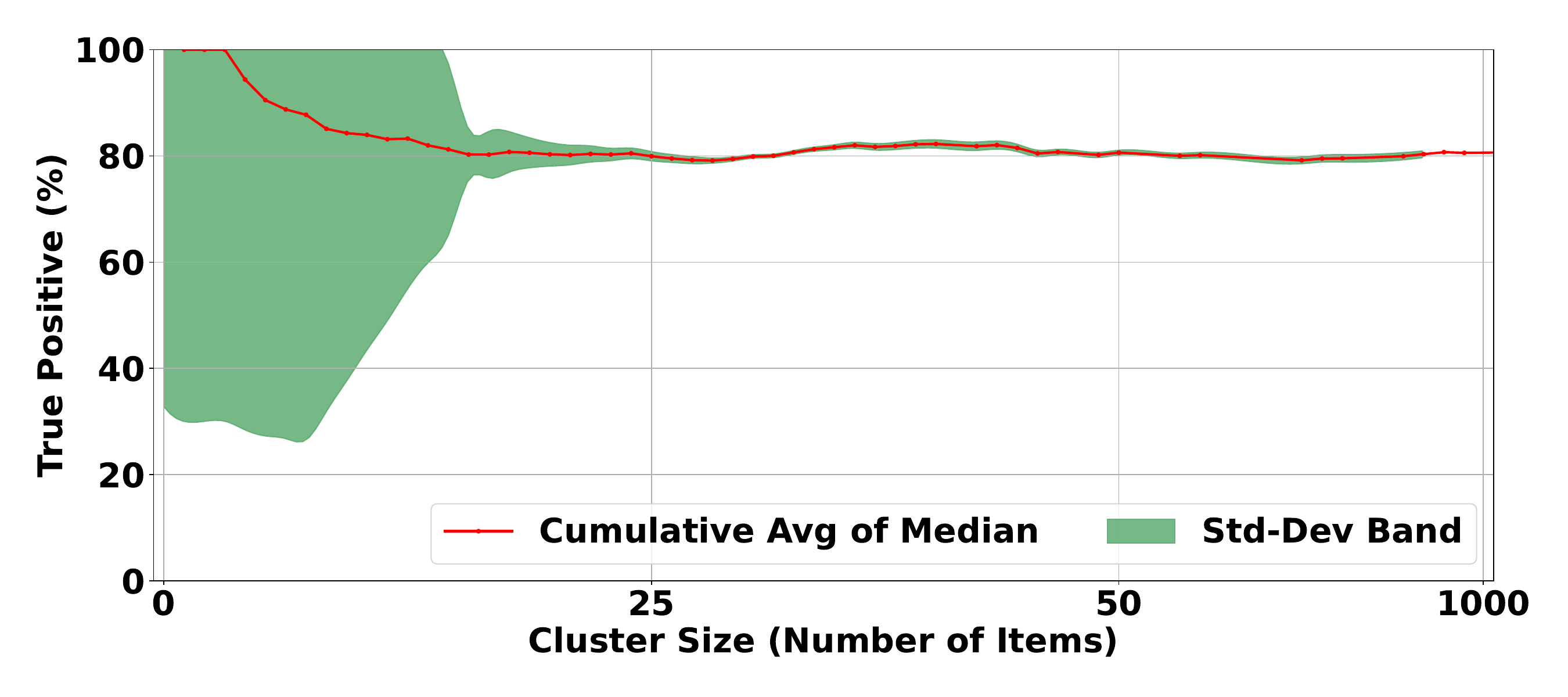} &
        \includegraphics[width=\linewidth]{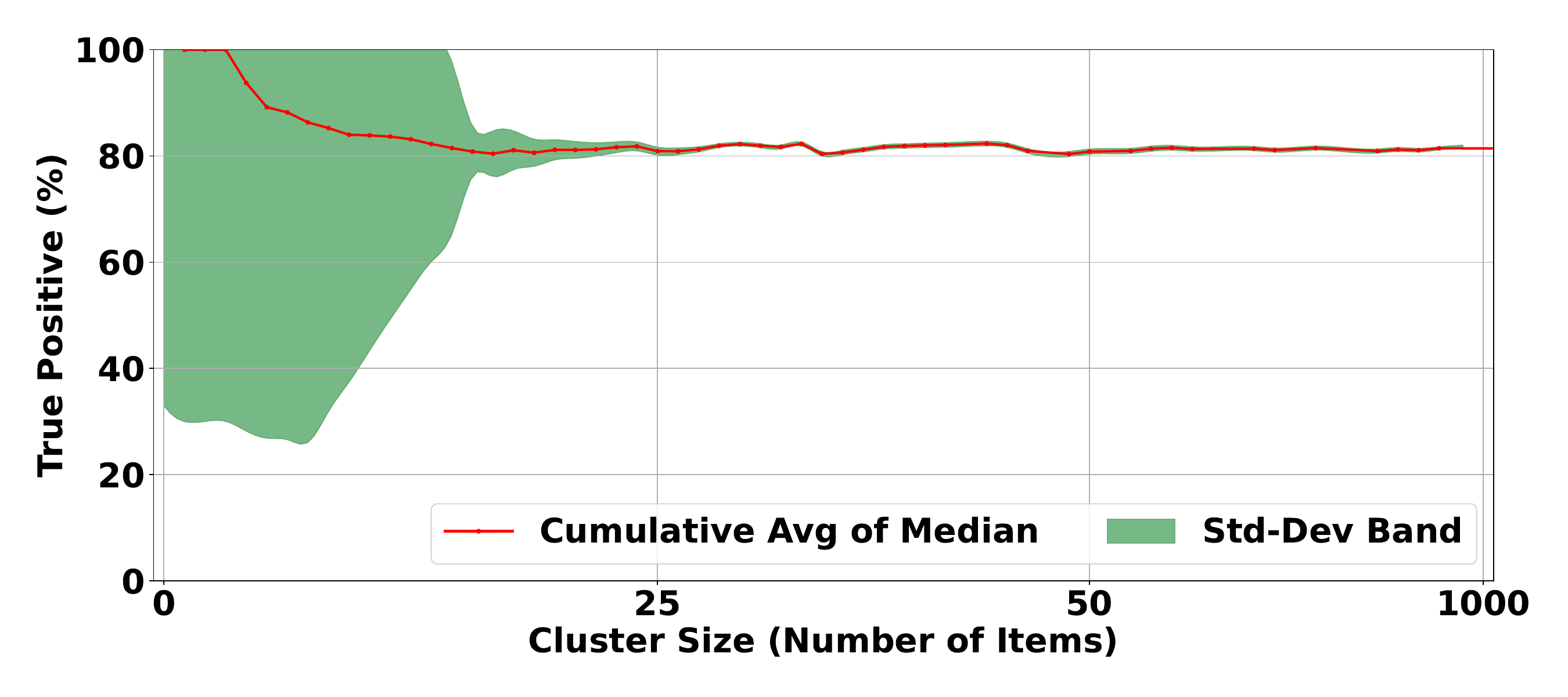} &
        \includegraphics[width=\linewidth]{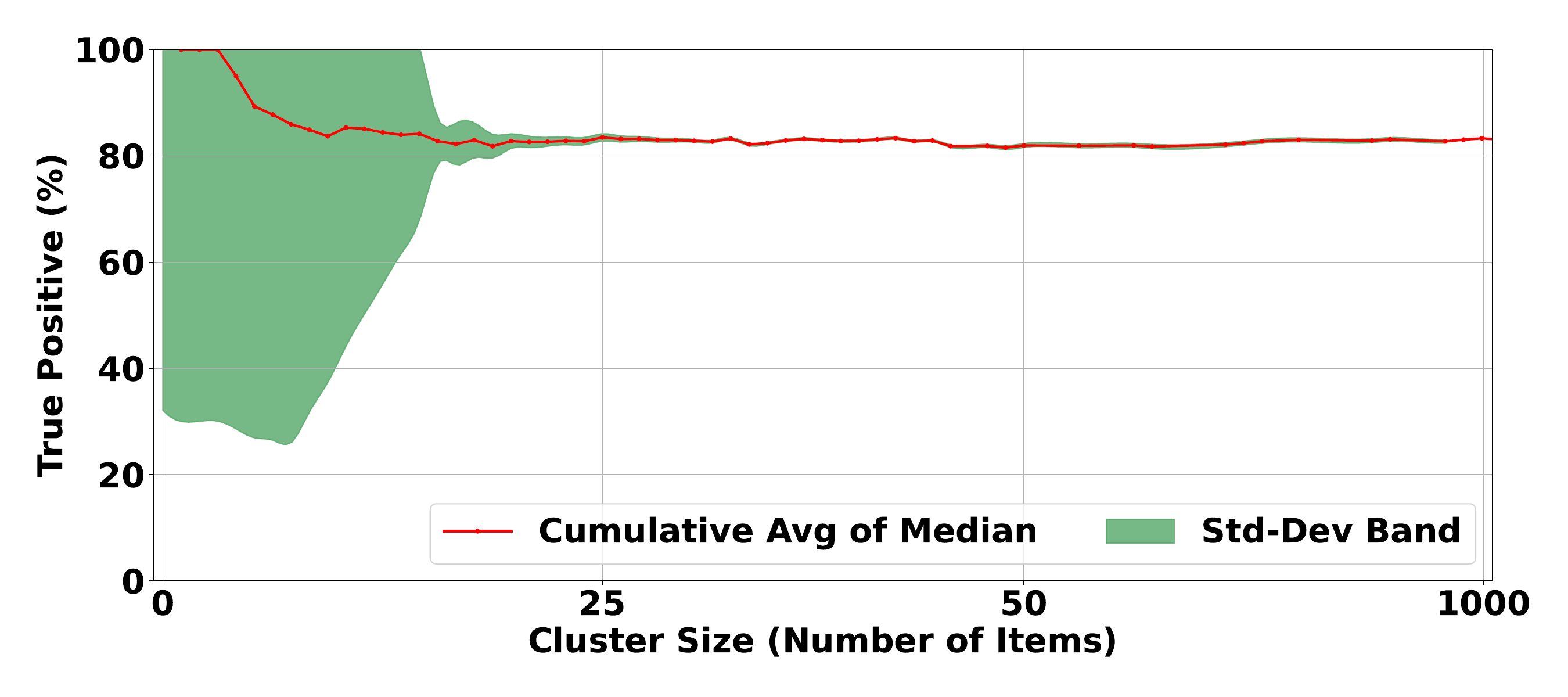} &
        \includegraphics[width=\linewidth]{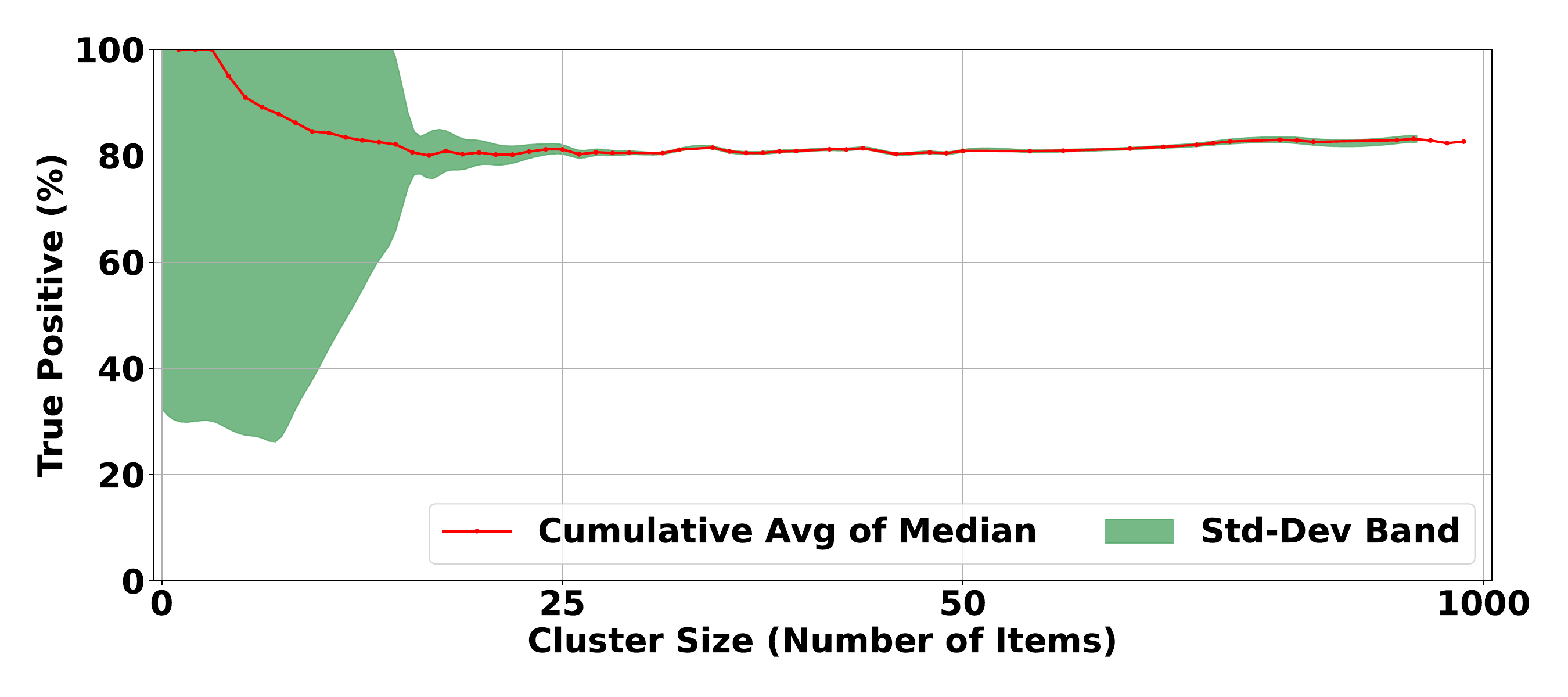} &
        \includegraphics[width=\linewidth]{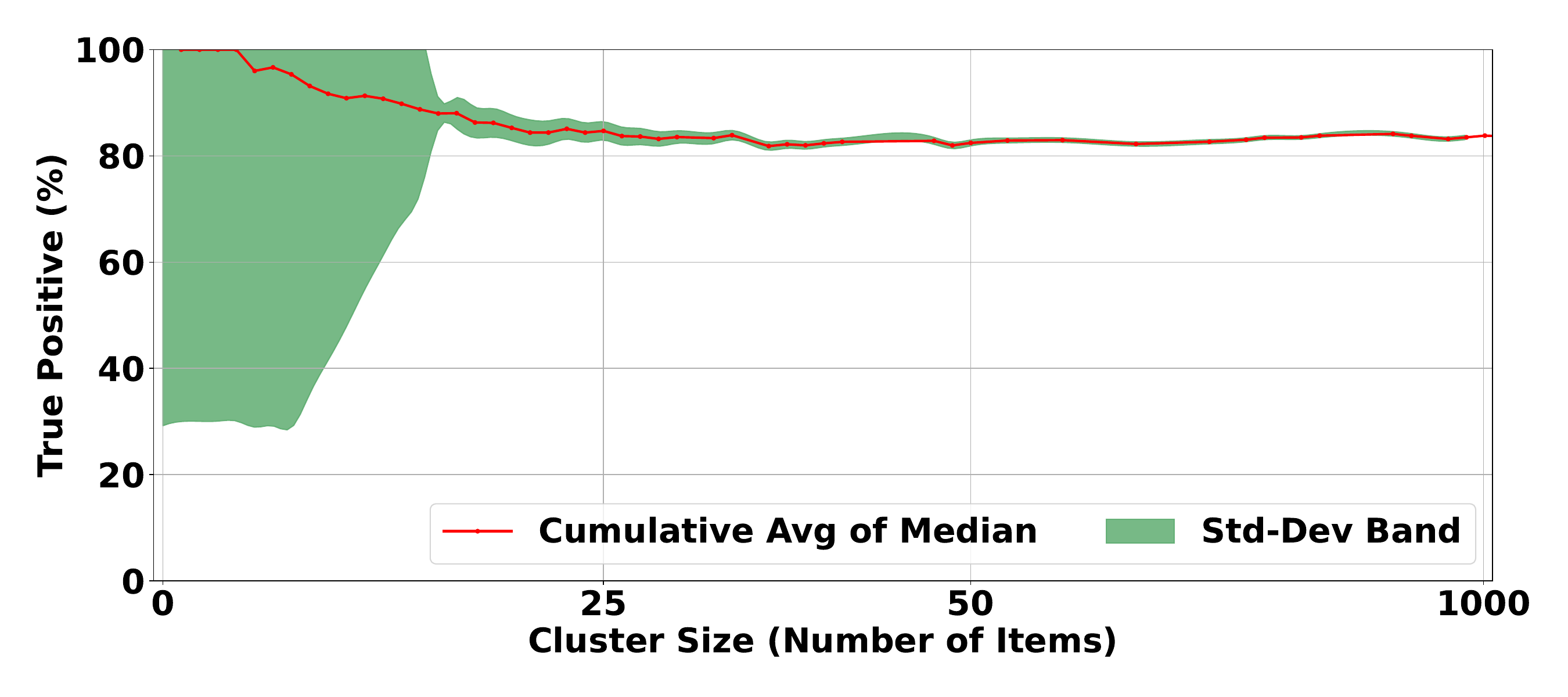} \\ 

        \multicolumn{5}{c}{\underline{\textbf{Sub-Optimal Heuristic}}} \\
        \multicolumn{5}{c}{\textbf{Average $K$}} \\    
        \includegraphics[width=\linewidth]{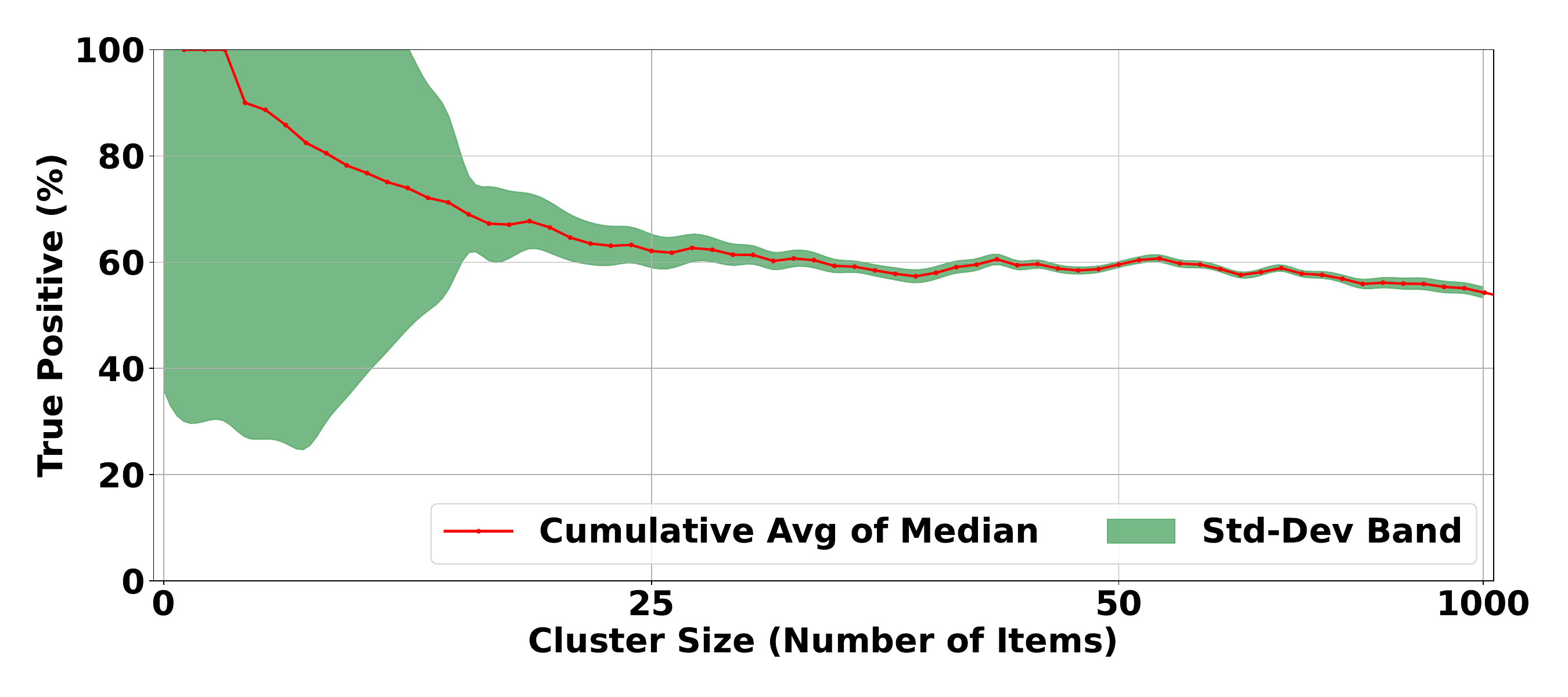} &
        \includegraphics[width=\linewidth]{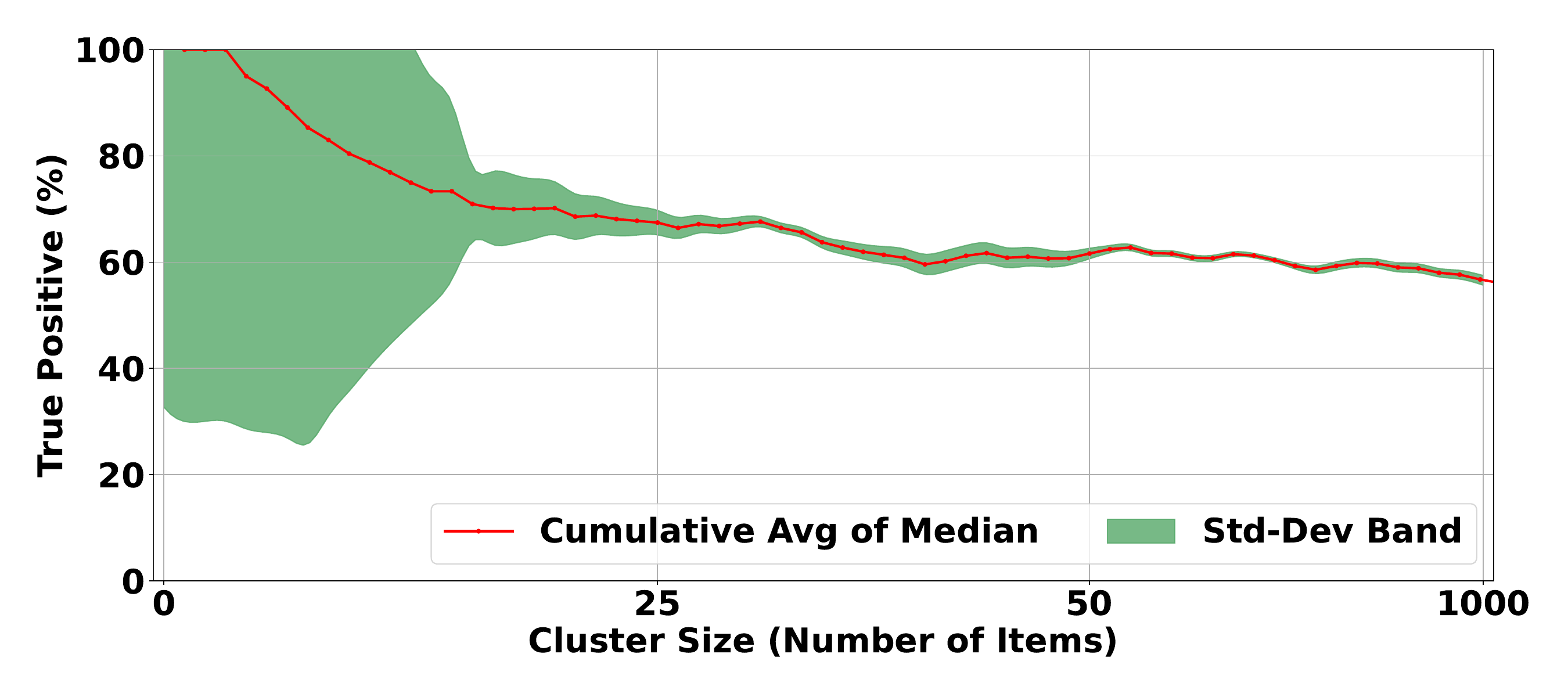} &
        \includegraphics[width=\linewidth]{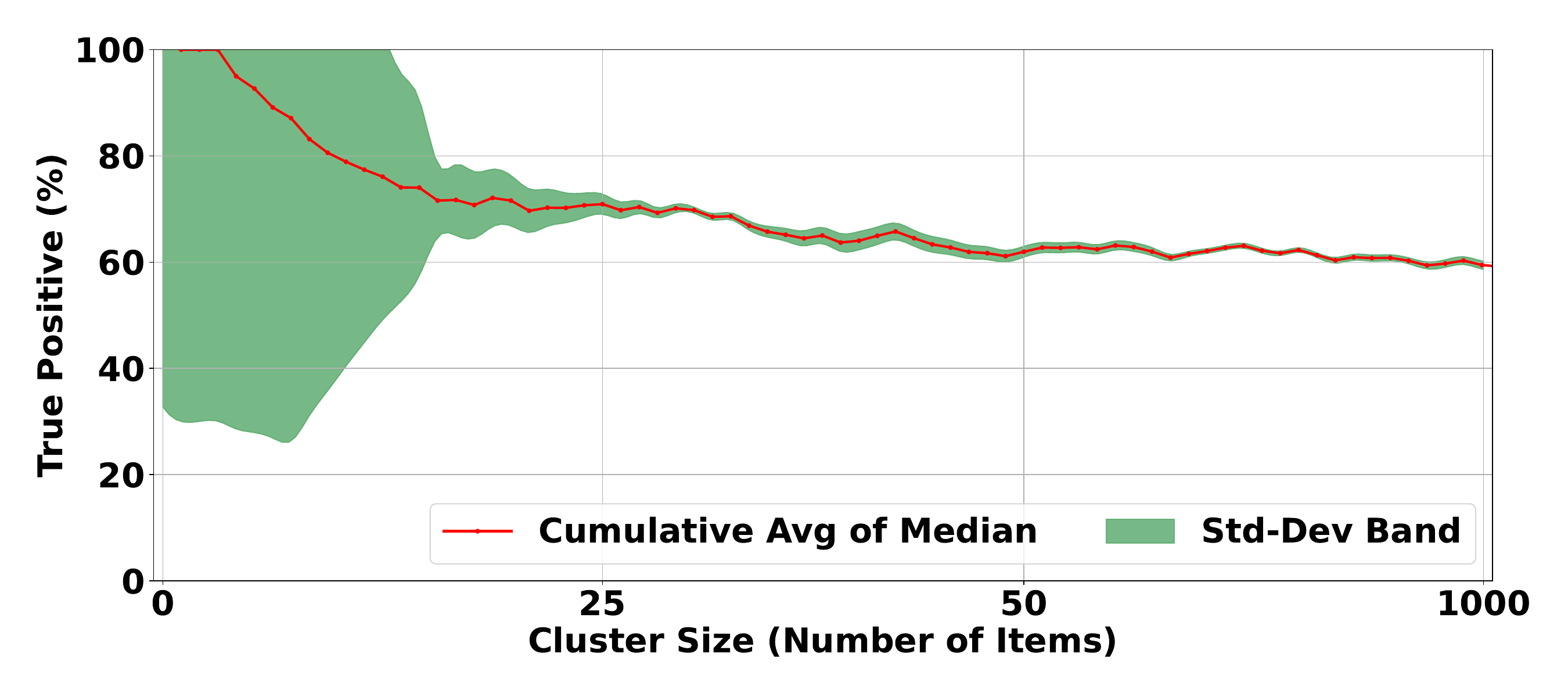} &
        \includegraphics[width=\linewidth]{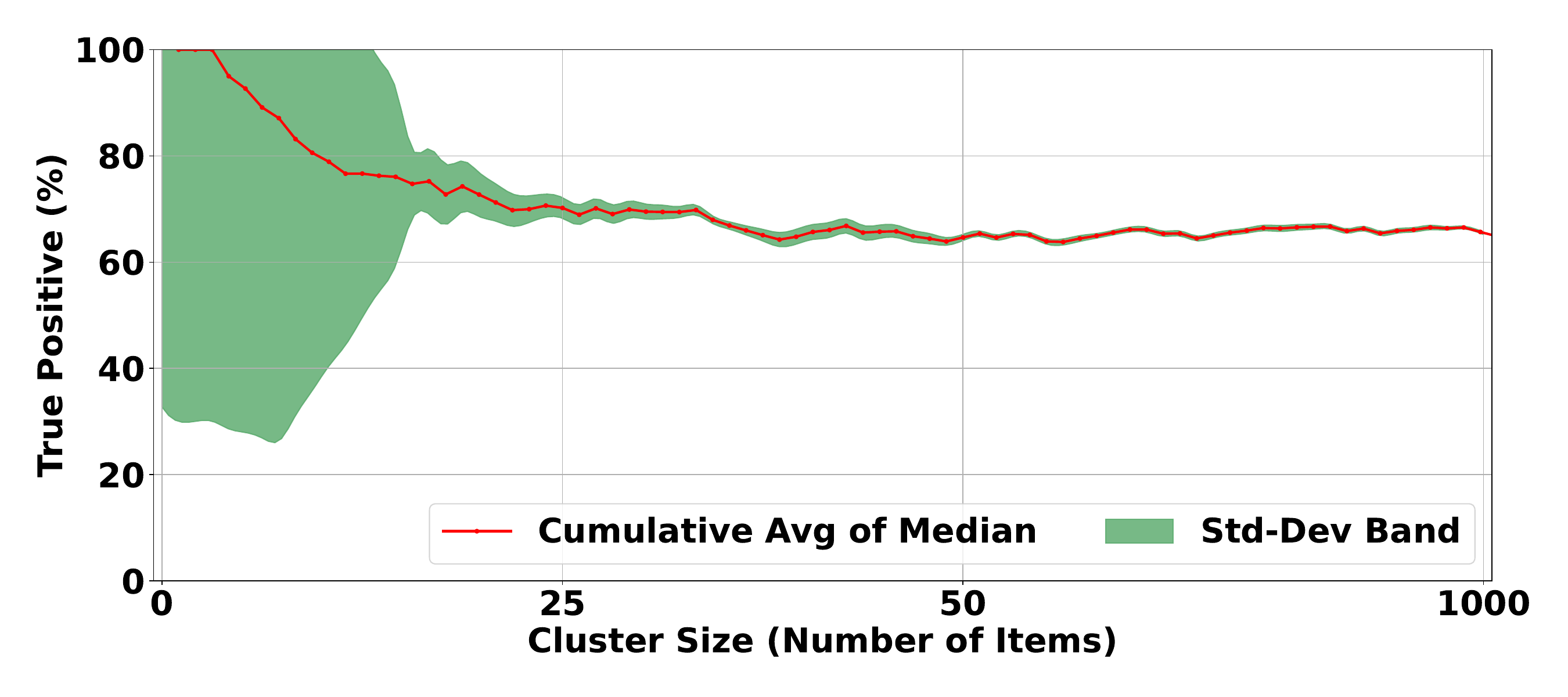} &
        \includegraphics[width=\linewidth]{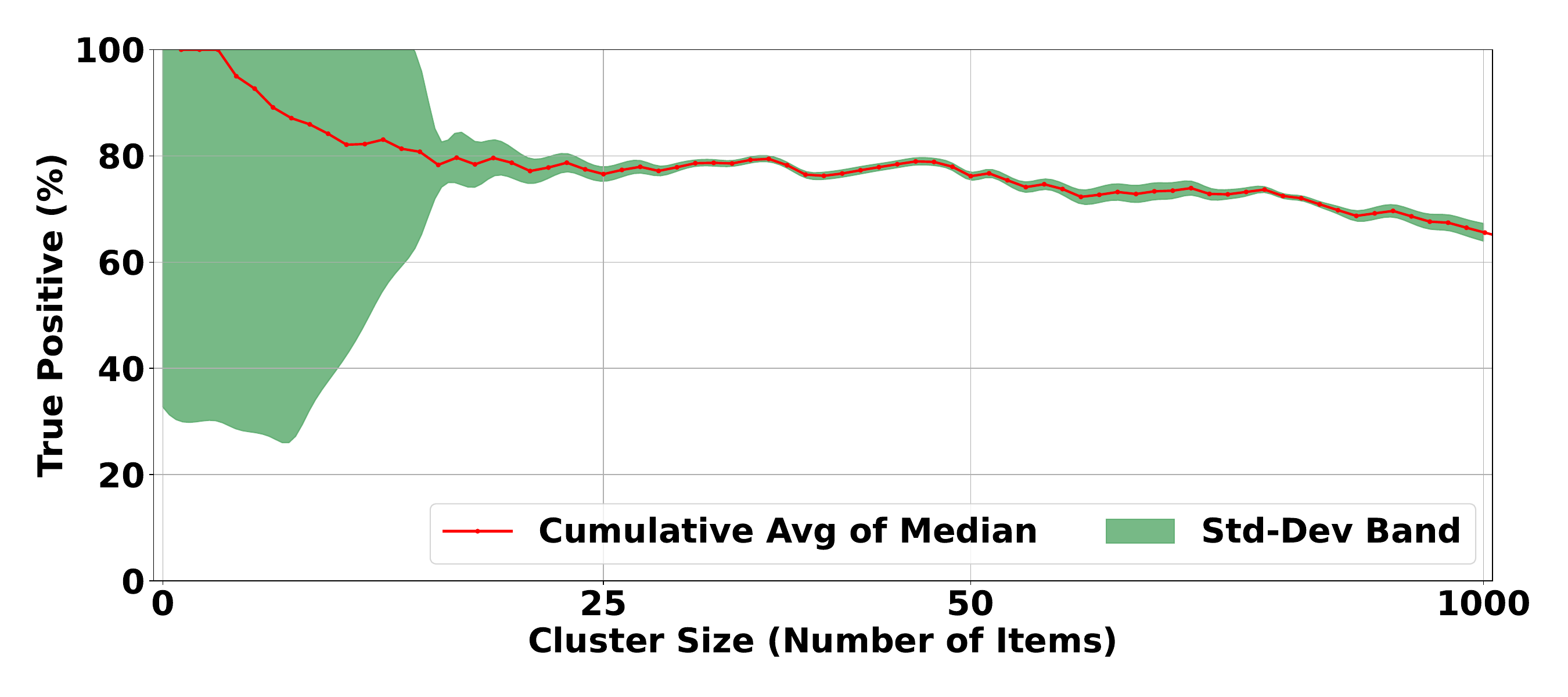} \\         
        \multicolumn{5}{c}{\textbf{Maximum $K$}} \\       
        \includegraphics[width=\linewidth]{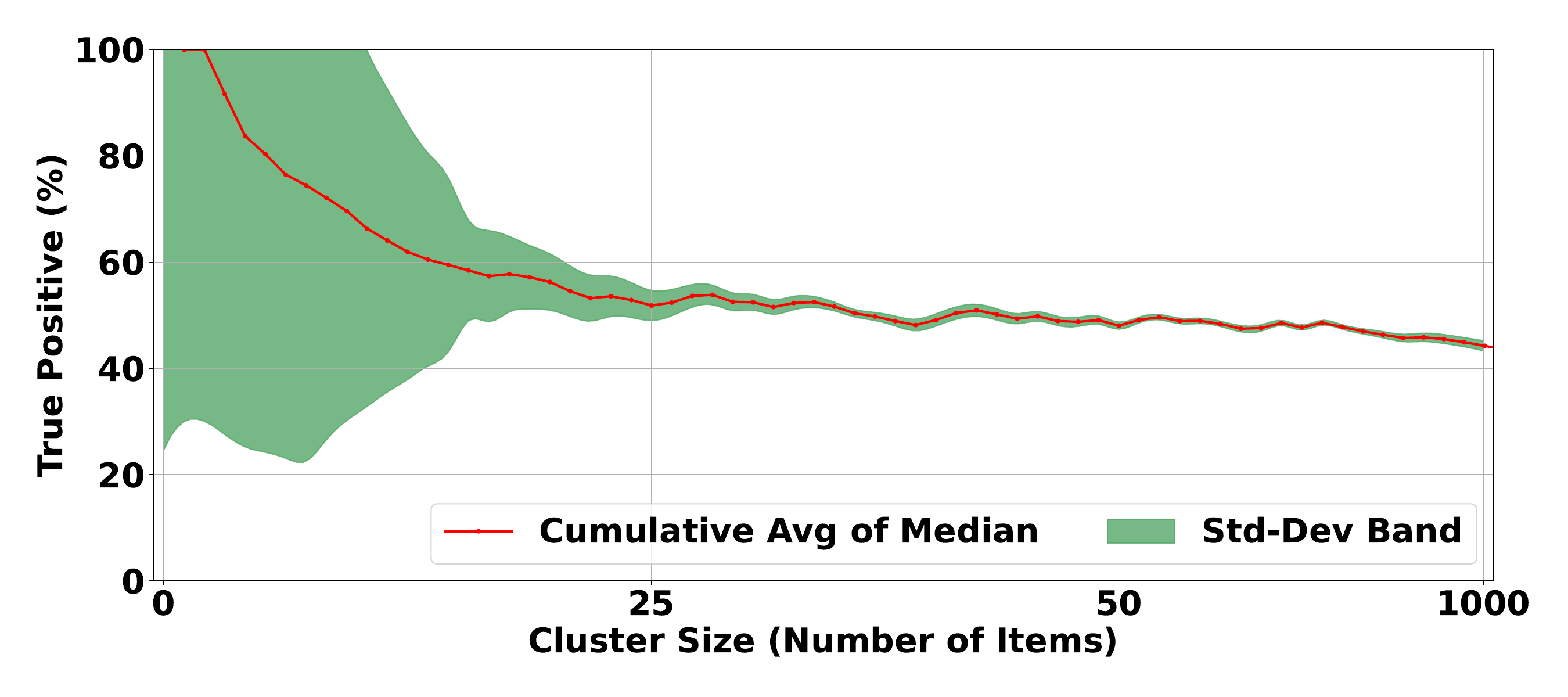} &
        \includegraphics[width=\linewidth]{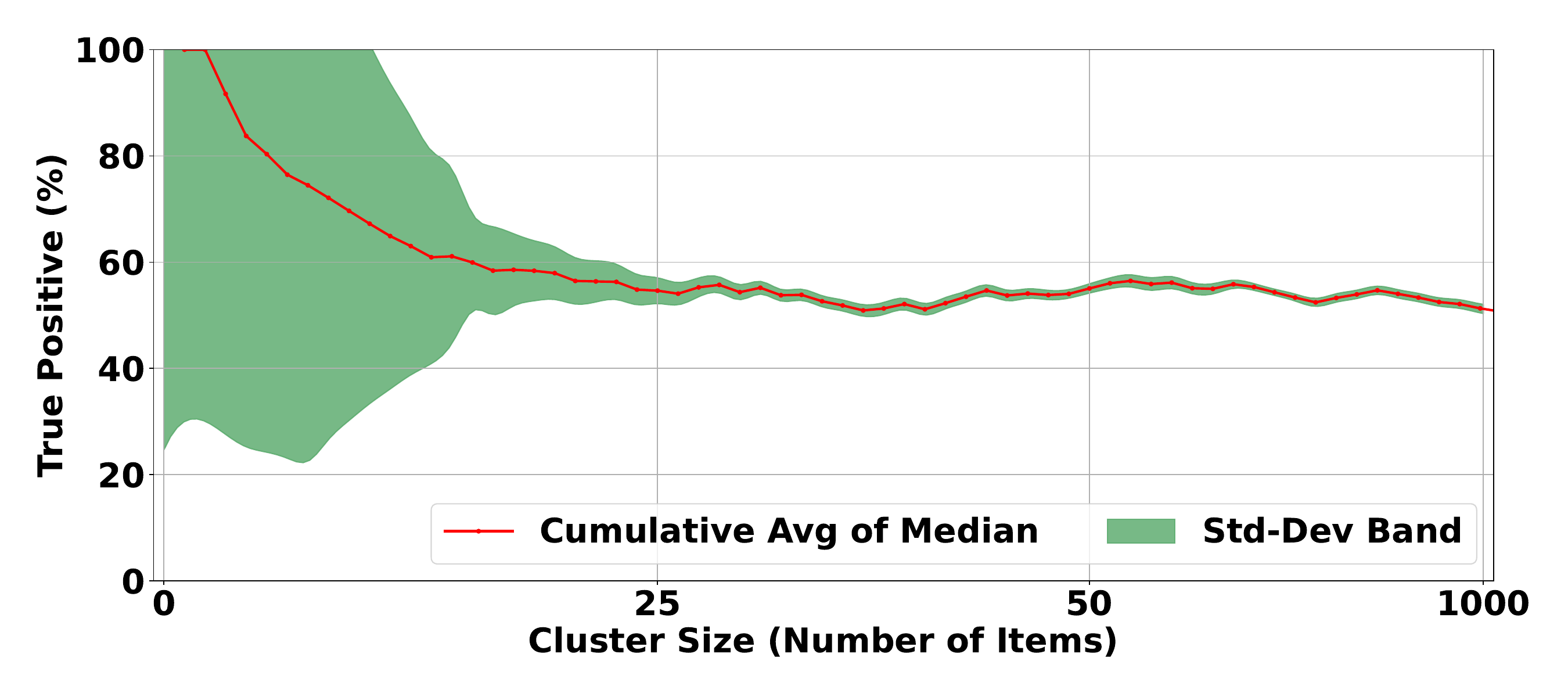} &
        \includegraphics[width=\linewidth]{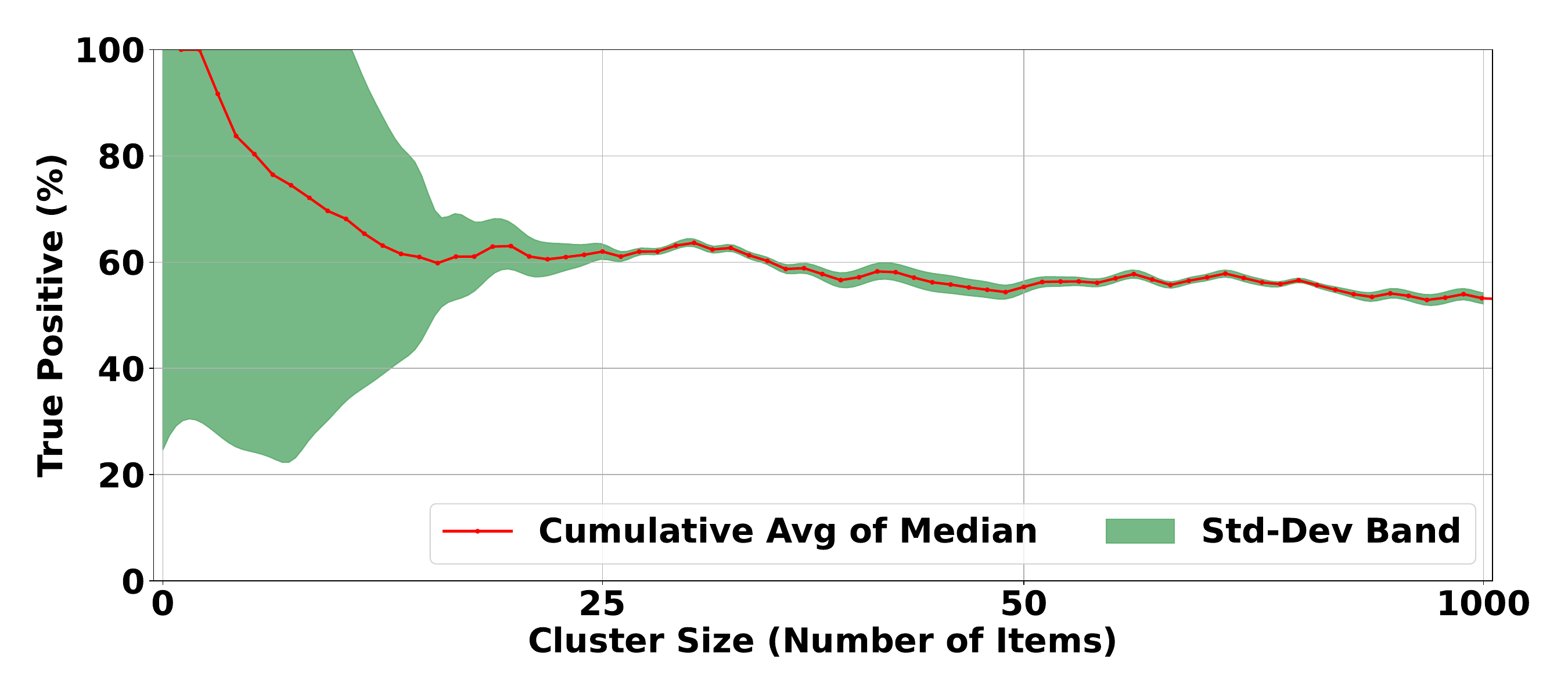} &
        \includegraphics[width=\linewidth]{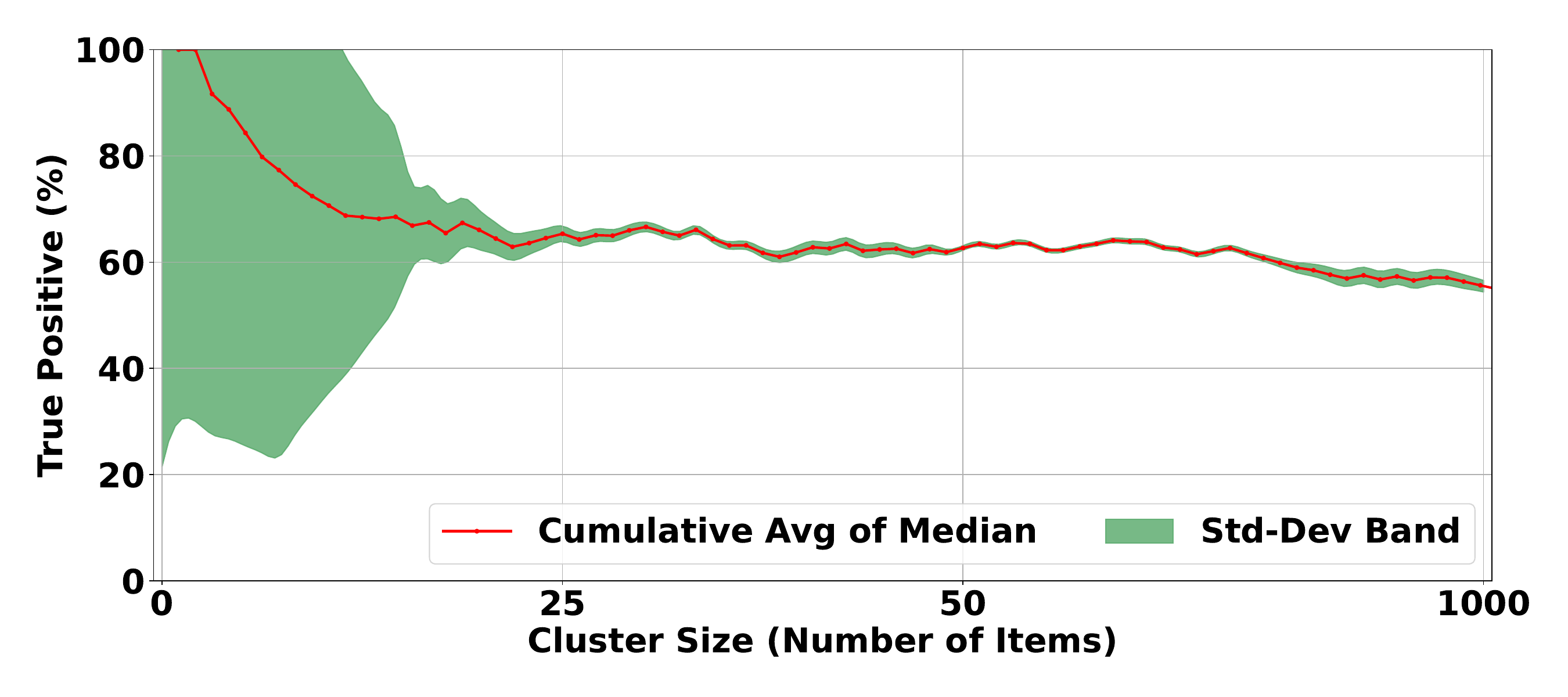} &
        \includegraphics[width=\linewidth]{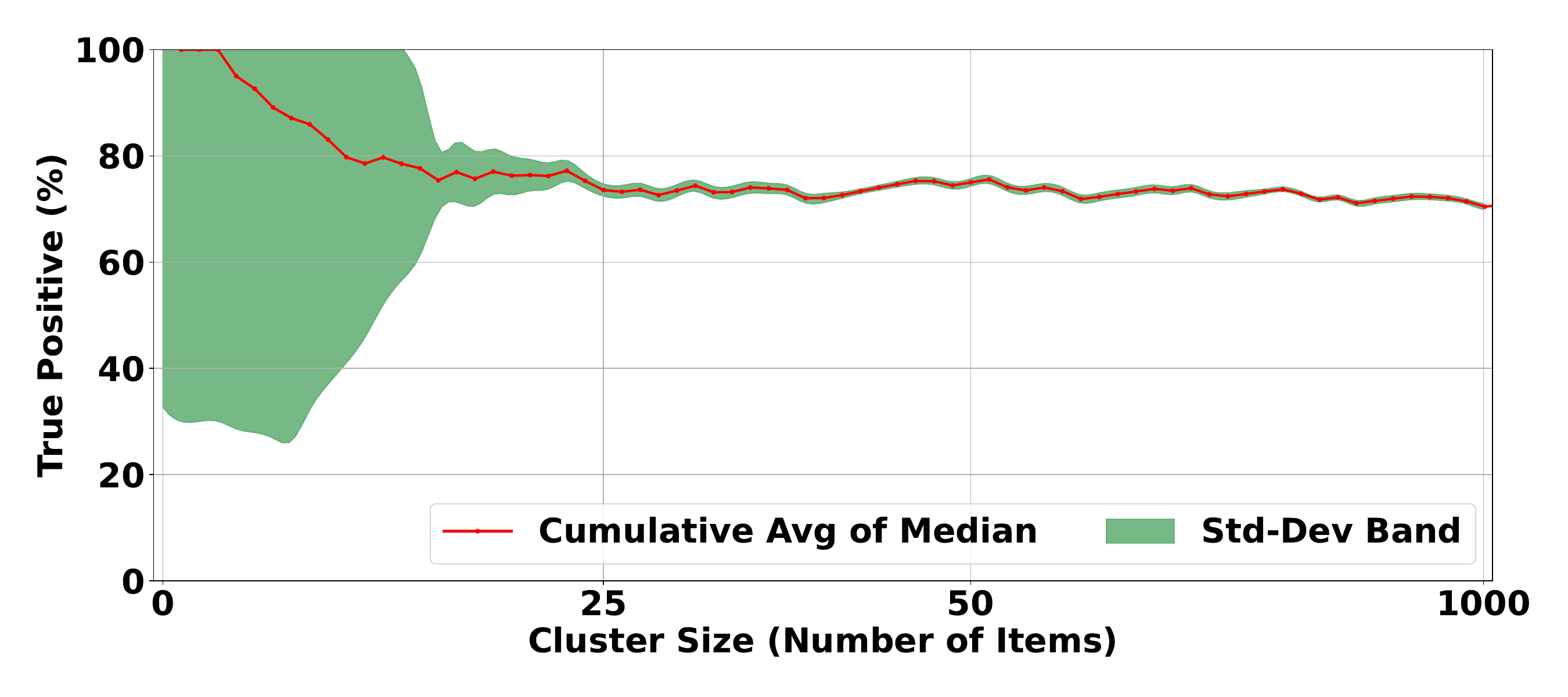} \\          
        \multicolumn{5}{c}{\textbf{Random $K$}} \\ 
        \includegraphics[width=\linewidth]{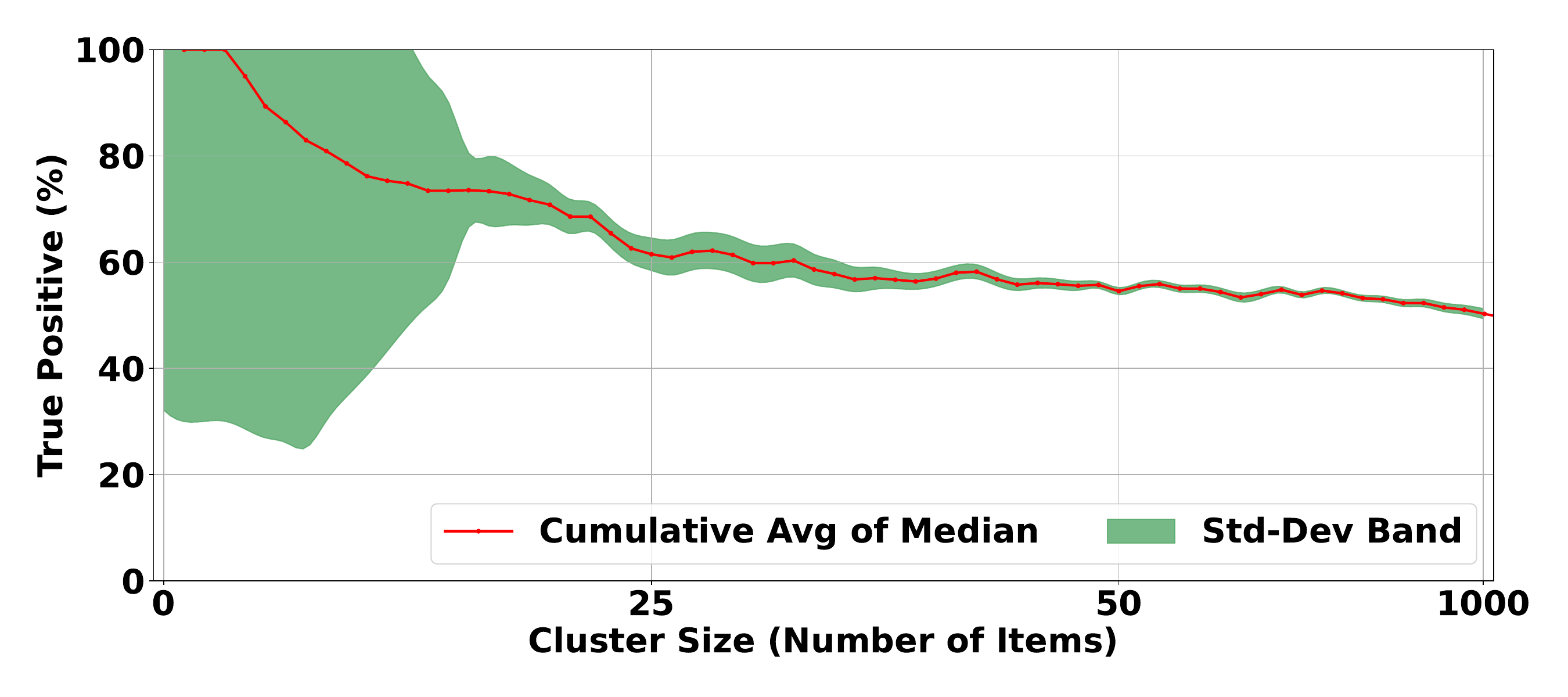} &
        \includegraphics[width=\linewidth]{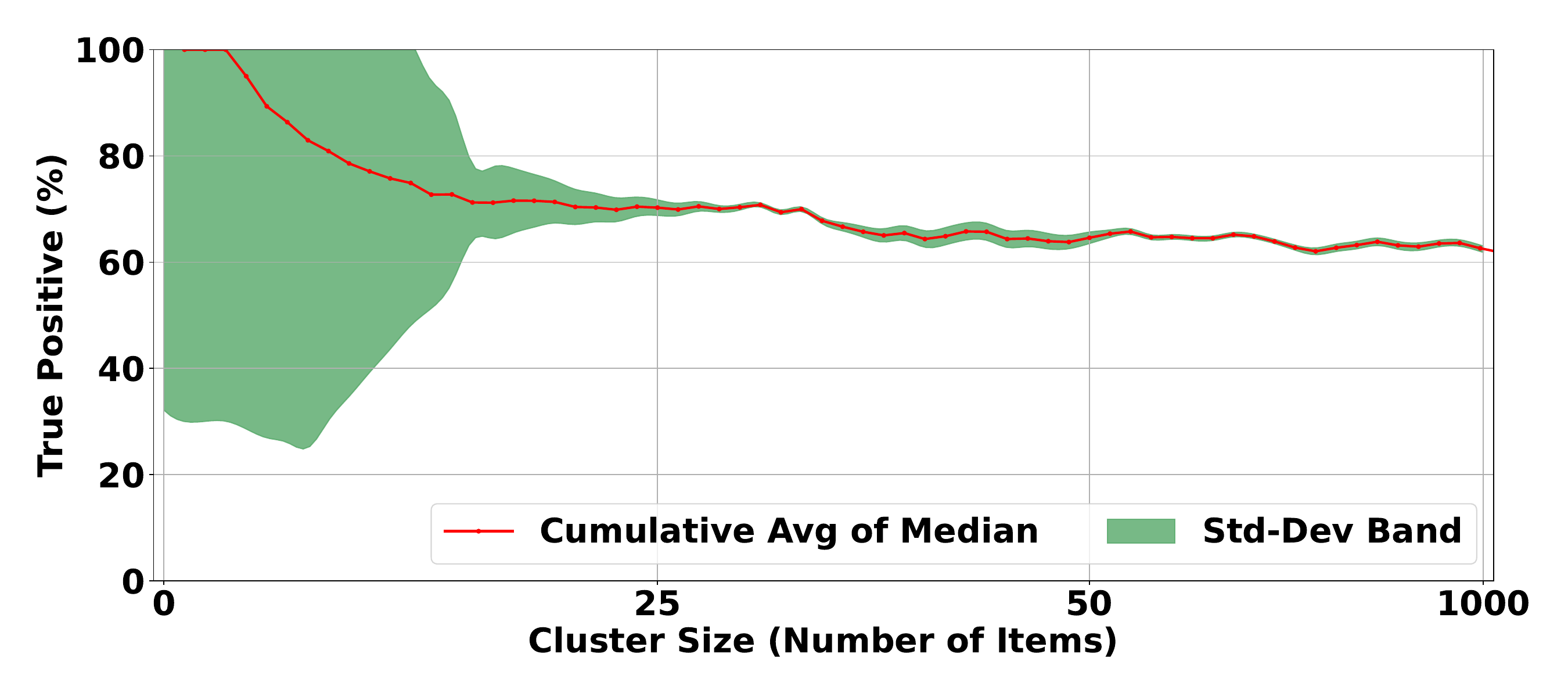} &
        \includegraphics[width=\linewidth]{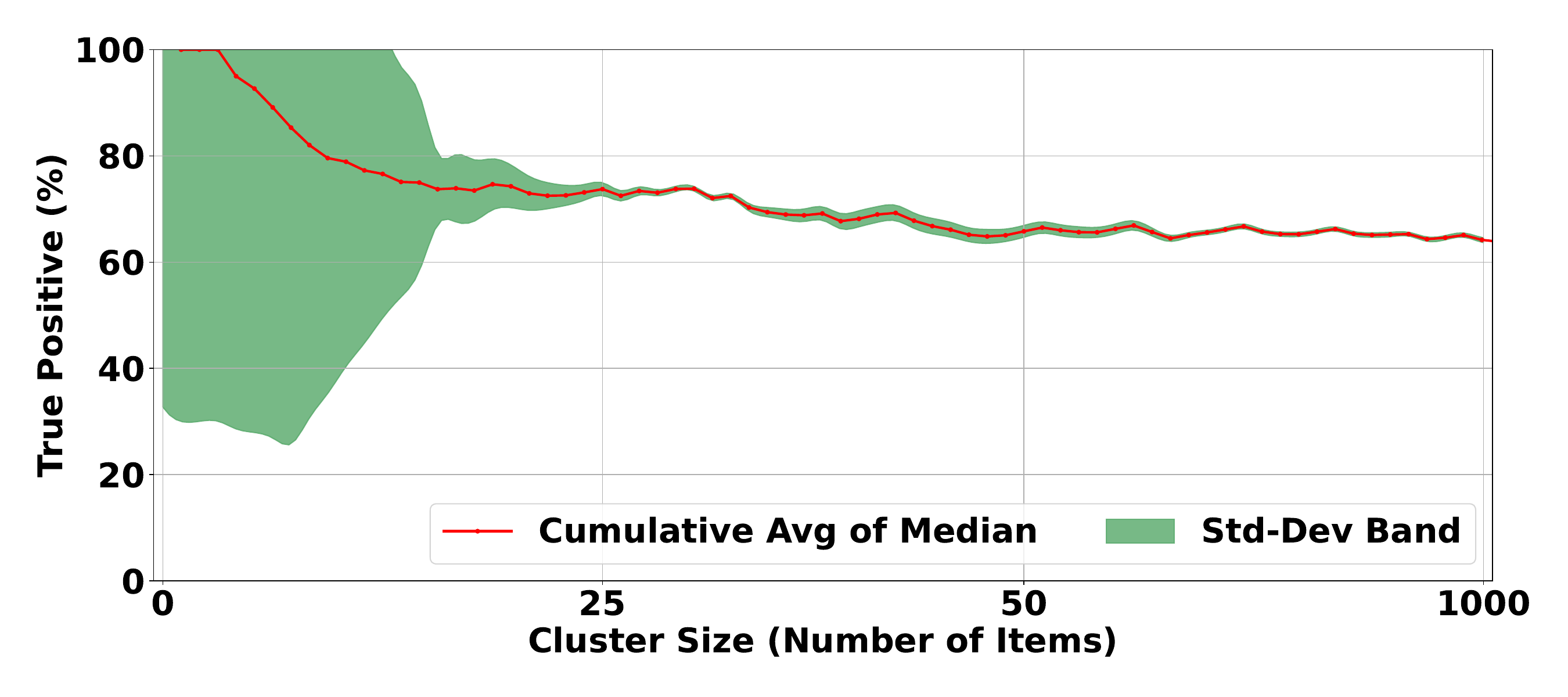} &
        \includegraphics[width=\linewidth]{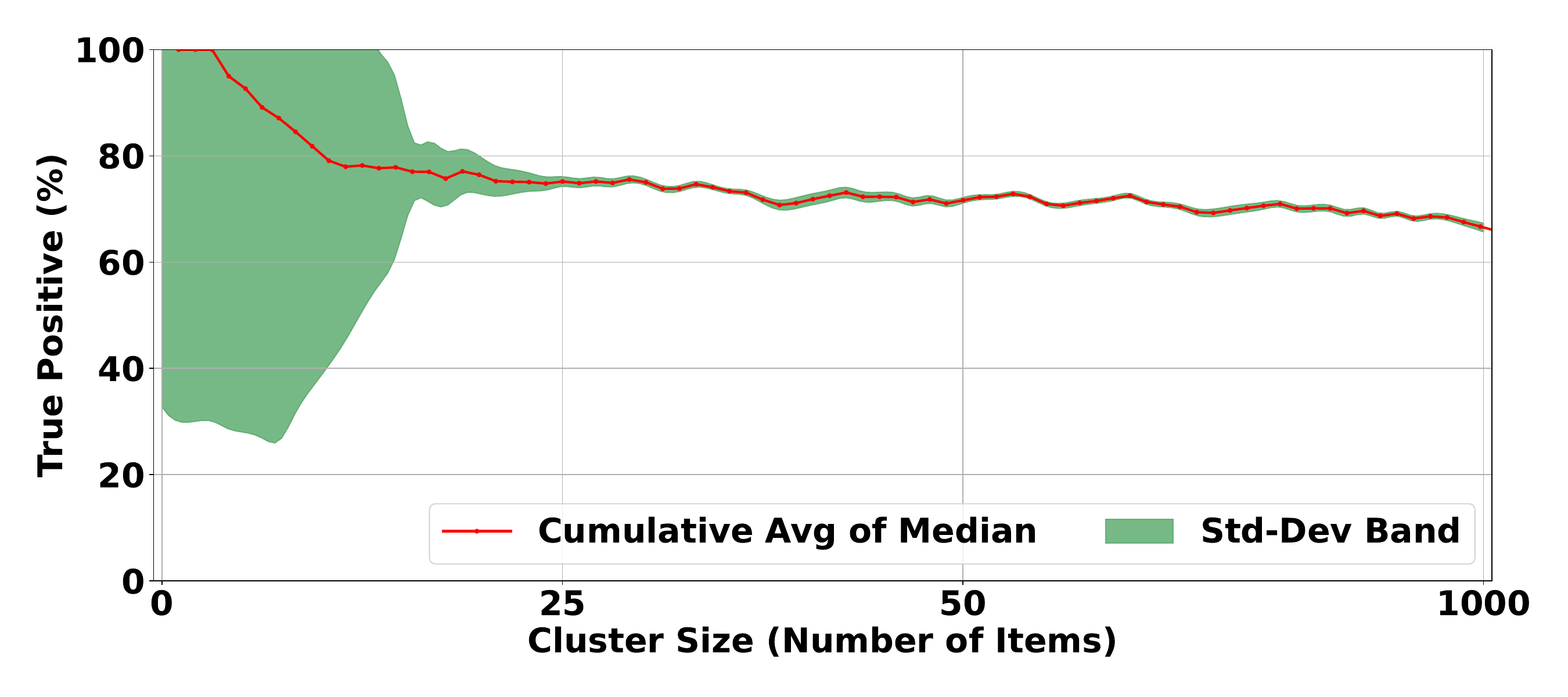} &
        \includegraphics[width=\linewidth]{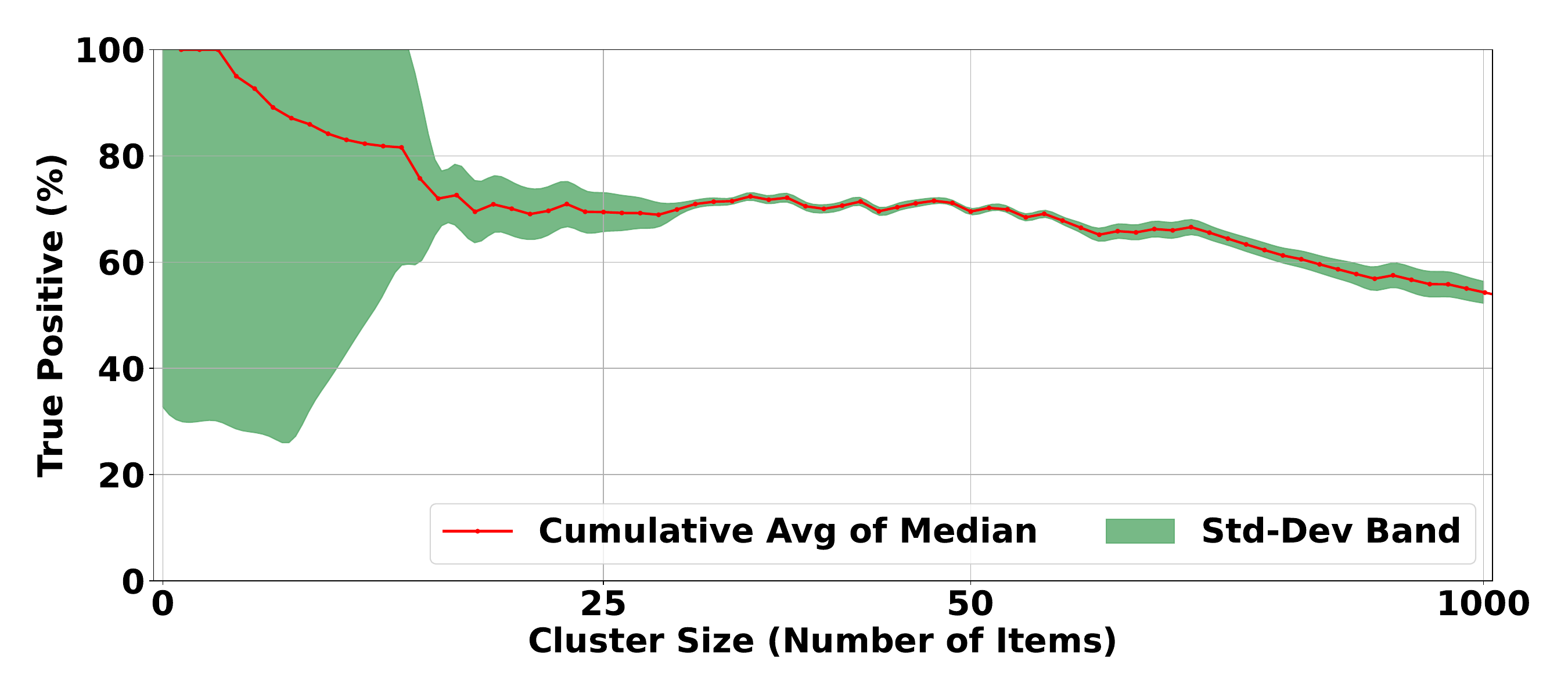} \\         
        
    \end{tabularx}%
    }
    \caption{SSdeep Threshold Variation with Retrained Bloom Filters}
    \label{tab:autoyara_autopyara_sdhash_variation}
\end{table*}
\clearpage



\end{document}